\documentclass[11pt]{cernrep}
\usepackage{graphicx}
\usepackage{here}
\usepackage{cite}
\usepackage{amssymb}
\usepackage{epsfig}

\def\beq{\begin{equation}}
\def\eeq{\end{equation}}
\def\beeq{\begin{eqnarray}}
\def\eeeq{\end{eqnarray}}
\def\nn{\nonumber}
\def\to{\rightarrow}
\newcommand\as{\alpha_{\mathrm{S}}}
\newcommand\msbar{\overline {\mbox {\rm {\small {MS}}}}}
\def\gsim{\mathrel{\rlap{\lower4pt\hbox{\hskip1pt$\sim$}}
    \raise1pt\hbox{$>$}}}         
\def\lsim{\mathrel{\rlap{\lower4pt\hbox{\hskip1pt$\sim$}}
    \raise1pt\hbox{$<$}}}         
\def\bom#1{{\mbox{\boldmath $#1$}}}
\def\pdf{pdf}
\def\pdfs{pdf's}
\def\ee{$e^+ e^-$}
\def\pp{$pp$ }
\def\ppb{$p\bar{p}$}
\def\ra{\rightarrow}

\newcommand\ep{\epsilon}
\newcommand\sss{\scriptscriptstyle}

\def\lra{\longrightarrow}
\def\bann{\begin{eqnarray*}}
\def\eann{\end{eqnarray*}}

\newcommand\ptg{p_{{\sss T}\gamma}}
\newcommand\mug{\mu_\gamma}
\newcommand\etag{\eta_\gamma}
\newcommand\phig{\phi_\gamma}
\newcommand\epc{\epsilon_c}
\newcommand\epg{\epsilon_\gamma}
\newcommand\muo{\mu_0}
\def\abs#1{\left| #1\right|}
\newcommand\aem{\alpha_{\rm em}}
\newcommand\lambdamsb{\Lambda_{\rm \sss \overline{MS}}^{(5)}}
\newcommand\ptj{p_{{\sss T}j}}
\newcommand\etaj{\eta_j}
\newcommand\ptmin{p_{\sss T}^{min}}
\newcommand\ptmax{p_{\sss T}^{max}}
\newcommand{\cK}{{\cal K}}
\def\gE{\gamma_{\mbox{\tiny E}}}
\def\mR{\mu_{\mbox{\tiny R}}}
\def\res{\,{\mbox{\scriptsize jet}}}
\def\om{\omega}
\def\glip{\gamma_{\mbox{\tiny L}}}
\def\alb{\bar\as}
\def\alom{\frac{\alb}{\om}}
\def\VEV#1{\overline{#1}}
\def\sig{\sigma}
\newcommand{\cO}[1]{{\cal O}\left(#1\right)}
\newcommand{\nf}{N_f}
\begin{document}
\title{QCD}
\author{{\bf Convenors}: S. Catani, M. Dittmar, D. Soper, W.J. Stirling, 
S. Tapprogge. \\
  {\bf Contributing authors}: 
S.~Alekhin, P.~Aurenche, C.~Bal\'azs, R.D.~Ball, G.~Battistoni, E.L.~Berger, T.~Binoth, 
R.~Brock, D.~Casey, G.~Corcella, V.~Del~Duca, A.~Del~Fabbro, A.~De~Roeck       
C.~Ewerz, D.~de~Florian, M.~Fontannaz, S.~Frixione, W.T.~Giele, M.~Grazzini,
J.P.~Guillet, G.~Heinrich, J.~Huston, J.~Kalk, A.L.~Kataev, K.~Kato,
S.~Keller, M.~Klasen, D.A.~Kosower, A.~Kulesza, Z.~Kunszt, A.~Kupco, V.A.~Ilyin, 
L.~Magnea, M.L.~Mangano, A.D.~Martin, K.~Mazumdar, Ph.~Min\'e, M.~Moretti,
W.L.~van~Neerven, G.~Parente, D.~Perret-Gallix, E.~Pilon, A.E.~Pukhov,
I.~Puljak, J.~Pumplin, E.~Richter-Was, R.G.~Roberts, G.P.~Salam, M.H.~Seymour,
N.~Skachkov, A.V.~Sidorov, H.~Stenzel, D.~Stump, R.S.~Thorne, D.~Treleani, W.K.~Tung,
A.~Vogt, B.R.~Webber, M.~Werlen, S.~Zmouchko.
}
\maketitle
\begin{abstract}
We discuss issues of QCD at the LHC including parton distributions, Monte Carlo
event generators, the available next-to-leading order calculations,
resummation, photon production, small $x$ physics, double parton scattering, and
backgrounds to Higgs production.
\end{abstract}

\section{INTRODUCTION}
\label{sec:intro;qcd}

It is well known that precision QCD calculations and their
experimental tests at a proton--proton collider
are inherently difficult.    
``Unfortunately'', essentially all physics aspects of the 
LHC, from particle searches beyond the Standard Model (SM)
 to electroweak precision measurements and 
studies of heavy quarks are connected to 
the interactions of quarks and gluons at large transferred momentum.
An optimal exploitation of the LHC is thus unimaginable 
without the solid understanding of many aspects of QCD and their 
implementation in accurate Monte Carlo programs.

This review on QCD aspects relevant for the LHC gives an overview
of today's knowledge, of ongoing theoretical efforts 
and of some experimental feasibility studies for the LHC.
More aspects related to the experimental feasibility and an
overview of possible measurements, classified according to final
state properties, can be found in Chapter~15 of Ref.~\cite{ATLAS-TDR;11qcd}.
It was impossible, within the time-scale of this Workshop,  
to provide accurate and quantitative
answers to all the needs for LHC measurements. 
Moreover, owing to the foreseen theoretical and experimental
progress, detailed quantitative studies of QCD will have necessarily 
to be updated just before the start of the LHC experimental program.
The aim of this review is to update Ref.~\cite{pastlhcproc;6qcd} and to provide
reference work for the activities required in  preparation of the LHC
program in the coming years.

Especially relevant for essentially all possible measurements at the LHC
and their theoretical interpretation is the knowledge of the parton (quark,
anti-quark and gluon) distribution functions (\pdfs),
discussed in Sect.~2.
Today's knowledge about quark and anti-quark distribution functions comes
from lepton-hadron deep-inelastic scattering (DIS) experiments and
from Drell-Yan (DY) lepton-pair production
in hadron collisions. Most information about the
gluon distribution function is extracted from hadron--hadron
interactions with photons in the final state.
The theoretical interpretation of a large number of experiments
has resulted in various sets of \pdfs\ which are the basis for 
cross section predictions at the LHC.
Although these \pdfs\ are widely used for LHC simulations, their
uncertainties are difficult to estimate and various quantitative
methods are being developed now (see Sects. $2.1 - 2.4$).

The accuracy of this
traditional approach to describe proton--proton interactions
is limited by the possible knowledge of the proton--proton 
luminosity at the LHC.
Alternatively, much more precise information might eventually
be obtained from an approach which considers the LHC directly
as a parton--parton collider at large transferred momentum.
Following this approach, the experimentally
cleanest and theoretically best understood reactions
would be used to normalize directly the LHC parton--parton luminosities
to estimate various other reactions.
Today's feasibility studies indicate that this approach might 
eventually lead to cross section accuracies, due to 
experimental uncertainties,  of about $\pm$ 1\%. 
Such accuracies require that in order to profit, 
the corresponding theoretical uncertainties have to be 
controlled at a similar level using 
perturbative calculations and the corresponding Monte Carlo
simulations. As examples, the one-jet inclusive cross section and the 
rapidity dependence of $W$ and $Z$ production are known
at next-to-leading order, implying a theoretical accuracy of
about 10~\%. To improve further, higher
order corrections have to be calculated.

Section~3 addresses the implementation of
QCD calculations in Monte Carlo programs, which are an 
essential tool in the preparation of physics data analyses.
Monte Carlo programs are composed of several building blocks,
related to various stages in the interaction: the hard
scattering, the production of additional parton radiation
and the hadronization. Progress is being made in the improvement
and extension of matrix element generators and in the
 prediction for the transverse momentum distribution
in boson production. Besides the issues of parton distributions
 and hadronization, another
non-perturbative piece in a Monte Carlo generator is the treatment of 
the minimum bias and underlying events. One of the important
issue discussed in the section on Monte Carlo generators is
the consistent matching of the various building
blocks. More detailed studies on Monte Carlo generators for 
the LHC will be performed in a foreseen topical workshop.

The status of higher order calculations and prospects for
further improvements are presented in Sect.~4. As mentioned
earlier, one of the essential ingredients for improving
the accuracy of theoretical predictions is the availability
of higher order corrections.
For almost all  processes of interest, containing a (partially) hadronic
final state, the next-to-leading order (NLO) corrections have been computed and allow to
make reliable estimates of production cross sections. However, 
to obtain an accurate estimate of the uncertainty,
the calculation of the next-to-next-to-leading order (NNLO) corrections is needed. These 
calculations are extremely challenging and once performed,
they will have to be matched with a corresponding increase
in accuracy in the evolution of the \pdfs.

Section~5 discusses the summations of logarithmically enhanced
contributions in perturbation theory. Examples of such 
contributions occur in the inclusive production of a final-state system 
 which carries a large fraction of the
available center-of-mass energy (``threshold resummation'')
or in case of the production of a system with high mass
at small transverse momentum (``$p_T$ resummation''). In
case of threshold resummations, the theoretical calculations for most
processes of interest have been performed at next-to-leading
logarithmic accuracy. Their importance is two-fold: firstly, 
the cross sections at LHC might be directly affected; secondly, 
the extraction of \pdfs\ from other reactions
might be influenced and thus the cross sections at LHC
are modified indirectly. For transverse momentum
resummations, two analytical methods are discussed.

The production of prompt photons (as discussed in Sect.~6)
can be used to put constraints on the gluon density in the
proton and possibly to obtain measurements of the strong
coupling constant at LHC. The definition of a photon
usually involves some isolation criteria (against 
hadrons produced close in phase space). This requirement
is theoretically desirable, as it reduces the dependence
of observables on the fragmentation contribution to
photon production. At the same time, it is useful from the
experimental point of view as the background due to 
jets faking a photon signature can be further reduced.
A new scheme for isolation is able to eliminate the 
fragmentation contribution.

In Sect.~7 the issue of QCD dynamics in the region of small $x$
is discussed. For semi-hard strong interactions, which are
characterized by two large, different scales, the cross sections
contain large logarithms. The resummation of these at 
leading logarithmic (LL) accuracy can be performed by the BFKL
equation. Available experimental data are however not 
described by the LL BFKL, indicating the present of large
sub-leading contributions and the need to include 
next-to-leading corrections.
Studies of QCD dynamics in this regime can be made not only
by using inclusive observables, but also through the study
of final state properties. These include the production
of di-jets at large rapidity separation (studying the
azimuthal decorrelation between the two jets) or the
production of mini-jets (studying their multiplicity).

An important topic at the LHC is  multiple
(especially double) parton scattering (described in Sect.~8), 
i.e. the simultaneous occurrence
of two independent hard scattering in the same interaction. 
Extrapolations to LHC energies, based
on measurements at the Tevatron show the importance 
of taking this process into account when small transverse momenta
are involved. Manifestations of double parton
scattering are expected in the production of four jet final
states and in the production of a lepton in association with
two $b$-quarks (where the latter is used as a final state for
Higgs searches).

The last section (Sect.~9) addresses the issue of the present knowledge
of background for Higgs searches, for final states containing
two photons or multi-leptons. For the case of di-photon final states
(used for Higgs searches with $90 < m_H < 140$~GeV), studies
 of the irreducible background are performed by
calculating the (single and double) fragmentation contributions to  
NLO accuracy and by studying the effects of soft gluon emission.
The production of rare five lepton final states could provide
valuable information on the Higgs couplings for $m_H > 200$~GeV,
awaiting further studies on improving the understanding of the
backgrounds.

During the workshop, no studies of 
diffractive scattering at the LHC have been performed. This topic is challenging both from
the theoretical and the experimental point of view. The study of
diffractive processes (with a typical signature of a leading proton
and/or a large rapidity gap) should lead to an improved understanding
of the transition between soft and hard process and of the 
non-perturbative aspects of QCD. From the experimental point of view,
the detection of leading protons in the LHC environment is 
challenging and requires adding additional detectors to ATLAS and
CMS. If hard diffractive scattering (leading proton(s) together
with e.g. jets as signature for a hard scattering) is to be studied
with decent statistical accuracy at large $p_T$, most of the 
luminosity delivered under normal running conditions has to be utilized.
A few more details can be found in Chapter~15 of  Ref.~\cite{ATLAS-TDR;11qcd},
some ideas for detectors in Ref.~\cite{TOTEM;1qcd}. Much more work remains
to be done, including a detailed assessment of the capabilities of the
additional detectors.

\subsection{Overview of QCD tools}

All of the processes to be investigated at the LHC involve QCD to some extent.
It cannot be otherwise, since the colliding quarks and gluons carry the QCD
color charge. One can use perturbation theory to describe the cross section for
an inclusive hard-scattering process,
\beq
\label{hardpro;1qcd}
h_1(p_1) + h_2(p_2) \to H(Q,\{ \dots \}) + X \;\;.
\eeq
Here the colliding hadrons $h_1$ and $h_2$ have momenta $p_1$ and $p_2$, $H$
denotes the triggered hard probe (vector bosons, jets, heavy quarks, Higgs
bosons, SUSY particles and so on) and $X$ stands for  any unobserved particles
produced by the collision. The typical scale $Q$ of the scattering process is
set by the invariant mass or the transverse momentum of the hard probe and the
notation $\{ \dots \}$ stands for any other measured kinematic variable of the
process. For example, the hard process may be the production of a $Z$ boson.
Then $Q=M_Z$ and we can take $\{\dots\} = y$, where $y$ is the rapidity of
the $Z$ boson. One can also measure the transverse momentum $Q_T$ of the the
$Z$ boson. Then the simple analysis described below applies if $Q_T \sim M_Z$.
In the cases $Q_T \ll M_Z$ and $M_Z \ll Q_T$, there are two hard scales in the
process and a more complicated analysis is needed. The case $Q_T \ll M_Z$ is of
particular importance and is discussed in 
Sects.~\ref{sec:corcella;3qcd}, \ref{sec:huston;3qcd} and
\ref{sec:resptdist;qcd}. 

The cross section for the process (\ref{hardpro;1qcd}) is
computed by using the factorization formula~\cite{Ellis:1991qj,Collins:1989gx}
\beeq
\sigma(p_1,p_2;Q, \{ \dots \} ) \!\!\!&=&\!\!\! 
\sum_{a,b} \int dx_1 \, dx_2 
\,f_{a/h_1}(x_1,Q^2)
\, f_{b/h_2}(x_2,Q^2) \; 
{\hat \sigma}_{ab}(x_1p_1,x_2p_2;Q, \{ \dots \};\as(Q))
\nn \\
\label{factfor;1qcd}
\!\!\!&+&\!\!\! {\cal O}\left( (\Lambda_{QCD}/Q )^p \right) \;\;.
\eeeq
Here the indices $a,b$ denote parton flavors,
\{$g,u,\bar u,d,\bar d,\dots$\}. The factorization formula (\ref{factfor;1qcd})
involves the convolution of the partonic cross section ${\hat \sigma}_{ab}$  
and the parton distribution functions 
$f_{a/h}(x, Q^2)$  of the
colliding hadrons. The term ${\cal O}\left( (\Lambda_{QCD}/Q )^p \right)$ on the right-hand side 
of Eq.~(\ref{factfor;1qcd}) generically denotes non-perturbative contributions
(hadronization effects, multiparton interactions, contributions of the soft
underlying event and so on).

Evidently, the \pdfs\ are of great importance to making
predictions for the LHC. These functions are determined from experiments. Some
of the issues relating to this determination are discussed in
Sect.~\ref{sec:pdf;qcd} In particular, there are discussions of the question
of error analysis in the determination of the \pdfs\ and there is
a discussion of the prospects for determining the \pdfs\ from LHC
experiments.

The partonic cross section ${\hat \sigma}_{ab}$ is computable as a power series
expansion in the QCD coupling $\as(Q)$:
\beeq
\label{pertex;1qcd}
\!\!\!\!\!\!
{\hat \sigma}_{ab}(p_1,p_2;Q, \{ \dots \}; \as(Q)) \!\!\!\!&=&\!\!\!\!
\as^k(Q) \left\{
{\hat \sigma}_{ab}^{(LO)}(p_1,p_2;Q, \{ \dots \}) \right. \nn \\
&~& \;\;\;\;\;\; \;\;\;\; + \,\as(Q) \;
{\hat \sigma}_{ab}^{(NLO)}(p_1,p_2;Q, \{ \dots \}) \nn \\
&~& \;\;\;\;\;\; \;\;\;\; +\left. \! \as^2(Q) \;
{\hat \sigma}_{ab}^{(NNLO)}(p_1,p_2;Q, \{ \dots \}) +
\cdots \right\} \,.
\eeeq
The lowest (or leading) order (LO) term ${\hat \sigma}^{(LO)}$ gives only 
a rough estimate
of the cross section. Thus one needs the next-to-leading order (NLO) term, which
is available for most cases of interest. A list of the available calculations is
given in Sect.~\ref{sec:nlo}. Cross sections at NNLO are not available at
present, but the prospects are discussed in Sect.~\ref{sec:nnlo}.

The simple formula (\ref{factfor;1qcd}) applies when the cross section being
measured is ``infrared safe.'' This means that the cross section does not
change if one high energy strongly interacting light particle in the
final state divides into two particles moving in the same direction or if one
such particle emits a light particle carrying very small momentum. Thus in
order to have a simple theoretical formula one does not typically measure the
cross section to find a single high-$p_T$ pion, say, but rather one measures
the cross section to have a collimated jet of particles with a given total
transverse momentum $p_T$. If, instead, a single high-$p_T$ pion (or, more
generally, a high-$p_T$ hadron $H$) is measured, 
the factorization formula has to include an additional convolution with the
corresponding parton fragmentation function $d_{a/H}(z, Q^2)$.
An example of a case where one needs a more
complicated treatment is the production of high-$p_T$ photons. This case is
discussed in Sect.~\ref{sec:photons;qcd}

\begin{figure}[t!]
\begin{center}
\mbox{\includegraphics[width=0.5\textwidth,clip]{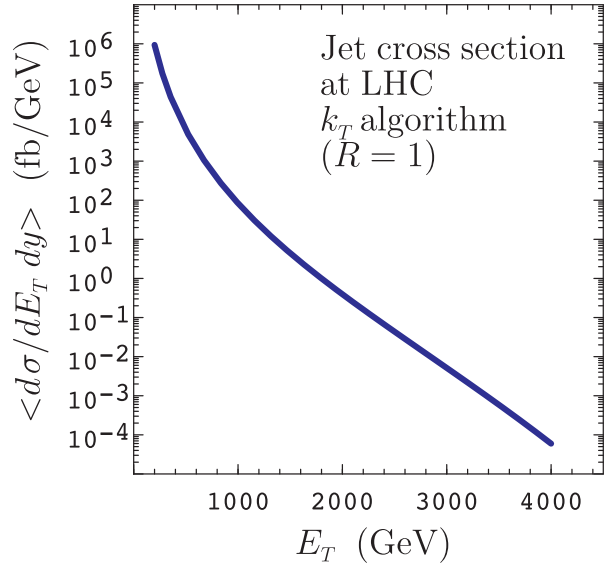}}
\end{center}
\vskip -0.8cm
\caption{Jet cross section at the LHC, averaged over the rapidity 
interval $-1<y<1$. The cross section is calculated at NLO using CTEQ5M 
partons with the renormalization and factorization
scales set to $\mu_R = \mu_F = E_T/2$. Representative values at
$E_T = $0.5, 1, 2, 3 and 4 TeV are
$(
6.2 \times 10^{3},
8.3 \times 10^{1},
4.0 \times 10^{-1},
5.1 \times 10^{-3},
5.9 \times 10^{-5} 
)$ fb/GeV with about 3\% statistical errors. 
}
\label{fig:jetxsect}
\end{figure}

As an example of a NLO calculation, we display in Fig.~\ref{fig:jetxsect}
the predicted cross section $d\sigma/dE_T\,dy$ at the LHC for the
inclusive production of a jet with transverse energy $E_T$ and rapidity
$y$ averaged over the rapidity interval $-1<y<1$. The calculation uses
the program in Ref.~\cite{Ellis:1992en} and the \pdf\ set CTEQ5M 
\cite{Lai:2000wy}. 
As mentioned above, the ``jets'' must be
defined with an infrared safe algorithm. Here we use the $k_T$
algorithm \cite{Catani:1993hr,Ellis:1993tq} with a joining parameter $R = 1$. 
The $k_T$ algorithm has better theoretical properties
than the cone algorithm that has often been used in hadron collider
experiments.

\begin{figure}[t!]
\centering
\includegraphics[width=0.6\textwidth,clip]{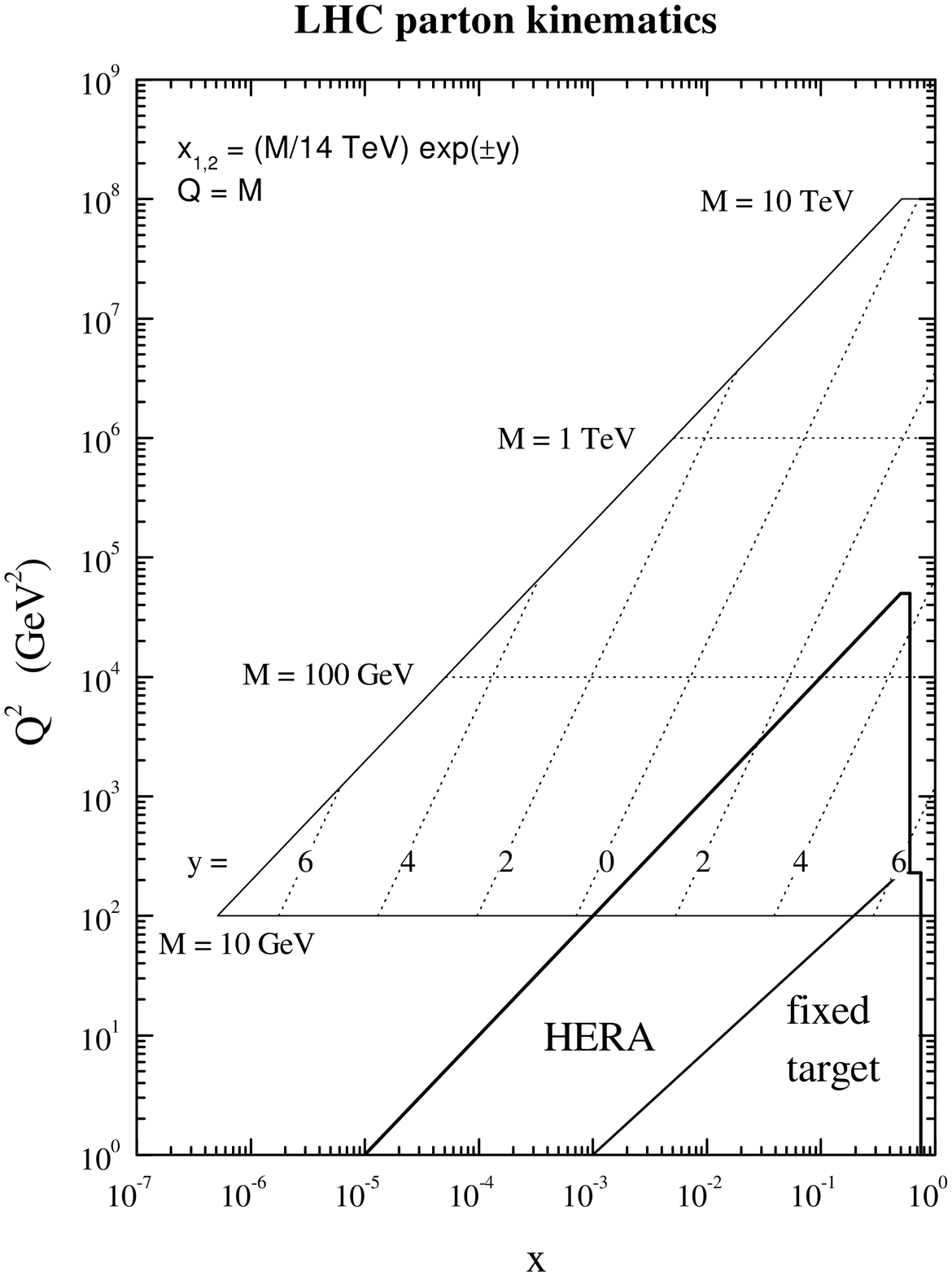}
\vskip-0.5cm
\caption{Values of $x$ and $Q^2$ probed in the production of an object
of mass $M$ and rapidity $y$ at the LHC, $\sqrt{s} = 14$~TeV.} 
\label{fig:LHCpartons}
\end{figure}

In Eq.~(\ref{factfor;1qcd}) there are integrations over the parton momentum
fractions $x_1$ and $x_2$. The values of $x_1$ and $x_2$ that dominate the
integral are controlled by the kinematics of the hard-scattering process. In
the case of the production of a heavy particle of mass $M$ and rapidity $y$,
the dominant values of the momentum fractions are $x_{1,2} \sim (M e^{\pm
y})/{\sqrt{s}}$, where $s=(p_1+p_2)^2$ is the square of the centre-of-mass energy
of the collision. Thus, varying $M$ and $y$ at fixed ${\sqrt{s}}$, we are
sensitive to partons with different momentum fractions. Increasing ${\sqrt{s}}$
the \pdfs\ are probed in a kinematic range that extends towards
larger values of $Q$ and smaller values of $x_{1,2}$. This is illustrated in
Fig.~\ref{fig:LHCpartons}. At the LHC, $x_{1,2}$ can
be quite small. Thus small $x$ effects that go beyond the simple formula
(\ref{factfor;1qcd}) could be important. These are discussed in
Sect.~\ref{sec:smallx;qcd}


%
\begin{figure}[t!]
\begin{center}
\mbox{\includegraphics[width=0.6\textwidth,clip]{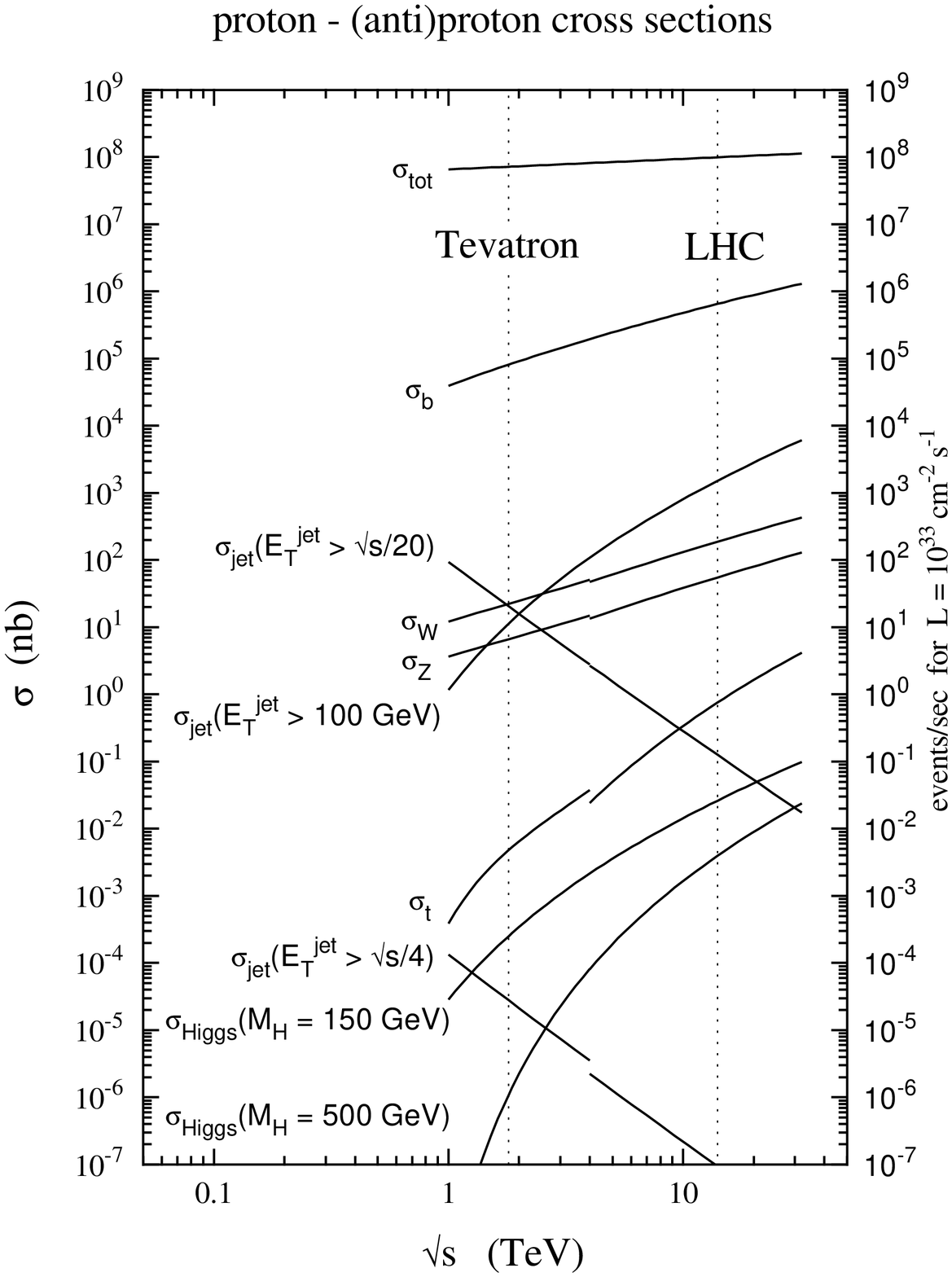}}
\end{center}
\vskip -0.8cm
\caption{Cross sections for hard scattering versus $\sqrt{s}$.
The cross section values at $\sqrt{s} = 14$~TeV are:
$\sigma_{\rm tot} = 99.4$~mb,
$\sigma_{\rm b} = 0.633$~mb,
$\sigma_{\rm t} = 0.888$~nb,
$\sigma_{\rm W} = 187$~nb,
$\sigma_{\rm Z} = 55.5$~nb,
$\sigma_{{\rm H}}(M_{\rm H} =150\;{\rm GeV}) = 23.8$~pb,
$\sigma_{{\rm H}}(M_{\rm H} =500\;{\rm GeV}) = 3.82$~pb,
$\sigma_{{\rm jet}}(E_T^{\rm jet}> 100\;{\rm GeV}) = 1.57\; \mu$b,
$\sigma_{{\rm jet}}(E_T^{\rm jet}> \sqrt{s}/20) = 0.133$~nb,
$\sigma_{{\rm jet}}(E_T^{\rm jet}> \sqrt{s}/4) = 0.10$~fb.
All except the first of these are calculated using the latest MRST
\pdfs\ \protect\cite{Martin:1999ww}.}
\label{fig:FNALandLHCxsect}
\end{figure}

In Fig.~\ref{fig:FNALandLHCxsect} we plot NLO cross sections for a selection
of hard processes versus $\sqrt{s}$. The curves for the lower values of $\sqrt{s}$
 are for $p\bar p$ collisions, as at the Tevatron, while the curves for the
higher values of $\sqrt{s}$ are for $p  p$ collisions, as at the LHC. An
approximation (based on an extrapolation of a standard Regge parametrization) 
to the total cross section is also displayed. We see that the
cross sections for production of objects with a fixed mass or jets with a
fixed transverse energy $E_T$ rise with $\sqrt{s}$. This is because the
important $x_{1,2}$ values decrease, as discussed above, and there are more
partons at smaller $x$. On the other hand, cross sections for jets with
transverse momentum that is a fixed fraction of $\sqrt{s}$ fall with $\sqrt{s}$.
This is (mostly) because the partonic cross sections $\hat \sigma$ fall with
$E_T$ like $E_T^{-2}$.

The perturbative evaluation of the factorization formula (\ref{factfor;1qcd})
is based on performing power series expansions in the QCD coupling $\as(Q)$.
The dependence of $\as$ on the scale $Q$ is logarithmic and it is given
by the renormalization group equation~\cite{Ellis:1991qj}
\beq
\label{rgeq;1qcd}
Q^2 \frac{d \as(Q)}{d Q^2} = \beta(\as(Q)) = - b_0 \,\as^2(Q) - b_1 \,\as^3(Q)
+ \cdots \;\;,
\eeq
where the first two perturbative coefficients are
\beq
\label{betas;1qcd}
b_0 = 
\frac{33 - 2 N_f}{12 \pi} \;\;, \quad
b_1 = 
\frac{153 - 19 N_f}{24 \pi^2} \;\;,
\eeq
and $N_f$ is the number of flavours of light quarks (quarks whose
mass is much smaller than the scale $Q$). The third and fourth coefficients
$b_2$ and $b_3$ of the $\beta$-function are also
known~\cite{Tarasov:1980au,vanRitbergen:1997va}. If we include only the LO
term, Eq.~(\ref{rgeq;1qcd}) has the exact analytical solution
\beq
\label{loas;1qcd}
\as(Q) = \frac{1}{b_0 \ln (Q^2/\Lambda_{QCD}^2)} \;\;,
\eeq
where the integration constant $\Lambda_{QCD}$ fixes the absolute size of the 
QCD coupling. From Eq.~(\ref{loas;1qcd}) we can see that a change of the scale
$Q$ by an arbitrary factor of order unity (say, $Q \to Q/2$) induces a variation
in $\as$ that is of the order of $\as^2$. This variation in  uncontrollable
because it is beyond the accuracy at which  Eq.~(\ref{loas;1qcd})
is valid. Therefore, in LO of perturbation theory the size of $\as$ is
not unambiguously defined. 

The QCD coupling $\as(Q)$ can be precisely defined only starting from
the NLO in perturbation theory. To this order, the 
renormalization group equation (\ref{rgeq;1qcd}) has no exact analytical solution.
Different approximate solutions can differ by higher-order corrections and
some (arbitrary) choice has to be made. Different choices can eventually be 
related to the definition of different renormalization schemes. The most popular
choice~\cite{Caso:1998tx} is to use the $\msbar$-scheme to define
renormalization and then to use the following approximate solution of the two
loop evolution equation to define $\Lambda_{QCD}$: 
\beq
\label{nloas;1qcd}
\as(Q) = \frac{1}{b_0 \ln (Q^2/\Lambda_{\msbar}^2)} 
\left[ 1 - \frac{ b_1 \ln \big[ \ln (Q^2/\Lambda_{\msbar}^2) \big]}
{b_0 \ln (Q^2/\Lambda_{\msbar}^2)} + 
{\cal O}\!\left( \frac{\ln^2 \big[ \ln (Q^2/\Lambda_{\msbar}^2)\big]}
{\ln^2 (Q^2/\Lambda_{\msbar}^2)} \right) \right]
\;\;.
\eeq
Here the definition of $\Lambda_{QCD}$ 
$(\Lambda_{QCD}=\Lambda_{\msbar})$
is contained in the fact that there is no term
proportional to $1/\ln^2(Q^2/\Lambda_{QCD}^2)$. 
In this expression there are $N_f$
light quarks. Depending on the value of $Q$, one may want to use different
values for the number of quarks that are considered light. Then one must match
between different renormalization schemes, and correspondingly change the value
of $\Lambda_{\msbar}$ as discussed in Ref.~\cite{Caso:1998tx}. The constant
$\Lambda_{\msbar}$ is the one fundamental constant of QCD that  must be determined
from experiments. Equivalently, experiments can be used to determine the value
of $\as$ at a fixed reference scale $Q=\mu_0$. It has  become standard to
choose $\mu_0=M_Z$. The most recent determinations of $\as$
lead~\cite{Caso:1998tx} to the world average  $\as(M_Z) = 0.119 \pm 0.002$. In
present applications to hadron collisions, the value of $\as$ is often varied
in the wider range $\as(M_Z)=0.113-0.123$ to conservatively estimate
theoretical uncertainties.

The parton distribution functions $f_{a/h}(x,Q^2)$ at any fixed scale $Q$
are not computable in perturbation theory. However, their scale dependence is
perturbatively controlled by the DGLAP evolution 
equation~\cite{Gribov:1972ri,Gribov:1972rt,Altarelli:1977zs,Dokshitzer:1977sg}
\beq
\label{evequa;1qcd}
Q^2 \frac{d \,f_{a/h}(x,Q^2)}{d Q^2} = 
\sum_{b} \int_{x}^1 \frac{dz}{z} \, P_{ab}(\as(Q^2), z) \,f_{a/h}(x/z,Q^2)
\;\;.
\eeq
Having determined $f_{a/h}(x, Q_0^2)$ at a given input scale $Q = Q_0$, the
evolution equation can be used to compute the \pdfs\
at different perturbative scales $Q$ and larger values of $x$.

The kernels $P_{ab}(\as, z)$ in Eq.~(\ref{evequa;1qcd}) are the
Altarelli--Parisi (AP) splitting functions. They depend on the parton flavours
$a,b$ but do not depend on the colliding hadron $h$ and thus they are
process-independent.  The AP splitting  functions can be computed as a power
series expansion in $\as$:
\beq
\label{apexp;1qcd}
P_{ab}(\as, z) = \as P_{ab}^{(LO)}(z) + \as^2 P_{ab}^{(NLO)}(z)
+ \as^3 P_{ab}^{(NNLO)}(z) + {\cal O}(\as^4) \;\;.
\eeq
The LO and NLO terms $P_{ab}^{(LO)}(z)$ and $P_{ab}^{(NLO)}(z)$ in the 
expansion are 
known~\cite{Floratos:1977au,Floratos:1979ny,Gonzalez-Arroyo:1979df,
Gonzalez-Arroyo:1980he,Curci:1980uw, Furmanski:1980cm,Floratos:1981hs}. 
These first two terms (their explicit expressions are collected 
in Ref.~\cite{Ellis:1991qj}) are used
in most of the QCD studies. Partial 
calculations~\cite{Larin:1994vu,Larin:1997wd} 
of the next-to-next-to-leading order (NNLO)
term $P_{ab}^{(NNLO)}(z)$ are also available (see Sects.~\ref{sec:vogt;2qcd}, 
\ref{sec:kataev;2qcd} and \ref{sec:nnlo}).

As in the case of $\as$, the definition and the evolution of the \pdfs\
depends on how many of the quark flavors are considered
to be light in the calculation in which the parton distributions are used.
Again, there are matching conditions that apply. In the currently popular sets
of parton distributions there is a change of definition at $Q = M$, where $M$
is the mass of a heavy quark.

The factorization on the right-hand side of Eq.~(\ref{factfor;1qcd}) in terms
of (perturbative) process-dependent partonic cross sections and 
(non-perturbative) process-independent \pdfs\ involves
some degree of arbitrariness, which is known as factorization-scheme
dependence. We can always `re-define' the \pdfs\ by
multiplying (convoluting) them by some process-independent perturbative
function. Thus, we should always specify the factorization-scheme used to
define the \pdfs. The most common scheme is the $\msbar$
factorization-scheme~\cite{Ellis:1991qj}. An alternative scheme, known as DIS
factorization-scheme~\cite{Altarelli:1979ub}, is sometimes used. Of course,
physical quantities cannot depend on the factorization scheme. Perturbative
corrections beyond the LO to partonic cross sections and AP splitting
functions are thus factorization-scheme dependent to compensate the
corresponding dependence of the \pdfs.  In the 
evaluation of hadronic cross sections at a given perturbative order,  the
compensation may not be exact because of the presence of yet uncalculated
higher-order terms. Quantitative studies of the factorization-scheme dependence
can be used to set a lower limit on the size of missing higher-order 
corrections. 

The factorization-scheme dependence is not the only signal of the uncertainty
related to the computation of the factorization formula (\ref{factfor;1qcd}) by
truncating its perturbative expansion at a given order. Truncation
leads to additional  uncertainties and, in particular, to a dependence on the
renormalization and factorization scales.  The renormalization scale $\mu_R$ is
the scale at which the QCD coupling $\as$ is evaluated. The factorization scale
$\mu_F$ is  introduced to separate the bound-state effects  (which are embodied
in the \pdfs) from the perturbative interactions 
(which are embodied in the partonic cross section) of the partons. In
Eqs.~(\ref{factfor;1qcd}) and (\ref{pertex;1qcd}) 
we took $\mu_R =\mu_F = Q$. On physical grounds
these scales have to be of the same order as $Q$, but their value cannot be
unambiguously fixed. In the general case, the right-hand side of
Eq.~(\ref{factfor;1qcd}) is modified by introducing explicit dependence on
$\mu_R, \mu_F$ according to the replacement 
\beeq
f_{a/h_1}(x_1,Q^2) \; f_{a/h_2}(x_2,Q^2) \!\!\!\!\!&& \!\!\!\!\!\!
{\hat \sigma}_{ab}(x_1p_1,x_2p_2;Q, \{ \dots \};\as(Q)) \nn \\
& \downarrow &  \nn \\
\label{mudep;1qcd}
f_{a/h_1}(x_1,\mu_F^2) \;\, f_{a/h_2}(x_2,\mu_F^2) \; \!\!\!\!\!&& \!\!\!\!\!\!
{\hat \sigma}_{ab}(x_1p_1,x_2p_2;Q, \{ \dots \}; \mu_R,\mu_F
;\as(\mu_R)) \;\;. 
\eeeq
The physical cross section $\sigma(p_1,p_2;Q, \{ \dots \} )$  
does not depend on the arbitrary scales $\mu_R, \mu_F$, but parton densities and
partonic cross sections  separately depend on these scales. The $\mu_R,
\mu_F$-dependence  of the partonic cross sections appears in their perturbative
expansion and compensates the $\mu_R$ dependence of $\as(\mu_R)$ and the
$\mu_F$-dependence of the \pdfs. The compensation would
be  exact if everything could be computed to all orders in perturbation theory.
However, when the quantities entering Eq.~(\ref{mudep;1qcd}) are evaluated at,
say, the $n$-th perturbative order, the result exhibits a residual $\mu_R,
\mu_F$-dependence, which is formally of the $(n+1)$-th order. That is, the
explicit $\mu_R, \mu_F$-dependence that still remains reflects the absence of
yet uncalculated higher-order terms. For this reason, the size of the $\mu_R,
\mu_F$ dependence is often used as a measure of the size of at least some of 
the uncalculated higher-order terms and thus as an estimator of the theoretical
error caused by truncating the perturbative expansion.

\begin{figure}[t!]
\begin{center}
\mbox{\includegraphics[width=0.5\textwidth,clip]{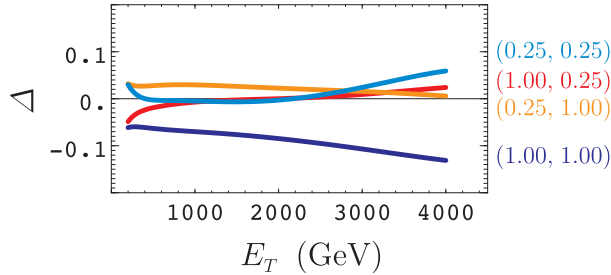}}
\end{center}
\vskip -0.8cm
\caption{Variation of the jet cross section with renormalization and
factorization scale. We show $\Delta$ defined in Eq.~(\ref{DeltaDef})
versus $E_T$ for four choices of  $\{\mu_{R}/E_T,
\mu_{F}/E_T\}$.}
\label{fig:mudepend}
\end{figure}

As an example, we estimate the theoretical error on the predicted jet cross
section in Fig.~\ref{fig:jetxsect}. We vary the renormalization scale $\mu_{R}$
and the factorization scale $\mu_{F}$. In Fig.~\ref{fig:mudepend}, we plot
\begin{equation}
\Delta(\mu_{R}/E_T,\mu_{F}/E_T)
= 
{\langle d\sigma(\mu_{R}/E_T,\mu_{F}/E_T)\,/dE_T\,dy\rangle
\over
\langle d\sigma(0.5, 0.5)/dE_T\,dy\rangle}
\label{DeltaDef}
\end{equation}
versus $E_T$ for four values of the pair $\{\mu_{R}/E_T,
\mu_{F}/E_T\}$, namely  $\{0.25,0.25\}$, $\{1.0,0.25\}$
$\{0.25,1.0\}$, and $\{1.0,1.0\}$.  We see about a 10\% variation in the
cross section. This suggests that the theoretical uncertainty is at least
10\%.

The issue of the scale dependence of the perturbative QCD calculations
has received attention in the literature and
various recipes have been proposed to choose `optimal' values of $\mu$
(see the references in~\cite{Caso:1998tx}). There is no compelling
argument that shows that these `optimal' values reduce the size of the yet
unknown higher-order corrections.
These recipes may thus be used to get more confidence on the central value of 
the theoretical calculation, but they cannot be used to reduce its theoretical
uncertainty as estimated, for instance, by scale variations around 
$\mu \sim Q$. The theoretical uncertainty ensuing from the 
truncation of the perturbative series can only be reduced by actually
computing more terms in perturbation theory.

We have so far discussed the factorization formula (\ref{factfor;1qcd}). We
should emphasize that there is another mode of analysis of the theory
available, that embodied in Monte Carlo event generator programs. In this type
of analysis, one is limited (at present) to leading order partonic hard
scattering cross sections. However, one simulates the complete physical
process, beginning with the hard scattering and proceeding through parton
showering via repeated one parton to two parton splittings and finally ending
with a model for how partons turn into hadrons. This class of programs, which
simulate complete events according to an approximation to QCD, are very
important to the design and analysis of experiments. Current issues in Monte
Carlo event generator and other related computer programs are discussed in
Sect.~\ref{sec:mcs;qcd}

%

\section{PARTON DISTRIBUTION FUNCTIONS\protect\footnote{Section
    coordinators: R. Ball, M. Dittmar and W.J. Stirling.}}
\label{sec:pdf;qcd}

Parton distributions (\pdfs) play a central role in hard scattering 
cross sections at the LHC. A precise knowledge of the \pdfs\
is absolutely vital for reliable predictions for signal and background 
cross sections. In many cases, it is the uncertainty in the input
\pdfs\ that dominates the theoretical error on the prediction.
Such uncertainties can arise both from the starting
distributions, obtained from a global fit to DIS,
DY
and other data, and from DGLAP evolution to the higher $Q^2$ scales typical
of LHC hard scattering processes. 

To predict LHC cross sections we will need accurate \pdfs\ over a wide 
range of $x$ and $Q^2$ (see Fig.~\ref{fig:LHCpartons}).
Several groups have made significant contributions to the determination of 
\pdfs\ both during and after the workshop. 
The MRST and CTEQ global analyses have 
been updated and refined, and small numerical problems have been corrected.
The `central' \pdf\ sets obtained from these global fits are, not surprisingly,
very similar,  and remain the best way to estimate central values
for LHC cross sections. Specially constructed variants of the central fits
(exploring, for example, different values of $\alpha_S$ or different
theoretical treatments of heavy quark distributions) allow the sensitivity
of the cross sections to some of the input assumptions.

A rigorous and global treatment of \pdf\ {\it uncertainties} remains elusive,
but there has been significant progress in the last few years, with several 
groups introducing sophisticated statistical analyses into 
quasi-global fits. While some of the more novel methods are still at a 
rather preliminary stage, it is hoped that over the next few years they 
may be developed into useful tools. 

One can reasonably expect that by LHC start-up time, the precision 
\pdf\ determinations will have improved from NLO to
NNLO. Although the complete  
NNLO splitting functions have not yet been calculated, several studies have 
made use of partial information (moments, $x\to 0,1$ limiting behaviour)
to assess the impact of the NNLO corrections.
 
At the same time, accurate measurements of Standard Model (SM)
cross sections at the LHC will further constrain the \pdfs. 
The kinematic acceptance of the LHC detectors allows a large 
range of $x$ and $Q^2$ to be probed. Furthermore, the wide variety
of final states and high parton-parton luminosities available
will allow an accurate determination of the gluon density and 
flavour decomposition of quark densities.

All of the above issues are discussed in the individual contributions that 
follow. Lack of space has necessarily restricted the amount of information
that can be included, but more details can always be found in the literature.


\subsection{MRS: \pdf\ uncertainties and $W$ and $Z$ 
production at the LHC\protect\footnote{Contributing authors: A.D.~Martin, 
R.G.~Roberts, W.J.~Stirling and R.S.~Thorne.}}
\label{mrssection}

There are several reasons why it is very difficult to derive overall
`one sigma' errors on parton distributions of the form $f_i \pm \delta f_i$.
In the global fit there are complicated correlations between
a particular \pdf\ at different
$x$ values, and between the different \pdf\ flavours. 
For example,  the
charm distribution is correlated with the gluon distribution, the gluon
distribution at low $x$ is correlated with the gluon at high $x$ via the
momentum sum rule, and so on. Secondly, many of the uncertainties in the input
data or fitting procedure are not `true' errors in the probabilistic sense.
For example,  the uncertainty in the high--$x$ gluon in the MRST
fits \cite{Martin:1998sq} derives from a subjective assessment of the impact 
of `intrinsic $k_T$' on the prompt photon cross sections 
included in the global fit.
\begin{figure}[t!]
\begin{center}
\mbox{\includegraphics[width=0.5\textwidth,clip]{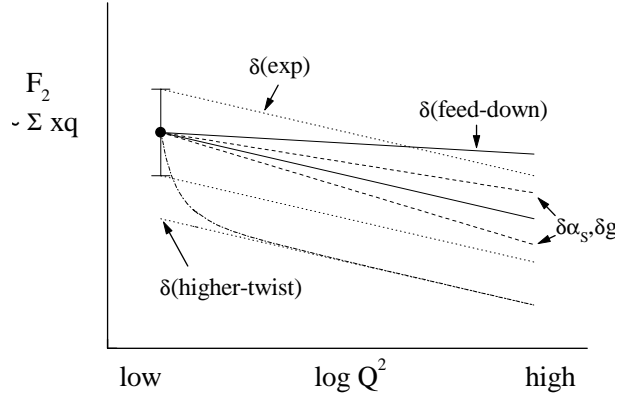}}
\end{center}
\vskip -0.8cm
\caption{Schematic representation of the various uncertainties
contributing to the prediction of a structure function
or parton distribution at high $Q^2$.}\label{fig:ecartoon}
\end{figure}
Despite these difficulties, several groups {\it have} attempted  to 
extract meaningful $\pm \delta f_i$ \pdf\ errors 
(see \cite{Alekhin:1999ck,Alekhin:1999kt} and 
Sects.~\ref{sec:keller;2qcd},\ref{sec:alekhin;2qcd}). 
Typically, these analyses focus on subsets 
of the available DIS and other data, which are statistically `clean', i.e.
free from undetermined systematic errors. As a result, various
aspects of the \pdfs\ that are phenomenologically important, 
the flavour structure of the sea and the sea and gluon distributions 
at large $x$ for example, are either only weakly constrained
or not determined at all.  

Faced with the difficulties in trying to formulate {\it global}
\pdf\ errors, one can adopt a more pragmatic approach to the problem
by making a detailed assessment
of the \pdf\ uncertainty for a {\it particular}  cross section of interest.
This involves determining which partons contribute and 
at which $x$ and $Q^2$ values,
and then systematically tracing back to the data sets
that constrained the distributions in the global fit. Individual \pdf\ sets
can then be constructed to reflect the uncertainty in the particular partons
determined by a particular data set.

We have recently performed such an analysis for $W$ 
and $Z$ total cross sections
at the Tevatron and LHC \cite{Martin:1999ww}.
The theoretical
technology for calculating these is very robust. The total cross sections are
known to NNLO in QCD perturbation theory
\cite{Matsuura:1990ba,Hamberg:1991np,vanNeerven:1992gh}, and the 
input electroweak parameters
($M_{W,Z}$, weak couplings, etc.) are known to high accuracy. 
The main theoretical uncertainty therefore derives from the input
\pdfs\ and, to a lesser extent, from $\alpha_S$.\footnote{The two are of course
correlated, see for example \cite{Martin:1998sq}.}

\begin{figure}[t!]
\begin{center}
\includegraphics[width=9cm,bb=60 150 520 660]{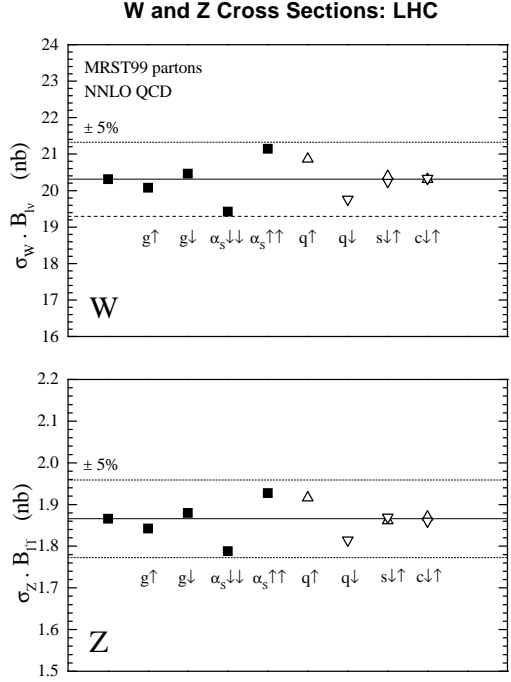} 
\end{center}
\vskip-0.5cm
\caption{Predictions for the $W$ and $Z$ total cross sections    
times leptonic branching ratio in     
$p  p$ collisions at  $14$~TeV
using the various MRST parton sets from Ref.~\cite{Martin:1999ww}.
The error bars on the default MRST prediction
correspond to a scale variation of $\mu = M_V/2 \to 2 M_V$, $V=W,Z$.
} \label{fig:LHCwdat}     
\end{figure}

For the hadro-production of a heavy object like a $W$ boson, with mass $M$ and
rapidity $y$,  leading-order
kinematics give   $x = M\exp(\pm y)/\sqrt{s}$
and $Q=M$. For example, a $W$ boson ($M = 80$~GeV)
produced at rapidity $y=3$ at the LHC corresponds to the annihilation of quarks
with $x=0.00028$ and $0.11$, probed at $Q^2 = 6400$~GeV$^2$.
Notice that $u,d$ quarks with these $x$ values are already
more or less directly `measured' in deep
inelastic scattering (at HERA and in fixed--target experiments
respectively), but at much lower $Q^2$, see Fig.~\ref{fig:LHCpartons}.
Therefore the first two
important sources of uncertainty in the \pdfs\ relevant to $W$ production
are
\begin{itemize}
\item[(i)] the uncertainty in the DGLAP evolution, which except at high $x$ comes
mainly from the gluon and $\alpha_S$;
\item[(ii)] the uncertainty in the quark distributions from measurement errors
on the structure function data used in the fit.
\end{itemize}
This is illustrated in Fig.~\ref{fig:ecartoon}.\footnote{The `feed-down'
 error represents a possible anomalously large contribution at $x \approx 1$
affecting the evolution at lower $x$. It is not relevant, however,
for $W$ production at the Tevatron or LHC.}
Only $75\%$ of the total $W$ cross section at the LHC arises from the 
scattering
of $u$ and $d$  (anti)quarks. Therefore also potentially important is
\begin{itemize}
\item[(iii)] the uncertainty in the input strange ($s$) and 
charm ($c$) quark distributions, which are relatively
poorly determined at low $Q^2$ scales.
\end{itemize}

In order to investigate these various effects we have constructed ten variants
of the standard MRST99 distributions \cite{Martin:1999ww} that 
probe approximate  $\pm1\sigma$ 
variations in the gluon, $\alpha_S$, the overall quark normalisation, and the $s$ and $c$ \pdfs.
The corresponding predictions for the $W$ total cross section at the LHC are shown 
in Fig.~\ref{fig:LHCwdat}.
Evidently the largest variation comes from the effect of varying $\alpha_S(M_Z^2)$, in this case
by $\pm 0.005$ about the central value of $0.1175$. The higher the value of $\alpha_S$,
the faster the (upwards) evolution, and the larger the predicted $W$ cross section.
The effect of a $\pm 2.5\%$ normalisation error, as parameterised
by the $q\updownarrow$ \pdfs, is also significant. 
The uncertainties in
the input  $s$ and $c$ distributions get washed out  
by evolution to high $Q^2$,
and turn out to be numerically unimportant.

In conclusion, we see from Fig.~\ref{fig:LHCwdat} that  $\pm 5\%$ 
represents a conservative
error on the prediction of $\sigma(W)$ at LHC. We arrive at this result
without recourse to complicated statistical analyses
in the global fit. It is also reassuring that the 
latest (corrected) CTEQ5  prediction \cite{Lai:2000wy} is very close 
to the central MRST99 prediction, see Fig.~\ref{fig:WprodA} below.
Finally, it is important to stress that the results of our analysis represent
a `snap-shot' of the current situation. As further data are added to 
the global fit in 
coming  years, the situation may change. However it is already 
clear that LHC $W$ and $Z$ cross sections
can already be predicted with high precision, and their measurement will
therefore provide a fundamental test of the SM.

\subsection{CTEQ: studies of \pdf\
uncertainties\protect\footnote{Contributing authors: R.~Brock, D.~Casey, J.~Huston, 
J.~Kalk, J.~Pumplin, D.~Stump and W.K.~Tung.}}
\label{cteqsection}

\noindent\underline{Status of Standard Parton Distribution Functions}

The widely used \pdf\ sets all have been updated
recently, driven mainly by new experimental inputs. Largely due to differences
in the choices of these inputs (direct photon vs. jets) and their theoretical
treatment, the latest MRST \cite{Martin:1999ww} and 
CTEQ \cite{Lai:2000wy} distributions have
noticeable differences in the gluon distribution for $x>0.2$. Details are
described in the original papers.

\begin{figure}[t!]
\centering
\mbox{\includegraphics[width=0.4\textwidth,clip]{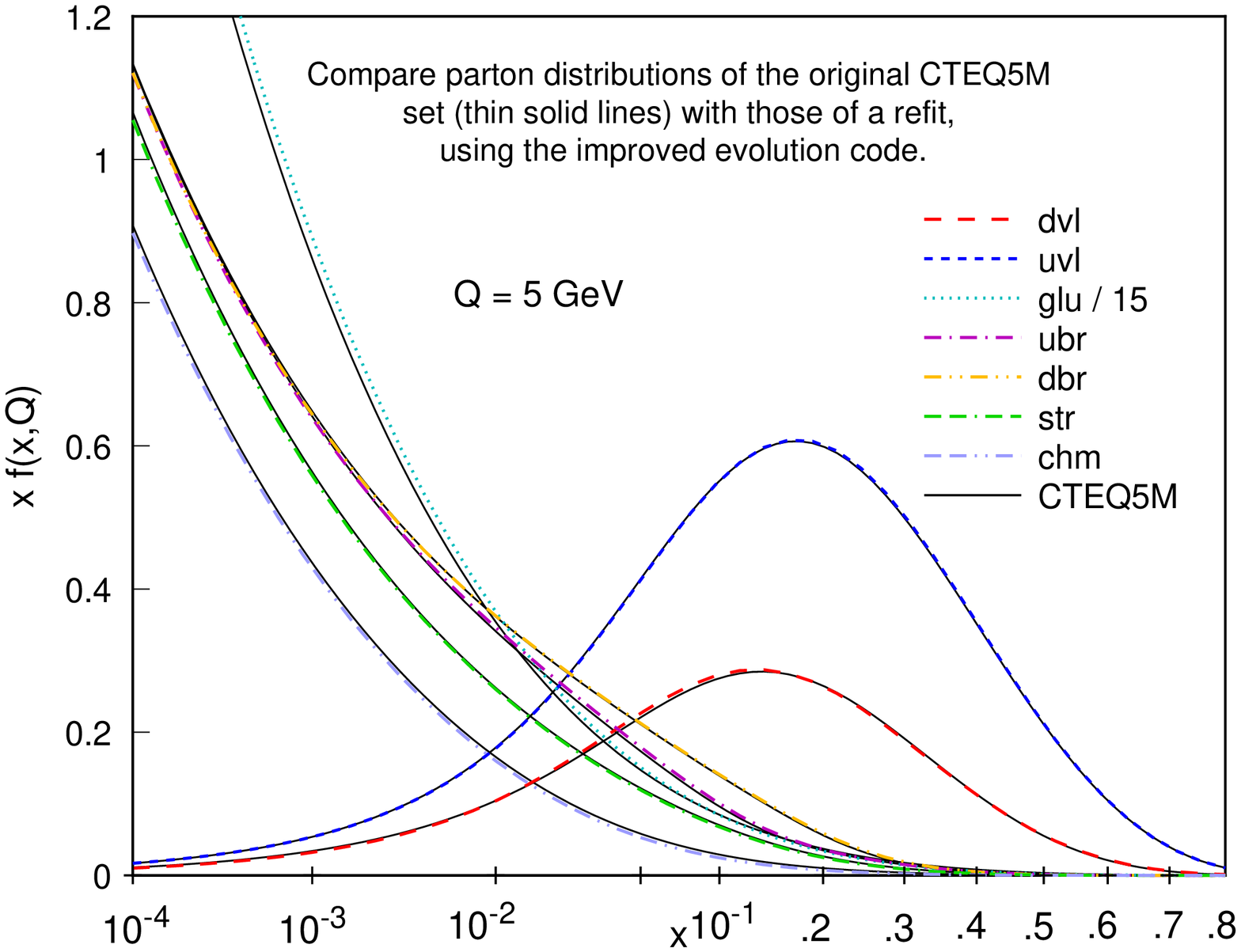}}
\hspace{2cm}
\mbox{\includegraphics[width=0.4\textwidth,clip]{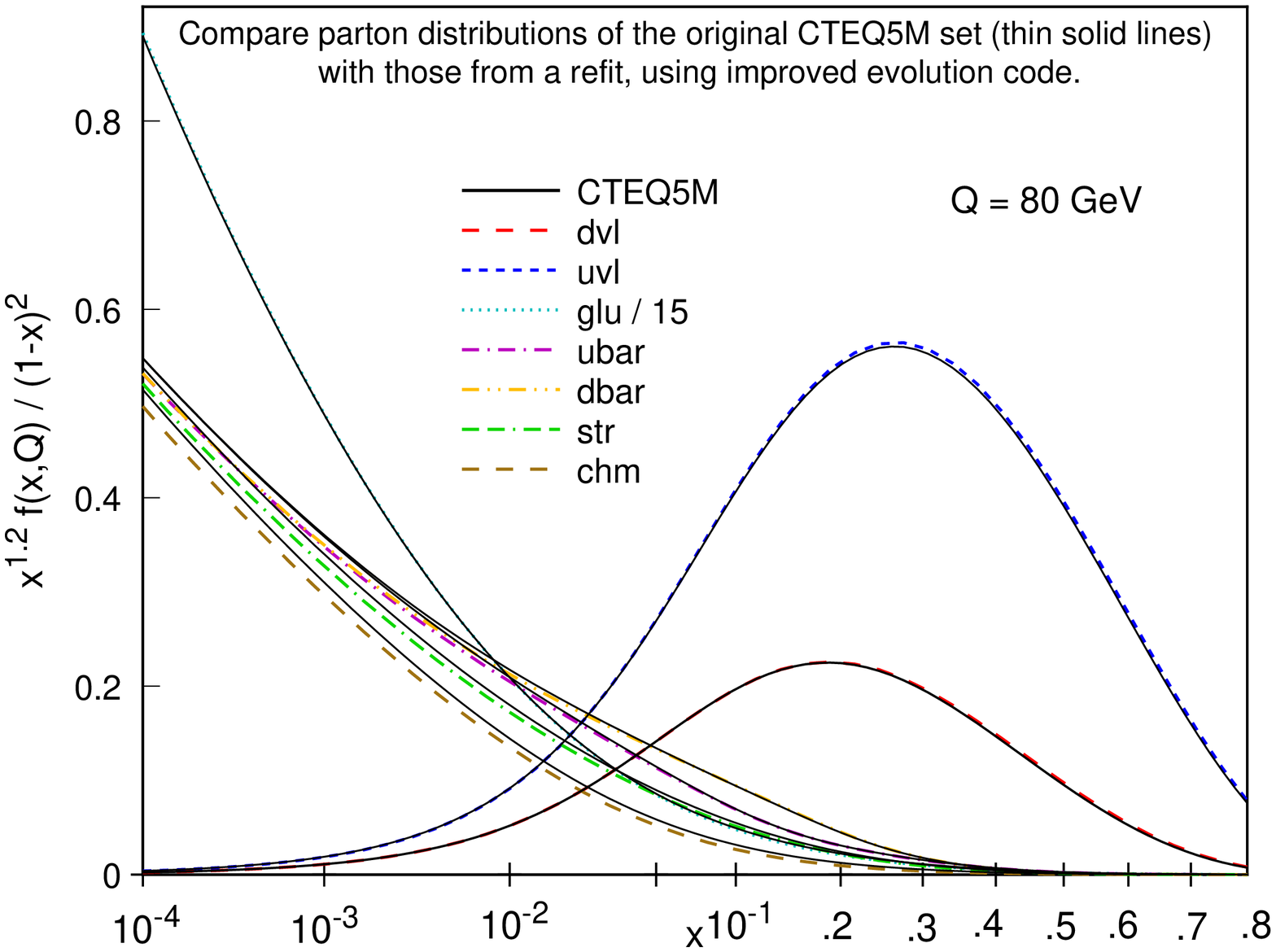}}
\vskip-0.5cm
 \caption{Comparison of CTEQ5M (original) and CTEQ5M1 (revised)
 distributions at two energy scales.}
 \label{fig:CteqA}
\end{figure}

The accuracy of modern DIS measurements and the expanding $(x,Q)$ range in
which \pdfs\ are applied require accurate QCD evolution calculations. Previously
known differences in the QCD evolution codes have now been corrected; all
groups now agree with established results \cite{Blumlein:1996rp} with good precision.
The differences between updated \pdfs\ obtained with the improved evolution code
and the original ones are generally small; and the differences between the
physical cross sections based on the two versions of \pdfs\ are insignificant ,
by definition, since both have been fitted to the same experimental data sets.
However, accurate predictions for physical processes not included in the global
analysis, especially at values of $(x,Q)$ beyond the
current range, can differ and require the improved \pdfs. Figs.~\ref{fig:CteqA}%
a,b compare the \pdf\ sets CTEQ5M (original) and CTEQ5M1 (updated) at scales $%
Q=5$ and $80$ GeV respectively. 

\begin{figure}[t!]
\centering
\mbox{\includegraphics[width=0.4\textwidth,clip]{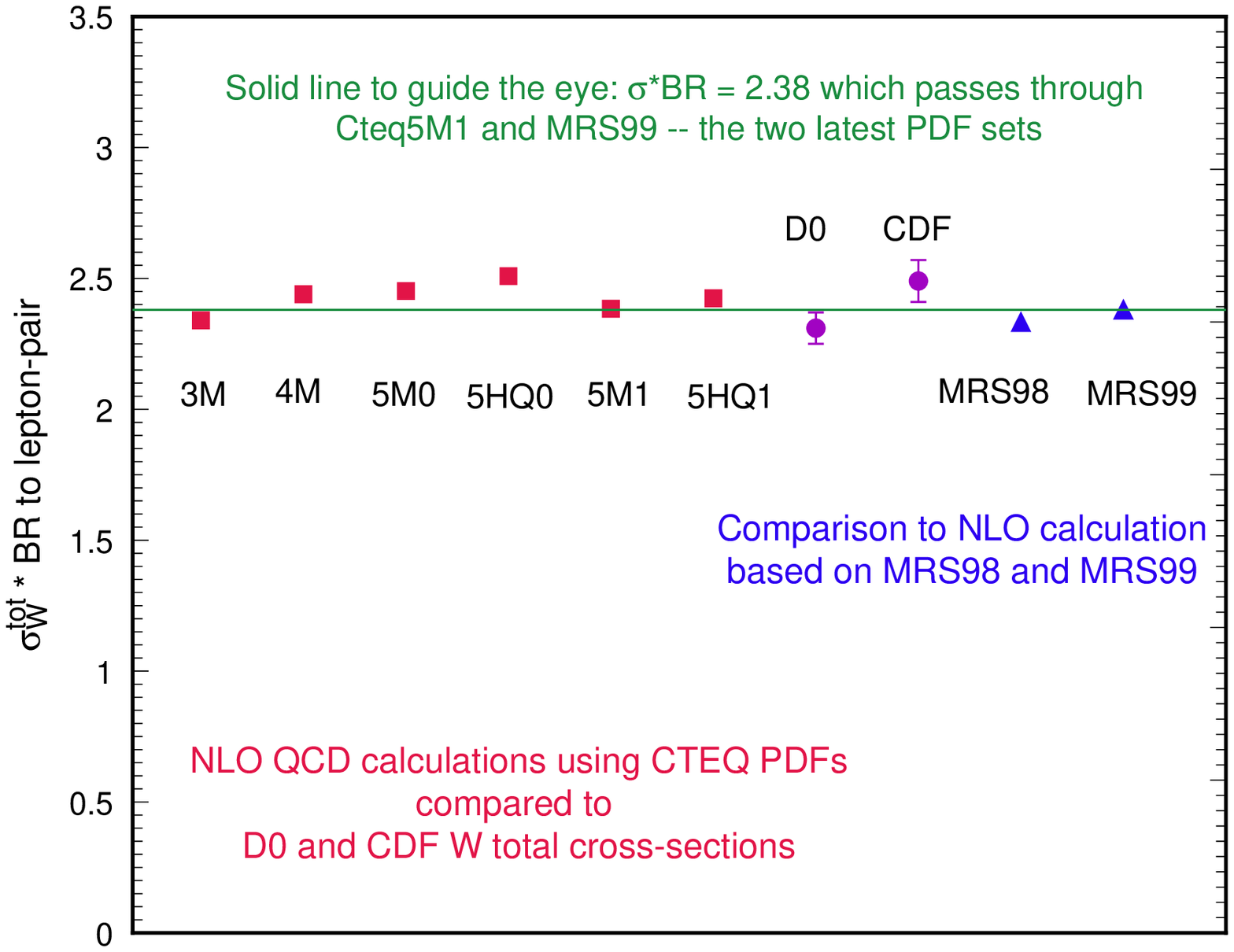}}
\hspace{2cm}
\mbox{\includegraphics[width=0.4\textwidth,clip]{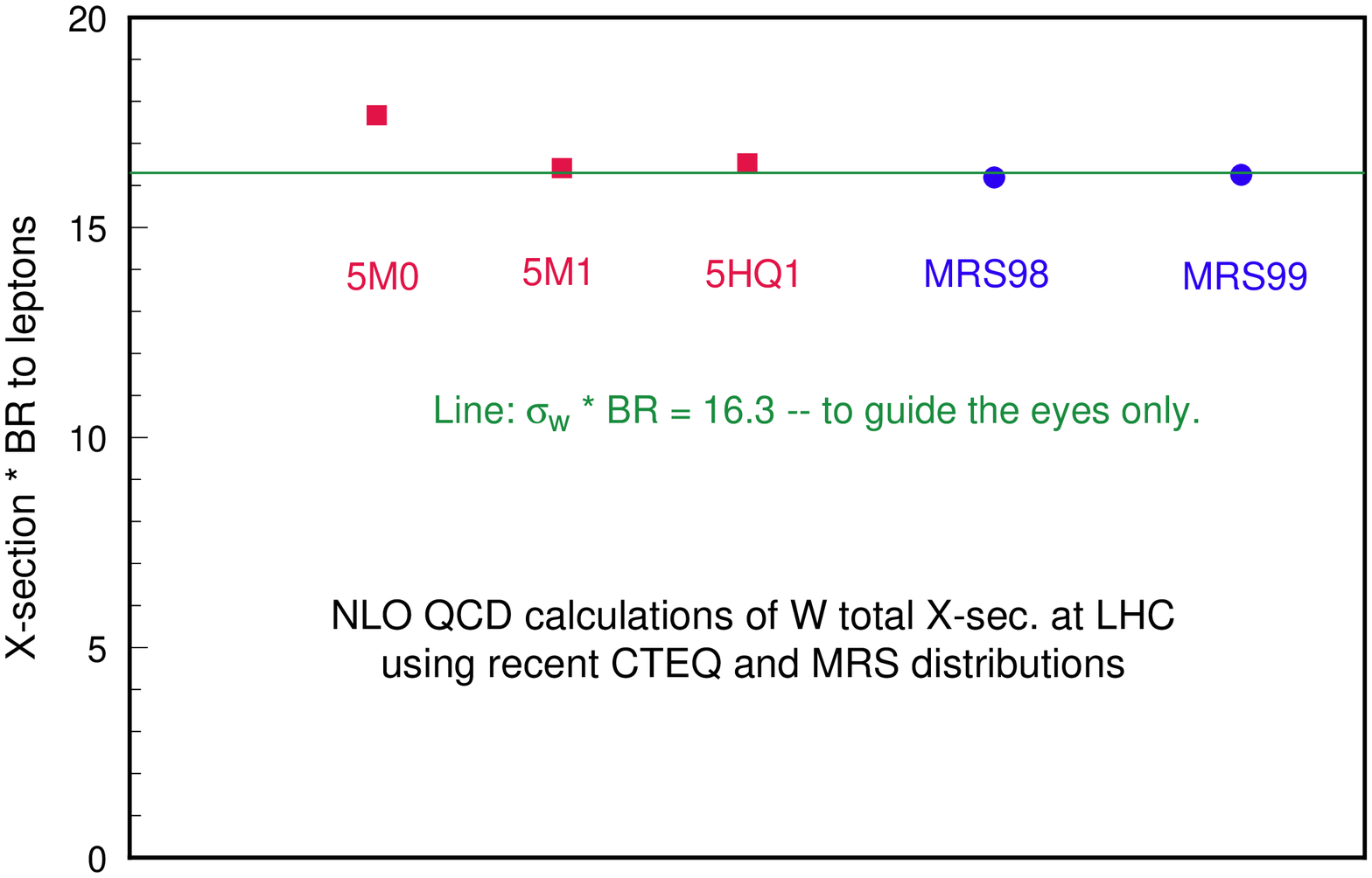}}
%
\vskip-0.5cm
 \caption{Predicted W production cross section, using various historical
and recent parton distribution sets.}
 \label{fig:WprodA}
\end{figure}

\noindent A comparison of the predicted $W$ production cross sections at the
Tevatron and at LHC, using the historical CTEQ parton distribution sets, as
well as the most recent MRST sets are given in Figs.~\ref{fig:WprodA}.
We see that the predicted values of $\sigma_W$ agree very well.
However, the spread of $\sigma_W$ from different ``best fit'' \pdf\ sets does
not give a quantitative measure of the uncertainty of $\sigma_W$!

\noindent\underline{Studies of \pdf\ Uncertainties}

It is important to quantify the uncertainties of physics predictions due to
imprecise knowledge of the \pdfs\ at future colliders (such as the LHC): these
uncertainties may strongly affect the error estimates in precision SM
measurements as well as the signal and backgrounds for new physics
searches.

Uncertainties of the \pdfs\ themselves are strictly speaking unphysical, since
\pdfs\ are not directly measurables. They are renormalization and factorization
scheme dependent; and there are strong correlations between different flavours
and different values of $x$ which can compensate each other in physics
predictions. On the other hand, since \pdfs\ are universal, if one can obtain
meaningful estimates of their uncertainties based on global analysis of
existing data, they can then be applied to all processes that are of interest
for the future.

An alternative approach is to assess the uncertainties on {\em specific
physical predictions} for the full range (i.e.\ the ensemble) of \pdfs\ allowed
by available experimental constraints which are used in current global
analyses, without explicit reference to the uncertainties of the parton
distributions themselves. This clearly gives more reliable estimates of the
range of possible predictions on the physical variable under study. The
disadvantage is that the results are process-specific; hence the analysis has
to be carried out for each process of interest.

In this short report, we present first results from a systematic study of
both approaches. In the next section we focus on the $W^{\pm }$
production cross section, as a proto-typical case of current interest. A
technique of Lagrange multiplier is incorporated in the CTEQ global analysis to
probe its range of uncertainty at the Tevatron and the LHC. This method is
directly applicable to other cross sections of interest, e.g.\ Higgs production.
We also plan to extend it for studying the uncertainties of $W$-mass measurements
in the future. In the following section we describe a Hessian study of the
uncertainties of the non-perturbative \pdf\ parameters in general, followed by
application of these to the $W^{\pm }$ production cross section study and a
comparison of this result with that of the Lagrange-multiplier approach.

\noindent
First, it is important to note the various \textbf{sources of
uncertainty} in \pdf\ analysis.

\newenvironment{Simlis}[2][$\bullet$]
{\begin{list}{#1}
 {
  \settowidth{\labelwidth}{#1}
  \setlength{\labelsep}{0.5em}
  \setlength{\leftmargin}{#2}
  \setlength{\rightmargin}{0em}
  \setlength{\itemsep}{0ex}
  \setlength{\topsep}{0ex}
 }
}
{\end{list}}
\begin{Simlis}{0em}
\item
\textbf{Statistical errors} of experimental data. These vary over a wide range,
but are straightforward to treat.

\item
\textbf{Systematic experimental errors} within each data set typically arise
from many sources, some of which are highly correlated. These errors can be
treated by standard methods \emph{provided} they are precisely known, which
unfortunately is often not the case -- either because they are not randomly
distributed or their estimation may involve subjective judgements. Since
strict quantitative statistical methods are based on idealized assumptions,
such as random errors, one faces an important trade-off in \pdf\ uncertainty
analysis. If emphasis is put on the ``rigor'' of the statistical method, then
most experimental data sets can not be included the analysis (see
Sect.~\ref{sec:keller;2qcd}).
If priority is placed on using the maximal experimental constraints from
available data, then standard statistical methods need to be supplemented by
physical considerations, taking into account existing experimental and
theoretical limitations. We take the latter tack.

\item
\textbf{Theoretical uncertainties} arise from higher-order PQCD corrections,
resummation corrections near the boundaries of phase space, power-law (higher
twist) and nuclear target corrections, etc.

\item
Uncertainties of \pdfs\ due to the \textbf{parametrization of
the non-perturbative \pdfs}, $f_{a}(x,Q^2_{0}),$ at some low energy scale $%
Q_{0}.$ The specific functional form used introduces implicit correlations
between the various $x$-ranges, which could be as important, if not more so,
than the experimental correlations in the determination of $f_{a}(x,Q^2)$
for all $Q.$

\end{Simlis}

In view of these considerations, the preliminary results reported here can
only be regarded as the beginning of a continuing effort which will be
complex, but certainly very important for the next generation of collider
programs.

\noindent\underline{\it The Lagrange multiplier method}

Our work uses the standard CTEQ5 analysis tools and results \cite{Lai:2000wy} as
the starting point. The ``best fit'' is the CTEQ5M1 set. There are 15
experimental data sets, with a total of $\sim$ 1300 data points; and 18
parameters {$a_i, i=1,\dots,18$} for the non-perturbative initial parton
distributions. A natural way to find the limits of a physical quantity $X$,
such as $\sigma_{W}$ at $\sqrt{s}=1.8$\,TeV, is to take $X$ as one of the
search parameters in the global fit and study the dependence of $\chi^{2}$ for
the 15 base experimental data sets on $X$.

Conceptually, we can think of the function $\chi^{2}$ that is minimized in the
fit as a function of \{$a_{1}$-$a_{17},X$\} instead of \{$a_{1}$-$a_{18}$\}.
This idea could be implemented directly in principle, but Lagrange's method of
undetermined multipliers does the same thing in a more efficient way. One
minimizes
\begin{equation}
F(\lambda)=\chi^{2}+\lambda X(a_{1},\dots,a_{18})
\end{equation}
for fixed $\lambda$. By minimizing $F(\lambda)$ for many values of $\lambda$,
we map out $\chi^{2}$ as a function of $X$.

\begin{figure}[t!]
\centering
\mbox{\includegraphics[width=0.4\textwidth,clip]{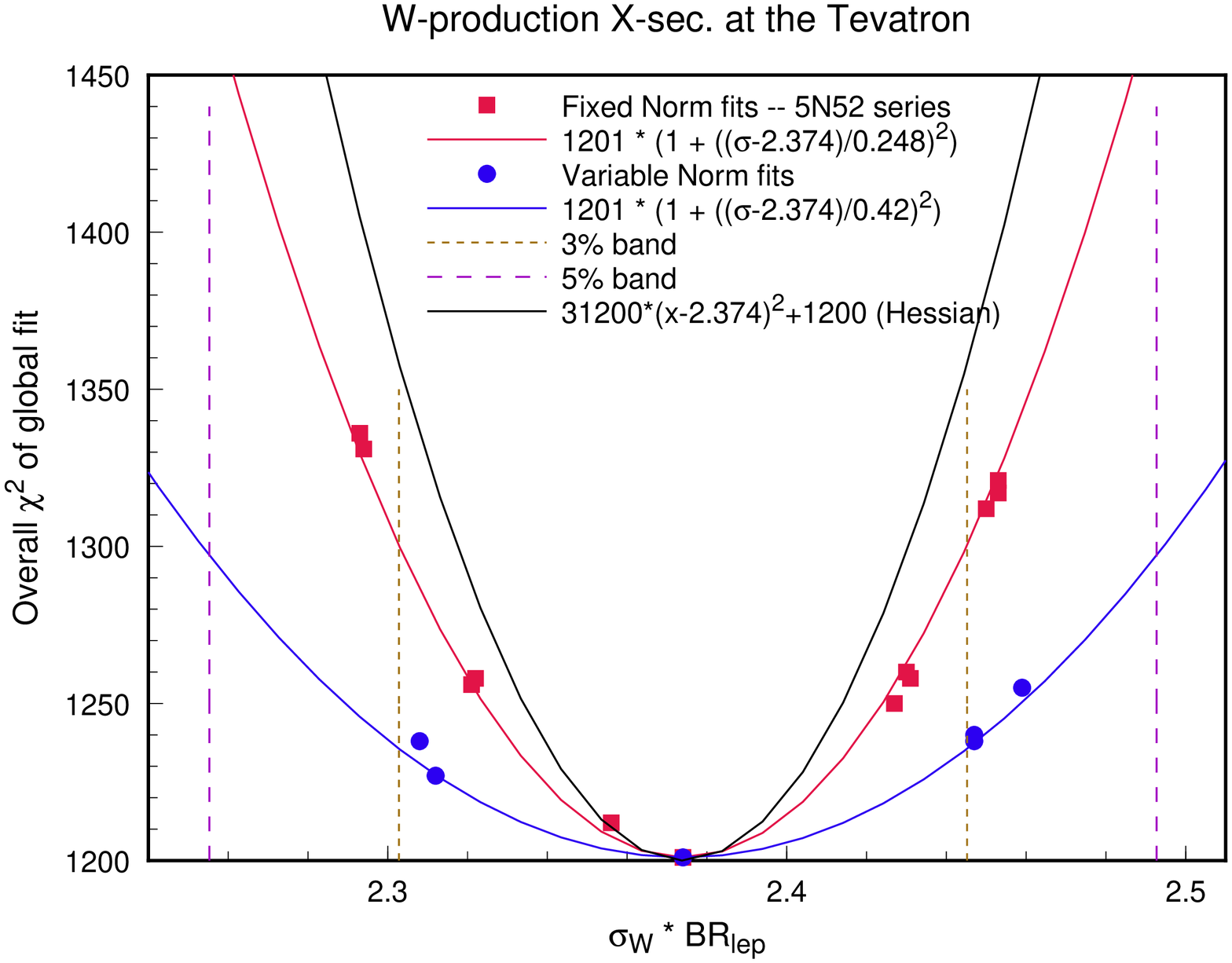}}
\hspace{2cm}
\mbox{\includegraphics[width=0.4\textwidth,clip]{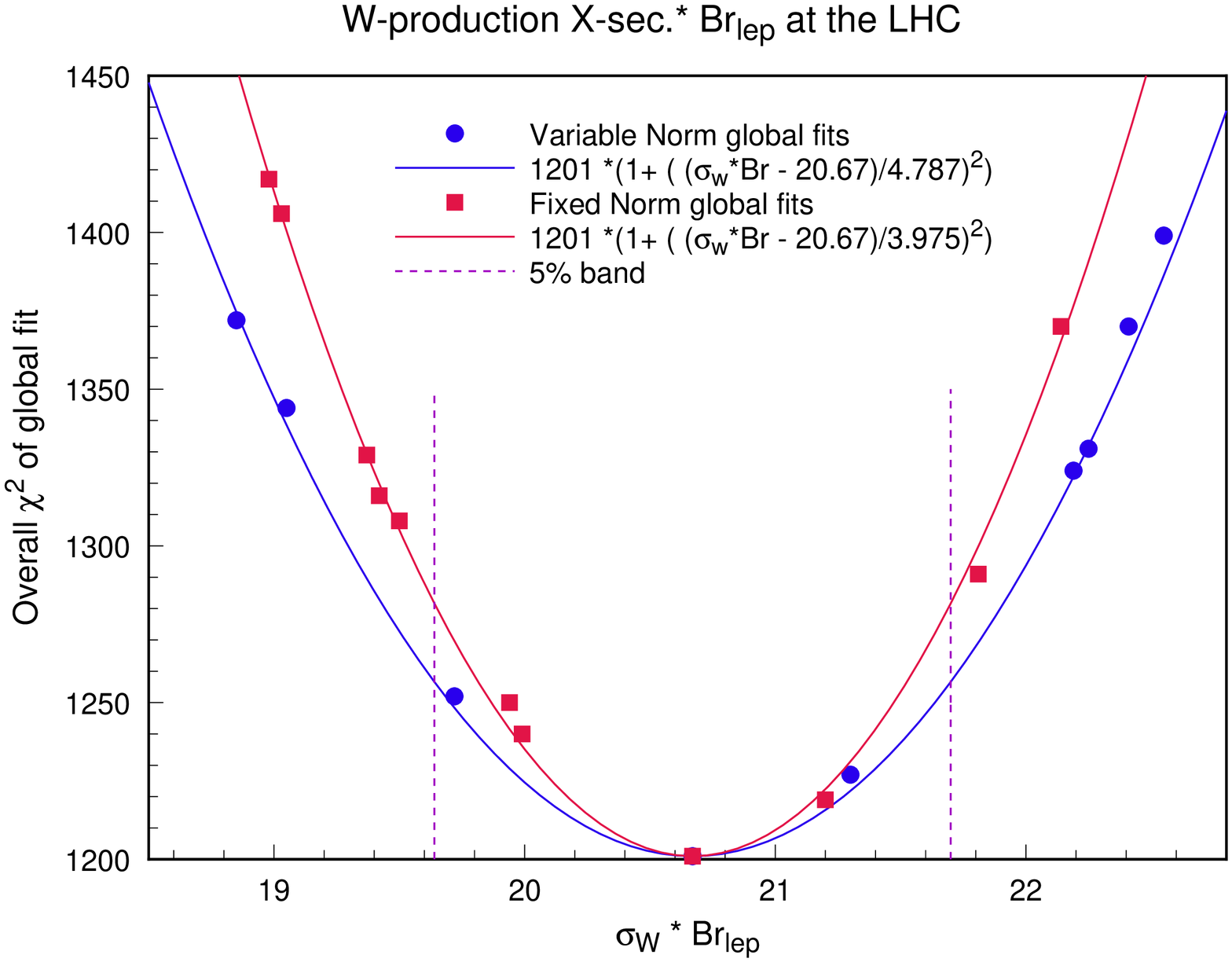}}
\vskip-0.5cm
 \caption{$\chi^2$ of the base experimental data sets vs. the W
 production cross section at the Tevatron and LHC.}
 \label{fig:WprodB}
\end{figure}

Figs.~\ref{fig:WprodB}a,b show the $\chi^{2}$ for the 15 base experimental
data sets as a function of $\sigma_{W}$ at the Tevatron and the LHC energies
respectively. Two curves with points corresponding to specific global fits are
included in each plot\footnote{The third line in Figs.~\ref{fig:WprodB}a
refers to results of the next section.}: one obtained with all experimental
normalizations fixed; the other with these included as fitting parameters
(with the appropriate experimental errors). 
We see that the
$\chi^2$'s for the best fits corresponding to various values of the $W$
cross section are close to being parabolic, as expected. Indicated on the
plots are 3\% and 5\% ranges for $\sigma_W$. The two curves for the Tevatron
case are farther apart than for LHC, reflecting the fact that the $W$-production
cross section is more sensitive to the quark/anti-quark distributions and
these are tightly constrained by existing DIS data.

The important question is: how large an increase in $\chi^{2}$ should be taken
to define the likely range of uncertainty in $X$. The elementary statistical
theorem that $\Delta\chi^{2}=1$ corresponds to 1 standard deviation of the
measured quantity $X$ relies on assuming that the errors are Gaussian,
uncorrelated, and with their magnitudes correctly estimated. Because these
conditions do not hold for the full data set (of 1300 points from 15 different
experiments), this theorem cannot be naively applied
quantitatively.\footnote{As shown by Giele {\it et.al.} \cite{Giele:1998gw}, taken
literally, only one or two selected experiments satisfy the standard
statistical tests.} We plan to examine in some detail how well the fits along
the parabolas shown in Fig.\ref{fig:WprodB}a,b compare with the individual
precision experiments included in the global analysis, in order to arrive at
reasonable quantitative estimates on the uncertainty range for the $W$
cross section. In the meantime, based on past (admittedly subjective)
experience with global fits, we believe a $\chi^2$ difference of 40-50
represents a reasonable estimate of current uncertainty of parton
distributions. This implies that the uncertainty of $\sigma_{W}$ is about 3\%
at the Tevatron, and 5\% at the LHC. These estimates certainly need to be put
on a firmer basis by the on-going detailed investigation mentioned above.

\noindent\underline{\it The Hessian matrix method}

The Hessian matrix is a standard procedure for error analysis.
At the minimum of $\chi^{2}$, the first derivatives with respect to the
parameters $a_{i}$ are zero, so near the minimum $\chi^2$ can be approximated by
\begin{equation}
\chi^{2} = \chi^{2}_{0} + \frac{1}{2}\sum_{i,j}F_{ij}y_{i}y_{j}
\end{equation}
where $y_{i}=a_{i}-a_{0i}$ is the displacement from the minimum, and $F_{ij}$
is the \emph{Hessian}, the matrix of second derivatives. It is natural to
define a new set of coordinates using the complete orthonormal set of
eigenvectors of the symmetric matrix $F_{ij}$. These vectors can be ordered by
their eigenvalues $e_{i}$. The eigenvalues indicate the uncertainties for
displacements along the eigenvectors. For uncorrelated Gaussian statistics,
the quantity $\ell_{i}=1/\sqrt{e_{i}}$ is the distance in the 18 dimensional
parameter space that gives a unit increase in $\chi^{2}$ in the direction of
eigenvector $i$.

From calculations of the Hessian we find the eigenvalues vary over a wide
range. There are ``steep" directions of $\chi^{2}$ -- combinations of
parameters that are well determined -- e.g.\ parameters for $u$ and $d$,
which are well-constrained by DIS data. There are also ``flat"
directions where $\chi^{2}$ changes little over large distance in $a_{i}$
space, some of them associated with the gluon distribution. These flat
directions are inevitable in global fitting, because as the data improve it
makes sense to maintain enough flexibility for $f_a(x,Q^2_0)$ to be determined by
the available experimental constraints. The Hessian method gives an analytic
picture of the region in parameter space around the minimum, hence allows us to
identify the particular degrees of freedom which need further experimental
input in future global analyses.

We have
calculated how the $W$ cross section $\sigma_{W}$ varies along the
eigenvectors of the Hessian. Details will be described elsewhere.
This provides another way to calculate the relation between the minimum $%
\chi^{2}$ for the base experimental data sets and the value of $\sigma_{W}$.
The results are shown as the third line in Fig.~\ref{fig:WprodB}a. We see that
there is approximate agreement between this method and the Lagrange multiplier
method. Armed with the Hessian, one can in principle make similar calculations
on other physical cross sections without having to do repeated global fits as
in the Lagrange multiplier method.  The latter, however, gives more reliable
bounds for each individual process.

\noindent\underline{Conclusion}

We have just begun the task of determining quantitative uncertainties for the
parton distribution functions and their physics predictions. The methods
developed so far look promising. Related work reported in this Workshop
(see \cite{Martin:1999ww,Giele:1998gw,Alekhin:1999za,LesHouches:2000}
and Sects.~\ref{mrssection},\ref{sec:keller;2qcd},\ref{sec:alekhin;2qcd})
share the same objectives, but have rather
different emphases, some of which are briefly mentioned in the text. These
complementary approaches should lead to eventual progress which is 
critical for the high-energy physics program at LHC, as well as at other colliders.

\subsection{Pdf uncertainties\protect\footnote{Contributing authors: W.T.~Giele,
S.~Keller and D.A.~Kosower.}}
\label{sec:keller;2qcd}
\noindent\underline{Introduction}

The goal of our work is to extract \pdfs\
from data with a quantitative estimation of the uncertainties.  There
are some qualitative tools that exist to estimate the uncertainties,
see e.g. \cite{Martin:1998sq}.  
These tools are clearly not adequate when the \pdf\
uncertainties become important.  One crucial example of a measurement
that will need a quantitative assessment of the \pdf\ uncertainty is the
planned high precision measurement of the mass of the $W$-vector boson
at the Tevatron.

The method we have developed in \cite{Giele:1998gw} is flexible and can
accommodate non-Gaussian distributions for the uncertainties associated
with the data and the fitted parameters as well as all their
correlations.  New data can be added in the fit without having to redo
the whole fit.  Experimenters can therefore include their own data
into the fit during the analysis phase, as long as correlation with
older data can be neglected.  Within this method it is trivial to
propagate the \pdf\ uncertainties to new observables, there is for
example no need to calculate the derivative of the observable with
respect to the different \pdf\ parameters.  The method also provides
tools to assess the goodness of the fit and the compatibility of new
data with current fit.  The computer code has to be fast as there is a
large number of choices in the inputs that need to be tested.

It is clear that some of the uncertainties are difficult to quantify
and it might not be possible to quantify all of them.  All the plots
presented here are for illustration of the method only, our results
are {\sl preliminary}.  At the moment we are not including all the
sources of uncertainties and our results should therefore be
considered as lower limits on the \pdf\ uncertainties.  Note that all
the techniques we use are standard, in the sense that they can be
found in books and papers on statistics \cite{D'Agostini:1995fv,D'Agostini:1999cm} and/or in Numerical
Recipes.

\noindent\underline{Outline of the Method}

We only give a brief overview of the method in this section.  More
details are available in \cite{Giele:1998gw}.  Once a set of core experiments is
selected, a large number of uniformly distributed sets of parameters
$\lambda \equiv\lambda_1, \lambda_2, \ldots, \lambda_{N_{par}}$ (each
set corresponds to one \pdf) can be generated.  The probability of each
set, $P(\lambda)$, can be calculated from the likelihood (the
probability) that the predictions based on $\lambda$ describe the
data, assuming that the initial probability distribution of the
parameters is uniform, see \cite{D'Agostini:1995fv,D'Agostini:1999cm}.

Knowing $P(\lambda)$, the probability of the possible values of any
observable (quantity that depends on $\lambda$) can be calculated using a
Monte Carlo integration.  For example, the average value and the \pdf\ 
uncertainty of an observable $x$ are given by:
\begin{displaymath}
\mu_{x}  =  \int \left(\prod_{i=1}^{N_{par}}  d\lambda_i \right) \, x(\lambda) P(\lambda),\qquad 
\sigma_{x}^2  =  \int \left(\prod_{i=1}^{N_{par}} d\lambda_i \right) \, (x(\lambda)- \mu_{x})^2 P(\lambda) 
\end{displaymath}
Note that the average value and the standard deviation represents the
distribution only if the latter is
a Gaussian.  The above is correct but computationally inefficient, 
instead we use a Metropolis algorithm to generate
$N_{pdf}$ unweighted \pdfs\ distributed according to $P(\lambda)$.
Then:
\begin{displaymath}
\mu_{x}  \approx  \frac{1}{N_{pdf}}\sum_{j=1}^{N_{pdf}} 
x\left(\lambda_j\right),\qquad\qquad 
\sigma_{x}^2  \approx  \frac{1}{N_{pdf}}\sum_{j=1}^{N_{pdf}} 
\left(x\left(\lambda_j\right)-\mu_{x}\right)^2\ .
\end{displaymath}
This is equivalent to importance sampling in Monte Carlo integration 
techniques and is very efficient.  
Given the unweighted set of \pdfs, a new experiment can be added to the
fit by assigning a weight (a new probability) to each of the \pdfs, using
Bayes' theorem.  The above summations become weighted.  There is
no need to redo the whole fit {\sl if} there is no correlation between
the old and new data.
  If we know how to calculate $P(\lambda)$ properly, the only 
uncertainty in the method comes from the Monte-Carlo integrations.

\noindent\underline{Calculation of $P(\lambda)$}
 
Given a set of experimental points $\{ x^e
\}=x^e_1,x^e_2,\ldots,x^e_{N_{obs}}$ the probability of a set of \pdf\
is proportional to the likelihood, the probability of the data given
that the theory is derived from that set of \pdf: 
$P(\lambda) \approx P(\{x^e\}|\lambda)$.
If all the uncertainties are Gaussian distributed, then it is well known that:
$P({x^e}|\lambda) \approx e^{-\frac{\chi^2}{2}}$,
where $\chi^2$ is the usual chi-square.  It is only in this case that
it is sufficient to report the size of the uncertainties and their
correlation.  When the uncertainties are not Gaussian distributed, it
is necessary for experiments to report the distribution of their
uncertainties and the relation between these uncertainties the theory
and the value of the measurements.  Unfortunately most of the time
that information is not reported, or difficult to extract from papers.
This is a very important issue that has been one of the focus of the
\pdf\ working group at a Fermilab workshop in preparation for run II
\cite{keller:4}.  In other words, experiments should always provide a way to
calculate the likelihood of their data given a theory prediction for
each of their measured data point ($P(\{x^e\}|\lambda)$).  This was
also the unanimous conclusion of a recent workshop on confidence
limits held at CERN \cite{keller:5}.  This is particularly crucial when
combining different experiments together: the pull of each experiment
will depend on it and, as a result, so will the central values of the
deduced \pdfs.  Another problem that is sometimes underestimated is the
fact that some if not all systematic uncertainties are in fact
proportional to the theory.  Ignoring this fact while fitting for the
parameters can lead to serious bias.

\begin{figure}[t!]
\begin{center}
\includegraphics[bb= 34 590 494 704]{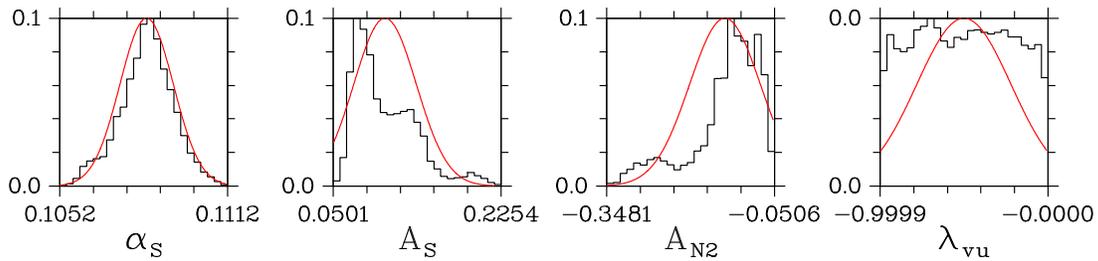}  
\end{center}
\vskip-0.5cm
\caption{Plot of the distribution (histogram) of four of the parameters.
  The first one is $\as$, the strong coupling constant at the
  mass of the $Z$-boson.  The line is a Gaussian distribution with
  same average and standard deviation as the histogram.}
  \label{keller:fig1}
\end{figure}

\noindent\underline{Sources of uncertainties} 

There are many sources of uncertainties beside the experimental
uncertainties.  They either have to be shown to be small enough to be
neglected or they need to be included in the \pdf\ uncertainties.  For
examples: variation of the renormalization and factorization scales;
non-perturbative and nuclear binding effects; the choice of functional
form of the input \pdf\ at the initial scale; 
accuracy of the evolution; Monte-Carlo uncertainties; and the
theory cut-off dependences.

\noindent\underline{Current fit}

Draconian measures were needed to restart from scratch and re-evaluate
each issue.  We fixed the renormalization and factorisation scales,
avoided data affected by nuclear binding and non-perturbative effects,
and use a MRS-style parametrization for the input \pdfs.  The evolution
of the \pdf\ is done by Mellin transform method, see \cite{Kosower:1998vj,Kosower:1997hg}.  All the
quarks are considered massless.  We imposed a positivity constraint on
$F_2$.  A positivity constraint on other ``observables'' could also be
imposed.

At the moment we are using H1 and BCDMS(proton) measurement of $F_2^p$
for our core set.  The full correlation matrix is taken into account.
{\em Assuming that all the uncertainties are Gaussian
  distributed}~\footnote{No information being given about the
  distribution of the uncertainties.} we calculate the $\chi^2
(\lambda)$ and $P(\lambda) \approx exp(-\chi^2/2)$.  We generated
50000 unweighted \pdfs\ according to the probability function.  For 532
data points, we obtained a minimum $\chi^2=530$ for 24 parameters.  We
have plotted in Fig.~\ref{keller:fig1}, 
the probability distribution of some of the
parameters.  Note that the first parameter is $\as$.  The value
is smaller than the current world average.  However, it is known that
the experiments we are using prefer a lower value of this parameter,
see \cite{keller:7}, and as already pointed out, our current uncertainties
are lower limits.  Note that the distribution of the parameter is not
Gaussian, indicating that the asymptotic region is not reached yet.
In this case, the blind use of a so-called chi-squared fitting
technique is not appropriate.  From this large set of \pdfs, it is
straightforward to plot, for example, the correlation between
different parameters and to propagate the uncertainties to other
observables.


\subsection{Uncertainties on \pdfs\ and parton-parton 
luminosities\protect\footnote{Contributing author: S. Alekhin.}}
\label{sec:alekhin;2qcd}

\begin{figure}[t!]
\begin{center}
\includegraphics[width=11.5cm,height=6.5cm]{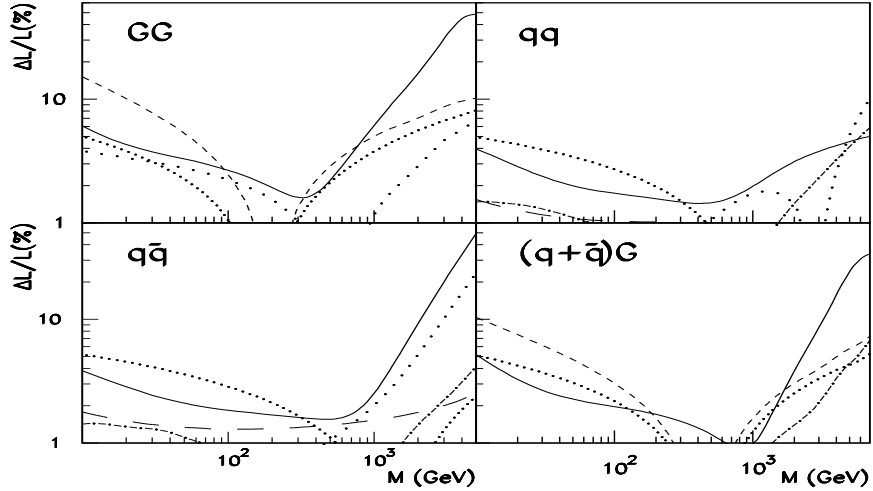}
\end{center}
\vskip-0.5cm
\caption{The relative uncertainties for selected set of 
parton luminosities (full lines: experimental errors (stat+syst);
short-dashed lines: RS; dotted-dashed lines: TS; 
sparse-dotted lines: DC; dense-dotted lines: MC; 
long-dashed lines: SS). Here $L_{\rm GG}$ is gluon-gluon 
luminosity; $L_{\rm qq}=L_{\rm uu}+L_{\rm dd}+L_{\rm du}$; 
$L_{\rm q\bar q}=L_{\rm u \bar d}+L_{\rm d \bar u}$; 
$L_{\rm (q+\bar q)G}=L_{\rm uG}+L_{\rm \bar uG}+L_{\rm dG}+L_{\rm \bar dG}$.}
\label{AL:lums}
\end{figure}

An important quantity for LHC physics is the uncertainty of 
\pdfs\ used for the cross section calculations. The modern widely 
used \pdfs\ parametrizations do not contain complete estimate 
of their uncertainties. This estimate is difficult
partially due to the lack of experimental information on the 
data points correlations, partially due to the fact that the theoretical uncertainties 
are conventional, and partially due to the fundamental 
problem of restoring the distribution from the finite number of 
measurements. These problems are not completely solved at the moment 
and a comprehensive estimate of the \pdfs\ uncertainties is 
not available so far. The study given below is based on the 
NLO QCD analysis of the world charged leptons 
DIS data of Refs.~\cite{Whitlow:1992uw,Benvenuti:1989rh,Benvenuti:1990fm,Arneodo:1997qe,Adams:1996gu,Aid:1996au,Derrick:1996hn} 
for proton and deuterium targets\footnote{More details 
of the analysis can be found in Ref.~\cite{Alekhin:1999ck}.}. 
The analysed data span the region $x=10^{-4}\div0.75$,
$Q^2=2.5\div5000$~GeV$^2$, $W\gsim 2$~GeV
and allows for precise determination of \pdfs\ at low $x$,
which is important for LHC since the most of accessible processes 
are related to small $x$. The data 
are accompanied by the information on point-to-point correlations
due to systematic errors. This allows the complete inference of 
systematic errors, that was performed using the covariance matrix 
approach, as in Ref. \cite{Alekhin:1999za}.
The \pdfs\ uncertainties due to the variation of the 
strong coupling constant $\as$
and the high twists (HT) contribution are 
automatically accounted for in the total
experimental uncertainties since $\as$ and HT are fitted\footnote
{The value of $\as(M_{\rm Z})=0.1165\pm0.0017({\rm stat+syst})$
is obtained, that is compatible with the world average.}. 
Other theoretical errors on \pdfs\ were estimated as the \pdfs\ variation 
after the change of different fit ansatzes:
\begin{itemize}
\item[{\bf RS}] -- the change of 
renormalization scale in the evolution equations from $Q^2$ to 
$4Q^2$. This uncertainty is evidently connected with the influence 
of NNLO corrections.
\item[{\bf TS}] -- the change of 
threshold value of $Q^2$ for the QCD evolution loops
with heavy quarks from $m_{\rm Q}^2$ to $6.5m_{\rm Q}^2$. The variation 
is conventional and was chosen following the arguments of
Ref.~\cite{Blumlein:1999mg}.
\item[{\bf DC}] -- the change of correction on nuclear effects in 
deuterium from the ansatz based on the Fermi motion model of Ref.~\cite{AL:WEST} 
to the phenomenological formula from Ref.~\cite{AL:GOMES}. Note that 
this uncertainty 
may be overestimated in view of discussions \cite{Melnitchouk:1999un,Yang:1999ew}
on the applicability of the 
model of Ref.\cite{AL:GOMES} to light nuclei.
 
\item[{\bf MC}] -- the change of c-quark mass by 0.25 GeV
(the central value is 1.5 GeV).
\item[{\bf SS}] -- the change of strange sea suppression factor by 0.1, 
in accordance with recent results by the NuTeV collaboration \cite{Adams:1999sx}
(the central value is 0.42). 
\end{itemize}
One can see that the scale of the theoretical errors is conventional 
and can change with improvements in the determination of the fit input  
parameters and progress in theory. Moreover, the uncertainties can 
be correlated with the uncertainties of the partonic cross sections,
e.g. the effect of RS uncertainty on \pdfs\ 
can be compensated by the NNLO correction to parton cross section. 
Thus the theoretical uncertainties should not be applied automatically 
to any cross section calculations, contrary to experimental ones.

\begin{figure}[t!]
\begin{center}
\includegraphics[width=9.5cm,height=4.0cm]{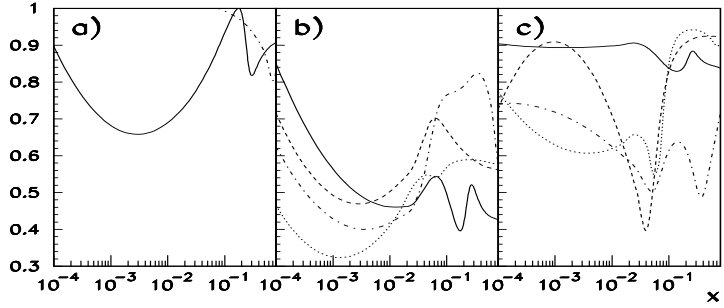}
\end{center}
\vskip-0.5cm
\caption{The ratios of the experimental \pdfs\ errors
calculated with some fitted parameters fixed to the
\pdfs\ errors calculated with all parameters released
($\as$ fixed -- a); HT fixed -- b)). 
The similar ratio for the systematic errors omitted/included
is also given -- c).
Full lines correspond to gluons, dashed ones -- to total sea, 
dotted ones -- to d-quarks, 
dashed-dotted ones -- to u-quarks.}
\label{AL:ahs}
\end{figure}

The \pdfs\ uncertainties have different importance for 
various processes. The limited space does not allow us
to review all of them. We give the figures for 
the most generic ones only.
The uncertainties of a specific cross section
due to \pdfs\ are entirely 
located in the uncertainties of the parton-parton luminosity $L_{ab}$, 
that is defined as 
\begin{displaymath}
L_{ab}(M)=\frac{1}{s}\int^1_{\tau}\frac{dx}{x} \; f_{a}(x,M^2) 
f_{b}(\tau/x,M^2),  
\end{displaymath}
where 
$M$ is the produced mass and $\tau=M^2/s$.
In Fig.~\ref{AL:lums} the uncertainties for selected set of 
parton luminosities calculated using the \pdfs\ 
from Ref.~\cite{Alekhin:1999ck} are given. 
The upper bound of $M$ was chosen so that the corresponding luminosity
is $\sim0.01$~pb. 
One can see that in general
at $M\gsim 1$~TeV experimental uncertainties dominate, while 
at $M\lsim 1$~TeV theoretical ones dominate. Of the latter 
the most important are the RS uncertainty for 
the gluon luminosity and MC uncertainty for the 
quark luminosities. At the largest $M$ the DC uncertainty 
for quark-quark luminosity is comparable with the experimental one.
In the whole the uncertainties do not exceed 10\% at $M\lsim 1$~TeV.
As for the quark-quark luminosity, its uncertainty is less than 10\% in the 
whole $M$ range. The uncertainties are not so large in view of the fact that 
only a small subset of data 
relevant for the \pdfs\ extraction was used in the analysis. 
Adding data on prompt photon production, DY process, and jet 
production can improve the \pdfs\ determination at large $x$. 
Meanwhile it is worth to note that high order QCD corrections are 
more important for these processes than for DIS
and the decrease of experimental errors due to adding data points
can be accompanied by the increase of theoretical errors.

\begin{figure}[t!]
\begin{center}
\includegraphics[width=9.5cm,height=6cm]{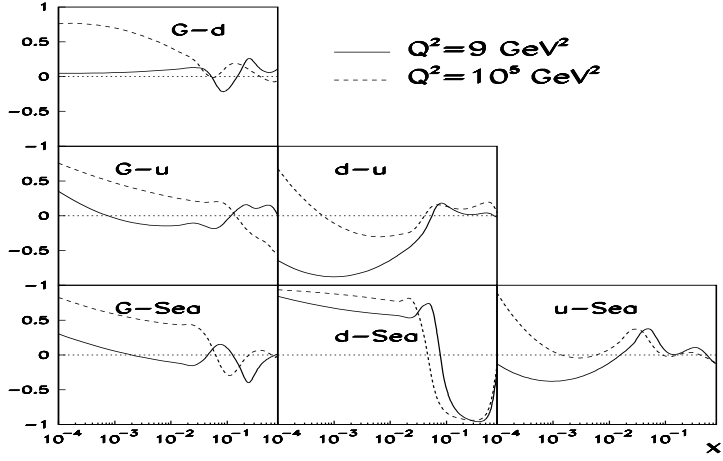}
\end{center}
\vskip-0.5cm
\caption{The \pdfs\ correlation coefficients.}
\label{AL:cor}
\end{figure}

As it was noted above, the experimental \pdfs\ errors by definition
include the statistical 
and systematical errors, as well as errors due to $\as$ and HT.
To trace the effect of $\as$ variation on the 
\pdfs\ uncertainties
the latter were re-calculated with $\as$ fixed at the value 
obtained in the fit. The ratios of obtained experimental \pdfs\ errors
to the errors calculated with $\as$ released are given in 
Fig.~\ref{AL:ahs}. It is seen that the $\as$ variation 
takes some effect on the gluon distribution errors only. Similar 
ratios for the HT fixed are also
given in Fig.~\ref{AL:ahs}. One can conclude, that the 
account of HT contribution have significant impact on the \pdfs\ errors.
Meanwhile it is evident that these ratios hardly depend on the scale
of \pdfs\ error and are specific for 
the analysed data set. For instance, in the analysis of CCFR data on the 
structure function $F_3$ no significant influence of HT on the \pdfs\
was observed \cite{Kataev:1999bp,Kataev:1999dp}. The contribution of systematic errors to  
the total experimental \pdfs\ uncertainties is also given  in Fig.~\ref{AL:ahs}: 
the systematic errors are most essential for the 
u- and d-quark distributions.

\begin{table}[t!] 
\begin{center}
\begin{tabular}{|c|c|c|c|c|c|c|}   \hline
                 & stat+syst & RS & TS & SS & MC & DC \\\hline
$\Delta L_{\rm W} (\%)$ &1.9 & 0.4 & 0.9 & 1.3 & 2.9  & 0.3\\\hline
$\Delta L_{\rm Z} (\%)$ &1.6 & 0.5 & 0.9 & 1.3 & 2.9  & 0.6\\\hline
$\Delta L_{\rm W/Z} (\%)$ & 0.5 & -- & -- & -- & -- & 0.3\\\hline
\end{tabular}
\caption{The uncertainties of the parton luminosities for $W/Z$
production cross sections and their ratios. Here
$L_{\rm W}=L_{\rm u\bar d}+L_{\rm d\bar u}$, 
$L_{\rm Z}=L_{\rm u\bar u}+L_{\rm d\bar d}$, and
$L_{\rm W/Z}=(L_{\rm u\bar d}+L_{\rm d\bar u})
/(L_{\rm u\bar u}+L_{\rm d\bar d})$.}
\end{center}
\label{AL:wz}
\end{table}

Except uncertainties itself the \pdf\ correlation are also important
(see Fig.~\ref{AL:cor}). The account of correlations can lead to 
cancellation of the \pdfs\ uncertainties in the calculated 
cross section. The luminosities uncertainties can also
cancel in the ratios of cross sections. An example of such 
cancellation is given in Table~\ref{AL:wz}, where the 
uncertainties of luminosities for the $W/Z$ production cross sections
and their ratios are given. 

The \pdf\ set discussed in this subsection can be obtained by the code
\cite{Alekhin:1999;3qcd}. The \pdfs\ are DGLAP evolved 
in the range $x=10^{-7}\div1$, $Q^2=2.5\div5.6\cdot10^{7}$~GeV$^2$.
The code returns the values of u-, d-, s-quark, 
and gluon distributions Gaussian-randomized with accordance of their 
dispersions and correlations including both experimental and 
theoretical ones.


\subsection{Approximate NNLO evolution of parton 
densities\protect\footnote{Contributing authors: W.L. van Neerven and A. Vogt.}}
\label{sec:vogt;2qcd}

In order to arrive at precise predictions of perturbative QCD for the
LHC, for example for the total \mbox{$W$-production} cross section 
discussed in Sects.~\ref{mrssection} and \ref{cteqsection}, 
the calculations need to be extended beyond 
%
%
the NLO. Indeed, the NNLO coefficient 
functions for the above cross section have been calculated some time ago 
\cite{Hamberg:1991np,vanNeerven:1992gh}. The same holds for the structure functions in 
DIS \cite{vanNeerven:1991nn,Zijlstra:1991qc,Zijlstra:1992kj,Zijlstra:1992qd} which form the 
backbone of the present information on the parton densities. On the 
other hand, the corresponding NNLO splitting 
functions have not been computed so far. Partial results are however 
available, notably the lowest four and five even-integer moments,
respectively, for the singlet and non-singlet combinations \cite
{Larin:1994vu,Larin:1997wd}. When supplemented by results on the leading \mbox{$ x\!
\rightarrow\! 0$} terms \cite{Blumlein:1996jp,Catani:1994sq,Fadin:1998py,Blumlein:1998em,Blumlein:1998pp} derived from small-$x$
resummations, these constraints facilitate effective para\-metrisations
\cite{vanNeerven:1999ca,NV2} which are sufficiently accurate for a wide range in $x$ 
(and thus a wide range of final-state masses at the LHC).       
In this section, we compile these expressions and take a brief look at 
their implications.  For detailed discussions the reader is referred to 
refs.~\cite{vanNeerven:1999ca,NV2}.

In terms of the flavour non-singlet (NS) and singlet (S) combinations 
of the parton densities (here $f_{q_f} \equiv q$ and $f_{g} \equiv g$),
\begin{eqnarray}
  q_{{\rm NS},ik}^{\pm} &\! =\! & q_i\pm\bar{q}_i - (q_k\pm\bar{q}_k)
  \:\: , \quad\quad
  q_{\rm NS}^V           \, =\,   \sum_{r=1}^{N_f} (q_r - \bar{q}_r)
  \: , \quad\quad 
  q_{\rm S}^{\,}         \, =\,   \left( \begin{array}{c} \!\Sigma\! \\
                                  \! g\! \end{array} \right) 
\end{eqnarray}
with $ \Sigma = \sum_{r=1}^{N_f} (q_r + \bar{q}_r)$, the evolution 
equations (\ref{evequa;1qcd})
%
%
consist of $\, 2N_f\! -\! 1\,$ scalar non-singlet equations and the 
$2\!\times\! 2$ singlet system. 
%
The LO and NLO splitting functions $P^{(LO)}(x)$
and $P^{(NLO)}(x)$ in Eq.~(\ref{apexp;1qcd}) are known for a long time.
For each of the NNLO functions $P^{(2)}(x)= (4 \pi)^3 P^{(NNLO)}(x) $ two approximate expressions 
(denoted by `$A$' and `$B$') are given below in the 
$\overline{\mbox{MS}}$ scheme, which span the estimated residual 
uncertainty. The central results are represented by the average 
$1/2\, (P_A^{(2)}+P_B^{(2)})$.

The NS$^+$ parametrisations \cite{vanNeerven:1999ca} read, using $\, \delta \equiv 
\delta (1\! -\! x)$, $\, L_1 \equiv \ln (1\! -\! x)$ and 
$\, L_0 \equiv \ln x$,
\begin{eqnarray}
\label{nsp}
 \lefteqn{ P^{(2)+}_{{\rm NS},A}(x) = 
 \frac{1137.897}{(1-x)_+} + 1099.754\, \delta - 2975.371\, x^2
 - 125.243 - 64.105\, L_0^2 + 1.580\, L_0^4 }
 \quad \\ & & \hspace*{-8mm} \mbox{}
 - N_f \left( \frac{184.4098}{(1-x)_+} + 180.6971\, \delta 
 + 98.5885\, L_1 - 205.7690\, x^2 - 6.1618 - 5.0439\, L_0^2 \right)
 + P^{(2)}_{{\rm NS}, N_f^2} 
 \:\: , \nonumber\\[1mm] 
 \lefteqn{ P^{(2)+}_{{\rm NS},B}(x) =
 \frac{1347.207}{(1-x)_+} + 2283.011\, \delta - 722.137\, L_1^2
 - 1236.264 - 332.254\, L_0 + 1.580\, (L_0^4 - 4 L_0^3) }
 \quad \nonumber \\ & & \hspace*{-8mm} \mbox{}
 - N_f \left( \frac{184.4098}{(1-x)_+} + 180.6971\, \delta 
 + 98.5885\, L_1 - 205.7690\, x^2 - 6.1618 - 5.0439\, L_0^2 \right)
 + P^{(2)}_{{\rm NS}, N_f^2} 
 \nonumber
\end{eqnarray}
with
\begin{eqnarray}
\label{PNS2}
  P^{(2)}_{{\rm NS}, N_f^2}(x) \: &=&
     \frac{1}{81} \bigg( \mbox{} - \frac{64}{(1-x)_+}
     - [ 204 + 192\, \zeta (3) - 320\, \zeta (2) ] \,\delta (1-x)
     + 64
  \nonumber \\ & & \mbox{}
     + \frac{x \ln x}{1-x}\, (96\,\ln x + 320)
     + (1-x) (48\, \ln^2 x + 352 \ln x + 384) \bigg) \:\: .
\end{eqnarray}
Here $\zeta (l)$ denotes Riemann's $\zeta$-function. Equation~(\ref{PNS2})
is an exact result, derived from large-$N_f$ methods \cite{Gracey:1994nn}.
The corresponding expressions for $ P^{(2)-}_{{\rm NS}}$ are
\begin{eqnarray}
\label{nsm}
 P^{(2)-}_{{\rm NS},A}(x) &\! =\! & P^{(2)+}_{{\rm NS},A}(x)
 + 20.687\, x^2 - 18.466 + 66.866\, L_0^2 -0.148\, L_0^4 
 \nonumber \\ & & \mbox{} + N_f \left(
 0.0163\, L_1 - 0.402\, x^2 + 0.4122\, - 1.4965L_0^2 \right)
 \:\: , \\[1mm]
 P^{(2)-}_{{\rm NS},B}(x) &\! =\! & P^{(2)+}_{{\rm NS},B}(x) 
 - 0.101\, L_1^2 + 1.508 + 4.775\, L_0 - 0.148\, (L_0^4 - 4 L_0^3) 
 \nonumber \\ & & \mbox{} + N_f \left( 
 0.0163\, L_1 - 0.402\, x^2 + 0.4122\, - 1.4965L_0^2 \right) 
 \:\: . \nonumber
\end{eqnarray}
The difference between $P^{(2)-}_{\rm NS}$ and $P^{(2)V}_{\rm NS}$ 
is unknown, but expected to have a negligible effect ($\ll 1\%$).  

The effective parametrisations for the singlet sector are 
given in Ref.~\cite{NV2}.
Besides the $1/x\, \ln x$ terms of $P_{qq}^{(2)}$, $P_{qg}^{(2)}$ and
$P_{gg}^{(2)}$ \cite{Catani:1994sq,Fadin:1998py}, only the $N_f^2$ contribution $ \propto 
1/[1-x]_+$ to $P_{gg}^{(2)}$ is exactly known here \cite{Bennett:1998ch}.

The evolution equations (\ref{evequa;1qcd}) are written for a factorization
scale $\mu_f=Q$. Their form can be straightforwardly 
generalized to include also the dependence on the renormalization scale $\mu_r$.

\begin{figure}[t!]
\centering
\includegraphics[width=0.75\textwidth,clip]{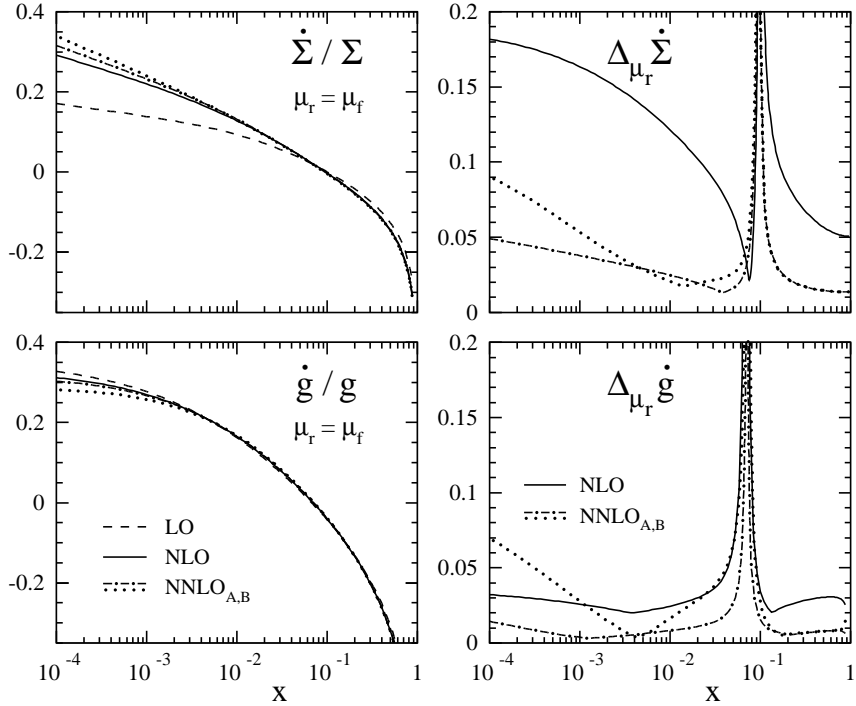}
\vskip-0.5cm
\caption{Left: The LO, NLO and approximate NNLO predictions for the 
 logarithmic derivatives $\dot{q}/q \equiv d \ln q/ d\ln \mu_f^2$ of
 the singlet quark and gluon densities, $q=\Sigma$ and $q=g$, at 
 $\mu_f^2\simeq 30\mbox{ GeV}^2$. Right: The relative scale uncertainty 
 $\Delta_{\mu_r}\dot{q}$ (defined in the text) of these NLO and NNLO 
 results. The number of flavours is $N_f = 4$.}
\label{nvfig}
\end{figure}

The expansion of Eq.~(\ref{evequa;1qcd}) is illustrated in the left part of
Fig.~\ref{nvfig} for $\,\mu_r\! =\!\mu_f$, $\,\as\! =\! 0.2$ 
and parton densities typical for $\,\mu_f^2 \simeq 30\mbox{ GeV}^2$. 
Under these conditions, the NNLO effects are small \mbox{($<\! 2\% $)} 
at medium and large $x$. This also holds for the non-singlet evolution 
not shown in the figure. The approximate character of the our results
for $P^{(2)}$
does not introduce relevant 
uncertainties at $x \gsim 2\cdot 10^{-3}$. The third-order corrections 
increase with decreasing $x$, reaching $(12 \pm 4)\% $ and $(\mbox{} -6
\pm 3)\% $, respectively, of the NLO predictions for $\dot{\Sigma}$ and 
$\dot{g}$ at $x = 10^{-4}$.

The renormalization-scale uncertainty of these results is shown in the
right part of Fig.~\ref{nvfig} in terms of $\Delta_{\mu_r}\dot{q}\equiv
(\dot{q}_{\rm max} - \dot{q}_{\rm min})/[2\, \dot{q}_{\rm average}]$, 
as determined over the range $ 0.5\, \mu_f \leq \mu_r \leq 2\, \mu_f$. 
Note that the spikes slightly below $x=0.1$ arise from $\dot{q}_{\rm 
average} \simeq 0$ and do not represent enhanced uncertainties. Thus 
the inclusion of the third-order terms in Eq.~(\ref{evequa;1qcd}), already 
in its approximate form,
leads to significant 
improvements of the scale stability, except for the gluon evolution 
below $x = 10^{-3}$.

\subsection{The NNLO analysis of the experimental data 
for $xF_3$  and the effects of high-twist power 
corrections\protect\footnote{Contributing authors: 
A.L.~Kataev, G.~Parente and A.V.~Sidorov.}}
\label{sec:kataev;2qcd}

During the last few years there has been considerable 
progress in calculations of the perturbative QCD corrections to 
characteristics of DIS. Indeed, the 
analytic expressions for the NNLO 
perturbative QCD corrections to the coefficient functions of structure functions 
$F_2$ \cite{vanNeerven:1991nn,Zijlstra:1991qc,Zijlstra:1992qd} and $xF_3$ \cite{Zijlstra:1992kj,Zijlstra:1994sh} are now known.
However, to perform the NNLO QCD fits of the concrete experimental data it 
is also 
necessary to know the NNLO expressions for the anomalous dimensions of the 
moments of 
$F_2$ and $xF_3$. At present, this information is available in the case of 
$n=2,4,6,8,10$ moments of $F_2$  \cite{Larin:1994vu,Larin:1997wd}.
The results of Refs.~\cite{vanNeerven:1991nn,Zijlstra:1991qc,
Zijlstra:1992kj,Zijlstra:1992qd,Larin:1994vu,Larin:1997wd,Zijlstra:1994sh} are  forming the  theoretical background 
for the study of the effects, contributing to scaling violation at the level of 
new theoretical precision, namely with taking into account the effects of the 
NNLO perturbative QCD contributions.

In the process of these studies it is rather instructive to include the 
available 
theoretical information on the effects of high-twist corrections, 
which could give rise to scaling violation of the form $1/Q^2$.
The development of the infrared renormalon (IRR) approach (for a review 
see Ref.~\cite{Beneke:1999ui}) and the dispersive method \cite{Dokshitzer:1996qm} (see also 
\cite{Shirkov:1997wi,Solovtsov:1999in}) made it possible to construct models for the power-suppressed 
corrections to DIS structure functions (SFs). Therefore, it became possible to include the predictions
of these models to the concrete analysis of the experimental data.

In this part of the Report the results of the series 
of works \cite{Kataev:1999bp,Kataev:1999dp,Kataev:1996vu,Kataev:1998nc} will be summarized. These works are 
devoted to the analysis of the experimental 
data of $xF_3$ SF of $\nu N$ DIS, obtained by the CCFR collaboration  
\cite{Seligman:1997mc}. They  have the aim to determine the NNLO 
values of  $\Lambda_{\overline{MS}}^{(4)}$ and  $\as(M_Z)$ with fixation of theoretical ambiguities 
due to uncalculated higher-order perturbative QCD terms and transitions 
from the case of $f=4$ number of active flavours to the case 
of $f=5$ number of active flavours. The second task was to extract the effects 
of the twist-4 contributions to   $xF_3$ \cite{Kataev:1999bp,Kataev:1998nc} 
and compare 
them with the IRR-model predictions of Ref.~\cite{Dasgupta:1996hh}.
Some estimates of the influence of the twist-4 corrections 
to the  constants of the initial parametrization of $xF_3$ 
\cite{Kataev:1999dp} are presented. These constants are   
related to the parton distribution parameters.

The analysis of Refs.~\cite{Kataev:1999bp,Kataev:1999dp,Kataev:1996vu,Kataev:1998nc} is based on reconstruction 
of the non-singlet (NS) SF $xF_3$ from the finite number of its  moments 
$M_n(Q^2)=\int_0^1x^{n-1}F_3(x,Q^2)dx$ using the Jacobi polynomial method, 
proposed in Ref.~\cite{Parisi:1979jv} and further developed 
in Refs.~\cite{Barker:1981wu,Chyla:1986eb,Krivokhizhin:1987rz,Krivokhizhin:1990ct}.
Within this method one has
\begin{equation}
xF_3(x,Q^2)=x^{\alpha}(1-x)^{\beta}\sum_{n=0}^{N_{max}}
\Theta_n^{\alpha,\beta}(x)\sum_{j=0}^{n}c_j^{(n)}(\alpha,\beta)M_{j+2}^{TMC}(Q^2)
\label{Jacobi}
\end{equation}
where  $\Theta_n^{\alpha,\beta}$ are the Jacobi polynomials,
$c_j^{(n)}(\alpha,\beta)$ are combinatorial coefficients given in terms of Euler 
$\Gamma$-functions and the $\alpha$, $\beta$-weight parameters.
In view of the reasons, discussed in Ref.~\cite{Kataev:1999bp} they were fixed to 
0.7 and 3 respectively, while  $N_{max}=6$ was taken.
Note, that 
the expressions for  Mellin moments were corrected by target mass 
contributions (TMC), taken into account as 
$M_{n}^{TMC}(Q^2)=M_n(Q^2)+(n(n+1)/(n+2))(M_{nucl}^2/Q^2) M_{n+2}(Q^2)$.
The QCD evolution of the moments is defined by the solution of the corresponding 
renormalization group equation 
\begin{equation}
\frac{M_n(Q^2)}{M_n(Q_0^2)}=exp\bigg[-\int_{A_s(Q_0^2)}^{A_s(Q^2)}
\frac{\gamma_{NS}^{(n)}(x)}{\beta(x)}dx\bigg]\frac{C_{NS}^{(n)}(A_s(Q^2))}
{C_{NS}^{(n)}(A_s(Q_0^2))}
\label{evol}
\end{equation}
The QCD running coupling constant enters this equation 
through $A_s(Q^2)=\as(Q^2)/(4\pi)$ and is defined as the expansion 
in terms of inverse powers of $ln(Q^2/\Lambda_{\overline{MS}}^{(4)~2})$-terms 
in the LO, NLO and NNLO.
The NNLO approximation of the coefficient functions of the  moments 
$C_{NS}^{(n)}(A_s(Q^2))=1+C^{(1)}(n)A_s(Q^2)+C^{(2)}(n)A_s^2(Q^2)$ 
were determined from the results of Ref.~\cite{Zijlstra:1992kj,Zijlstra:1994sh}.
The related anomalous dimension functions are defined as 
\begin{equation}
\mu\frac{\partial lnZ_n^{NS}}{\partial\mu}=\gamma_{NS}^{(n)}(A_s)=
\sum_{i\geq 0}\gamma_{NS}^{(i)}(n)A_s^{i+1}
\label{anom}
\end{equation}
where $Z_n^{NS}$ are the renormalization constants of the corresponding NS 
operators. The expression for the QCD 
$\beta$-function in the $\overline{MS}$-scheme is 
known analytically at the NNLO 
\cite{Tarasov:1980au,Larin:1993tp}. However, as was already mentioned, the 
NNLO corrections to $\gamma_{NS}^{(n)}$ are known at present only in the case 
of $n=2,4,6,8,10$ NS moments of $F_2$ SF of $eN$ DIS \cite{Larin:1994vu,Larin:1997wd}.
Keeping in mind that in these cases the difference between the NLO expressions 
for $\gamma_{NS,F_2}^{(1)}$ and $\gamma_{NS,xF_3}^{(1)}$ is rather small
\cite{Kataev:1996vu}, it was assumed that the similar feature is true at the NNLO also. The $xF_3$ fits of 
Refs.~\cite{Kataev:1999bp,Kataev:1999dp,Kataev:1996vu,Kataev:1998nc} 
were done within this approximation. The one more approximation, entering 
onto these analysis, was the estimation of the anomalous dimensions of odd 
moments with $n=3,5,7,9$ by means of smooth interpolation 
of the results of Refs.~\cite{Larin:1994vu,Larin:1997wd}, 
originally proposed in Ref.~\cite{Parente:1994bf}. In view  
of the basic role of the NNLO corrections to the coefficient functions of 
$xF_3$ moments, revealed  in the process of the concrete fits 
\cite{Kataev:1999bp,Kataev:1999dp,Kataev:1996vu,Kataev:1998nc}, 
it is expected that neither the calculations of the NNLO corrections to $xF_3$ 
odd anomalous dimensions (which are now in progress \cite{kataev:V}) 
and further interpolation to even values of $n$, nor the fine-tuning 
of the reconstruction method of Eq.~(\ref{Jacobi}), which depends on the values 
of $\alpha$, $\beta$ and $N_{max}$, will not affect significantly the accuracy 
of the main results of Refs.~\cite{Kataev:1999bp,Kataev:1999dp,Kataev:1998nc}.

\begin{figure}[t!]
\begin{center}
\includegraphics[width=11.5cm,height=6.5cm]{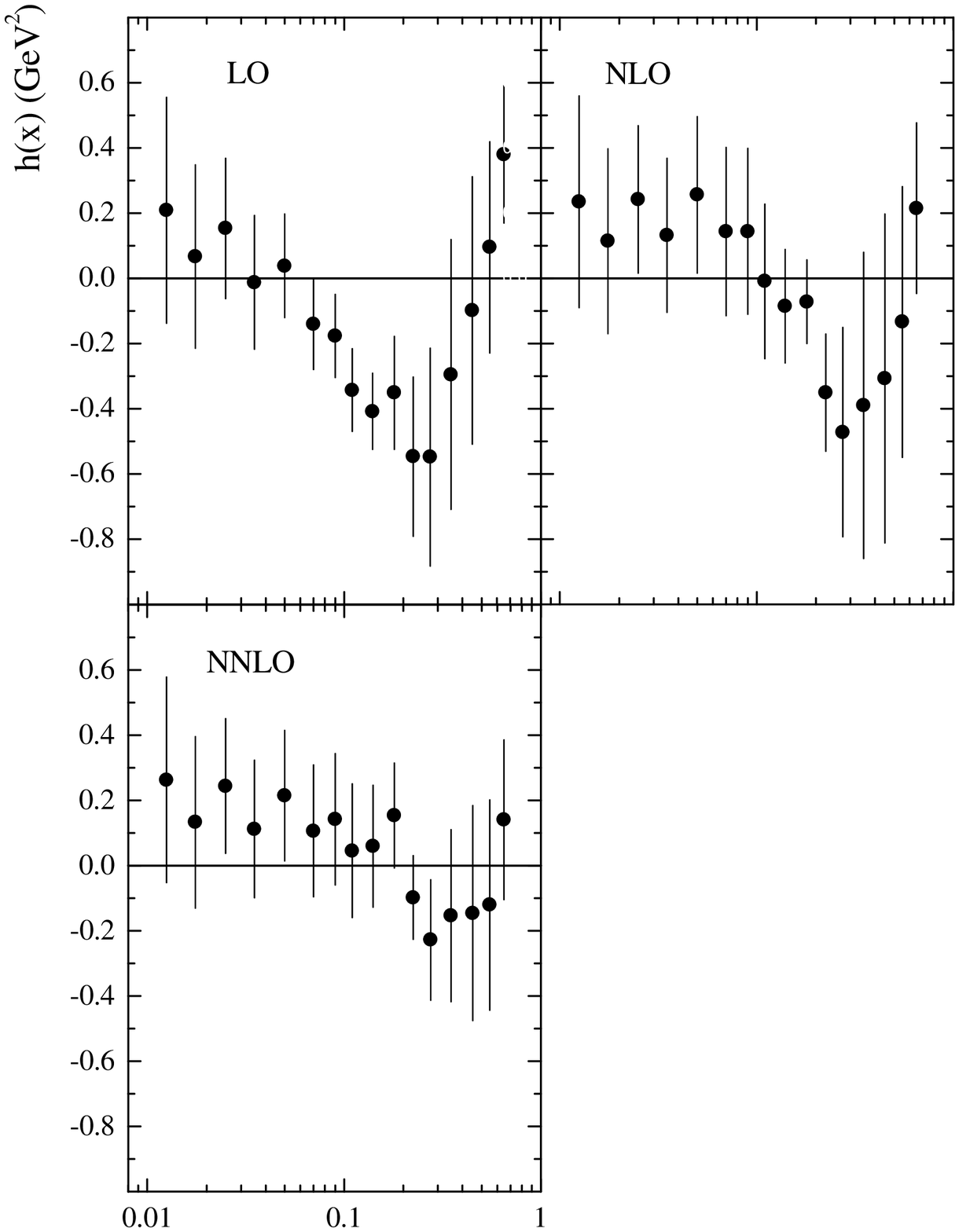}
\end{center}
\vskip-1cm
\caption{$h(x)$ extracted from CCFR'97 data for $xF_3$}
\label{kataev:fig1}
\end{figure}

The power corrections were included in the analysis using 
two different approaches. First, following the ideas of Ref.~\cite{Virchaux:1992jc}, the term 
$h(x)/Q^2$ was added onto the r.h.s. of Eq.~(\ref{Jacobi}). 
The function $h(x)$ was 
parameterized by a set of free constants $h_i$ for each $x$-bin of the 
analysed data. These constants were extracted from the concrete LO, NLO 
and NNLO fits. The resulting behaviour of $h(x)$ is presented in Fig.~\ref{kataev:fig1}, 
taken from Ref.~\cite{Kataev:1999bp}.
Secondly, the IRR model contribution 
$M_n^{IRR}=\tilde{C}(n)M_n(Q^2)A_2^{'}/Q^2$ was added into the 
reconstruction formula of Eq.~(\ref{Jacobi}), where $A_2^{'}$ is the free 
parameter and  
was estimated in Ref.~\cite{Dasgupta:1996hh}.  
The factor $M_n(Q_0^2)$ in the l.h.s. of Eq.~(\ref{evol})
was defined at the initial scale $Q_0^2$ using the parametrization 
$xF_3(x,Q_0^2)=A(Q_0^2)x^{b(Q_0^2)}(1-x)^{c(Q_0^2)}(1+\gamma(Q_0^2)x)$. 
In Table~\ref{kataev:tab1} the combined results of the fits of 
Refs.~\cite{Kataev:1999bp,Kataev:1999dp}
of CCFR'97 data are presented. The twist-4 terms were switched off and 
retained following the 
discussions presented above.

The comments on the extracted   behaviour of $h(x)$ (see Fig.~\ref{kataev:fig1})
are now in order.
Its $x$-shape, obtained from LO and NLO analysis of Ref.~\cite{Kataev:1999bp}
is in agreement with the IRR-model formula of Ref.~\cite{Dasgupta:1996hh}. Note also, 
that the combination of quark counting rules \cite{kataev:qqr,Brodsky:1973kr} 
with the results of Ref.~\cite{Berger:1979du,Gunion:1984ay} predict the following 
$x$-shape of $h(x)$: $h(x)\sim A_2^{'}(1-x)^2$. Taking into account the 
negative values of $A_2^{'}$, obtained in the process of LO and NLO fits 
(see Table~\ref{kataev:tab1}), one can conclude, that the related behaviour of $h(x)$ is 
in qualitative agreement with these predictions. Though a certain indication of the 
twist-4 terms survives even at the NNLO, the NNLO part of Fig.~\ref{kataev:fig1} demonstrates 
that the $x$-shape of $h(x)$ starts to deviate from the IRR model 
of Ref.~\cite{Dasgupta:1996hh}. Notice also, that within the statistical error bars 
the NNLO value of $A_2^{'}$ is indistinguishable from zero (see Table~\ref{kataev:tab1}). 
This feature might be related to the interplay between NNLO perturbative  
and $1/Q^2$ corrections. Moreover, at the used reference scale $Q_0^2=20~GeV^2$ 
the high-twist parameters cannot be defined independently from the 
effects of perturbation theory, which at the NNLO can mimic the contributions 
of higher-twists provided the experimental data is not precise enough and 
the value of $Q_0^2$ is not too small (for the recent discussion of this 
subject see Refs.~\cite{Alekhin:1999ck,Alekhin:1999kt}).
\begin{table}[t!]
\begin{center}
\begin{tabular}{|c|c|c|c|c|c|c|c|}
\hline
Order & $\Lambda_{\overline{MS}}^{(4)}$ & A & b & c & $\gamma$ & 
$A_2^{'}$[$GeV^2$]  
& $\chi^2$/points \\
\hline
LO   & 264$\pm$ 36 & 4.98$\pm$ 0.23 & 0.68$\pm$ 0.02 & 4.05$\pm$ 0.05 & 
0.96$\pm$ 0.18 & -- & 113.1/86 \\
     & 433$\pm$ 51 &  4.69$\pm$ 0.13 & 0.64$\pm$ 0.01 & 4.03 $\pm$ 0.04 & 
1.16$\pm$0.12 &  -0.33$\pm$ 0.12 & 83.1/86 \\
     &  331$\pm$162 &  5.33$\pm$1.33 & 0.69$\pm$0.08 & 4.21 $\pm$0.17 & 
1.15$\pm$0.94 & h(x) in Fig.~\ref{kataev:fig1} & 66.3/86 \\
\hline
NLO & 339$\pm$35 & 4.67$\pm$0.11 & 0.65$\pm$0.01 & 3.96$\pm$0.04 &
0.95$\pm$0.09 & -- & 87.6/86\\
    &  369$\pm$37 & 4.62$\pm$0.16 & 0.64$\pm$0.01 & 3.95$\pm$0.05 &
0.98$\pm$0.17 &  -0.12$\pm$0.06 & 82.3/86\\
    &  440$\pm$183 & 4.71$\pm$1.14 & 0.66$\pm$0.08 & 4.09$\pm$0.14 &
1.34$\pm$0.86  &  h(x) in Fig.~\ref{kataev:fig1} & 65.7/86\\
\hline
NNLO & 326$\pm$35 & 4.70$\pm$0.34 & 0.65$\pm$0.03 & 3.88$\pm$0.08 & 
0.80$\pm$0.28  & -- & 77.0/86\\
     &  327$\pm$35 & 4.70$\pm$0.34 & 0.65$\pm$0.03 & 3.88$\pm$0.08 & 
0.80$\pm$0.29  & -0.01$\pm$0.05 & 76.9/86\\
     &  372$\pm$133 & 4.79$\pm$0.75 & 0.66$\pm$0.05 & 3.95$\pm$0.19 &
0.96$\pm$0.57  & h(x) in Fig.~\ref{kataev:fig1} & 65.0/86 \\                      
\hline
\end{tabular}
\end{center}
\vskip-0.25cm
\caption{The results of the fits of the CCFR'97 data with the 
cut $Q^2>5~GeV^2$. The parameters $A$, $b$, $c$, $\gamma$ 
are normalized at $Q_0^2=20~GeV^2$, which is initial scale 
of the QCD evolution. Statistical errors are indicated.}
\label{kataev:tab1}
\end{table}

The results of Table~\ref{kataev:tab1} demonstrate, that despite the correlation 
of the NLO values $\Lambda_{\overline{MS}}^{(4)}$ with the values 
of the twist-4 coefficient $A_2^{'}$, the parameters of the adopted 
model for $xF_3 (x,Q_0^2)$ remain almost unaffected by the inclusion of the 
$1/Q^2$-term via the IRR-model of Ref.~\cite{Dasgupta:1996hh}. Thus, the corresponding parton 
distributions are less sensitive to twist-4 effects, than the NLO value of 
$\Lambda_{\overline{MS}}^{(4)}$. At the NNLO level the similar 
feature is related to already discussed tendency of the effective minimization 
of the $1/Q^2$- contributions to $xF_3$  (see also NNLO part of Fig.~\ref{kataev:fig1}).

For the completeness the NLO and NNLO values of $\as(M_Z)$, obtained in 
Ref.~\cite{Kataev:1999bp} from the results of Table~\ref{kataev:tab1}  with twist-4 terms modelled 
through the IRR approach are also presented:
\begin{eqnarray}
{\rm NLO}~~~\as(M_Z)&=& 0.120 \pm 0.003 (stat) \pm 0.005 (syst)^{+0.009}_{-0.007} 
\\ \nonumber 
{\rm NNLO}~~\as(M_Z)&=& 0.118 \pm 0.003 (stst) \pm 0.005 (syst)\pm 0.003
\label{alphas}
\end{eqnarray}
The systematical uncertainties in these results are determined by the 
pure systematical uncertainties of the CCFR'97 data for $xF_3$ \cite{Seligman:1997mc}.
The theoretical errors are fixed by variation of the factorization and 
renormalization scales \cite{Kataev:1999bp}. The incorporation into the 
$\overline{MS}$-matching formula for $\as$ \cite{Bernreuther:1982sg,Larin:1995va,Chetyrkin:1997sg} of the 
proposal of Ref.~\cite{Blumlein:1999mg} to vary the scale of smooth transition to the 
world with $f=5$ number of active flavours from $m_b^2$ to $(6.5m_b)^2$ 
was also taken into account. The theoretical uncertainties, presented in 
Eq.~(\ref{alphas}) are in agreement with the ones, estimated in Ref.~\cite{vanNeerven:1999ca} 
using the DGLAP equation. The NNLO value of $\as(M_Z)$ is in agreement 
with another NNLO result $\as(M_Z)=0.1172\pm0.0024$, which was obtained 
in Ref.~\cite{Santiago:1999pr} from the analysis of SLAC, BCDMS, E665 and HERA data for 
$F_2$  with the help of the Bernstein polynomial technique \cite{Yndurain:1978wz}.


\subsection{Measuring Parton Luminosities and Parton Distribution 
Functions at the LHC\protect\footnote{Contributing authors: M.~Dittmar, 
K.~Mazumdar and N.~Skachkov.}}
\label{dittmarsection}

The traditional approach for cross section calculations 
and measurements at hadron colliders 
uses the proton--proton luminosity, 
$L_{proton-proton}$, and the ``best'' known 
quark, anti-quark and gluon parton--distribution functions,
$PDF(x_1,x_2,Q^{2})$ to predict event rates $N_{events}$ for 
a particular parton parton process
with a calculable cross section 
$\sigma_{theory}(q,\bar{q},g \rightarrow X)$,  
using:
\begin{equation}
N_{events}(pp \rightarrow X) = L_{proton-proton} 
\times PDF(x_1,x_2,Q^{2}) \times \sigma_{theory}(q,\bar{q},g \rightarrow X).
\end{equation}
The possible quantitative accuracy of such comparisons
depends not only on the statistical errors, but also 
on the knowledge of $L_{proton-proton}$, the $PDF(x_1,x_2,Q^{2})$ and 
the theoretical and experimental uncertainties for 
the observed and predicted event rates for the studied process.

For many interesting reactions at the LHC one finds that 
statistical uncertainties become quickly negligible
when compared to today's uncertainties. 
Besides the technical difficulties to perform higher order calculations, 
limitations arise from the knowledge of the proton--proton luminosity and the 
parton distribution functions. Estimates for  
proton--proton luminosity measurements at the LHC 
assign typically uncertainties of $\pm$5\%. Similar uncertainties
are expected from the limited knowledge of parton distribution functions.    
Consequently, the traditional approach to cross section 
predictions and the corresponding measurements will be limited 
to uncertainties of at best $\pm$5\%.   

A more promising method \cite{Dittmar:1997md}, 
using only relative  cross section measurements, might lead eventually
to accuracies of $\pm$1\%.
The new approach starts from the idea that for high $Q^{2}$ processes 
one should consider the LHC as a parton--parton collider instead of 
a proton--proton collider. Consequently, one needs to determine 
the different parton--parton luminosities from
experimentally clean and theoretical well understood reactions.  

The production of the vector bosons $W^{\pm}$ and $Z^{0}$ with their 
subsequent leptonic decays fulfil these requirements. 
Taking today's experimental results, the vector boson masses 
are precisely known and their couplings to fermions
have been measured with accuracies of better than 1\%. 
Furthermore, $W^{\pm}$ and $Z^{0}$ bosons with leptonic 
decays have 1) huge cross sections (several nb's) 
and 2) can be identified over a large rapidity range with small backgrounds.

\begin{figure}[t!]
\begin{center}
\begin{tabular}{lr}
\begin{minipage}[t]{0.51\textwidth}
\includegraphics[width=\textwidth]{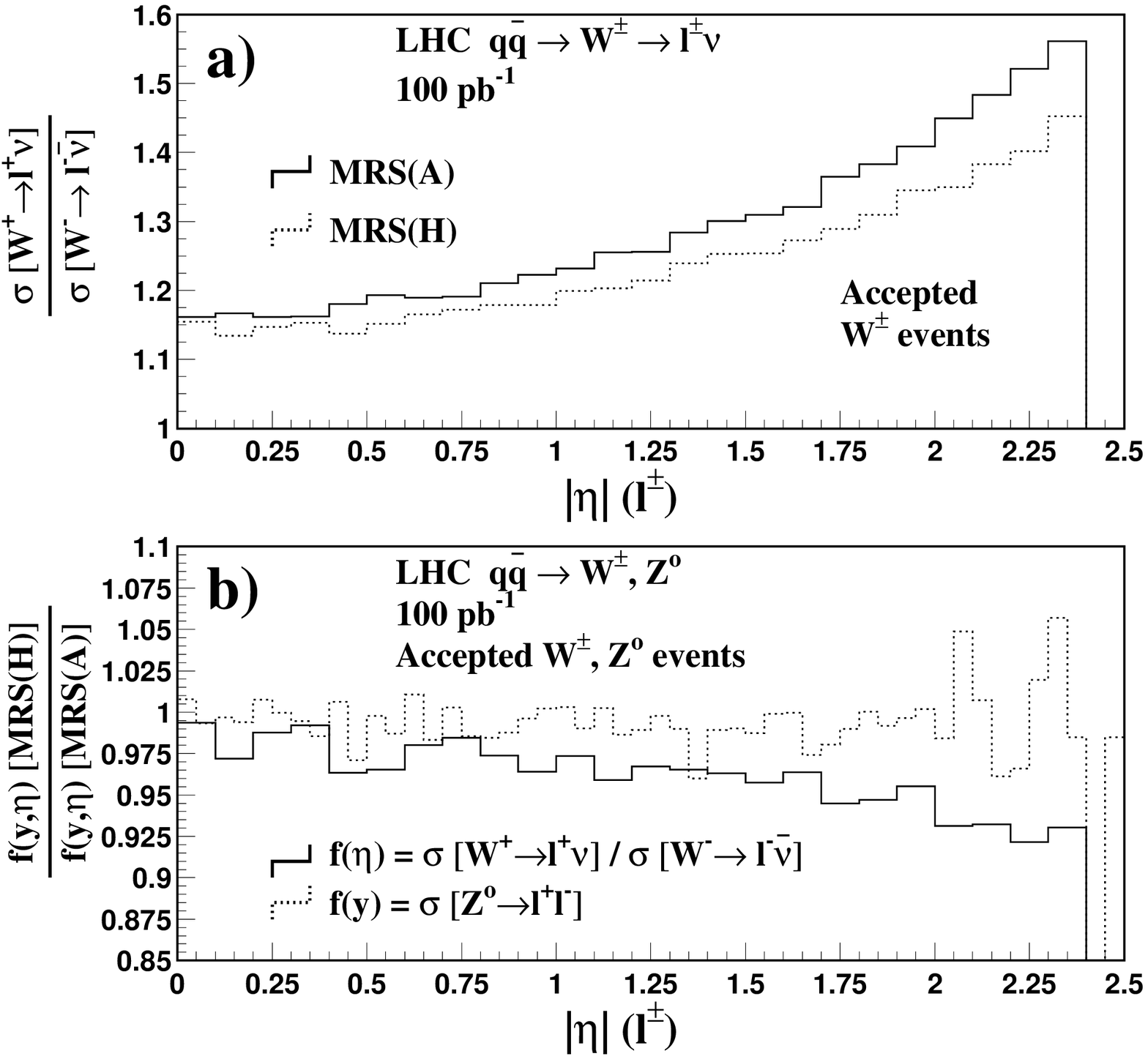} 
\vskip-0.5cm
\caption{a) The detected charged lepton cross section ratio,
$\sigma(\ell^{+} \nu)/\sigma(\ell^{-} \bar{\nu})$, originating
from the reaction
$q\bar{q} \rightarrow W^{\pm} \rightarrow \ell^{\pm} \nu$
as a function of the lepton pseudorapidity for
the MRS(H) and MRS(A) structure function parametrisation.
b) The relative changes for the charged lepton distributions between
the MRS(H) and MRS(A) parametrisations for $W^{+}$,  $W^{-}$
and for  $Z^{0}$ production~\cite{Dittmar:1997md}.
}
\label{fig:mrsmrsh}     
\end{minipage}
&
\begin{minipage}[t]{0.44\textwidth}
\includegraphics[width=\textwidth,bb=40 120 530 660]{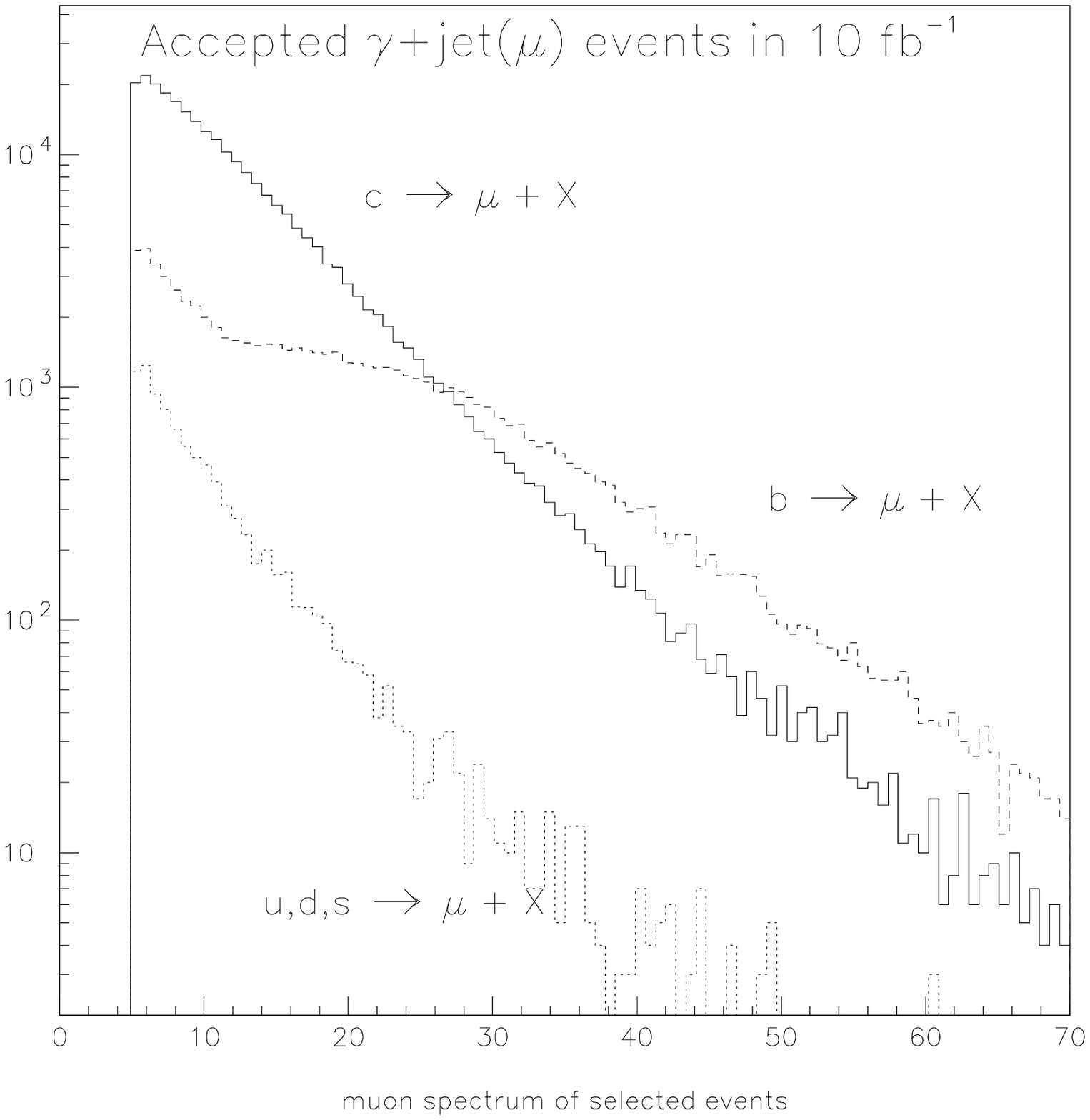} 
\vskip-0.5cm
\caption{The inclusive muon $p_{t}$ 
spectrum in selected photon--jet events originating from light and heavy 
quarks~\cite{Kajari}. 
Assuming standard b--lifetime tagging expectations from 
ATLAS or CMS one should reduce 
the b--flavoured jets by about a factor of 2, the charm--jets by 
a factor of 10 and the light quarks by roughly a factor of 50.
} 
\label{fig:ptmuspectrum}     
\end{minipage}
\\
\end{tabular}
  \end{center}
\vspace{-0.5cm}
\end{figure}

From the known mass and the number of 
``counted'' events as a function of the rapidity $Y$
one can use the relations $M^{2}= s x_{1} x_{2}$ and 
$Y = {1\over 2} ln \frac{x_1}{x_2}$
to measure directly the corresponding quark and anti-quark luminosities
over a wide $x$ range (see fig.\ref{fig:LHCpartons}). 
Simulation studies indicate that the leptonic $W$ and $Z$ decays 
can be measured with good accuracies up to lepton pseudorapidities 
$|\eta| < 2.5$, corresponding roughly to quark and anti-quark $x$ ranges 
between 0.0003 to 0.1.   
The sensitivity of $W$ and $Z$ production data at the LHC  
even to small variations of the \pdfs\ is indicated in 
Figure~\ref{fig:mrsmrsh}.

  
Once the quark and anti-quark luminosities are 
determined from the $W$ and $Z$ data over a 
wide $x$ range, SM event rates of 
high mass Drell--Yan lepton pairs and other processes dominated by 
quark--anti-quark scattering can be predicted.
The accuracy for such predictions is only limited by the theoretical 
uncertainties of the studied process relative to the one for 
$W$ and $Z$ production.

The approach can also be used to measure the gluon luminosity
with unprecedented accuracies.
Starting from gluon dominated ``well'' understood reactions within the SM,  
one finds that the cleanest experimental 
conditions are found for the production of high mass $\gamma$--Jet, 
$Z^{0}$--Jet and perhaps, $W^{\pm}$--Jet events.
However, the identification of these final states 
requires more selection criteria and includes an irreducible background of  
about 10--20\% from quark--anti-quark scattering.   
Some experimental observables to constrain the gluon luminosity
from these reactions have been investigated previously
\cite{Behner:1997es}. The study,
using rather restrictive selection criteria to select 
the above reactions with well defined kinematics,
indicated the possibility to extract the gluon luminosity function
with negligible statistical errors and systematics which might approach 
errors of about $\pm$1\% over a wide $x$ range.


Furthermore, the use of the different rapidity distributions  
for the Vector bosons and the associated jets 
has been suggested in \cite{Chiappetta:1999yi}. The proposed 
measurement of the rapidity asymmetry
improves the separation
between signals and backgrounds and should thus improve 
the accuracies to extract the gluon luminosity.

For this workshop, previous experimental simulations of photon--jet 
final states have been repeated with much larger Monte Carlo statistics 
and more realistic detector simulations
\cite{Bandourin}. These studies select events with exactly one jet 
recoiling against an isolated photon with a minimum $p_{t}$ of 40 GeV.
With the requirement that, in the plane transverse to the beam direction 
the jet is back--to--back with the photon, 
only the photon momentum vector and the jet angle 
needs to be measured. Using the selected kinematics, the mass of the 
photon--jet system can be reconstructed with good accuracy. 
These studies show that several million of photon--jet
events with the above kinematics will be detected 
for a typical LHC year of 10 fb$^{-1}$ and thus 
negligible statistical errors
for the luminosity and $x$ between 0.0005 to $\approx$ 0.2.
This $x$ range seems to be sufficient for essentially all 
high $Q^{2}$ reactions involving gluons.
In addition, it might however be possible 
using dedicated trigger conditions, to select events with 
photon $p_{t}$ as low as 10--20 GeV, which should enlarge the 
$x$ range to values as low as 0.0001. 
The above reactions are thus excellent candidates 
to determine accurately the parton luminosity 
for light quarks, anti-quarks and gluons. 

To complete the 
determination of the different parton luminosities one 
needs also to constrain the 
luminosities for the heavier {\bf $s$,$c$ and $b$} quarks.
The charm and beauty quarks can be measured from a quark flavour 
tagged subsample of the photon--jet final states.  
One finds that the photon--jet subsamples with charm or beauty flavoured jets 
are produced dominantly from the heavy quark--gluon scattering 
($c(b) g \rightarrow c(b) \gamma$). 
For this additional study of photon--jet final states, 
the jet flavour has been identified as being a charm or beauty jet, using  
inclusive high $p_{t}$ muons and in addition $b$-jet identification 
using standard lifetime tagging techniques \cite{Kajari}. 
The simulation indicates that clean photon--charm jet and 
photon--beauty jet 
event samples with high $p_{t}$ photons 
($>$40 GeV) and jets with inclusive high $p_{t}$ muons. 
The muon $p_{t}$ spectrum from the different initial quark flavours
is shown in Figure~\ref{fig:ptmuspectrum}. 

Assuming that inclusive muons with a minimum $p_{t}$ of 5--10 GeV
can be clearly identified, a PYTHIA Monte Carlo
simulation shows that a few 10$^{5}$ $c$--photon events and 
about 10$^{5}$ $b$--photon events per 10 fb$^{-1}$ LHC year should be 
accepted. These numbers correspond to statistical
errors of about $\pm$ 1\% for a $x_{c}$ and $x_{b}$ range 
between 0.001 and 0.1. However, without a much better understanding of 
charm and beauty fragmentation functions such measurements will 
be limited to systematic uncertainties of $\pm$~5--10\%.

Finally, the strange quark luminosity can be determined 
from the scattering of $s g \rightarrow W c$. The events would 
thus consist of $W^{\pm}$ charm--jet final states. 
Using inclusive muons to tag 
charm jets and the leptonic decays of $W$'s to electrons and muons 
we expect about an accepted event sample with a cross section of
2.1 pb leading to about 20k tagged events per 10 fb$^{-1}$ LHC year.
Again, it seems that the corresponding statistical errors 
are much smaller than the expected 
systematic uncertainties from the charm tagging
of $\pm$ 5--10\%.

In summary, we have identified and studied several final 
states which should allow to constrain the light quarks and anti-quarks 
and the gluon luminosities with statistical errors well below 1\% 
for an $x$ range between 0.0005 to at $\approx$ 0.2. 
However, experimental systematics for isolated charged leptons and photons, 
due to the limited knowledge of the 
detector acceptance and selection efficiencies 
will be the limiting factor which optimistically 
limit the accuracies to perhaps $\pm$1\% for light quarks and gluons.   
The studied final states with photon--jet events with tagged  
charm and beauty jets should allow to constrain experimentally  
the luminosities of $s$, $c$ and $b$ quarks and anti-quarks over 
a similar $x$ range and systematic uncertainties 
of perhaps 5--10\%. 

These promising experimental feasibility studies need now to be 
combined with the corresponding theoretical calculations and Monte Carlo
modelling. In detail one has to study how well 
uncertainties from scale dependence, $\as$ and higher order 
corrections change expected cross section ratios.
Figure~\ref{fig:LHCwdat} gives an example 
of today's uncertainties for $W$ and $Z$ cross sections 
at the LHC~\cite{Martin:1999ww}.
Similar estimates for all studied 
processes need to be done during the coming years in order to 
know the real potential of this approach 
to precision cross section measurements 
and their interpretation at the LHC.


\subsection{Lepton Pair Production at the LHC and the Gluon Density in the 
Proton\protect\footnote{Contributing authors: E.~L.~Berger and M.~Klasen.}}

The production of lepton pairs in hadron collisions $h_1h_2\rightarrow\gamma^*
X;\gamma^*\rightarrow l\bar{l}$ proceeds through an intermediate
virtual photon via $q {\bar q} \rightarrow \gamma^*$, and the subsequent 
leptonic decay of the virtual photon. Interest in this DY process is 
usually focused on lepton pairs with large mass $Q$ which justifies the application 
of perturbative QCD and allows for the extraction of the anti-quark density in 
hadrons \cite{Drell:1970wh}.  Prompt photon production 
$h_1h_2\rightarrow\gamma X$ can be calculated in perturbative QCD if the transverse 
momentum $Q_T$ of the photon is sufficiently large. Because the quark-gluon Compton 
subprocess is dominant, $g q \rightarrow \gamma X$, this reaction provides essential 
information on the gluon density in the proton at large $x$ 
\cite{Martin:1998sq}. 
Alternatively, the gluon density can be 
constrained from the production of jets with large 
transverse momentum at hadron 
colliders \cite{Lai:2000wy}.

In this report we exploit the fact that, along prompt photon production, lepton pair
production is dominated by quark-gluon scattering in the region $Q_T>Q/2$.
This realization means that new independent constraints on the gluon density 
may be derived from DY data in kinematical regimes that are accessible 
at the LHC but without the theoretical and experimental 
uncertainties present in the prompt photon case.



At LO, two partonic subprocesses contribute to the
production of virtual and real photons with non-zero transverse momentum:
$q\bar{q}\rightarrow\gamma^{(*)}g$ and $qg\rightarrow\gamma^{(*)}q$.
The cross section for lepton pair production is related to the cross section
for virtual photon production through the leptonic branching ratio of the
virtual photon $\alpha/(3\pi Q^2)$. The virtual photon cross section reduces
to the real photon cross section in the limit $Q^2\rightarrow 0$.

The NLO corrections arise from virtual one-loop
diagrams interfering with the LO diagrams and from real emission diagrams. At
this order $2 \rightarrow 3$ partonic processes with incident gluon pairs $(gg)$, 
quark pairs $(qq)$, and non-factorizable quark-anti-quark $(q\bar{q}_2)$ processes 
contribute also.  
An important difference
between virtual and real photon production arises when a quark emits a
collinear photon. Whereas the collinear emission of a real photon leads to a
$1/\epsilon$ singularity that has to be factored into a fragmentation
function, the collinear emission of a virtual photon yields a finite
logarithmic contribution since it is regulated naturally by the photon
virtuality $Q$. In the limit $Q^2\rightarrow 0$ the NLO virtual photon
cross section reduces to the real photon cross section if this logarithm is
replaced by a $1/\epsilon$ pole. A more detailed discussion can be found
in Ref.~\cite{Berger:1998ev,Berger:1999fm}.

The situation is completely analogous to hard
photo-production where the photon participates in the scattering in the initial
state instead of the final state. For real photons, one encounters an
initial-state singularity that is factored into a photon structure function.
For virtual photons, this singularity is replaced by a logarithmic dependence
on the photon virtuality $Q$ \cite{Klasen:1998jm}.

A remark is in order concerning the interval in $Q_T$ in which our analysis is 
appropriate.  In general, in two-scale situations, a series of logarithmic 
contributions will arise with terms of the type $\as^n \ln^n (Q/Q_T)$.  Thus, 
if either $Q_T >> Q$ or $Q_T << Q$, resummations of this series must be considered. 
For practical reasons, such as event rate, we do not venture into the domain 
$Q_T >> Q$, and our fixed-order calculation should be adequate.  On the 
other hand, the cross section is large in the region $Q_T << Q$.  In previous 
papers~\cite{Berger:1998ev,Berger:1999fm}, we compared our cross sections with available 
fixed-target and collider data on massive lepton-pair production, and we were able
to establish that fixed-order perturbative calculations, without resummation, 
should be reliable for $Q_T > Q/2$.  At smaller values of $Q_T$, non-perturbative 
and matching complications introduce some level of phenomenological ambiguity.  For 
the goal we have in mind, viz., constraints on the gluon density, it would appear 
best to restrict attention to the region $Q_T \geq Q/2$, but below $Q_T >> Q$.


We analyze the invariant cross section $Ed^3\sigma/dp^3$ averaged over the 
rapidity interval -1.0 $<y<$ 1.0.  We integrate the cross section over various 
intervals of pair-mass $Q$ and plot it as a function of the transverse 
momentum $Q_T$.  Our predictions are based on a NLO  
calculation \cite{Arnold:1991yk} and are evaluated in the $\overline{\rm MS}$ 
renormalization scheme. The renormalization and factorization scales are set to 
$\mu=\mu_R=\mu_F= \sqrt{Q^2+Q_T^2}$. If not stated otherwise, we use the CTEQ4M
parton distributions \cite{Lai:1997mg} and the corresponding value of
$\Lambda$ in the two-loop expression of $\as$ with four flavours (five if
$\mu>m_b$). The DY factor $\alpha/(3\pi Q^2)$ for the decay of the
virtual photon into a lepton pair is included in all numerical results.

\begin{figure}[t!]
\begin{center}
\begin{tabular}{lr}
\begin{minipage}[t]{0.47\textwidth}
\includegraphics[width=\textwidth,bb=60 100 495 725]{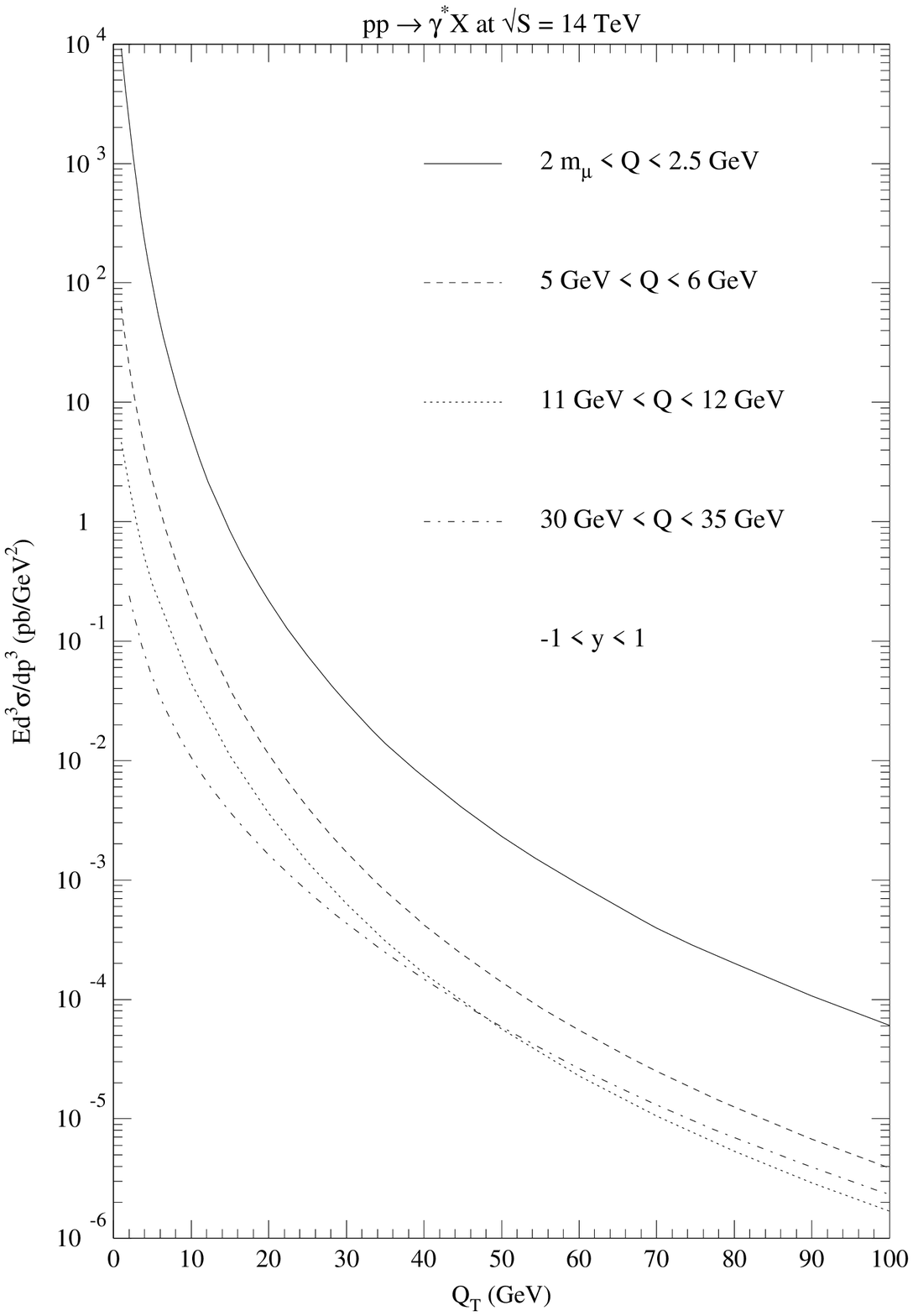} 
\vskip-0.5cm
\caption{Invariant cross section $Ed^3\sigma/dp^3$ as a function of $Q_T$
for $pp \rightarrow \gamma^* X$ at $\sqrt{s}=14$ TeV.
}
\label{fig:berger1}     
\end{minipage}
&
\begin{minipage}[t]{0.47\textwidth}
\includegraphics[width=\textwidth,bb=60 100 495 725]{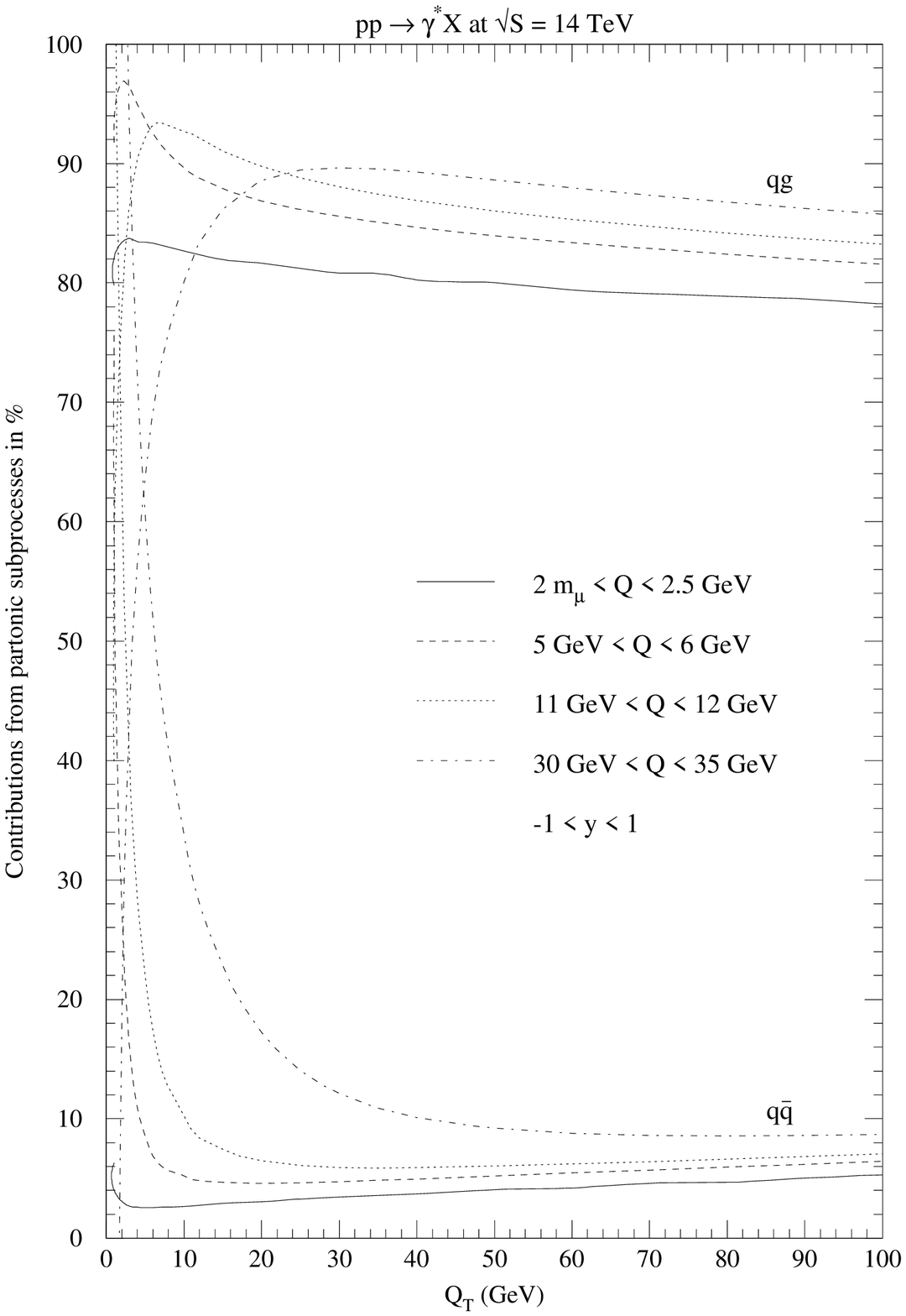} 
\vskip-0.5cm
\caption{Contributions from the partonic subprocesses $qg$ and $q\bar{q}$ to
the invariant cross section $Ed^3\sigma/dp^3$ as a function of $Q_T$
for $pp\rightarrow \gamma^* X$ at $\sqrt{s}$ = 14 TeV. The
$qg$ channel dominates in the region $Q_T > Q/2$.
}
\label{fig:berger2}     
\end{minipage}
\\
\end{tabular}
  \end{center}
\vspace{-0.5cm}
\end{figure}

In Fig.~\ref{fig:berger1} we display the NLO cross section for lepton pair
production at the LHC as a function of $Q_T$ for
four regions of $Q$ chosen to avoid resonances, {\it i.e.\ } from threshold to $2.5$ GeV,
between the $J/\psi$ and the $\Upsilon$ resonances, above the $\Upsilon$'s,
and a high mass region. The cross section falls both with the mass of the
lepton pair $Q$ and, more steeply, with its transverse momentum $Q_T$.
The initial LHC luminosity is expected to be 10$^{33}$ cm$^{-2}$ s$^{-1}$, or
10 fb$^{-1}$/year, and to reach the design luminosity of 10$^{34}$ cm$^{-2}$
s$^{-1}$ after three or four years. Therefore it should be possible to analyze
data for lepton pair production to at least $Q_T\simeq 100$ GeV where one
can probe the parton densities in the proton up to $x_T = 2Q_T/\sqrt{s}\simeq
0.014$. The UA1 collaboration measured the transverse momentum distribution
of lepton pairs at $\sqrt{s}=630$ GeV to $x_T=0.13$ \cite{Albajar:1988iq},
and their data agree well with our expectations 
\cite{Berger:1998ev,Berger:1999fm}.

The fractional contributions from the $qg$ and $q\bar{q}$ subprocesses 
through NLO are shown in Fig.~\ref{fig:berger2}. It is evident 
that the $qg$ subprocess is the most important subprocess as long as
$Q_T > Q/2$. 
%
%
The dominance of the $qg$ subprocess increases somewhat with $Q$,
rising from over 80 \% for the lowest values of $Q$ to about 90 \%
at its maximum for $Q \simeq$ 30 GeV.
Subprocesses other than those initiated by the $q\bar{q}$ and
$q g$ initial channels are of negligible import.

\setcounter{footnote}{0}

The full uncertainty in the gluon density is not known.  We estimate the 
sensitivity of LHC experiments to the gluon density in the proton from
the variation of different recent parametrizations. We choose the latest
global fit by the CTEQ collaboration (5M) as our point of reference
\cite{Lai:2000wy} and compare results to those based on their preceding analysis 
(4M) \cite{Lai:1997mg} and on a fit with a higher gluon density (5HJ) intended to
describe the CDF and D0 jet data at large transverse momentum.  We also compare 
to results based on global fits by MRST \cite{Martin:1998sq}, who provide three 
different sets with a central, higher, and lower gluon density, and to GRV98
\cite{Gluck:1998xa}\footnote{In this set a purely perturbative generation of
heavy flavours (charm and bottom) is assumed. Since we are working in a massless
approach, we resort to the GRV92 parametrization for the charm contribution
\cite{Gluck:1992ng} and assume the bottom contribution to be negligible.}.

\begin{figure}[t!]
\begin{center}
\includegraphics[width=0.45\textwidth,bb=60 100 495 725]{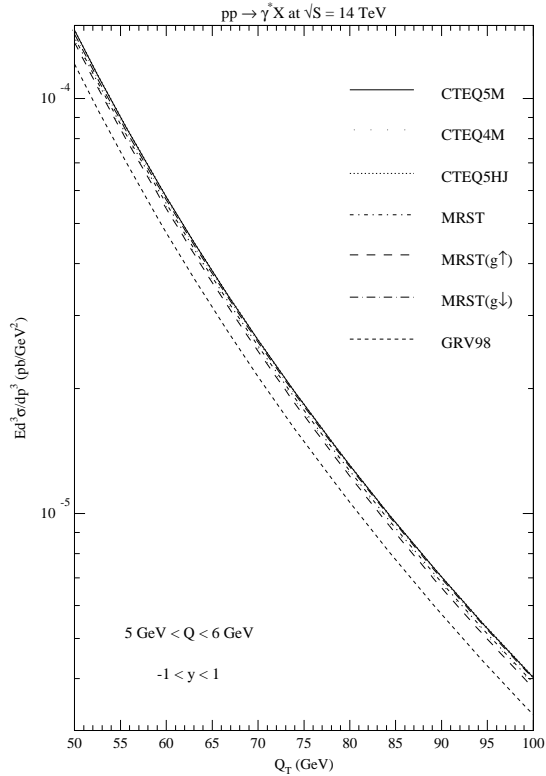} 
\end{center}
\vskip-0.5cm
\caption{Invariant cross section $Ed^3\sigma/dp^3$ as a function of $Q_T$
for $pp \rightarrow \gamma^* X$ at $\sqrt{s}=14$ TeV in the
region between the $J/\psi$ and $\Upsilon$ resonances. The largest differences
from CTEQ5M are obtained with GRV98 (minus 18 \%).
}
\label{fig:berger3}     
\end{figure}

In Fig.~\ref{fig:berger3} we plot the cross section 
for lepton pairs with mass between the
$J/\psi$ and $\Upsilon$ resonances at the LHC in the region
between $Q_T=50$ and 100 GeV ($x_T = 0.007 \dots 0.014$). For the CTEQ
parametrizations we find that the cross section increases from 4M to 5M by 5
\% and does not change from 5M to 5HJ in the whole $Q_T$-range. The largest
differences from CTEQ5M are obtained with GRV98 (minus 18 \%).

The theoretical uncertainty in the cross section can be estimated by varying
the renormalization and factorization scale $\mu_R=\mu_F$ about the central
value $\sqrt{Q^2+Q_T^2}$. 
In the region between the $J/\psi$ and $\Upsilon$ resonances,
the cross section drops from $\pm 39\%$ (LO) to
$\pm 16\%$ (NLO) when $\mu$ is varied over the 
interval interval $0.5 < \mu/\sqrt{Q^2+Q_T^2} < 2$.   
The $K$-factor ratio (NLO/LO) is approximately 1.3 at 
$\mu/\sqrt{Q^2+Q_T^2} = 1$.

We conclude that the hadroproduction of low mass lepton pairs is an advantageous
source of information on the parametrization and size of the gluon density.
With the design luminosity of the LHC, regions of $x_T \simeq 0.014$ should
be accessible. The theoretical uncertainty has been estimated from the scale
dependence of the cross sections and found to be small at NLO.


\section{MONTE CARLO EVENT GENERATORS\protect\footnote{Section
    coordinator: D. Perret-Gallix.}}
\label{sec:mcs;qcd}
The event generation package is the first link of the event 
simulation/reconstruction software suite
which is central to any experimental data analysis. 
Physics results are obtained by a direct comparison
of simulated and observed data. Therefore, precision analyses rely 
on an accurate and detailed implementation of the underlying physics model 
in the generation of
signal as well as background processes.

An event generator is built from various pieces whose object and nature are
quite different. Some are perturbative: 
the hard-scattering matrix element (ME) 
which can be calculated exactly, the parton shower (PS) which approximates,
through the evolution equations, the
initial parton conditions and final-state jet structure, and some 
are 
non-perturbative and probabilistic like the parton distribution in the 
composite 
initial particles and the fragmentation of the final partons. 
The main 
difficulty in writing event generator programs lies on the consistent 
matching of those different components.

Several multi-process parton shower event generators 
(PSEG) have been developed to cover the 
physics programme at \ee, \pp or \ppb~colliders:
{\sc Pythia}~\cite{Sjostrand:1994yb}, {\sc Herwig}~\cite{Marchesini:1992ch,Marchesini:1996vc,Corcella:1999qn}, {\sc Isajet}~\cite{Paige:1998xm,Baer:1999sp}.
These Monte Carlo programs provide an accurate description 
of jet physics at existing high-energy colliders,
which allow the simulation of a large variety of final-state 
processes within and beyond the SM. 
These programs have been essential to 
demonstrate the impressive LHC potential on many different 
and detailed physics questions, to develop new analysis strategies and  
also to optimize the performance of the LHC experiments.  
 
Nevertheless, the increasing potential of very accurate 
measurements at the LHC and the sensitivity to exotic 
physics processes using specific and rare kinematics demand for 
the implementation of higher-order processes and thus
a rethinking of the organisation and probably an extensive rewriting
of many specific Monte Carlo generators.
\par
In the first section, we list the major points of 
concern or pending issues in the development of event generators for the 
LHC physics. The next section discusses the present treatments of minimum bias 
and underlying events. 
The following two contributions address 
the implementation of transverse momentum effects in boson production.
The last three sections present a short description of some of the currently 
available ME generators.
\subsection{QCD event generators: major issues\protect\footnote{
Contributing authors: V.A. Ilyin, D. Perret-Gallix and A.E. Pukhov.}}
\label{sec:intromc;qcd}

\subsubsection{Multi-particle final states}
\noindent{\it Matrix element}

\noindent PSEG are essentially limited to the simulation of 
$2 \ra 2$ processes\footnote{
$n\ra m$ represents processes where $n$ initial particles
decays or scatter to produce $m$ particles in the final state. 
} 
based on analytic matrix element expressions. 
However,
the LHC center of mass energy is large enough to open many high
multiplicity channels. 
In addition, new particle searches
in the Higgs and Susy sectors require the simulation of $2\ra 4$, 
$2\ra 6$ or even $2 \ra 10$
jet processes\footnote{In R-parity non-conserving models.}
for which
a precise knowledge of the SM background processes is 
mandatory.

QCD multi-jets events $pp \ra n_1$ jets and 
$pp \ra Z/W + n_2$ jets
have been computed at LO, for $n_1 \leq 6$ 
by using the {\sc SPHEL} approximation~\cite{Kunszt:1988it}
(i.e. assuming all helicity amplitudes give similar contributions), and
for $n_1 \leq 6$ ({\sc Njets})~\cite{Berends:1991vx} and 
$n_2 \leq 4$ ({\sc Vecbos})~\cite{Berends:1991ax} by
using exact recurrence relations~\cite{Berends:1988me}.

In the PSEG, partonic final states are mimicked through the PS
mechanism based on the leading logarithmic (LL) approximation.
It properly describes parton radiations only in the soft and 
collinear region leading to a crude estimate of the multi-parton dynamics
of the event.
The remedy for a better multi-parton event generator is two-fold:
({\it i}) to improve the simulation of the PS
    by introducing ME corrections 
    (see Sects.~\ref{sec:corcella;3qcd} and \ref{sec:huston;3qcd}), 
({\it ii}) to implement the complete multi-parton hard scattering ME  
process.

The evaluation of ME for multi-particle QCD processes has been reviewed 
in~\cite{Mangano:1991by}. A powerful technique is
the use of helicity amplitudes in the massless limit~\cite{DeCausmaecker:1982bg,Berends:1982uq,Berends:1986qf}. 
Recent developments in this direction were done in~\cite{Dittmaier:1999nn} 
where the Weyl-van-der-Waarden spinor calculus
was generalized to the
massive fermions.
At this level of complexity where so many sub-processes must be calculated,
the analytic hand-made approach becomes literally intractable 
unless stringent
approximations are imposed, as the narrow width approximation, massless
fermions, 
averaging/summing over initial/final helicity state or selecting only a 
subset of gauge invariant diagrams. 

A more systematic approach is needed:
({\it i})  to provide all required channels, 
({\it ii}) to allow for a detail study of finite width effects and
helicity and color correlations, 
({\it iii}) to generate complete ME expressions in order to match 
    the experimental precision.
For example, the LHC statistics will allow to measure
the top quark mass with negligible uncertainty.
This implies that both top quark and
$W$ finite widths must be taken into account in the evaluation of 
the interference between signal and background diagrams.

The automatic Feynman diagram generator packages, largely used for the
\ee~physics analysis, generate complete and approximation-free tree-level 
ME codes, in principle, for any final state multiplicity and with a 
higher reliability level than hand written procedures\footnote{
The packages automatically
generate checks for gauge invariance and gauge independence.}. 
They are gradually upgraded to \pp physics. 
{\sc Grace}~\cite{Tanaka:1991wn,grace:3qcd,intro:3qcd}, {\sc Madgraph}~\cite{Stelzer:1994ta}, 
{\sc Alpha}~\cite{Caravaglios:1995cd} 
and {\sc Phact}~\cite{Ballestrero:1995jn,Accomando:1997es}
are based on tree level helicity amplitude algorithms in arbitrary 
massive gauge theories. 
The evaluation is purely numerical and the code size scales 
linearly with the number of external particles.
In {\sc Alpha} (see Sect.~\ref{sec:alpha;3qcd}), an iterative algorithm, based on 
Green functional methods, evaluates the 
amplitudes for any given Lagrangian and leads to more compact expressions
allowing, for example,  the generation of $gg/q\bar{q} \ra n $ with 
$n \leq 9$~\cite{Caravaglios:1999yr}. 
The {\sc CompHEP}~\cite{Pukhov:1999gg,intro:3qcd} package is based on the squared 
amplitude technique.
Here, the size of the ME code grows exponentially with the number of
external particles, but it produces more powerful 
symbolic expressions. This method
has shown good efficiency for the evaluation $2 \ra 3,4$ processes,
comparable to the helicity amplitude algorithms.

However, the completeness of the
automatically produced matrix elements and the poor 
optimization of the code (when compared to hand coding)
often translate into computationally intensive 
and memory hungry expressions, sometimes reaching the limit
of computability on conventional workstations.

The development effort is focused on two directions:
({\it i}) to improve the code efficiency by the introduction of new 
     computational algorithm, by a 
better optimization and by the ``automated'' introduction of 
approximations,
({\it ii}) to develop code taking advantage of massively parallel 
systems~\cite{Yuasa:1997kx,Yuasa:1997fa}.

\vspace{0.5cm}
\noindent{\it Multi-dimensional integration}

\noindent
The cross section computation and the event generation stage are based
on the multi-dimensional integration procedure. It needs to be focused
to the phase space region where the amplitude is large. 
The amplitude behavior on those regions can be sharp and multi-variate
due to complex singularity patterns.
Integration packages including {\sc Vegas}~\cite{Lepage:1978sw,numrec:3qcd}, 
{\sc Bases/Spring}~\cite{Kawabata:1986yt,Kawabata:1995th}, 
{\sc MILXy}~\cite{tkachov:3qcd}, 
{\sc Foam}~\cite{Jadach:1999sf} use self-adapting techniques
based on {\it importance} and/or {\it stratified sampling}.
However, a faster integration convergence is obtained by providing
the integration algorithm with information on the location and
behavior of the singularities. This is usually done by the
so-called ``kinematics'' routine 
performing the mapping of the integration
variables to the physics parameters. Not yet fully automated~\cite{Ilin:1996gy}, 
it is aiming by appropriate variable transforms at smoothing the singularities
and reducing their dimensionality.

For many important processes, it is impossible to match all singularities
within a single set of variable transforms (e.g. $pp\to u\bar u d\bar d$
with $W$,$Z$ decays and $t$-channel singularities).
In those cases, one relies on a {\it
multichannel} algorithm~\cite{Berends:1995xn,Kovalenko:1997qp} 
where each peaking structure has its own appropriate mapping.

\vspace{0.5cm}
\noindent{\it Interface to the PSEG package }

\noindent The implementation of automatically-produced hard-process ME
in PSEG is a delicate but essential task
to benefit from the implementation of the complex QCD machinery reproducing
the initial and final states.

The ultimate goal is to embed the full ME with its appropriate kinematics
mapping into the kernel of the PSEG through some automated procedure.
Although some progress has been achieved toward this end, a simpler approach is to  
generate parton level event sample using a program dedicated to a given
ME, then let them fragment through
the PS and hadronization scheme of the selected PSEG. For example,
in {\sc Pythia} the routines {\tt PYUPIN} and {\tt PYUPEV} are available for 
the implementation of externally produced event processes. Similar facilities
exist or can be implemented in other PSEG. This technique already used by 
the LHC experiments (see section~\ref{sec:comphep;3qcd}) may
raise consistent parameter and parton distribution bookkeeping issues.

\subsubsection{Heavy-quark production and parton shower}
Keeping the fermion masses at their on-shell value,
although making the expressions more complex,
is always a good practice to get rid of the propagator 
pole divergence.
At LHC, from a phenomenological point of view, light $u$, $d$ and $s$ quark masses 
can be neglected, but heavy $c$, $b$ and $t$ quark should be implemented
not only to reproduce threshold effects, but
also for a correct treatment of spin correlation and NLO corrections. 
Beside the basic $t$-quark physics studies, the heavy-quark event 
generation plays an 
important role as the dominant background to
the Higgs search ($W/Z b\bar{b}$, $t\bar{t} + 2 jets$ and $t\bar{t}t\bar{t}$,
$b\bar{b}b\bar{b}$, $b\bar{b}t\bar{t}$).
Those computations require the use of multi-particle massive ME
as developed in the automatic approach.

The simulation of the PS developed by a massive quark is similar 
to the massless case above an angular cut-off of $\theta = m_q/E_q$, while
below no radiation is emitted.
This is true only in the soft and collinear region,
if the physics observable is sensitive to high-$p_T$ effects (e.g. top mass
reconstruction) full massive radiative heavy-quark decay ME 
(i.e. $t\ra bWg$) must be embedded
in the PS code~\cite{Corcella:1998rs,Andre:1998vh}.

\subsubsection{Color and helicity implementation} 
Color and spin effects are important at LHC. Color correlations beside driving
the fragmentation of partons lead to color reconnection effects
acting on the local event
multiplicity. Spin effects in the top physics, for example, provide a 
useful handle on the nature of the couplings~\cite{Mahlon:1999gz}.

The procedure to assign helicity and color to the initial/final partons 
requires similar implementations in an event generator.
For $2\ra 2$ processes, the number of possible color flows is small
and can be handled easily through an overall factor for the single
diagram case and through a slightly more elaborated treatment when dealing
with  the interference of 2 diagrams with different color 
flows~\cite{Ellis:1987bv}.
For higher multiplicity~\cite{Fujimoto:1997wj}, in the super-symmetric QCD~\cite{Odagiri:1998ep} 
and in the $R$-parity violated
processes~\cite{Dreiner:1999qz}, the selection of the color final state is more 
involved.
In the helicity amplitude approach, each diagram must be decomposed over a
color flow reference base. The cross sections
for all possible color/helicity combinations 
($8^{n_g} \times 3^{n_q} \times 2^{n_g+n_q}$)
are then evaluated. 
Adding more final-state particles drastically increases the number
of cross section computations.

\subsubsection{NLO and NNLO corrections} 
In QCD, talking about corrections concerning the NLO
and NNLO contributions is an understatement.
Higher-order computations are very important not only due to the rather
large coupling constant $\alpha_s$ inducing substantial corrections,
but mainly because they reduce the renormalization and factorization scale
dependence.
Furthermore, analysis or 
experimental-cut dependencies (like the cone-size dependence in jet analysis) are better 
reproduced when higher-order corrections are included.
Roughly speaking if one can say that NLO is the first order giving a sensible
perturbative result, NNLO can be seen as the error estimate 
on this result.

In principle, computing NLO matrix elements is straightforward using loop
integral reduction techniques, but the number of involved diagrams and their
complexity have lead to the development of automatic coding 
programs like FeynArt/FeynCalc Formcalc/Looptools \cite{Kublbeck:1990xc,Mertig:1991an,Hahn:1999wr} or 
{\sc Grace} (see Sect.~\ref{sec:grace;3qcd}). The latter is geared to provide
1-loop $n$-body final-state ME while, in practice,  a maximum of $n=4$ and further
approximations are imposed by computational limitations.

But the main problem lies in the cancellation of soft and collinear
infinities present at NLO precision. 
Fully inclusive computations generate the so-called $K$-factor as a global 
scaling factor, but detailed analyses need phase-space
dependent corrections.
Two techniques (see the general discussion in Sect.~\ref{sec:focal;qcd})
have been developed to handle the
cancellations:
the phase-space slicing method~\cite{Giele:1992vf} and the subtraction 
method~\cite{Frixione:1996ms,Catani:1997vz}. In the former, the cancellation
is performed by approximate integration within regions delimited by
some unphysical cut-off 
(the approximation becomes better as the cut-off
becomes smaller),
in the latter the divergent terms are replaced by a suitable 
analytically-integrable expression plus its finite difference with the original expression.
For these two approaches, Monte-Carlo integration techniques are used, allowing for a 
precise implementation of the experimental cuts. 
These NLO programs (see Sect.~\ref{sec:focal;qcd}) 
can be seen as ``pseudo-event generators''. 
Phase space points 
(pseudo-events) after being tested against the cuts have their corresponding 
weights accumulated to form the observable. Single or multi differential 
distributions can be 
built in one go. But two issues prevent the use of these packages as true event
generators: {\it (i)} the handling of negative weighted events and {\it (ii)} the
interface to the PS and fragmentation stage.
No definite scheme currently exists to properly implement 
LO+NLO processes in a stochastic event generator.

The negative weighted events arise from the virtual corrections cancelling the
soft and collinear divergences. Several attempts are on trial.
One approach is to treat those events as the usual
positive weighted events and to observe
the cancellation only after the reconstruction stage where the
experimental resolution will have introduced a natural cut-off. 
This implies the generation, the simulation and the reconstruction of many
events which finally cancel, not contributing to the statistical significance
and therefore leads to unstable results.
More advanced attempts have been based on a re-weighting of event generated by 
showering from the LO matrix 
elements ~\cite{Seymour:1995df,Corcella:2000gs,Miu:1999ju,Mrenna:1999dy,Corcella:1998rs,Andre:1998vh}.
Recently, a modified subtraction method is exercised to built NLO event 
generators~\cite{Friberg:1999fh,Collins:2000qd} by point-by-point cancellation of the 
singularities. This approach looks quite encouraging although final implementations
have not been realized yet. 

The second problem is the matching of a NLO ME to the PS. A consistent
approach would be to interface a NLO ME to next-to-leading logarithmic (NLL)
order parton
shower, but no such algorithm exist yet (see Sect.~\ref{sec:ps;3qcd}) and therefore
one has to find the least damaging approach to connect NLO ME and LL PS 
and final hadronization. 
Basically the ordered evolution PS variable should be matched to
the ME regularization parameter. Remaining double counting effects will be removed
by the rejection algorithm for each event topology~\cite{Friberg:1999fh}.

\subsubsection{Parton shower}\label{sec:ps;3qcd}
In hadronic collision, the parton showering occurs both in the initial and in
the final state. In the latter, the high-virtuality partons are evolved using
the DGLAP equations down to quasi-real objects ready to 
undergo final hadronization.
The initial partons selected from the parton distribution functions with a 
relative momentum fraction $x$ and virtuality $Q^2$ follow a
backward evolution~\cite{Sjostrand:1985xi,Bengtsson:1986gz,Marchesini:1988cf}
to bring back the 
virtuality down to values compatible with the confinement of partons in a fast
hadron (cloud of quasi-real particles). In this process, gluons and quarks
are emitted (absorbed in the backward-evolution time frame) 
by quark radiation or gluon splitting. This radiation contributes
 to the final-state multiplicity (beam remnants). 
In addition, the parton acquires
a transverse momentum and the full kinematic of the initial centre-of-mass
of the hard scattering will be uniquely defined (see Sects.~\ref{sec:corcella;3qcd}
and \ref{sec:huston;3qcd}).

The parton shower model implemented in the PSEG is essentially
a LL approximation, even if some NLL 
corrections  have been added through exact energy-momentum 
conservation,
angular ordering and 
`optimal scheme' definition
for $\as$~\cite{Catani:1991rr}. 
The dominant logarithmic singularities are resummed in the Sudakov form factors.

As seen in the previous section, the need for a NLL parton shower is high.
The problem is that resumming
higher-order correction breaks one major ``raison d'\^etre'' of the
PS: the universality. At LL level, the hard scattering and 
the parton showering are 2 independent processes
(factorization between the short an the long range) 
and the success of the PSEG
is based on this feature. Incorporating higher-order corrections may break
universality and each type of hard scattering process may require a specific NLL PS
evaluation (see also the last paragraph in Sect.~\ref{sec:grace;3qcd}). 

\subsubsection{Multi-parton scattering}
PSEG for rare events usually include single-scattering processes only.  
At the LHC, one expect, due to the unitary bound, multi-parton interactions 
to give important
contributions to several processes~\cite{DelFabbro:1999tf,Kulesza:1999zh}. 
As an example the cross section
for the production of four jets with double-parton collisions dominates the
single-scattering process when the minimum of the produced jets 
transverse momenta is $p_{t_{min}}<20\,$GeV (see Sect.~\ref{sec:doubpar;qcd}).
These processes, observed by CDF~\cite{Abe:1997bp,Abe:1997xk}, are
largely discussed in Sect.~\ref{sec:doubpar;qcd}, in the Bottom Production
Chapter of this Report and in the 
ATLAS TDR~\cite{ATLAS-TDR;11qcd}.
Information related to the PSEG implementation of
multi-parton scattering can be found in 
the {\sc Pythia}~\cite{Sjostrand:1994yb} and 
{\sc Herwig v6.1}~\cite{Corcella:1999qn} manuals.

Under the simplifying assumptions of no correlation between the
longitudinal-momentum fractions of the initial 
partons, and of the process-independence of parton correlations in 
transverse-momentum space, double-parton 
interactions are easyly implementable into PSEG codes, in terms of a single
universal parameter $\sigma_{eff}$ (see Sect.~\ref{sec:doubpar;qcd}).
However, none of those 
hypotheses can be taken for granted. It is therefore important to 
implement those effects in PSEG programs by using different
dynamical models.
In addition to their contributions to
the background to new particle searches,
the multi-parton interactions at the LHC can provide insights on the 
dynamical structure of the 
hadrons~\cite{Calucci:1998rw,Calucci:1999yz,Calucci:1999uw}.
\subsubsection{Standardization and language issues}
The availability of several independent event generation packages 
although aiming at similar scopes is a big advantage for the experimental 
community.
It makes possible comparative checks and leads to a deeper understanding of 
the various approximations used and implementation dependent issues.

However, one must strongly stress that the definition of a
common interface scheme between the event generators and
the simulation/analysis experimental packages would be extremely valuable.
Such a standardization would cover the following issues:
({\it i}) parameter naming convention, ({\it ii}) parameter database management,
({\it iii}) event output format, ({\it iv}) event sample database. 

Although the standardization scheme can already be exercised on the
existing Fortran PSEG, it takes its full meaning with the current transition
to the object oriented (OO) methodology.
The 
maintenance issue\footnote{Maintenance here means much more than a mere bug
correcting process, it refers to the ability to implement new physics 
models, processes or features on request.} 
 of those large and complex packages over the long expected lifetime
of the LHC experiments is the main reason for
using the OO technology, but the built-in object modularity opens
the door to
a finer grained standardization at least to the level of the interfaces 
of the main procedures
(random number generator, diagram generation,
diagram display, matrix element code, integrator, parton shower,
fragmentation, structure functions). This would allow the building of event
generators using procedures from various origins.   
Most of the PSEG package developers have endorsed C++ as the language
for the future developments. Design and implementation
studies are already in progress~\cite{Lonnblad:1999cq,garren:3qcd,Krauss:1999fc}.  

On these last issues, the setting up of a dedicated
working group with all concerned authors and users would be quite timely.

\subsection{Minimum bias and underlying events\protect\footnote{
Contributing author: B.R.~Webber.}}

A crucial area of physics for the LHC is the structure of finals states in
soft minimum-bias collisions and the soft underlying event in hard
processes. At present very little is understood about these matters on
the basis of QCD starting from first principles. The three principal
event generators in use for LHC physics, {\sc Isajet}, {\sc Herwig} and 
{\sc Pythia},
use quite different models for this type of physics, although each
uses basically the same model to generate both minimum-bias and
underlying events.

\begin{figure}[t!]
\begin{center}
\includegraphics[width=0.5\textwidth]{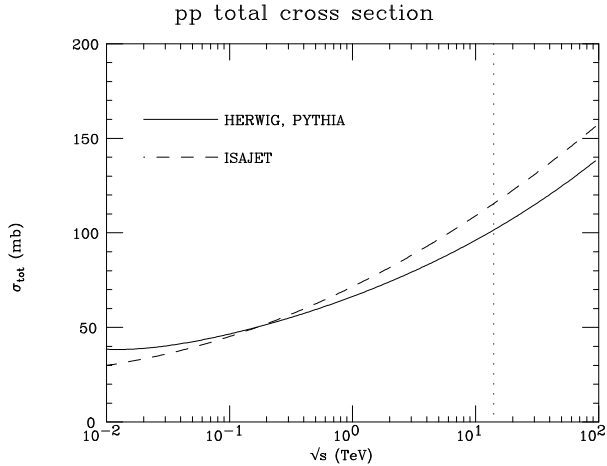}
\vskip-0.5cm
\caption{The $pp$ total cross section according to the parametrizations used in
{\sc Herwig}, {\sc Pythia} and {\sc Isajet}.
\label{fig:bryan;3qcd}}
\end{center}
\end{figure}

Simulation of minimum-bias events starts with a parametrization
of the total cross section. {\sc Herwig} and {\sc Pythia} both use the
Donnachie-Landshoff fit \cite{Donnachie:1992ny}
$$
\sigma_{tot} = 21.70 s^{0.0808} + 56.08 s^{-0.4525}
$$
(where $\sigma$ is in mb and $\sqrt{s}$ in GeV), whereas
{\sc Isajet} uses a $\log^2 s$ form:
$$
\sigma_{tot} = 25.65\left[1+0.0102\log^2(s/1.76)\right]\;.
$$ 
Notice (see Fig.~\ref{fig:bryan;3qcd})
that, although smaller asymptotically, the {\sc Isajet} value
is larger at LHC energies.

To model soft final states, {\sc Herwig} uses the UA5 minimum-bias
Monte Carlo \cite{Alner:1987is}, adapted to its own cluster fragmentation
model. See the {\sc Herwig} manuals \cite{Corcella:1999qn} for further details. 
The model
is based on a negative binomial parametrization of the overall charged
multiplicity. This
has the property of generating large multiplicity fluctuations with
long range in rapidity, in addition to short-range correlations due to
cluster decay.  For true minimum-bias simulation, the soft events
generated by {\sc Herwig} should be mixed with an appropriate fraction
of QCD hard-scattering events.
For the underlying event in hard collisions, the same model is used
to simulate a soft collision between beam clusters containing the
spectator partons.

The minimum-bias/underlying event model used in {\sc Isajet} is based on a
mechanism of multiple
Pomeron exchange \cite{Abramovsky:1973fm}, with a fluctuating number of
`cut Pomerons' acting as sources of final-state hadrons. Each cut Pomeron
fragments directly into hadrons according to the {\sc Isajet} independent
fragmentation model, with the fragmentation axis along the beam direction.
The model again produces large long-range multiplicity fluctuations,
but short-range correlations are weak due to the absence of clustering. 

In {\sc Pythia} a multiple interaction model is used to generate hard, soft and
underlying events in a unified manner. Multiple interactions are discussed in
more detail below. The number $n$ and distribution $P(n)$ of interactions
per event is controlled by the minimum transverse momentum allowed in each
interaction and, optionally, by a model for the impact parameter profile.
Long-range fluctuations may be somewhat weaker in this model, with
short-range correlations somewhere between the two other generators.
In minimum-bias events the choice $n=0$ can occur, in which case a
two-string fragmentation model linking a quark in each beam proton to
a diquark in the other is used.

A study of energy-flow correlations between well-separated phase-space
regions would be helpful in understanding the underlying event and in
separating its contribution from that of the hard
subprocess \cite{Marchesini:1988hj}.  Such a study is currently
being undertaken by the CDF Collaboration.

\subsection{Matrix-element corrections to vector boson production and 
transverse-momentum distributions\protect\footnote{
Contributing authors: G. Corcella and M.H. Seymour.}}\label{sec:corcella;3qcd}

Vector boson production will be a fundamental process to test QCD
and the SM of the electroweak interactions.
Monte Carlo event generators~\cite{Sjostrand:1994yb,Marchesini:1992ch,Marchesini:1996vc}
simulate the initial-state radiation 
in vector boson production processes in the soft/collinear approximation,
but can have `dead zones' in phase space, where no parton emission is allowed.
The radiation in the dead zone is physically suppressed, since 
it is not soft or collinear logarithmically enhanced, but not complete absent 
as nevertheless happens in standard PS algorithms. 
Matrix-element corrections to the {\sc Herwig}
simulation of Drell--Yan processes have been implemented
in~\cite{Corcella:2000gs} 
following the method described in~\cite{Seymour:1995df}: 
the dead zone is populated by the
using of 
the exact first-order amplitude and the cascade in the already-populated 
phase-space region is corrected using the exact matrix element every time an 
emission is capable of being the hardest so far.
A somewhat different procedure is followed to implement matrix-element 
corrections to the {\sc Pythia} event generator~\cite{Miu:1999ju,Mrenna:1999dy}:
the PS probability 
distribution is applied over the whole phase space, the previous algorithm 
having a cut $q_T<m_V$ on the vector boson $V$ transverse momentum to avoid 
double counting,
and the exact ${\cal O}(\alpha_S)$ matrix element is used only to generate the 
closest branching to the hard vertex.
\begin{figure}[t!]
\begin{center}
\begin{tabular}{lr}
\begin{minipage}[t]{0.47\textwidth}
    \includegraphics[width=\textwidth,clip]{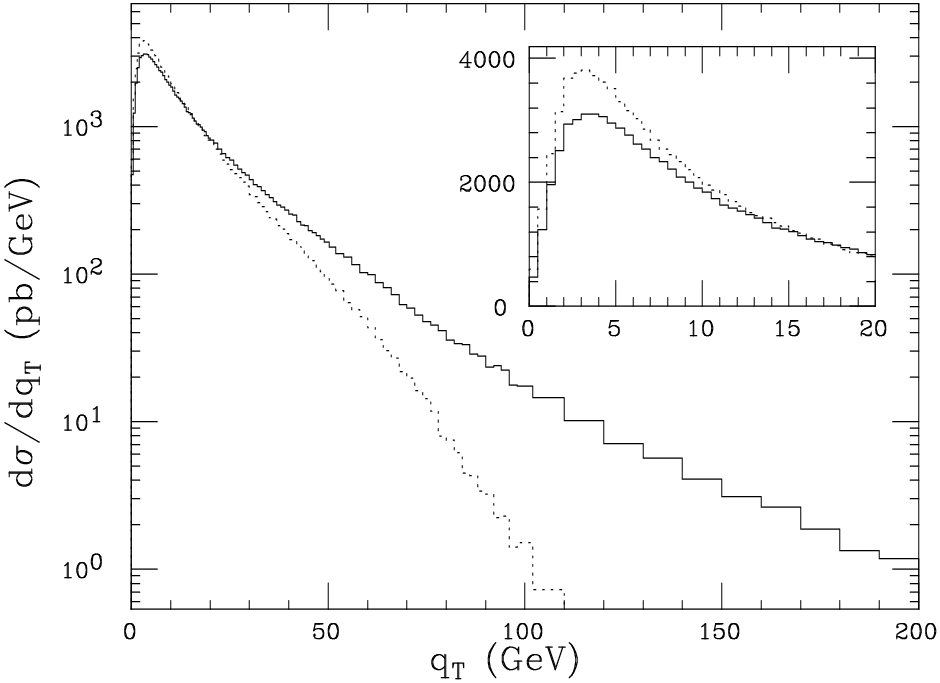}
\vskip-0.5cm
    \caption{$W$ transverse momentum distribution at the LHC, according to 
{\sc Herwig} before (dotted line) and after matrix-element corrections (solid).
    \label{fig:qtlhc;3qcd}}
\end{minipage}
&
\begin{minipage}[t]{0.47\textwidth}
    \includegraphics[width=\textwidth,clip]{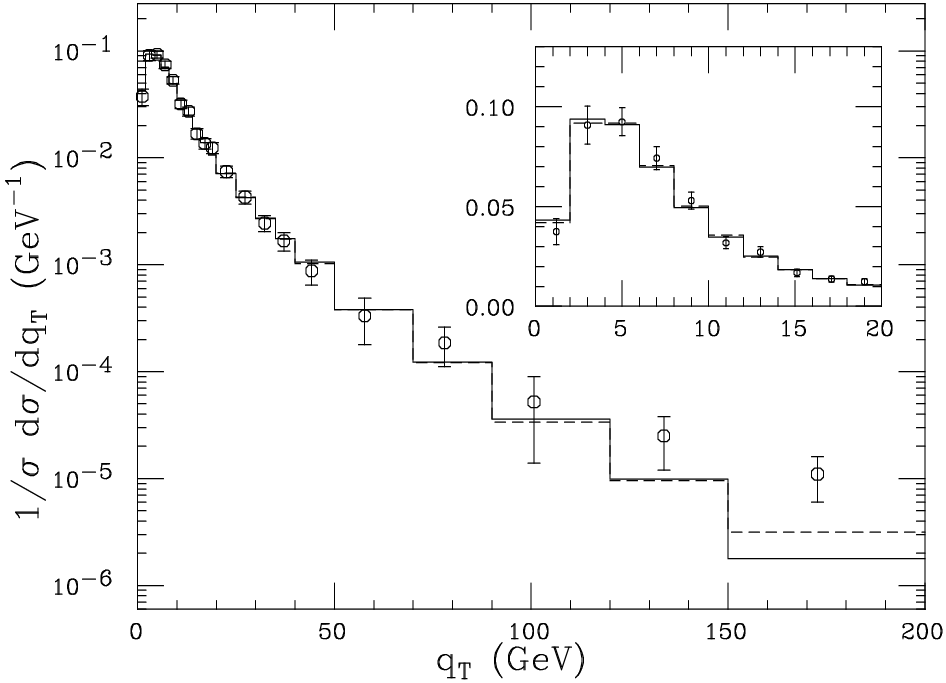}
\vskip-0.5cm
    \caption{Comparison of the D\O\ data with {\sc Herwig} 6.1 for 
$q_{T{\mathrm{int}}}=0$ (solid) and 1 GeV (dashed).
    \label{fig:d0;3qcd}}
\end{minipage}
\end{tabular}
  \end{center}
\vspace{-0.5cm}
\end{figure}
Referring hereinafter to the {\sc Herwig} event generator, in Fig.~\ref{fig:qtlhc;3qcd} 
the distribution of the $W$ transverse momentum $q_T$ 
is plotted at the LHC by running {\sc Herwig} 5.9, the latest public version, 
and {\sc Herwig} 6.1~\cite{Corcella:1999qn}, 
the new version including matrix-element corrections to vector boson 
production, for an intrinsic transverse momentum $q_{T{\mathrm{int}}}=0$, 
its default value. 
A big difference can be seen at large $q_T$, where the 6.1 version
has many more events which are generated via the exact ${\cal O}(\alpha_S)$
amplitude. In the PS soft/collinear approximation, on the contrary, 
$q_T$ is constrained to be $q_T<m_W$.
A suppression can be seen at small $q_T$, due to the fact 
that, even though we are providing the Monte Carlo shower with the 
tree-level ${\cal O}(\alpha_S)$ matrix-element corrections, virtual 
contributions are missing and, by default, 
we still get the total leading-order cross section. No next-to-leading order 
parton shower algorithm is presently available.

In Fig.~\ref{fig:d0;3qcd} some recent D\O\ data~\cite{Abbott:1998jy}
on the $W$ $q_T$ spectrum at the Tevatron is compared with the {\sc Herwig} 6.1
results, which are corrected for detector smearing effects.
A good agreement is found after hard and soft matrix-element corrections;
the options $q_T=0$ and 1 GeV are investigated, but no relevant effect is 
visible after detector corrections, which have been shown
in~\cite{Corcella:2000gs} to be pretty strong.
\begin{figure}[t!]
\begin{center}
\begin{tabular}{lr}
\begin{minipage}[t]{0.47\textwidth}
    \includegraphics[width=\textwidth,clip]{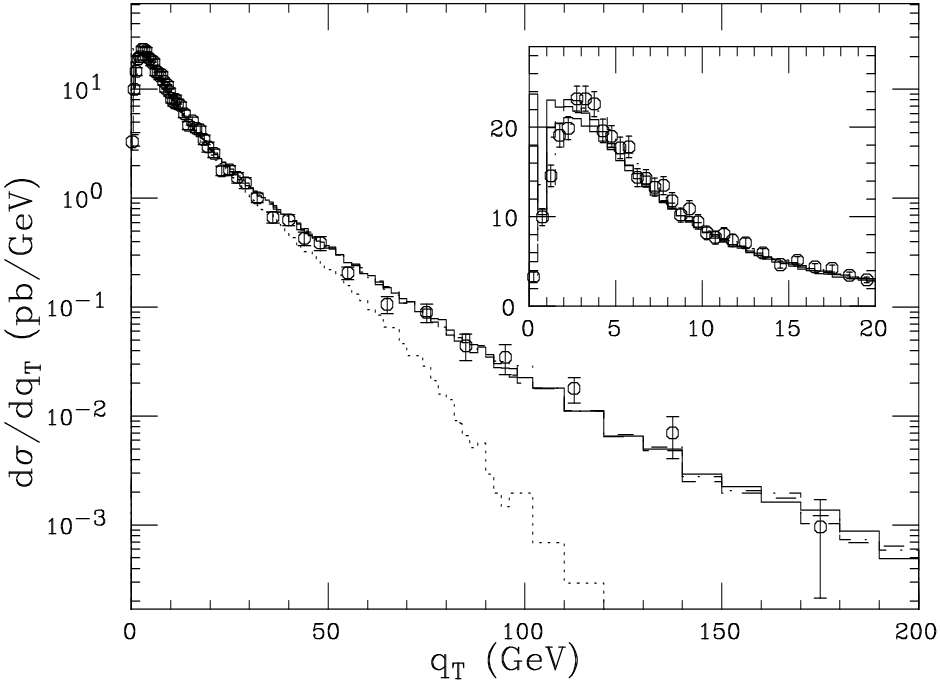}
\vskip-0.5cm
    \caption{Comparison of the CDF data on $Z$ production with {\sc Herwig} 5.9 
    (dotted line) and {\sc Herwig} 6.1 for 
    $q_{T{\mathrm{int}}}=0$ (solid), 1 GeV (dashed) and 2 GeV (dash-dotted).
    \label{fig:cdf;3qcd}}
\end{minipage}
&
\begin{minipage}[t]{0.47\textwidth}
    \includegraphics[width=\textwidth,clip]{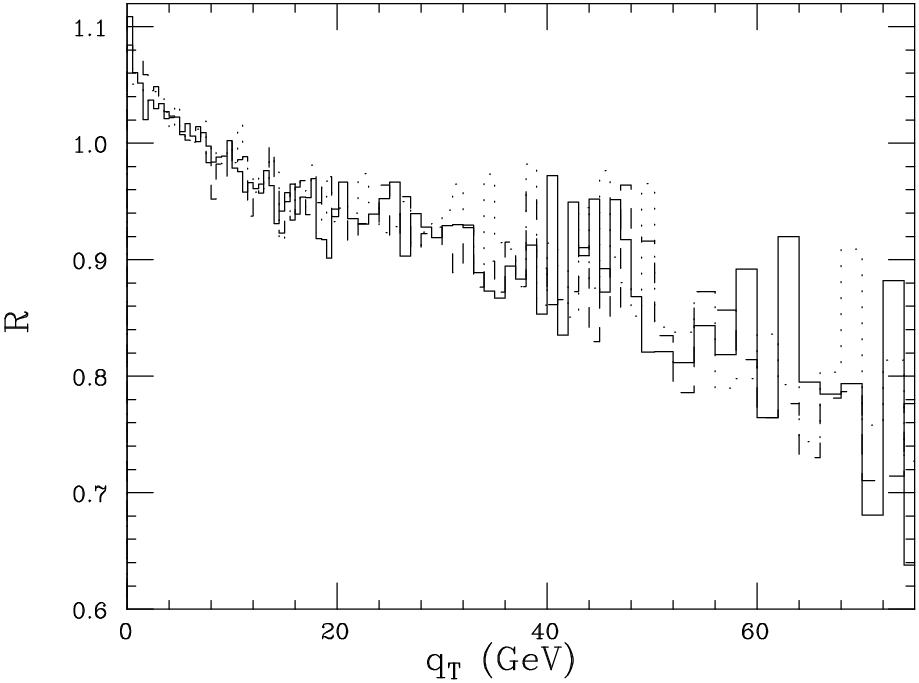}
\vskip-0.5cm
    \caption{Ratio of the $W$ and the $Z$ $q_T$ distributions, 
    according to {\sc Herwig} 6.1 for 
    $q_{T{\mathrm{int}}}=0$ (solid), 1 GeV (dashed) and 2 GeV (dotted).
    \label{fig:ratio;3qcd}}
\end{minipage}
\\
\end{tabular}
  \end{center}
\vspace{-0.5cm}
\end{figure}

In Fig.~\ref{fig:cdf;3qcd}, we compare {\sc Herwig} with some CDF 
data~\cite{Affolder:2000jh}
on $Z$ production, already corrected for detector 
effects, which are however much smaller than the $W$ case.
We consider the options $q_{T{\mathrm{int}}}=0$, 1 and 2 GeV. 
The overall agreement is good, with a crucial role of matrix-element 
corrections to fit in with the data at large $q_T$. 
At low $q_T$, the best fit is the one corresponding to 
$q_{T{\mathrm{int}}}=2$~GeV.
Even though, as can be seen from Fig.~\ref{fig:cdf;3qcd},
the $Z$ distribution is strongly dependent on the intrinsic 
transverse momentum at low $q_T$, in~\cite{Corcella:1999dh} and in 
Fig.~\ref{fig:ratio;3qcd} it is shown that the ratio of the $W$ and $Z$ 
differential cross sections, both normalized to one, is roughly independent of 
$q_{T{\mathrm{int}}}$, which means that the effect of a non-zero 
$q_{T{\mathrm{int}}}$ is approximately the same for both $W$ and $Z$ spectra.
This ratio is one of the main inputs for the experimental analyses and the
fact that it is not strongly dependent on unknown non-perturbative effects is
good news for studies on the $W$ mass measurement.

\begin{figure}[t!]
\begin{center}
\begin{tabular}{lr}
\begin{minipage}[t]{0.47\textwidth}
    \includegraphics[width=\textwidth,clip]{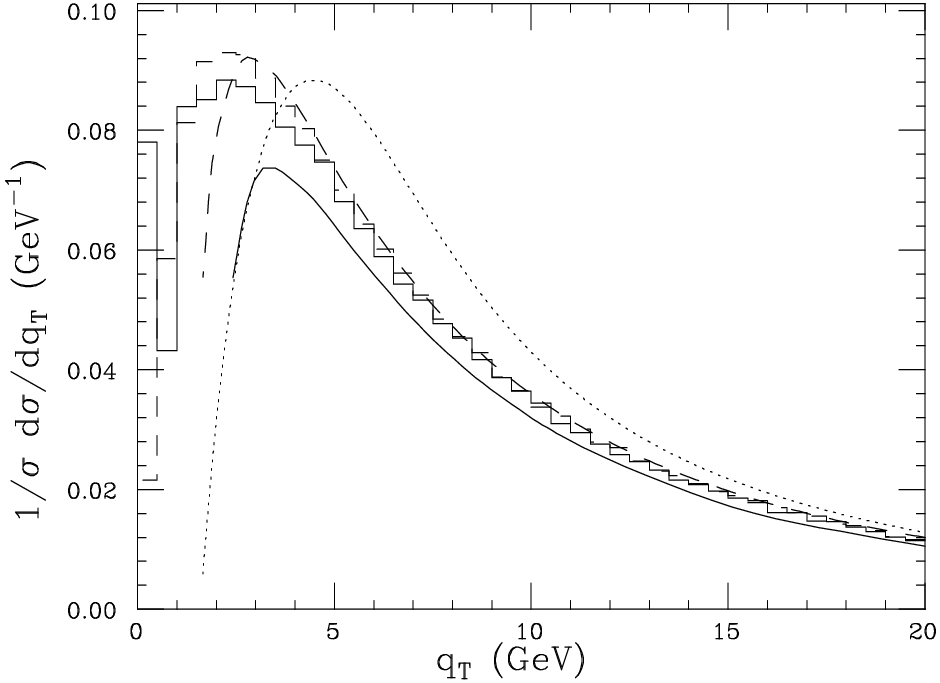}
\vskip-0.5cm
    \caption{The $W$ $q_T$ distribution in the low $q_T$ range at the
Tevatron, according to {\sc Herwig} 6.1, for $q_{T{\mathrm{int}}}=0$ (solid 
histogram) and 1 GeV (dashed histogram), compared with the resummed results
of \cite{Frixione:1999dw} in $q_T$- (solid line) and $b$-space (dotted line) and of 
\cite{Ellis:1998ii} in the $q_T$-space. \label{fig:resum1;3qcd}}
\end{minipage}
&
\begin{minipage}[t]{0.47\textwidth}
    \includegraphics[width=\textwidth,clip]{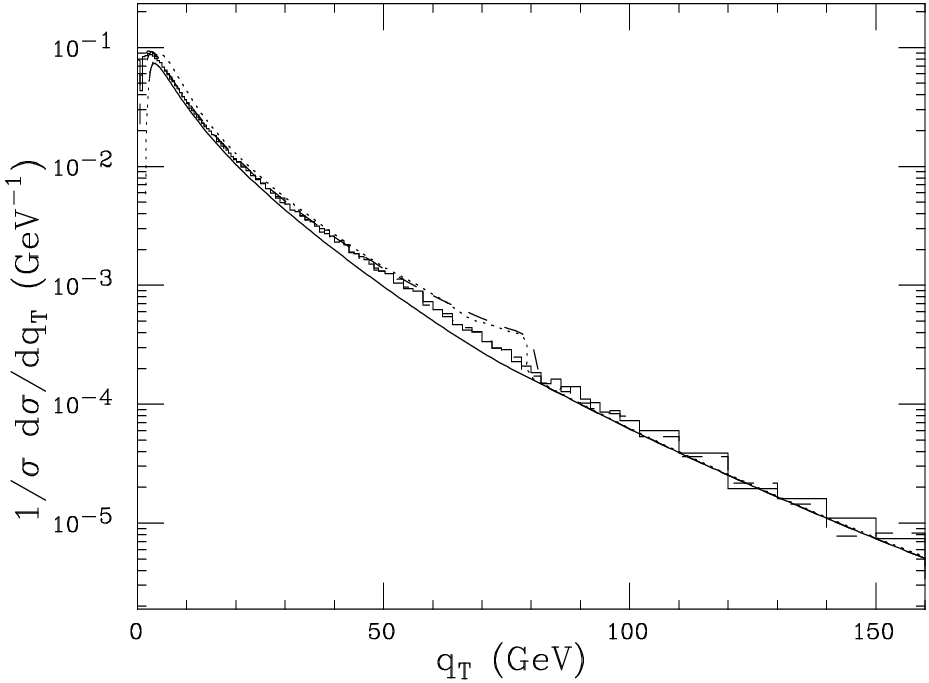}
\vskip-0.5cm
 \caption{As in Fig.~\ref{fig:resum1;3qcd}, but over the whole $q_T$ spectrum.
\label{fig:resum2;3qcd}}
\end{minipage}
\\
\end{tabular}
  \end{center}
\vspace{-0.5cm}
\end{figure}

It is also worthwhile comparing the {\sc Herwig} 6.1 $q_T$ distributions with 
some available calculations which resum the logarithms $l=\log(m_V/q_T)$,
$m_V$ being the vector boson mass, in a Sudakov-like exponential form factor
(see Sect.~\ref{sec:rescal;qcd} for a review of theoretical aspects of 
Sudakov resummation).
Such logarithms are large in the low $q_T$ range.
In~\cite{Corcella:2000gs} the Monte Carlo 
results are compared with the resummation approaches of~\cite{Frixione:1999dw},
where all terms down to the next-to-leading logarithmic order
$\approx \alpha_S^n l^n$ are kept in the Sudakov exponent, both
in $q_T$- and impact parameter $b$-space, and of~\cite{Ellis:1998ii}, where 
the authors expand the Sudakov exponent and keep in the differential cross 
section all terms down to the order $\approx \alpha_S^n l^{2n-3}$, which are 
next-to-next-to-leading logarithms after the expansion of the form factor.
Such resummations are also matched to the
exact first-order result in~\cite{Corcella:2000gs}. 
In Figs.~\ref{fig:resum1;3qcd} and \ref{fig:resum2;3qcd}
the $W$ $q_T$ distributions are plotted according to {\sc Herwig} 6.1 and the 
resummed calculations at small $q_T$ and over the whole $q_T$ range 
respectively. The overall agreement at low $q_T$ is reasonable and the {\sc Herwig}
plots lie well within the range of the 
resummed approaches. At large $q_T$ the matching of the resummed calculations 
to the exact ${\cal O}(\alpha_S)$ result works well only for the approach
of~\cite{Frixione:1999dw} in the $q_T$-space, as we have a continuous 
distribution at the point $q_T=m_W$, the other distributions showing a 
step due to uncompensated contributions of order $\alpha_S^2$ or higher.

In~\cite{Corcella:2000gs}, it is also shown that matrix-element corrections
to vector boson production
have a negligible effect on rapidity distributions, the latest version {\sc Herwig} 
5.9 agreeing well with the CDF data on the $Z$ rapidity.
The implemented hard and large-angle gluon radiation has nevertheless 
a marked impact on jet distributions both at the Tevatron and LHC, 
as many more events with high transverse energy jets are now generated.
While these analyses are performed assuming that the produced vector
boson decays into a lepton pair, the implementation of matrix-element 
corrections to the {\sc Herwig} simulation of the hadronic $W$ decay $W\to q\bar q'$ 
is in progress, however it is expected to be a reasonably straightforward 
extension of the corrections already applied to the process 
$Z\to q\bar q$. Furthermore, the method applied to improve the initial-state 
shower for $W/Z$ production could be extended to many other processes which are 
relevant for the LHC. Among these, we expect that the implementation of 
matrix-element corrections to top and Higgs production may have a 
remarkable phenomenological effect at the LHC. This is in progress as well.

\subsection{
A comparison of the predictions from Monte Carlo programs and 
transverse momentum resummation\protect\footnote{
Contributing authors: C. Bal\'azs, J. Huston and I. Puljak.}}
\label{sec:huston;3qcd}
%
        For many physical quantities, the predictions from PS Monte
Carlo programs should be nearly as precise as those from analytic theoretical 
calculations. This is expected, among others, for calculations which resum logs
with the transverse momentum of partons initiating the hard scattering
(resummed calculations are described in Sect.~\ref{sec:rescal;qcd}).
In the recent literature, most calculations of this kind are either based on 
or originate from the formalism developed by J. Collins, D. Soper, and G. 
Sterman~\cite{Collins:1985kg}, which we choose as the analytic `benchmark' of this Section.
In this case, both the Monte Carlo and analytic calculations
should accurately describe the effects of the emission of 
multiple soft gluons from the incoming partons,
an all orders problem in QCD. The initial state soft gluon emission can affect
the kinematics of the final state partons. This may have an impact on the 
signatures of physics processes at both the trigger and analysis levels and thus
it is important to understand the reliability of such predictions. The best 
method for testing the reliability is the direct comparison of the predictions
to experimental data. If no experimental data is available for certain 
predictions, then some understanding of the reliability may be gained from
the comparison of the predictions from the two different methods.


Parton showering resums primarily the leading logarithms, which are universal, 
i.e. process independent, and depend only on the given initial state.
In this lies one of the strengths of Monte Carlos, since parton showering can
be incorporated into a wide variety of physical processes.
As discussed in Sect.~\ref{sec:rescal;qcd}, an analytic calculation, 
in comparison, 
can resum all large logarithms, since all (in principle) are included in the
Sudakov exponent given in Eq.~(\ref{eq:abseries;5qcd}).

If we try to interpret parton showering in the same language as resummation, 
which is admittedly risky, then we can say that the Monte Carlo Sudakov 
exponent always contains terms analogous to $A^{(1)}$ and $B^{(1)}$
in Eq.~(\ref{AB;5qcd}). It was shown in 
Ref.~\cite{Catani:1991rr} that a suitable modification of the
Altarelli--Parisi splitting function, or equivalently the strong coupling
constant $\alpha_s$, also effectively approximates the $A^{(2)}$
coefficient.~\footnote{Reference \cite{Catani:1991rr} deals only with the 
high-$x$ (or $\sqrt{\tau}$) region, but the same results apply to the 
small-$p_T$ region in transverse momentum distributions.}

Both Monte Carlo and analytic calculations describe the effects of the 
emission of multiple soft gluons from the incoming partons, an all orders
problem in QCD. The initial state soft gluon emission affects the
kinematics of the final state partons, which, in turn, may have an impact 
on the signatures of physics processes at both the trigger and analysis 
levels. Thus it is important to understand the reliability of such
predictions. The best method for testing the reliability is the direct
comparison of the predictions to experimental data. If no experimental
data is available for certain predictions, then some understanding of the
reliability may be gained from the comparison of the predictions from the
two different methods.

In particular, one quantity which should be well--described by both
calculations is the transverse momentum ($p_T$) of the final state 
electroweak boson in a subprocess such as $q\overline{q} \to WX$, $ZX$ 
or $gg \to H X$, where most of the $p_T$ is provided by initial state 
parton showering. The parton showering supplies the same sort of transverse 
kick as the soft gluon radiation in a resummation calculation. 
This correspondence between the Sudakov form factors in resummation and 
Monte Carlo approaches may seem trivial, but there are many
subtleties in the relationship between the two approaches relating 
to both the arguments of the Sudakov factors as well as the impact of 
sub--leading logs~\cite{Mrenna:1999dy,Affolder:2000jh,Corcella:2000gs}.

At a point in its evolution corresponding to (typically) the virtuality of a
few GeV$^2$, the parton shower is cut off and the effects of gluon emission at
softer scales must be parameterized and inserted by hand. This is similar to
the (somewhat arbitrary) division between perturbative and non--perturbative
regions in a resummation calculation. The parametrization is typically
done with a Gaussian smearing similar to that used for the non--perturbative
$k_T$ in a resummation program. In general, the value for the non--perturbative
$\langle k_T \rangle$ needed in a Monte Carlo program will depend on the
particular kinematics being investigated.~\footnote{Note that this is 
unlike the case of the resummation calculations in 
Refs.~\cite{Collins:1985kg,Balazs:1997xd,Balazs:2000wv}, where
the non--perturbative physics is determined from fits to fixed
target data and then automatically evolved to the kinematic regime of
interest.}

A value for the average non--perturbative $k_T$ greater than 1 GeV does 
not imply that there is an anomalous intrinsic $k_T$ associated with the 
parton size; rather, this amount of $\langle k_T \rangle$ needs to be 
supplied to provide what is missing in the truncated parton shower. 
If the shower is cut off at a higher virtuality, more
of the `non--perturbative' $k_T$ will be needed.
\subsubsection{Vector boson production and comparison with {\sc Pythia} and 
{\sc Resbos}}
\begin{figure}[t!]
  \begin{center}
    \includegraphics[width=0.6\textwidth,clip]{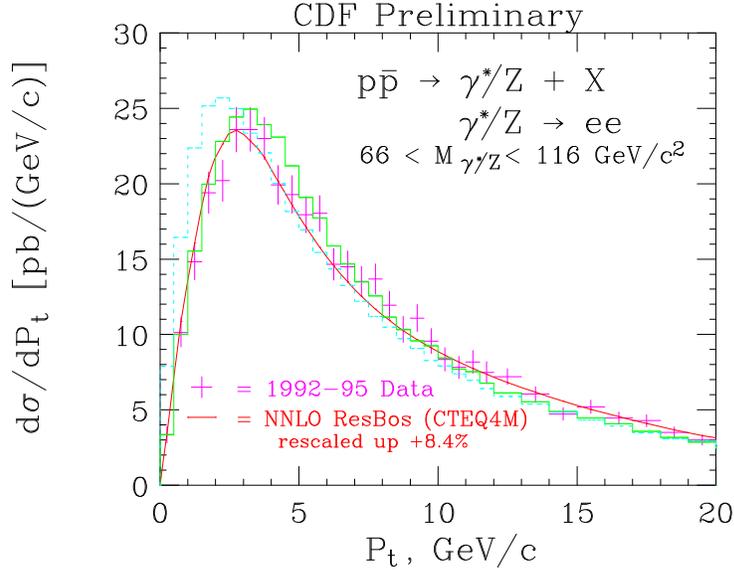}
\vskip-0.5cm
    \caption{The $Z^0$ $p_T$ distribution (at low $p_T$) from CDF for 
             Run 1 compared to predictions from ResBos and from {\sc Pythia}. 
             The two {\sc Pythia} predictions use the default (rms) value 
             for the non--perturbative $k_T$ (0.44 GeV) and the value that 
             gives the best agreement with the shape of the data (2.15 GeV).}
    \label{fig:run1_ee_pt;5qcd}
  \end{center}
\end{figure}

The (resolution corrected) $p_T$ distribution for $Z^0$ bosons (in the low 
$p_T$ region) for the CDF experiment \cite{Affolder:2000jh} is shown in 
Figure~\ref{fig:run1_ee_pt;5qcd} \cite{Balazs:2000sz}, compared to both the resummed prediction 
from ResBos\cite{Balazs:1997xd}, and to two predictions from {\sc Pythia} (version 6.125). 
One {\sc Pythia} prediction uses the default (rms)\footnote{For a Gaussian 
distribution, $k_T^{rms}=1.13\langle k_T \rangle$.} value of  intrinsic 
$k_T$ of 0.44 GeV and the second a value of 2.15 GeV per incoming parton. 
The latter value was found to give the best agreement for {\sc Pythia} with the
data.~\footnote{See Sect.~\ref{sec:corcella;3qcd} and Fig.~\ref{fig:cdf;3qcd} 
for comparisons of the CDF $Z^0$ $p_T$ data with {\sc Herwig}.}
All of the predictions use the CTEQ4M parton distributions~\cite{Lai:1997mg}.
Good agreement is observed between ResBos, {\sc Pythia} and the CDF data.
\subsubsection{Higgs boson production and comparison with {\sc Pythia}}
\begin{figure}[t!]
\begin{center}
\vspace*{-2.cm}
\includegraphics[width=12cm]{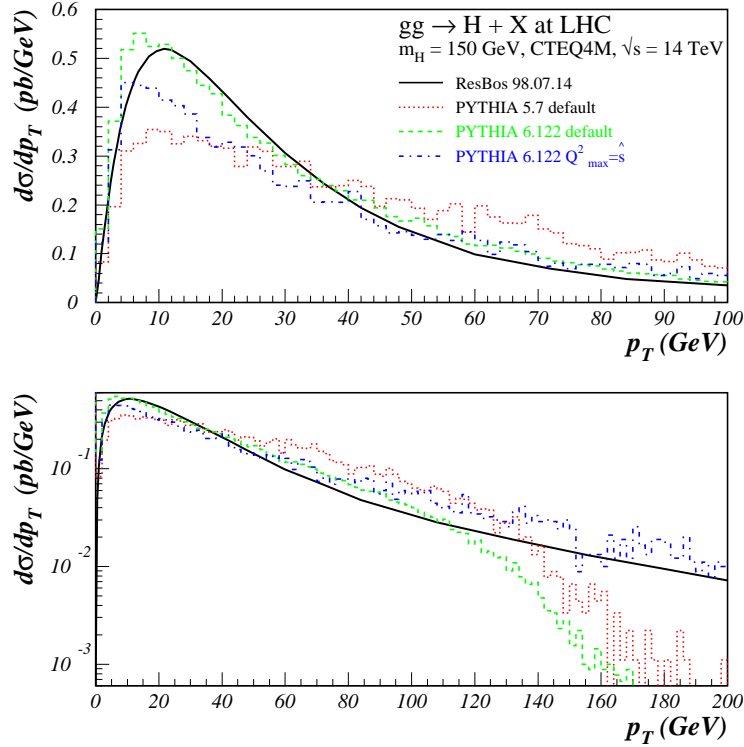}
\end{center}
\vspace{-1cm}
\caption{
A comparison of predictions for the Higgs $p_T$ distribution at the LHC from 
ResBos and from two recent versions of {\sc Pythia}. The ResBos and {\sc Pythia} 
predictions have
been normalized to the same area. 
\label{fig:resbos_pythia;3qcd}}
\end{figure}

A comparison of the Higgs $p_T$ distribution at the 
LHC~\cite{Balazs:2000sz}\footnote{
A more complete comparison of Monte Carlo and resummation treatments of
Z and Higgs 
production at both the Tevatron and the LHC can be found in 
Ref.~\cite{les-houches;11qcd}.
}, for a Higgs mass of 
150 GeV, is shown in Figure~\ref{fig:resbos_pythia;3qcd}, for 
ResBos~\cite{Balazs:2000wv} and the two 
recent versions of {\sc Pythia}. {\sc Pythia} has been rescaled 
to agree with the normalization of ResBos to allow for a better shape comparison. 
Note that the peak of the resummed distribution is at $p_T \approx$ 
11 GeV (compared to about 3 GeV for $Z^0$ production at the Tevatron). 
This is  partially due to  the larger mass (150 GeV compared to 90 GeV), 
but is primarily because of the larger color factors associated with 
initial state gluons ($C_A = 3$) rather than quarks ($C_F = 4/3$), and also 
because of the larger phase space for initial state gluon emission at the LHC. 
%
The newer version of {\sc Pythia} agrees well with ResBos at low to 
moderate $p_T$, but falls below the resummed prediction at high $p_T$. 
This is easily understood: ResBos switches to the NLO Higgs + jet matrix 
element~\cite{deFlorian:1999zd} at high $p_T$ while the default {\sc Pythia} can generate the 
Higgs $p_T$ distribution only by initial state gluon radiation, using as 
maximum virtuality the  Higgs mass squared. High $p_T$ Higgs production is 
another example where a $2 \to 1$ Monte Carlo calculation with parton 
showering can not completely reproduce the exact matrix element 
calculation, without the use of matrix element corrections
as already discussed in section~\ref{sec:corcella;3qcd}. The high $p_T$ 
region is better reproduced if the maximum virtuality $Q_{max}^2$ is set 
equal to the squared partonic center of mass energy, $s$, rather than 
$m_H^2$. This is equivalent to applying the PS to all of phase 
space. However, this has the consequence of depleting the low $p_T$ region 
as `too much' showering causes  events to migrate out of the peak.  The 
appropriate scale to use in {\sc Pythia} (or any Monte Carlo) depends on 
the $p_T$ range to be probed.  If matrix element information is used to 
constrain the behavior, the correct high $p_T$ cross section can be 
obtained while still using the lower scale for showering. The 
incorporation of matrix element corrections to Higgs production (involving 
the processes $gq \to qH$,$q{\overline{q}} \to gH$, $gg \to gH$) is the 
next logical project for the Monte Carlo experts, in order to accurately 
describe the high $p_T$ region.

The older version of {\sc Pythia} produces too many Higgs events at 
moderate $p_T$ (in comparison to ResBos). 
Two changes have been implemented in the newer version. The first change 
is that a cut is placed on the combination of $z$ and $Q^2$ values in a 
branching: $\hat{u} = Q^2-\hat{s}(1-z) < 0$, where $\hat{s}$ refers to the 
subsystem of the hard scattering plus  the shower partons considered to 
that point.  The association with $\hat{u}$ is relevant if the branching 
is interpreted in terms of a $2 \to 2$ hard scattering. The corner of 
emissions that do not respect this requirement occurs when the $Q^2$ value 
of the space-like emitting parton is little changed and the $z$ value of 
the branching is close to unity. This effect is mainly for the hardest 
emission (largest $Q^2$). The net result of this requirement is a 
substantial reduction in the total amount of gluon 
radiation~\cite{pythiaman:3qcd}~\footnote{Such branchings are kinematically allowed, but 
since matrix element corrections would assume initial state partons to 
have $Q^2=0$, a non-physical $\hat{u}$ results (and thus  no possibility 
to impose matrix element corrections). The correct behavior is beyond the 
predictive power of LL Monte-Carlos.}. In the second change, the 
parameter for the minimum gluon energy emitted in space-like showers is 
modified by an extra factor roughly corresponding to the $1/\gamma$ factor 
for the boost to the hard subprocess frame~\cite{pythiaman:3qcd}. The effect of 
this change is to increase the amount of gluon radiation. Thus, the two 
effects are in opposite directions but with the first effect being 
dominant. 

This difference in the $p_T$ distribution  between the two versions,
5.7 and 6.1, of 
{\sc Pythia} could have an impact on the analysis strategies for Higgs 
searches at the LHC~\cite{denegri:3qcd}.  
For example, for the CMS simulation of the Higgs search and the decay
into two photons it is envisaged to optimize the efficiency 
and the mass resolution for 
the high-luminosity running phase using charged particles 
with relatively large $p_{t}$, 
which balance the Higgs $p_{T}$ spectrum. 
These associated charged particles will allow to 
distinguish the Higgs event vertex from other vertices 
of unrelated proton--proton interactions 
with good accuracy. 
The efficiency of such an analysis 
strategy depends obviously on the knowledge of 
the Higgs $p_{T}$ spectrum and is thus somewhat 
sensitive to the used Monte Carlo parametrisation. 
\begin{figure}[t!]
\begin{center}
\vspace*{-2.cm}
\includegraphics[width=12cm]{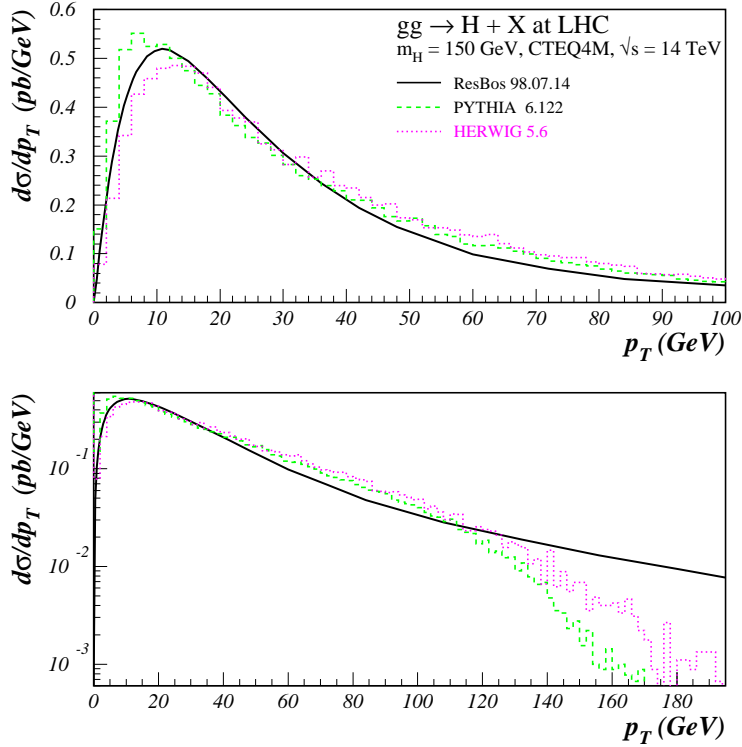}
\end{center}
\vspace{-1cm}
\caption{
 A comparison of predictions for the Higgs $p_T$ distribution at the
 LHC from 
ResBos, two recent versions of {\sc Pythia} and {\sc Herwig}. The ResBos, 
{\sc Pythia} and {\sc Herwig}
 predictions have been normalized to the same area.
\label{fig:comparison_lhc;3qcd}}
\end{figure}
\subsubsection{Comparison with {\sc Herwig}}
The variation between versions 5.7 and 6.1 of {\sc Pythia} gives an 
indication of the uncertainties due to the types of choices that can be 
made in Monte Carlos. The requirement that $\hat{u}$ be negative for all 
branchings is a choice rather than an absolute requirement.  Perhaps the 
better agreement of version 6.1 with ResBos is an indication that the 
adoption of the $\hat{u}$ restrictions was correct. Of course, there may 
be other changes to {\sc Pythia} which would also lead to better agreement 
with ResBos for this variable. 

Since there are a variety of choices that can be made in Monte Carlo 
implementations, it is instructive to compare the predictions for the 
$p_T$ distribution for Higgs production from ResBos and {\sc Pythia} with 
that from {\sc Herwig} (version 5.6, also using the CTEQ4M parton 
distribution functions). The {\sc Herwig} prediction is shown in 
Figure~\ref{fig:comparison_lhc;3qcd} along with the {\sc Pythia} and ResBos 
predictions, all normalized to the ResBos 
prediction\footnote{The normalization factors (ResBos/Monte Carlo) are {\sc 
   Pythia} (both versions)(1.61) and {\sc Herwig} (1.76). Figures of
the absolutely normalized predictions from ResBos,  {\sc 
Pythia}  and {\sc Herwig} for the $p_T$ distribution of 
the Higgs at 
the LHC can be found in Ref.~\cite{Balazs:2000sz}.}. 
In all cases, the CTEQ4M parton distribution was used. The predictions 
from {\sc Herwig} and {\sc Pythia} 6.1 are very similar, with the {\sc 
Herwig} prediction matching the ResBos shape somewhat better at low $p_T$. 
%
An understanding of the signature for Higgs boson production at either the 
Tevatron or LHC depends upon the understanding of the details of soft 
gluon emission from the initial state partons.  This soft gluon emission 
can be modelled either in a Monte Carlo or in a resummation calculation, 
with various choices possible in both implementations.  A comparison of the 
two approaches is useful to understand the strengths and weaknesses of 
each. The data from the Tevatron that either exists now, or will exist in 
Run 2, will be extremely useful to test both approaches.
 
In contrast to the case for Z production at the Tevatron, the 
Higgs cross section at the LHC is not particularly sensitive to the non--perturbative 
$k_T$ added at the scale $Q_0$. In the evolution to the hard scatter scale 
$Q$, the $k_T$ is `radiated away', given the enhanced gluon radiation 
probability present for a $gg$ initial state. For a more thorough discussion
of the comparison between analytic methods and parton showers, 
see Ref.~\cite{Balazs:2000sz}.

\subsection{{\sc CompHEP} for LHC\protect\footnote{
Contributing authors:  V.A.Ilyin and A.E.Pukhov.}}
\label{sec:comphep;3qcd}
The {\sc CompHEP} package is available from:
{\tt http://theory.npi.msu.su/\verb|~|comphep/}.
A version adapted to the LHC physics {\sc CompHEP} V.33~\cite{Pukhov:1999gg}, 
including executable Linux modules
is available at CERN from:
{\tt /afs/cern.ch/cms/physics/COMPHEP-Linux}.

The current {\sc CompHEP} version performs all calculation at tree level (LO).
Three issues must be discussed as they open several setting options:
a) the parton distributions, b) the QCD scale, and  
c) the running strong coupling.

In {\sc CompHEP} v.33, the following parton distribution sets
are implemented:  MRS(A') and MRS(G) \cite{Martin:1995ws},  CTEQ4l and 
CTEQ4m~\cite{Lai:1997mg}. Note that CTEQ4l is a LO parametrization, while in all others the
evolution of parton distributions is treated at NLO. 
Dedicated routines are available to allow the addition of any other 
defined parton distribution (e.g. CTEQ5).

As discussed in Sect.~\ref{sec:intro;qcd},
the factorization theorem states that the parton distribution 
depends not only on Bjorken variable $x$ but also on its virtuality $Q^2$ or,
equivalently, on the factorization scale.
This parameter is related to the energy (or momentum) scale 
which characterizes the hard subprocess, but it cannot be unambiguously fixed
(see Sect.~\ref{sec:intro;qcd}).
Therefore it can be
experimentally tuned. It can be set by the user for each specific QCD
process as either {\it fixed} or {\it running}. In the latter case, $Q^2$
can be set to any linear combination squared of the external
particles momenta (e.g. $(p_1-p_3)^2$, $(p_1-p_3-p_4)^2$, 
$(p_3+p_4)^2$ \ldots where initial and outgoing momenta enter with opposite signs).

In {\sc CompHEP} V.33, the QCD coupling $\as$ can be computed at 
LO, NLO or NNLO precision. All the corresponding
formulas are written in terms of $\Lambda_{\msbar}^{(6)}$, the fundamental QCD
scale for $N_f=6$ flavours of massless quarks (see Sect.~\ref{sec:intro;qcd}
and~\cite{Caso:1998tx}). 
In {\sc CompHEP}, to evaluate a QCD process, one first 
fixes the $\as$ normalization point (e.g. a popular normalization point  
is the mass of $Z$ boson, $Q=M_Z$) to which correspond an experimental
fit (e.g. $\as^{NLO}(M_Z)=0.118$). 
Then, the corresponding $\Lambda_{\msbar}^{(N_f)}$ ($N_f=5$ at $Q=M_Z$)
can be deduced from the $\as$ expression at the selected precision 
order. The {\sc CompHEP} input parameter $\Lambda_{\msbar}^{(6)}$ 
is then obtained 
from $\Lambda_{\msbar}^{(N_f)}$.
Finally, the choice of the QCD scale $Q$ determines 
$\as$ and the factorization scale for the \pdfs.
Therefore, complete LO calculations of LHC processes are made available 
for a consistent phenomenological analysis of the influence of higher order 
contributions.
\subsubsection{{\sc CompHEP-Pythia} interface}
An interface between {\sc CompHEP} and {\sc Pythia} can be found in:\\
{\tt /afs/cern.ch/cms/physics/COMPPYTH}.

A library of {\sc CompHEP} based partonic event generators for LHC 
processes has been initiated and various samples of event are available at:
{\tt /afs/cern.ch/cms/physics/PEVLIB}
for $Zb \bar b$,
$Wb \bar b$, $t\bar t b\bar b$ and some others. Unweighted event sample files,
located in the corresponding directories (see the files {\sf README} for
details) when handled by the {\sc CompHEP-Pythia} interface code, generate 
complete LHC events, ready to be fed to the detector simulation software.
For example, the $Zb \bar b$ process can be found 
in:
{\sf /afs/cern.ch/cms/physics/PEVLIB/Z\_b\_b}.
The file {\sf \_\_pevZbb} contains about 200K unweighted events. 
Each event is represented by the Lorentz 
momenta of all external particles. In the current version of the package,
there is no color information associated to the events. 
Thus, only the {\it Independent Fragmentation Model} can be invoked. 
One can always require the Lund model option for the fragmentation, 
as long as the corresponding color strings have been set by an external
algorithm in the routine {\tt PYUPEV}. 
The same remark applies also to the final
state radiations (FSR), which are, by default, switched off 
in {\sc CompHEP-Pythia} interface although initial state radiations (ISR)
are switched on. 
In the upcoming version of the {\sc CompHEP} package~\cite{newCompHEP:3qcd} 
color strings will be generated
from the matrix element factors allowing for the use of
the Lund fragmentation model.

\subsection{{\sc Grace} for LHC\protect\footnote{
Contributing author: K. Kato.}}\label{sec:grace;3qcd}
The URL of web page for the {\sc Grace} system is
{\tt http://www-sc.kek.jp/minami/}
where the latest information, the reports and manuals~\cite{Tanaka:1991wn,grace:3qcd},
the {\sc Grace} version.2 and the other 
products are available.

The automated system allows us to create event generators for 
complicated processes which are hard to calculate by hand.
For instance the process $gg \rightarrow b\bar{b}b\bar{b}$
has been calculated without any approximation (e.g. accounting for 
massive fermions) 
by use of the {\sc Grace} system~\cite{Tanaka:1991wn,grace:3qcd}.

The intrinsic function of the {\sc Grace} system
is to generate the amplitude for a specified
parton interaction. The system has been tested for many reactions 
and it was confirmed to be able to manage 2-body to 6-body
final state processes. The interface with the \pdfs, 
PS and the fragmentation
tools will be implemented in the coming versions
(see for example {\sc Grape} for
$ep\rightarrow \ell\bar{\ell}X$~\cite{Abe:1998km}).
For the parton showering and the fragmentation, two kinds of approach
can be followed.
The first is, like in {\sc Grape}  a 2 step procedure:
the {\sc Bases/Spring} package including \pdfs\ is used for the integration
over the phase space and for the generation of
unweighted events. If the ``kinematics'' code is
appropriate, {\sc Spring} generates events
with high efficiency and writes the four-vectors
of the final-state particles on a temporary file. Then
the generated momenta are passed to {\sc Pythia} for PS and fragmentation.
The other approach is more convenient but more complex.
Here
the code including the kinematics
and the generated matrix element is prepared
so that {\sc Pythia} can drive them directly.
This type of interface is tested till now only for the processes
whose final state consist of 2-, 3- and 4-bodies.

The {\sc Grace} system can automatically
deal with one-loop processes (NLO)
for the electroweak and QED-like QCD interactions.
For the final two-body processes the performance has been shown to be 
good. The application to the multi-body final states,
however, would be limited because of the huge CPU time 
required when the code is used as event generators. 
For such cases a practical use of the generated code 
will be to evaluate the cross sections and to give the
distribution of several physical quantities rather than
providing event generators.

\def\bar{\overline}
As mentioned the contributions beyond LO are crucial for 
a detailed QCD study. Since the PS method is based on the
renormalization group equation, it works as a bridge between the ``hard'' 
parton collision and the fragmentation. This bridge is built on the solid 
and reliable ground of perturbative QCD. In other words the parton 
shower provides an unambiguous theoretical understanding of $pp(\bar{p})$ 
interactions except for the ``soft'' component which cannot be controlled 
by the perturbative QCD. However, the PS in LL order is not 
enough. One of the shortcomings is as follows. The \pdfs\ for the initial 
state, products of elaborated works, are parameterized according to 
the NLO QCD formulas. On the other hand the corresponding PS, 
implemented in the existing programs like {\sc Pythia}, is evolved using only 
the LL algorithm at least in their current status. Then the systematic 
summation of large logarithms up to NLL order must solve this annoying 
situations. Though the basic technology has been already established and 
known for many years~\cite{Kalinowski:1981we,Kalinowski:1981ju,Kato:1987sg}, its implementation is not a trivial task 
as simply imagined. First it is process-dependent. Once the idea evolves 
and is realized as one of the environments of {\sc Grace}, it should 
allow more precise prediction for LHC. Thus this must be the biggest 
issue to us.

\subsection{{\sc Alpha} for LHC\protect\footnote{
Contributing authors:  M.L. Mangano and M. Moretti.}}\label{sec:alpha;3qcd}

As discussed in the introduction to this Section, the ability to
evaluate production rates for multi-jet final states will be
fundamental at the LHC to study a large class of processes, within and
beyond the SM. As was also discussed in the Sect.~\ref{sec:intromc;qcd}, 
a necessary
feature of any multi-jet calculation is the possibility to properly
evolve the purely partonic final state, for which exact fixed-order
perturbative calculations can be performed, into the observable
hadronic final state. This evolution is best performed using shower
Monte Carlo calculations. The accurate description of color-coherence
effects, furthermore, requires as noticed in the introduction a
careful bookkeeping of the contribution to the matrix elements of all
possible color configurations. The goal of the algorithm~\cite{Caravaglios:1999yr}
described in this Section is to allow the effective calculation of
multi-parton matrix elements, allowing the separation, to the leading
order in $1/N_c^2$ ($N_c=3$ being the number of colors), of the
independent color configurations. This technique allows an
unweighting of the color configurations, and allows the merging of
the parton level calculation with the {\sc Herwig} Monte Carlo.

The key element of the strategy is the use of the algorithm {\tt
ALPHA}, introduced in Ref.~\cite{Caravaglios:1995cd} for the evaluation of
arbitrary multi-parton matrix elements. This algorithm determines the
matrix elements from a (numerical) Legendre transform of the effective
action, using a recursive procedure which does not make explicit use
of Feynman diagrams.  The algorithm has a complexity growing like a
power in the number of particles, compared to the factorial-like
growth that one expects from naive diagram counting.  This is a
necessary feature of any attempt to evaluate matrix elements for
processes with large numbers of external particles, since the number
of Feynman diagrams grows very quickly beyond any reasonable
value. For example, this calculation allowed~\cite{Caravaglios:1999yr} the
evaluation of the matrix elements for the production of 8-gluon final
states.  The number of Feynman diagrams which describe this process
exceeds 7 billion.

The interface of the parton level scattering matrix element with the PS
requires the capability to reconstruct the appropriate
color flow for a given event.
The strategy to deal with this issue is described in detail in 
\cite{Caravaglios:1999yr}.
The following points have to be noticed:
\begin {enumerate}
\item
Dual amplitudes~\cite{Mangano:1987vj,Berends:1987cv,Mangano:1988xk} can be easily evaluated using the {\tt
ALPHA} algorithm.  Since the dual amplitudes $A$ are independent of
the number $N_c$ of colors, they can be calculated exactly by taking
$N_c$ sufficiently large.

\item
With an appropriate choice for the color of the external partons, the
full amplitude is proportional to a single dual amplitude.
\end {enumerate}

We explicitly calculated $n$-gluon dual amplitudes using the
large-$N_c$ Lagrangian. The correctness of the calculation was checked 
for $n$ up to 11 by comparing the results for maximally helicity
violating (MHV) amplitudes~\cite{Parke:1986gb} (e.g. $g^+ g^+ \to g^+
\cdots g^+$) with the analytic expressions known exactly for arbitrary
$n$~\cite{Mangano:1987vj,Berends:1987cv,Mangano:1988xk,Berends:1988me}.  The input of the numerical evaluation
of the matrix element is a string containing the total number of
gluons, their helicity state, and their momenta. From these data, the
amplitude is evaluated automatically.

The prescription to correctly generate the parton-shower
associated to a given event in the large-$N_c$ limit is therefore the
following:
\begin{enumerate}
\item Calculate the $(n-1)!$ dual amplitudes corresponding to all
possible planar color configurations.
\item Extract the {\em most likely} color configuration for this
event on a statistical basis, according to the relative contribution
of the single configurations to the total event
weight~\footnote{Defining $w_i=\vert A_i \vert^2$ for each color flow
$i$, and $W_i=\sum_{k=1,\dots,i} \, w_k/\sum_{k=1,\dots,n} \, w_k$,
the $j$-th color structure will be selected if $W_{j-1} \le \xi <
W_j$, for a random number $\xi$ uniformly distributed over the
interval $[0,1]$.}. Since each dual amplitude is gauge invariant, the
choice of color-configurations is also a gauge-invariant operation.
\item Develop the PS out of each initial and final-state
parton, starting from the selected color configuration. This step can
be carried out by feeding the generated event to a Monte-Carlo program
such as {\sc Herwig}, which is precisely designed to {\em turn partons
into jets} starting from an assigned color-ordered configuration.
\end{enumerate}                                                        
Notice that, if the dual amplitudes are evaluated for a specific
helicity configuration, {\sc Herwig} will also include
spin-correlation effects in the evolution of the parton
shower~\cite{Collins:1988cp,Knowles:1988vs,Marchesini:1988cf,Marchesini:1992ch,Marchesini:1996vc}.

As a result, use of the dual-amplitude representation of a multi-gluon
amplitude allows to accurately describe not only the large-angle
correlations in multi-jet final states, but also the full shower
evolution of the initial- and final-state partons with the same
accuracy available in {\sc Herwig} for the description of 2-jet
final states.

In alternative to the above prescription, one can use {\tt ALPHA} to
calculate the matrix elements for external states with assigned
colors. Since these states are all orthogonal, such an approach is
particularly efficient if one wants to use a Monte Carlo approach to
the summation over all possible color states.  The program will then
extract through a standard unweighting (at the leading order in
$1/N_c^2$) a specific color flow from all possible color flows
contributing to a given orthogonal color state.  This color flow is
then suitable as an initial condition for the shower evolution.
Further details can be found in~\cite{Caravaglios:1999yr}. At this time, the program
is only available in its parton-level form, and allows the calculation
of matrix elements for $gg\to g\dots g$ and $q \bar{q} \to g\dots g$
processes, with up to 8 final-state gluons. A full version including
the interface with {\sc Herwig} is being prepared.


\section{AVAILABLE NLO CALCULATIONS AND PROSPECTS AT NNLO\protect\footnote{Section
    coordinators: V. Del Duca, D. Soper and W.J. Stirling.}}
\label{sec:focal;qcd}

%
%

\subsection{Available NLO calculations of multijet processes\protect\footnote{
Contributing authors: V.~Del Duca and S.~Frixione.}}
\label{sec:nlo}

QCD calculations of multijet\footnote{For the sake of brevity, in this
section we will term as multijet any kind of (partly) hadronic final
state.} processes beyond LO in the strong coupling
constant $\as$ are quite involved.  Nowadays we know (see below)
how to perform in general calculations of the
NLO corrections to multijet processes, and almost
every process of interest has been computed to that accuracy. Instead,
the calculation of the NNLO
corrections is still at an organisational stage and represents a main
challenge.  Why should we perform calculations which are technically
so complicated ?

The general motivation is that the calculation of the NLO corrections
allows us to estimate reliably a given production rate, while the NNLO
corrections allow us to estimate the theoretical uncertainty on the
production rate. That comes about because higher-order corrections
reduce the dependence of the cross section on the renormalization
scale, $\mu_R$, and for processes with strongly-interacting incoming
particles the dependence on the factorization scale, $\mu_F$, as well.

An example is the determination of $\as$ from
event shape variables in $e^+ e^- \rightarrow 3$ jets~\cite{Burrows:1996db,
Schmelling:1996wm,Catani:1997rn,Bethke:2000rw}.
The calculation of the
NNLO contributions to this process would be needed to further reduce 
the theoretical
uncertainty in the determination of $\as$.
An additional motivation for performing calculations at NNLO is to obtain
a more accurate theoretical determination of signal and QCD background
to Higgs production (for further details, see Sect.~\ref{sec:backg;qcd}).

In recent years it has become clear how to construct 
general-purpose algorithms for the calculation of multijet processes
at NLO accuracy.
The crucial point is to organise the cancellation of the infrared
(i.e. collinear and soft) singularities of the QCD amplitudes
in a universal, i.e. process- and observable-independent, way. 
The universal terms in a NLO calculation are given by the tree-level 
collinear~\cite{Gribov:1972ri,Lipatov:1975qm,Altarelli:1977zs,
Dokshitzer:1977sg} and soft~\cite{Yennie:1961ad,Bassetto:1983ik,
Berends:1989zn} 
functions, and by the universal structure of the poles of the one-loop
amplitudes~\cite{Giele:1992vf,Kunszt:1992tn,Kunszt:1994mc}. 
The universal NLO terms and the 
process-dependent amplitudes are combined into 
effective matrix elements, which are devoid of singularities. The various NLO 
algorithms (phase-space 
{\em slicing}~\cite{Giele:1992vf,Giele:1993dj,Glover:1995vz,Keller:1999tf} and
{\em subtraction} method~\cite{Frixione:1996ms,Nagy:1997bz,Frixione:1997np,
Catani:1996jh,Catani:1997vz}) provide different methods to
construct the effective matrix elements. These
can be integrated in four dimensions, in practice almost always numerically, 
due to the complexity of the integrand. 
The integration can be performed with arbitrary experimental acceptance cuts.

We now outline how to perform a NLO calculation of a generic physical
observable. As is well known from Bloch-Nordsieck and
Kinoshita-Lee-Nauenberg theorems, QCD (like QED) does not have an
infinite-resolution power; any attempt to compute the kinematical
properties of a fixed number of final-state quarks and gluons results
in an infrared-divergent cross section. In order to obtain finite
quantities, all the partonic subprocesses which contribute to the same
order in $\as$ to the squared amplitude have to be included in the
computation, regardless of the number of final-state particles. In
addition, one is forced to consider variables which are inclusive
enough to be {\em infrared safe}. Roughly speaking, an observable 
is said to be infrared safe when its value, computed with the kinematical
variables of the final-state partons, does not change abruptly when
a soft gluon is emitted, or a parton splits almost collinearly
into a pair of partons. More technically, an infrared-safe observable
must have a smooth limit (that is, must behave continuously) in the
following three configurations: {\it a)} when a gluon in the final
state gets soft; {\it b)} when two partons in the final state tend
to get collinear to each other; {\it c)} when an initial-state
parton emits collinearly another parton.

At NLO (assuming that the LO cross section gets contributions from
the $n$-parton amplitudes), this implies that one has simply to 
consider two contributions, denoted as virtual and real.
The former is the product of the $n$-parton one-loop amplitudes
with the $n$-parton tree amplitudes, while the latter is the square
of the $(n+1)$-parton tree amplitudes. In order to deal with finite quantities 
in the intermediate steps of the calculation, we adopt dimensional 
regularization -- i.e. we change the dimensionality of space-time 
to $d=4-2\ep$. Thus, we can schematically write the virtual and real
contributions to the cross section as follows:
\beq
\left(\frac{d\sigma}{dx}\right)_{\sss V}=\frac{1}{2\ep}\,\delta(1-x)\, ,
\qquad \qquad
\left(\frac{d\sigma}{dx}\right)_{\sss R}=\frac{1}{1-x}\,;
\eeq
here, $1-x$ represents the radiated energy. So, $x=1$ means no radiation, 
and $x=0$ is the maximum of radiation. The relevant physical quantity
will be the average value $<F>$ of a certain function $F(x)$; for example,
we can think of $F$ as being the product of theta functions representing 
a histogram bin. Then, the NLO contribution to $<F>$ is
\beeq
<F>_{\sss\rm NLO}&=&
\int_0^1 dx\,\left(\frac{d\sigma}{dx}\right)_{\sss V}\!F(x)+
\int_0^1 dx\,(1-x)^{-2\ep}\,\left(\frac{d\sigma}{dx}\right)_{\sss R}\!F(x)
\\
&=&\frac{1}{2\ep}\int_0^1 dx\,\delta(1-x)\,F(x)+
\int_0^1 dx\,(1-x)^{-1-2\ep}\,F(x)
\label{Fav}
\\
&=&\frac{1}{2\ep}\,F(1)+<F>_{\sss R}\,.
\label{Fres}
\eeeq
The factor $(1-x)^{-2\ep}$ in the real contribution comes from the 
necessity of performing the computation in $d$ dimensions,
in order to regulate the divergences arising when performing the
integration over the phase space. As it is apparent from eq.~(\ref{Fav}),
the most difficult task is the computation of the real contribution.
In practice, the form of $F(x)$ is too complicated to perform an
analytical integration. On the other hand, we cannot proceed
straightforwardly, and compute the integral numerically; in fact,
the integral is divergent in the limit $\ep\to 0$, and the pole in
$1/\ep$ will exactly cancel that explicitly displayed in the virtual 
contribution (provided that $F$ describes an infrared-safe quantity).

Two strategies have been developed to tackle this problem.
In the framework of the {\em slicing} method, the real contribution
is rewritten as follows:
\beq
<F>_{\sss R}=\int_0^{1-\delta} dx\,\frac{F(x)}{(1-x)^{1+2\ep}}
+\int_{1-\delta}^1 dx\,\frac{F(x)}{(1-x)^{1+2\ep}},
\eeq
where $\delta$ is an arbitrary parameter, $0<\delta\le 1$. The first term on 
the right hand side of this equation is free of divergences ($F(x)$ is regular
in the limit $x\to 1$); in this term, one can therefore set $\ep=0$, and 
compute the integral with standard numerical methods. On the other hand, 
the second term is still divergent for $\ep\to 0$; however, if $\delta$ 
is small enough, one can approximate $F(x)$ with $F(1)$ (that is, with 
the first term of its Taylor expansion around $x=1$). Therefore
\beeq
<F>_{\sss R}&=&\int_0^{1-\delta} dx\,\frac{F(x)}{1-x}
+F(1)\int_{1-\delta}^1 dx\,\frac{1}{(1-x)^{1+2\ep}}
+{\cal O}(\delta)
\label{approxslc}
\\
&=&\int_0^{1-\delta} dx\,\frac{F(x)}{1-x}
-\frac{\delta^{-2\ep}}{2\ep}F(1)
+{\cal O}(\delta)\,.
\label{resslc}
\eeeq
Eq.~(\ref{resslc}) can now be substituted into eq.~(\ref{Fres}).
Expanding eq.~(\ref{resslc}) in powers of $\ep$, keeping only the 
terms which do not vanish in the limit $\ep\to 0$, and neglecting
the contributions of the terms of ${\cal O}(\delta)$, we see that
the pole terms in $1/\ep$ cancel, and one is left with a finite result:
\beq
<F>_{\sss\rm NLO}^{slicing}=\int_0^{1-\delta} dx\,\frac{F(x)}{1-x}
+F(1)\log\delta.
\label{FNLOslc}
\eeq
At a first glance, this expression is seemingly puzzling: the
parameter $\delta$ is arbitrary, and the physical results should not
depend on it. However, it is easy to see that the upper bound of the
integral gives a contribution behaving (approximately) like
$-F(1)\log\delta$. It has to be stressed that the slicing method is
based on the approximation performed in eq.~(\ref{approxslc}); for
this approximation to hold, it is crucial that $\delta$ is as small as
possible; otherwise, the terms collectively denoted with ${\cal O}(\delta)$ 
in eq.~(\ref{resslc}) are not negligible. On the other hand, 
in practical computations, the integral in eq.~(\ref{FNLOslc})
is performed numerically; due to the divergence of the integrand for
$x\to 1$, $\delta$ cannot be taken too small, because of the loss of
accuracy of the numerical integration. Thus, the value of $\delta$ is
a compromise between these two opposite requirements, being neither
too small nor too large. Of course, ``small'' and ``large'' are
meaningful only when referred to a specific computation. Therefore,
when using the slicing method, it is mandatory to check that the
physical results are stable against the variation of the value of
$\delta$, chosen in a suitable range. In principle, this check would
have to be performed for each observable $F$ computed; in practice,
only one observable is checked, generally chosen to be rather
inclusive (such as a total rate).

Another possibility to compute $<F>_{\sss R}$ is given by the 
{\em subtraction} method. One writes
\beq
<F>_{\sss R}=\int_0^1 dx\,\frac{F(x)-F(1)\theta(x-1+x_c)}{(1-x)^{1+2\ep}}
+F(1)\int_0^1 dx\,\frac{\theta(x-1+x_c)}{(1-x)^{1+2\ep}},
\label{subt}
\eeq
where $x_c$ is an arbitrary parameter $0<x_c\le 1$. The first term on the 
right hand side of this equation is convergent, and we can set $\ep=0$.
The second term is formally identical to the one appearing in
eq.~(\ref{approxslc}). Notice, however, that no approximation has
been made in eq.~(\ref{subt}); the price to pay is a more complicated
expression for the first integral. Proceeding as before, we get:
\beq
<F>_{\sss\rm NLO}^{subt}=\int_0^1 dx\,\frac{F(x)-F(1)\theta(x-1+x_c)}{1-x}
+F(1)\log x_c.
\label{FNLOsubt}
\eeq
This equation has to be compared to eq.~(\ref{FNLOslc}); although the
two are quite similar, there are two important differences that have
to be stressed. Firstly, the parameter $x_c$ introduced in the subtraction
method does not need to be small (actually, in the original
formulation of the method $x_c$ was not even introduced, which 
corresponds to set $x_c=1$ here). This is due to the fact that in
the subtraction method no approximation has been performed in the
intermediate steps of the computation. This in turn implies the
second point: there is no need to check that the physical results
are independent of the value of $x_c$, since this is true by
construction. 

The universal algorithms previously mentioned allow the computation
of any infrared-safe observable in a straightforward manner; the matrix 
elements do not need any algebraic manipulation, and can be computed in 
four dimensions. It is therefore relatively easy to construct computer
codes, accurate to NLO in QCD, that are flexible enough to become
a useful tool in the analysis of the experimental data. In the following,
we will list the codes which are of direct interest for the physics
of high-energy hadronic collisions. We do not intend to give a complete
list of references to the papers relevant for the calculation of a
given production process~\footnote{Further details on codes involving the 
production of a single vector
boson and of a Higgs boson can be found in Sect.~\ref{sec:photons;qcd} and 
\ref{sec:backg;qcd}, respectively.}, 
but rather only to quote the computer codes
which will have a chance to be used by the experimental collaborations at 
the LHC. Most of the codes listed here are available as free software.

\begin{itemize}

\item Jets
\vskip.15cm
\begin{itemize}
\item S.D.~Ellis, Z.~Kunszt and D.E.~Soper~\cite{Kunszt:1992tn,Ellis:1992en}, 
{\em subtraction}, computes one- and two-jet observables.
\item W.T.~Giele, E.W.N.~Glover and D.A.~Kosower (JETRAD)~\cite{Giele:1993dj},
{\em slicing}, computes one- and two-jet observables.
\item S.~Frixione~\cite{Frixione:1997np}, {\em subtraction},
computes one- and two-jet observables.
\item W.~Kilgore and W.T.~Giele~\cite{Kilgore:1999qg},
{\em slicing}, computes three-jet observables.
\end{itemize}
\vskip.2cm

\item Single Isolated Photon (plus one jet)
\vskip.15cm
\begin{itemize}
\item H.~Baer, J.~Ohnemus and J.F.~Owens~\cite{Baer:1990ra}, 
{\em slicing}, fragmentation contribution computed to LO accuracy.
\item L.E.~Gordon and W.~Vogelsang~\cite{Gordon:1994ut}, analytical 
integration over the variables of the recoiling partons: no information 
on the accompanying jet; dependence on the isolation variables treated to
logarithmic approximation.
\item S.~Frixione~\cite{Frixione:1998hn}, {\em subtraction}, only effective
with the isolation prescription of ref.~\cite{Frixione:1998jh}.
\item M.~Werlen (PHONLL) [http://home.cern.ch/\~{}monicaw/phonll.html],
{\em slicing}, based on ref.~\cite{Aurenche:1988fs,Aversa:1989vb}.
\end{itemize}
\vskip.2cm

\item Isolated-Photon Pairs
\vskip.15cm
\begin{itemize}
\item B.~Bailey, J.F.~Owens and J.~Ohnemus~\cite{Bailey:1992br},
{\em slicing}, fragmentation contributions computed to LO accuracy.
\item C.~Balazs, E.L.~Berger, S.~Mrenna and 
C.P.~Yuan~\cite{Balazs:1998xd}, {\em slicing},
resummation effects included, fragmentation 
contributions computed with parton shower methods.
\item T.Binoth, J.Ph.~Guillet, E.~Pilon and M.Werlen 
(DIPHOX)~\cite{Binoth:1999qq}, {\em slicing}, all contributions
computed to NLO accuracy.
\end{itemize}
\vskip.2cm

\item Single Heavy Vector Boson (plus one jet)
\vskip.15cm
\begin{itemize}
\item W.T.~Giele, E.W.N.~Glover and D.A.~Kosower (DYRAD)~\cite{Giele:1993dj},
{\em slicing}.
\end{itemize}
\vskip.2cm

\item Single Heavy Vector Boson plus one photon
\vskip.15cm
\begin{itemize}
\item U.~Baur, T.~Han, J.~Ohnemus~\cite{Baur:1993ir,Baur:1998kz}, 
{\em slicing}.
\item D.~de~Florian and A.~Signer~\cite{DeFlorian:2000sg}, {\em subtraction},
includes spin correlations in the decay of the bosons; fragmentation 
contributions computed to LO accuracy.
\end{itemize}
\vskip.2cm

\item Heavy Vector Boson Pairs
\vskip.15cm
\begin{itemize}
\item U.~Baur, T.~Han, J.~Ohnemus and J.F.~Owens~\cite{Ohnemus:1991za,
Ohnemus:1991kk,Ohnemus:1991gb,Baur:1995aj,Baur:1996uv}, {\em slicing}.
\item S.~Frixione, B.~Mele, P.~Nason and G.~Ridolfi~\cite{Mele:1991bq,
Frixione:1992pj,Frixione:1993yp}, {\em subtraction}.
\item J.M.~Campbell and R.K.~Ellis (MCFM)~\cite{Campbell:1999ah}, 
{\em subtraction}, includes spin correlations in the decay of the bosons.
\item L.~Dixon, Z.~Kunszt and A.~Signer~\cite{Dixon:1999di}, {\em subtraction},
includes spin correlations in the decay of the bosons.
\end{itemize}
\vskip.2cm

\item Higgs Boson at large transverse momentum (plus one jet)
\vskip.15cm
\begin{itemize}
\item D.~de~Florian and M.~Grazzini and Z.~Kunszt~\cite{deFlorian:1999zd},
{\em subtraction}, computes Higgs-boson production in the infinite 
top-quark-mass limit.
\end{itemize}
\vskip.2cm

\item Heavy Quarks
\vskip.15cm
\begin{itemize}
\item M.~Mangano, P.~Nason and G.~Ridolfi~\cite{Mangano:1992jk}, 
{\em subtraction}, computes single-inclusive distribution and correlations 
betweeen $Q$ and $\bar{Q}$.
\end{itemize}

\end{itemize}

Since the universal algorithms accomplish the task of cancelling the
infrared divergences of the virtual and real contributions in a process-%
independent way, the remaining work that has to be performed to
calculate a production rate at NLO is the computation of the
appropriate tree and one-loop amplitudes. As we said previously,
to compute $n$-jet production at NLO, two sets of amplitudes
are required: {\it a}) $n$-particle production amplitudes at tree
level and one loop; {\it b}) $(n+1)$-particle production
amplitudes at tree level. If the one-loop amplitudes are regularised 
through dimensional regularisation, it suffices at NLO to compute them 
to ${\cal O}(\epsilon^0)$.

Efficient methods based on the color decomposition~\cite{Mangano:1991by,
Bern:1991ux,DelDuca:1999ha,DelDuca:1999rs} of an amplitude in color-ordered
subamplitudes, which are then projected onto the helicity states of the
external partons, have largely enhanced the ability of computing 
tree~\cite{Mangano:1991by} and one-loop~\cite{Bern:1996je} amplitudes.
Accordingly, tree amplitudes with up to seven massless
partons~\cite{Mangano:1991by,Berends:1990hf,Kuijf:1991kn} and with a vector boson and up to five massless partons~\cite{Berends:1989yn}
have been computed analytically. In addition, efficient
techniques to evaluate numerically tree multi-parton amplitudes have been 
introduced~\cite{Draggiotis:1998gr,Caravaglios:1999yr}
(see Sect.~\ref{sec:mcs;qcd} for a description of available numerical codes), 
and have been used to compute tree amplitudes
with up to eleven massless partons~\cite{Caravaglios:1999yr}.
The calculation of one-loop amplitudes can be reduced to the
calculation of one-loop $n$-point scalar integrals~\cite{Passarino:1979jh,vanNeerven:1984vr,vanOldenborgh:1990wn}.
The reduction method~\cite{Passarino:1979jh} allowed the computation
of one-loop amplitudes with four massless 
partons~\cite{Ellis:1986er} and with a vector boson and three massless 
partons~\cite{Ellis:1981wv}.
However, one-loop scalar integrals present infrared divergences, 
induced by the massless external legs.
For one-loop multi-parton amplitudes, the infrared divergences hinder 
the reduction methods of 
ref.~\cite{Passarino:1979jh,vanNeerven:1984vr,vanOldenborgh:1990wn}.
This problem has been overcome in ref.~\cite{Bern:1993em,Bern:1994kr}.
Accordingly, one-loop amplitudes with five massless 
partons~\cite{Bern:1993mq,Bern:1995fz,Kunszt:1994tq} and with a vector boson and four massless 
partons~\cite{Bern:1996fj,Bern:1998sc,Bern:1997ka,Glover:1997eh,Campbell:1997tv} have been computed analytically. 
The reduction procedure of ref.~\cite{Bern:1993em,Bern:1994kr} 
has been generalised
in ref.~\cite{Binoth:1999sp}, where it has been shown that any 
one-loop $n$-point scalar integral, with $n> 4$, can be reduced
to box scalar integrals.
The calculation of one-loop multi-parton amplitudes thus can be
pushed a step further in the near future.

\subsection{Prospects for NNLO calculations\protect\footnote{
Contributing authors: V.~Del~Duca and G.~Heinrich.}}
\label{sec:nnlo}

Eventually, a procedure similar to the one followed at NLO will permit the 
construction of general-purpose algorithms at NNLO accuracy.
It is mandatory then to fully investigate the infrared structure of the
matrix elements at NNLO. The universal pieces needed to organise the 
cancellation of the infrared singularities are given by the tree-level 
triple-collinear~\cite{Campbell:1998hg,Catani:1999nv,DelDuca:1999ha}, 
double-soft~\cite{Berends:1989zn,Catani:2000ss} and 
soft-collinear~\cite{Campbell:1998hg,Catani:2000ss} functions, 
by the one-loop 
splitting~\cite{Bern:1998sc,Bern:1999ry,Kosower:1999xi,Kosower:1999rx} and 
eikonal~\cite{Bern:1998sc} functions,
and by the universal structure of the poles of the two-loop
amplitudes~\cite{Catani:1998bh}. These universal pieces have yet to be 
assembled together, to show the cancellation of the infrared divergences
at NNLO.

Then to compute $n$-jet production at NNLO, three sets of amplitudes
are required: {\it a}) $n$-particle production amplitudes at tree
level, one loop and two loops; {\it b}) $(n+1)$-particle production
amplitudes at tree level and one loop; {\it c}) $(n+2)$-particle
production amplitudes at tree level. 
In dimensional regularisation at NNLO, the two-loop amplitudes 
need be computed to ${\cal O}(\epsilon^0)$, while the one-loop amplitudes 
must be evaluated to ${\cal O}(\epsilon^2)$~\cite{Bern:1998sc,Bern:1995ix}. 
The main challenge is the calculation of the two-loop amplitudes.
At present, the only amplitude known at two loops is the one for $V 
\leftrightarrow q\bar q$~\cite{Kramer:1987sg,Matsuura:1988wt,Hamberg:1991np}, 
with $V$ a massive vector boson,
which depends only on one kinematic variable. It has been used to evaluate
the NNLO corrections to Drell-Yan production~\cite{Hamberg:1991np,
vanNeerven:1992gh} and to 
deeply inelastic scattering (DIS)~\cite{Zijlstra:1992qd,Zijlstra:1992kj}.
Two-loop computations for configurations involving two kinematic 
variables, which are needed in the case of parton-parton scattering,
exist only in the special cases of maximal 
supersymmetry~\cite{Bern:1997nh}, and of maximal helicity
violation~\cite{Bern:2000dn}. The latter contributes only beyond NNLO. 
One of the main obstacles 
for configurations involving two kinematic variables
is the analytic computation of the two-loop four-point functions
with massless external legs, where significant progress has just been 
achieved. 
These consist of planar double-box 
integrals~\cite{Smirnov:1999gc,Smirnov:1999wz}, 
non-planar double-box integrals~\cite{Tausk:1999vh},
single-box integrals with a bubble insertion on one of the 
propagators~\cite{Anastasiou:1999cx} and 
single-box integrals with a vertex correction~\cite{Anastasiou:1999bn}. 
Finally, processes such as
$e^+ e^- \rightarrow 3\, jets$ and $p\,p\to H\, jet$
sport configurations involving three kinematic variables and require
the analytic computation of two-loop four-point functions  with a massive 
external leg. Some of the required two-loop four-point functions of this kind
have been derived recently~\cite{Gehrmann:1999as}.
Another obstacle is the color decomposition of
two-loop amplitudes, which is not generally known yet. 
Substantial progress is expected in the near future on all the 
issues outlined above, which should make the present note soon outdated.

Finally, we mention that in the factorization of collinear 
singularities for strongly-interacting incoming particles,
the evolution of the \pdfs\ in the jet cross section should be 
determined to an accuracy matching the one of the parton cross section.
For hadroproduction of jets computed at NLO, one needs the NLO AP splitting
functions for the evolution of the \pdfs\ 
(see Eqs.~(\ref{evequa;1qcd}) and (\ref{apexp;1qcd})).
Accordingly, for hadroproduction at NNLO the evolution 
of the \pdfs\ should be computed using the NNLO AP splitting functions.
Except for the 
lowest five (four) even-integer moments of the NNLO non-singlet (singlet)
AP splitting functions~\cite{Larin:1994vu,Larin:1997wd},
no calculation of the NNLO evolution of the \pdfs\ exists yet. 
Some NNLO analyses based on the finite set of known moments 
have been performed for the DIS structure functions
$xF_3$ and $F_2$ (see Sects.~\ref{sec:vogt;2qcd} and \ref{sec:kataev;2qcd}
and Ref.~\cite{Santiago:1999pr}).
Furthermore, in ref.~\cite{vanNeerven:1999ca} a quantitative assessment of the 
importance of the yet unknown higher-order terms has been performed, 
with the conclusion that they should be numerically significant only for 
Bjorken $x$ smaller than $10^{-2}$. 

The computation of the evolution kernels of the 
\pdfs\ at NNLO accuracy is a major challenge in QCD.
The NLO computation was performed with two different methods, one using
the  operator product expansion (OPE) in a covariant
gauge~\cite{Floratos:1977au,Floratos:1979ny,Gonzalez-Arroyo:1979df,
Gonzalez-Arroyo:1980he,Floratos:1981hs}, the other using the 
light-cone axial (LCA) gauge with principal value prescription~
\cite{Curci:1980uw,Furmanski:1980cm}.
However, the prescription used in ref.~\cite{Curci:1980uw,Furmanski:1980cm} 
has certain shortcomings. Accordingly, the calculation has been repeated 
in the LCA gauge using a prescription~\cite{Heinrich:1998kv,Bassetto:1998uv} 
which makes it amenable to extensions beyond NLO, whereas the principal value
prescription does not seem to be applicable beyond NLO~\cite{drag}.
On the other hand, using the OPE method, there had been a problem with 
operator mixing in the singlet sector, which  has been 
fixed~\cite{Hamberg:1992qt,Collins:1994ee,Matiounine:1998ky} only recently, 
and the result finally coincides with  
the one obtained in the LCA gauge in ref.~\cite{Furmanski:1980cm}.
Thus the result for the AP splitting functions at NLO accuracy is fully
under control. Recent proposals for their calculation beyond  NLO
include extensions of the OPE technique, which have been used to recompute the 
NNLO corrections to DIS~\cite{Moch:1999eb}, and a
computation based on combining universal 
gauge-invariant collinear pieces~\cite{uwer}.


\section{SUMMATIONS OF PERTURBATION THEORY\protect\footnote{Section
    coordinator: L. Magnea.}}
\label{sec:rescal;qcd}

\subsection{Summations of logarithmically-enhanced 
contributions\protect\footnote{  
Contributing author: S. Catani.}}
\label{sec:introres;qcd}

The calculation of hard--scattering cross sections in hadron collisions
requires the knowledge of partonic cross sections ${\hat \sigma}$, 
as well as that of parton densities (see the factorization formula in 
Eq.~(\ref{factfor;1qcd})). The partonic cross sections 
${\hat \sigma}(p_1,p_2;Q, \{Q_1, \dots \}; \mu^2)$ are 
usually computed by truncating their perturbative expansion at a fixed order 
in $\as$, as in Eq.~(\ref{pertex;1qcd}). However, fixed--order calculations
are quantitatively reliable only when all the kinematical scales
$Q, \{Q_1, \dots \}$ are of the same order of magnitude. 
When the hard--scattering process involves two (or several) very different 
scales, say $Q \gg Q_1$, the $n$-th term in Eq.~(\ref{pertex;1qcd}) can 
contain double-- and single--logarithmic contributions of the type
$(\as L^2)^n$ and $(\as L)^n$ with $L=\ln (Q/Q_1) \gg 1$. These terms spoil
the reliability of the fixed--order expansion and have to be summed to all 
orders, systematically improving on the logarithmic accuracy of the expansion.

Typical examples of such large logarithms are the terms $L= \ln Q/Q_0$ 
related to the 
evolution of parton densities (and parton fragmentation functions)
from a low input scale $Q_0$ to the hard--scattering scale $Q$. 
These logarithms are produced by collinear radiation from the colliding 
partons and give {\em single--logarithmic} contributions. They never explicitly
appear in the calculation of the partonic cross section, because they are 
systematically (LO, NLO and so forth) resummed in the evolved parton densities 
$f_{a/h}(x,Q^2)$ and parton fragmentation functions $d_{a/H}(x,Q^2)$
by using DGLAP equations (\ref{evequa;1qcd}).

A different sort of large logarithm, $L=\ln {\sqrt{s}}/Q$, arises when the 
centre--of--mass energy ${\sqrt{s}}$ of the collision is much larger than the 
hard scale $Q$. These small--$x$ $(x=Q/{\sqrt{s}})$ logarithms are 
produced by multiple gluon radiation over the wide rapidity range 
that is available at large energy. For sufficiently inclusive
processes in singlet channels these give {\it single--logarithmic} (LLx)
contributions that can be calculated by using the BFKL 
equation \cite{Fadin:1975cb, Lipatov:1976zz,
Kuraev:1976ge, Kuraev:1977fs, Balitsky:1978ic}. 
The subleading (NLLx) contributions have also been calculated recently
\cite{Fadin:1998py,Ciafaloni:1998gs} and turn out to be very large. This is 
understood to be due to contamination by collinear logarithms of 
$Q^2/Q_0^2$, which must be simultaneously resummed to obtain reliable 
predictions at small $x$ \cite{Salam:1998tj,Ciafaloni:1999iv}. Various resummation
procedures have been suggested, and will be briefly discussed in 
Sects.~\ref{sec:ressmallx;qcd} and \ref{sec:smallx;qcd} Unfortunately there are as yet no 
substantial phenomenological analyses which use these resummations.
The resummation of small--$x$ logarithms will be important for the accurate 
determination of the behaviour of singlet parton 
densities $f_{a/h}(x,Q^2)$ at small values
of the parton momentum fraction $x$, and thus for making
reliable predictions of any process that is sensitive to the 
hard--scattering of low--momentum partons
(for example $b$--quark production\footnote{See the Bottom Production Chapter 
of this Report.} and inclusive production of 
low--$E_T$ jets and prompt photons at the LHC).
The BFKL equation is however also relevant for understanding  
the structure of final states, for example when there are jets with
large rapidity intervals, or diffractive processes 
with large rapidity gaps. These more general aspects of 
small--$x$ physics are discussed in Sect.~\ref{sec:smallx;qcd}

Yet another class of large logarithms is associated to the bremsstrahlung
spectrum of soft gluons. Since soft gluons can be radiated collinearly,
they give rise to {\em double--logarithmic} contributions to the partonic 
cross section, which takes the form
\beq
\label{logxs;5qcd}
{\hat \sigma} \sim \as^k ~{\hat \sigma}^{(LO)} 
\left\{ 1 + \sum_{n=1}^{\infty} \as^n \left( C_{2n}^{(n)} L^{2n} +  
C_{2n-1}^{(n)} L^{2n-1}
+ C_{2n-2}^{(n)} L^{2n-2} + \dots \right) \right\} \;\;.
\eeq
Double--logarithmic terms due to soft gluons arise in all the 
kinematic configurations where the contributions of real and virtual partons
are highly unbalanced (see Ref.~\cite{Bassetto:1983ik} and references therein).

When partons (particles or jets) with low momentum fraction $z$ are directly
triggered in the final state, the r\^ole of (real) soft radiation is evidently
enhanced. The low--momentum region of the fragmentation spectra of particles 
and subjets in jet final--states is thus particularly sensitive to the 
resummation of small--$z$ logarithms. 
The calculations based on the resummation of these
logarithms are probably the perturbative predictions that are most sensitive
to the coherence properties~\cite{Bassetto:1983ik, Dokshitzer:1991wu} of QCD.
Detailed studies of fragmentation processes have been performed
in $e^+e^-$ annihilation, DIS and at the Tevatron (see the recent review
in Ref.~\cite{Webber:1999ui}). Although this topic is not included in these 
proceedings, similar studies at the LHC would certainly be valuable. 

In different kinematic configurations, (real) radiation in the
final state can instead be strongly inhibited.
For instance, this happens in the case of transverse momentum distributions
at low transverse momentum, in the case of hard--scattering production near
threshold or when the structure of the final state
is investigated with high resolution (internal jet structure, shape variables).

Soft--gluon resummation for jet shapes has extensively been
studied and applied to hadronic final states produced by $e^+e^-$ annihilation
\cite{Catani:1993ua, Catani:1997rn, Mangano:1999sz}. 
Applications to hadron--hadron collisions have just begun to 
appear~\cite{Seymour:1994by, Seymour:1998kj,Forshaw:1999iv}
and have a large, yet uncovered, potential (from $\as$ determinations to 
studies of  non--perturbative dynamics). Future studies of this topic are 
certainly warranted.

Threshold logarithms, $L = \ln (1-x)$, occur when the tagged final state
produced by the hard scattering is forced to carry a very large fraction
$x$ ($x \to 1$) of the available centre--of--mass energy $\sqrt{s}$. 
Outstanding examples of hard processes near
threshold are DIS at large $x$ (here $x$ is the Bjorken variable), production
of DY lepton pairs or di-jets with large total
invariant mass $Q=M_{ll}$ or $M_{jj}$ ($x = Q/{\sqrt{s}}$), 
production of $W$, $Z$ and Higgs bosons ($x = M_{W,Z,H}/{\sqrt{s}}$),
production of heavy quark--anti-quark pairs ($x = 2m_Q/{\sqrt{s}}$), inclusive
production of single jets and single photons at large transverse energy $E_T$ 
($x = 2E_T/{\sqrt{s}}$).

Transverse--momentum logarithms, $L = \ln Q^2/{\bf p}_{T}^2$,
occur in the distribution of transverse momentum ${\bom p}_{T}$
of systems with high mass $Q$ $(Q \gg p_{T})$ that are produced with
a vanishing ${\bom p}_{T}$ in the LO subprocess. Examples of such systems
are DY lepton pairs, lepton pairs produced by $W$ and $Z$ decay, heavy 
quark--anti-quark pairs, photon pairs and Higgs bosons. 

Studies of soft--gluon resummation for transverse--momentum distributions
at low transverse momentum and hard--scattering production near
threshold were pioneered two decades ago 
\cite{Parisi:1979se, Curci:1979bg, Dokshitzer:1980hw, Amati:1980ch,
Parisi:1980xd, Curci:1980am, Curci:1981yr, Ciafaloni:1981nm, Ellis:1981sj,
Rakow:1981uh, Chiappetta:1982mw}.
The physical bases for a systematic all--order summation of the 
soft--gluon contributions are dynamics and kinematics 
factorizations~\cite{Sterman:1995aj, Catani:1997rb}. The first factorization
follows from gauge invariance and unitarity: in the soft limit multigluon
amplitudes fulfil factorization formulae given in terms of universal
(process independent) soft contributions. The second factorization regards 
kinematics and strongly depends on the actual cross section to be evaluated.
When phase--space kinematics is factorizable, resummation
is analytically feasible in the form of a {\em generalized exponentiation}
of the universal soft contributions that appear in the factorization formulae
of QCD amplitudes.

Typically, phase--space factorization does not occur in the space 
of the kinematic variables where the cross section is defined. 
It is thus necessary to introduce a conjugate space to overcome phase space 
constraints. This is the case for hard--scattering production near threshold,
where the relevant kinematical constraint is (one--dimensional) energy
conservation, which can be factorized performing a Laplace (or Mellin)
transformation
(see Sect.~\ref{sec:threshold;qcd}). Analogously, the relevant kinematical 
constraint for ${\bom p}_{T}$--distributions is (two--dimensional) 
transverse--momentum conservation and it can be factorized by performing 
a Fourier transformation (see Sect.~\ref{sec:resptdist;qcd}). In the 
conjugate space, the logarithms $L$ of the relevant ratio of momentum scales
are replaced by logarithms $\tilde{L}$ of the conjugate variable.

The resummed cross section is thus typically of the form
\beq
\label{resxs;5qcd}
{\hat \sigma}_{\rm res.} = \as^k \; \int_{\rm inv.} {\hat \sigma}^{(LO)}
\cdot C \cdot S \;,
\eeq
where the integral $\int_{\rm inv.}$ denotes the inverse transformation from
the conjugate space where resummation is actually carried out. The factor $C$ 
contains all constant contributions in the limit ${\tilde L} \to \infty$.
The singular dependence on ${\tilde L}$ is entirely {\it exponentiated} in the
effective form factor $S$: 
\beq
\label{resff;5qcd}
S = \exp \left\{ {\tilde L} \;g_1(\as(\mu) {\tilde L}) + 
g_2(\as(\mu) {\tilde L};\mu^2) + \as(\mu) \;g_3(\as(\mu) {\tilde L};\mu^2) + 
\dots \right\} \;\;.
\eeq
The structure of the exponent is formally analogous to that of the fixed--order
expansion of the partonic cross sections (see Eq.~(\ref{pertex;1qcd})). 
The function $L \,g_1$ resums all the leading logarithmic (LL) contributions
$\as^n L^{n+1}$, while $g_2$ contains the next--to--leading logarithmic 
(NLL) terms $\as^n L^n$ and so forth. Note that the NLL terms are formally 
suppressed by a power of $\as$ with respect to the LL ones, and the same is 
true for the successive classes of logarithmic terms\footnote{This has to be 
contrasted with the tower expansion sketched on the right--hand side of 
Eq.~(\ref{logxs;5qcd}). Within the framework of the tower expansion that sums 
the double-logarithmic terms $(\as L^2)^n$, then the terms 
$\as^n L^{2n-1} \sim \as L (\as L^2)^{n-1}$ and so forth, the ratio of two 
successive towers is, roughly speaking, of the order of 
$\as L$. More precisely, the tower expansion allows us to formally extend the
applicability of perturbative QCD to the region $\as L^2 \lsim 1$, and the
exponentiation in Eq.~(\ref{resff;5qcd}) extends it to the wider region 
$\as L \lsim 1$.
}. Thus, this logarithmic 
expansion is as systematic as the fixed--order expansion in 
Eq.~(\ref{pertex;1qcd}).

In general, a resummed expression such as  Eq.~(\ref{resxs;5qcd}) must be 
properly combined with the best available fixed--order result.
Using a shorthand notation, this is achieved by writing the partonic 
cross section ${\hat \sigma}$ as 
\beq
\label{resmatch;5qcd}
{\hat \sigma} = {\hat \sigma}_{\rm res.} + {\hat \sigma}_{\rm rem.} \;.
\eeq
The term ${\hat \sigma}_{\rm res.}$ embodies the all--order resummation,
while the remainder ${\hat \sigma}_{\rm rem.}$ contains no large logarithmic
contributions. The latter has the form
\beq
\label{remxs;5qcd}
{\hat \sigma}_{\rm rem.} = {\hat \sigma}^{({\rm f.o.})} -
\left[ \,{\hat \sigma}_{\rm res.} \,\right]^{({\rm f.o.})} \;\;,
\eeq
and it is obtained from ${\hat \sigma}^{({\rm f.o.})}$,
the truncation of the perturbative expansion for ${\hat \sigma}$ at a given
fixed order (LO, NLO, \dots), by subtracting the corresponding truncation
$\left[ {\hat \sigma}_{\rm res.}\right]^{({\rm f.o.})}$ of the resummed part.
Thus, the expression on the right--hand side of Eq.~(\ref{resmatch;5qcd}) 
includes soft--gluon logarithms to all orders and it is {\it matched} to 
the exact (with no logarithmic approximation) fixed--order calculation. 
It represents an improved perturbative calculation that is everywhere as 
good as the fixed--order result, and much better in the kinematics regions 
where the soft--gluon logarithms become large ($\as L \sim 1$). 
Eventually, when $\as L \gsim 1$, the resummed perturbative contributions 
are of the same size as the non--perturbative contributions and the effect 
of the latter has to be implemented in the resummed calculation.

Using a matched NLL+NLO calculation as described above, we can 
consistently  introduce a precise definition (say $\msbar$)
of $\as(\mu)$ and investigate the theoretical accuracy of the calculation
by studying its dependence on the renormalization/factorization scale $\mu$.

Resummed calculations for hadron collisions near threshold and for 
$p_{T}$--distributions are discussed in  
Sects.~\ref{sec:threshold;qcd} and \ref{sec:resptdist;qcd}, respectively.
Some overviews can also be found in Ref.~\cite{les-houches;11qcd}. We refer 
the reader to Sects.~\ref{sec:corcella;3qcd} and \ref{sec:huston;3qcd} 
for comparisons of resummed calculations 
with parton shower event generators.

\subsection{Threshold resummations\protect\footnote{  
Contributing author: L. Magnea.}}
\label{sec:threshold;qcd}

Large logarithms arise
in any inclusive
cross section for the production of an object with a large mass $Q$,
whenever the partonic energy $\sqrt {\hat{s}}$ available for the process 
is close to $Q$, the production threshold. The physical mechanism responsible 
for these logarithms is simple. Close to threshold the phase space for the 
emission of gluon radiation in the final state is kinematically restricted; 
soft real radiation is, however, responsible for the cancellation of infrared 
divergences associated with virtual gluon exchange; whenever radiation is 
inhibited, the cancellation is partially spoiled: finite but large 
contributions are left over, in the form of logarithms of the ratio of the 
two relevant energy scales, $\ln [(\hat{s} - Q^2)/\hat{s}]$. Close to 
partonic threshold these logarithms become large
and must be resummed.
Processes for which this resummation is relevant are 
ubiquitous, as noted in the previous subsection. 
Techniques to perform threshold resummations have been developed and 
progressively extended for well over a decade; references in which these 
techniques are explained is some detail include 
\cite{Sterman:1987aj,Catani:1989ne,Catani:1991rp,Sterman:1995fz,Catani:1996yz,
Contopanagos:1997nh,Kidonakis:1997gm,Bonciani:1998vc}; here we will briefly review the basic 
theoretical issues, and sketch the status of phenomenological applications 
of relevance to the LHC.

As described in the introduction to the present Section, the resummation of 
threshold logarithms in performed in Mellin space.
To illustrate the structure of a typical resummation of threshold logarithms, 
let us concentrate on the simplest and best known example: the DY 
cross section. In this case the resummed formula for the Mellin transform of 
the partonic cross section, in the DIS factorization scheme, takes the form
\cite{Sterman:1987aj,Catani:1989ne}
\beq
\hat{\sigma}_{{\rm res.}} (N, Q^2) = 
C(\as (Q^2)) \exp \left[ E(N, Q^2) \right]~~,
\label{DYexp;5qcd}
\eeq
where the function $C$ collects terms independent of the Mellin variable $N$,
while the exponent can be written as
\beq
E(N, Q^2) = - ~2 ~\int_0^1 d z ~\frac{z^N - 1}{1 - z} \left[ 
B(\as((1 - z) Q^2)) + \int_{(1 - z)^2 Q^2}^{(1 - z) Q^2} 
\frac{d q^2}{q^2} A(\as(q^2)) \right]~~.
\label{DYres;5qcd}
\eeq
Equations (\ref{DYexp;5qcd}) and (\ref{DYres;5qcd}) resum, in principle, all 
logarithms of $N$ to all orders in perturbation theory, in the sense that all 
such logarithms exponentiate and are calculable from the functions $A$ and 
$B$, for which Feynman rules can be derived. In practice, the functions 
$A$ and $B$ are known only to two loops, so that the resummation can 
explicitly be performed only for leading and next--to--leading logarithms
{\it in the exponent}. Performing the integrals in $E(N, Q^2)$, after 
expansion of the running couplings in terms of $\as(Q^2)$ to the desired 
accuracy, yields in general an expression of the form
\beq
E(N, Q^2) = \ln N ~g_1 (\as \ln N) + g_2 (\as \ln N) +
\sum_{k = 1}^\infty \as^k ~g_{k + 2} (\as \ln N)~~,
\label{DYNLL;5qcd}
\eeq
where the functions $g_1$ and $g_2$ are known in terms of the coefficients
$A^{(1)}$, $A^{(2)}$ and $B^{(1)}$ of the perturbative expansion of the 
functions $A$ 
and $B$, together with the one-- and two--loop coefficients of the QCD 
$\beta$ function. The (unknown) function $g_3$, giving the NNL logarithms, 
would require the determination of $A^{(3)}$, as well as $B^{(2)}$ and the 
three--loop $\beta$ function.

Several comments are necessary in order to introduce the practical 
applications of resummed formulas such as Eq.~(\ref{DYres;5qcd}).

\begin{itemize}

\item At the present level of accuracy (NLL) the dependence on the 
renormalization scale and on the factorization scheme is under control.
A change in renormalization scale shifts the function $g_2$ by an amount 
proportional to the derivative of the function $g_1$. A change in 
factorization scheme changes both $g_1$ and $g_2$, because it affects the 
way in which the DIS process is subtracted from DY to construct a 
finite cross section, however the change is well understood and both 
functions can be translated from one scheme to another 
\cite{Catani:1991rr,Contopanagos:1997nh}.

\item To understand the effects of resummation, one should keep in mind that
it is performed at the level of the {\it partonic} cross section. One 
consequence of this fact is that resummation generically {\it enhances}
the cross section, although one might expect a Sudakov {\it suppression}, 
since the probability of having a nearly radiation-less hard scattering is 
exponentially suppressed. This is easily understood in the DIS scheme: there
one computes the (factorized) partonic DY cross section by taking 
the ratio of the DY process to the square of the DIS process, since
there are two partons in the DY initial state. In this ratio, the 
denominator is Sudakov suppressed twice as much as the numerator, resulting
in an overall Sudakov enhancement.

\item The fact that the resummed partonic cross section must be folded with 
parton distributions to extract a physical prediction also means that the
effects of resummation are felt quite far away from the {\it hadronic} 
threshold. In fact, given a hadronic centre--of--mass energy $S$, the typical
partonic energy available for the production process will be $<\hat{s}> =
<x_1 x_2> S$, where $x_1$ and $x_2$ are the momentum fractions of the
scattered partons. Clearly $\hat{s}$ becomes close to threshold long before 
$S$ does.

\item The resummed partonic cross section by construction contains a subset
of the finite order perturbative calculations available for the process at 
hand. One should then work with a ``matched'' cross section, as described in 
the previous subsection (see Eqs.~(\ref{resmatch;5qcd}) and (\ref{remxs;5qcd})). 

\item The alert reader will have noticed that Eq.~(\ref{DYres;5qcd}), although 
well--defined order by order if the running couplings depending on variable 
arguments are re-expanded in terms of a fixed large scale, is actually 
ill--defined in the leading--logarithm (of $Q^2$) approximation, because the
integration contour runs over the Landau pole. This is a general feature of
most known resummations of perturbation theory: in fact, perturbation theory
is pointing us to its own limitations, and to the need to include information
concerning the non--perturbative structure of QCD \cite{Beneke:1999ui}. 
This fact has two consequences. 
On the one hand, it is possible to exploit partial resummations 
such as Eq.~(\ref{DYres;5qcd}) to estimate the size of the first relevant 
non--perturbative corrections: in the case of the DY process, two
independent approaches \cite{Beneke:1995pq,Dokshitzer:1996qm} lead to the 
conclusion that the first power correction to Eq.~(\ref{DYexp;5qcd}) is 
${\cal O}((N/Q)^2)$. On the other hand, experience has shown that the 
necessary inversion of the Mellin transform back to momentum space 
can generate unjustified (and stronger) power corrections that are not 
present in the original resummed expression. Methods to circumvent 
this problem have been developed \cite{Catani:1996yz}, so that 
Eq.~(\ref{DYexp;5qcd}) can be used confidently, with a definite understanding 
of the size of expected corrections.

\item In the general case of colored final states, a comparatively simple 
expression for the resummed cross section, such as Eq.~(\ref{DYres;5qcd}), is 
not available to all logarithmic orders, because the corresponding evolution 
equations are in matrix form, and their solution involves a scale--dependent 
mixing of color tensors. To NLL accuracy, however, a simple exponentiation
can still be achieved, by diagonalizing a matrix of anomalous dimensions in
the space of available color configurations \cite{Kidonakis:1997gm,
Bonciani:1998vc}. This results in a matrix of exponentials, each similar 
to Eq.~(\ref{DYres;5qcd}), with two new color--dependent functions of the 
running coupling. These new functions also carry the necessary dependence 
on the angles between incoming and outgoing colored partons.

\item It should be emphasized that further improvements are possible, and in 
some cases have already been achieved. In the case of 
the DY process, the terms independent of $N$ contained in the factor
$C$ in Eq.~(\ref{DYexp;5qcd}) can also be resummed: in the DIS scheme, they 
contain the absolute value of the ratio of the time-like to the space-like 
Sudakov form factor, which is known to exponentiate \cite{Magnea:1990zb}. 
Methods to resum classes of terms of the form $\ln N/N$ have 
recently been suggested \cite{Kramer:1998iq}.
Finally, a technique to resum 
simultaneously threshold logarithms and recoil enhancements in single 
particle inclusive cross sections has been introduced \cite{Laenen:2000de}.
\end{itemize}

Turning to practical applications, we observe that resummations of threshold 
logarithms have been performed to NLL accuracy for most of the processes of 
interest at the LHC, ranging from DIS and DY \cite{Sterman:1987aj,
Catani:1989ne,Magnea:1991qg,Catani:1991rr,Contopanagos:1997nh,Akhoury:1998gs}
to Higgs boson \cite{Kramer:1998iq} production, to include 
more recently studies of processes with hard colored particles in the final 
states, such as heavy quark 
\cite{Kidonakis:1997gm,Bonciani:1998vc,Kidonakis:1999ze},
prompt photon 
\cite{Laenen:1998qw,Catani:1998tm,Catani:1999hs,Kidonakis:1999hq},
$W$ boson \cite{Kidonakis:1999ur} and di-jet \cite{Kidonakis:1998bk}
production; applications of the formalism to quarkonium production have 
been proposed \cite{Cacciari:1999sy}.
Detailed phenomenological calculations, however, are presently available 
only for a subset of these processes. 

It is important to note that at the LHC threshold resummation can 
be important for two reasons. On one side, it can directly be applied to 
LHC processes through the corresponding partonic cross sections. On the 
other side, it can be applied to the lower--energy processes that are 
typically used to determine the parton densities, and thus it can 
indirectly affect LHC predictions through the use of (evolved) parton 
distributions reevaluated in this manner. 

We shall illustrate the phenomenological effects
of the application of these techniques with few examples, which will serve
to point out another relevant feature of NLO+NLL calculations: their 
increased stability with respect to scale variations.

\begin{figure}[t!]
  \begin{center}
    \includegraphics[width=0.5\textwidth,clip]{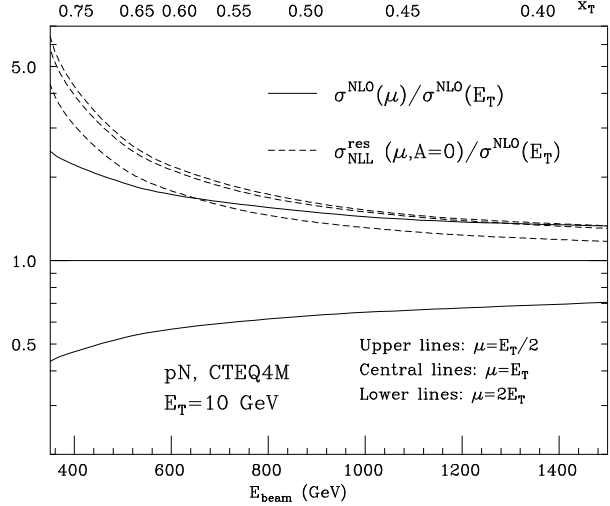}
\vskip-0.5cm
    \caption{Scale dependence of $d\sigma/dE_T$ for single prompt--photon 
     production in $pN$ collisions. The solid lines represent the NLO result 
     for different choices of $\mu=\mu_R=\mu_F$ ($\mu=E_T/2$ and $2E_T$), 
     normalized to the result for $\mu=E_T$. The dashed lines represent the 
     NLO+NLL results for different choices of $\mu$, normalized to NLO result 
     for $\mu=E_T$. See Ref.~\cite{Catani:1999hs} for details.}
    \label{fig:gammabeam;5qcd}
  \end{center}
\end{figure}

As discussed in Sect.~\ref{sec:pdf;qcd}, present data and NLO calculations do 
not constrain very well the determination of the parton distributions at large 
values of the parton momentum fraction $x$. This is particularly true for the
gluon density $f_g(x,Q^2)$ at $x \gsim 10^{-1}$ and $Q \sim 5-10~{\rm GeV}$.
The uncertainty on $f_g$ in this kinematic region propagates (although with
a reduced overall size) to smaller values of $x$ and larger values of $Q^2$
in LHC processes. Threshold resummation can help to extract parton 
distributions at large $x$ with more confidence than is at present in NLO 
analyses. 
Consider, for instance, the production of prompt photons with high transverse 
energy $E_T$ at fixed--target experiments. This process is very sensitive 
to the behaviour of the gluon density at large $x$ $(x \sim x_T = 
2 E_T/{\sqrt{s}})$. The corresponding theoretical calculations at fixed 
perturbative order, however, are not very accurate, as can be argued by 
studying their dependence on the factorization/renormalization scale $\mu$. 
When NLL resummation is applied~\cite{Catani:1999hs}, the scale dependence 
of the calculation is highly reduced and the resummed NLL contributions 
lead to large corrections at high $x_T$ (and smaller corrections at lower 
$x_T$). The scale dependence of the theoretical cross section in $pN$ 
collisions is shown in Fig.~\ref{fig:gammabeam;5qcd} as a function of 
$E_{\rm beam}$, the energy of the proton beam. Fixing $\mu_R=\mu_F=\mu$ and 
varying $\mu$ in the range $E_T/2 < \mu < 2E_T$ with $E_T=10~{\rm GeV}$ and 
$E_{\rm beam}=530~{\rm GeV}$ (this corresponds to the largest value of $x_T$
that is reachable by the E706 kinematics~\cite{Apanasevich:1998hm}),
the cross section
varies by a factor of $\sim 6$ at LO (the result of the LO calculation
is not shown in the plot), by a factor of $\sim 4$ at NLO and
by a factor of $\sim 1.3$ after NLL resummation. The central value (i.e. with
$\mu=E_T$) of the NLO cross section increases by a factor of $\sim 2.5$ after
NLL resummation. As expected, the size of these effects is reduced by 
decreasing $x_T$ (e.g. by increasing ${\sqrt{s}}$ at fixed $E_T$). This 
(extreme) example clearly illustrates how NLO+NLL resummed calculations 
can improve the present NLO determinations of parton distributions.
The method of Ref.~\cite{Laenen:2000de} can also be applied to investigate
the relevance of recoil effects in prompt-photon production.
  
NLL resummations of threshold logarithms are now available for all the most 
important processes (DIS, DY, and prompt--photon production) used to 
determine the parton densities via global fits. It is thus possible to 
consistently~\cite{Catani:1998tm, Sterman:2000pu} take into account 
all threshold effects affecting the different hadronic cross sections.
Preliminary studies~\cite{Sterman:2000pu,Vogt:1999xa,CMNunpub} suggest
that NLO+NLL fits are not likely to make drastic differences
in the parton densities that are strongly constrained by DIS data, at least
so long as the region of small $Q$ $(Q \sim {\rm few~GeV})$ is avoided at very
large $x$. At the same time, they suggest that resummed fits can make some
difference where the \pdfs\ are not so well known (gluon density at large $x$
and quark densities at larger values of $x$). In particular,
NLO+NLL fits, if implemented, are likely to reduce scale dependence,
and thus further improve our confidence in the theoretical predictions
for LHC cross sections.

As for direct effects of NLL threshold resummation at the LHC, we briefly
discuss top pair production, which is currently the best studied process in 
LHC kinematics~\cite{Bonciani:1998vc}. One could argue that 
threshold resummation effects in this case should
not be expected, since at the LHC we have $x=2 m_t/\sqrt{s} \sim 0.03$. 
This would however be incorrect since, as explained above, partonic 
threshold can be, on average, quite far from hadronic threshold. 
In the case of top production at the LHC, the dominant partonic subprocess
is gluon fusion. The gluon density is steeply falling at large $x$ and
quite large at small $x$, so that the average momentum fraction of gluons 
entering the partonic hard subprocess is relatively small and $\hat s \ll s$.
As a consequence, the effect of NLL resummation is still visible at the LHC:
the NLO+NLL resummed cross section is larger than the NLO estimate by about 
$5\%$. Moreover, NLL resummation reduces the scale dependence of 
the cross section by approximately a factor of two (from about $10\%$ to 
about $5\%$). This can be relevant, because the uncertainty due to the 
present knowledge of the parton densities is estimated to be twice as 
large. We refer the reader to the Top Physics Chapter of this Report for 
full details.

Other topical LHC processes are Higgs production, DY production of $W, Z$ and
lepton pairs, as well as production of high--$E_T$ jets. 
Since the Higgs mass $M_H$ is expected to be of the same order as the 
top-quark mass, Higgs production will be dominated by gluon fusion. 
Thus, the effects of threshold resummation on this process should be at 
least as important as for top-pair production. The results of 
Ref.~\cite{Kramer:1998iq}, based on the expansion at NNLO of threshold 
resummation, support this conclusion. Complete quantitative studies to 
NLL accuracy are not yet available and would be valuable.
The production of $W$ and $Z$ at the LHC is less close to threshold than 
top production. Moreover, its dominant partonic subprocess is 
$q{\bar q}$ annihilation. The large--$x$ behaviour of the quark densities
is less steep than that of the gluon density, and 
soft--gluon radiation from initial--state quarks is depleted by 
the colour charge factor $C_F/C_A \sim 1/2$ with respect to radiation from
gluons. Thus, the effects of threshold resummation on $W, Z$ production
should be small. Their size could however increase in the case of
production of high--mass (say, $Q \gsim 1~\rm{TeV}$) DY lepton pairs.
The inclusive production of high--$E_T$ jets and di-jets with large invariant 
mass at the Tevatron and at the LHC can be sensitive to threshold logarithmic
contributions. Nonetheless, phenomenological analyses to NLL accuracy are not
available for these processes. An important conceptual reason for that is 
the fact that the cone algorithms used so far to experimentally define jets 
are not infrared and collinear safe~\cite{Seymour:1998kj, Kilgore:1997sq}. 
Although their unsafety may show up only at some high order in perturbation 
theory, it prevents all--order summations. The future use \cite{procrunII;5qcd}
of safe algorithms, 
such as the $k_{\perp}$-algorithm~\cite{Catani:1993hr, Ellis:1993tq} and the 
improved cone algorithm studied at the Workshop on Physics at the Tevatron 
in Run II, will overcome this problem. For the definition of different jet 
algorithms, we refer the reader to 
Ref.~\cite{procrunII;5qcd}.

\subsection{Resummation of transverse momentum distributions\protect\footnote{  
Contributing authors: A.~Kulesza and W.~J.~Stirling.}}
\label{sec:resptdist;qcd}

The description of vector and scalar boson production properties, in
particular their transverse momentum ($p_T$) distribution, is likely to be
one of the most investigated topics at the LHC, especially in the context
of Higgs searches. To obtain a reliable theoretical prediction for the
$p_T$ distribution, the corrections due to soft gluon radiation have to be
taken into account. At small transverse momentum the $p_T$ distribution is
dominated by large logarithms $\ln(Q^2/p_T^2)$, which are directly related
to the emission of gluons by the incoming partons. Therefore, at sufficiently
small $p_T$, fixed--order perturbation theory breaks down and the logarithms
must be resummed. The origin of the large logarithms is visible already at
leading--order: in fact, the contribution from real emission
diagrams for $q\bar{q} \lra Vg$ contains a term of the form $\as C_F
\ln\left({Q^2 /p_T^2}\right)/(\pi p_T^2) $. When more gluons are emitted,
the logarithmic divergence becomes stronger. It can be shown that in the
approximation of {\it soft and collinear} gluons with strongly ordered
transverse momenta $k_T$, i.e.
\beq
k_{T,1}^2 \ll k_{T,2}^2 \ll \ldots \ll k_{T,n}^2 \lsim p_T^2 \ll Q^2
\label{strong;5qcd}
\eeq
the dominant contributions to the $q \bar{q} \lra V X $ cross section can be
resummed, giving a so--called Sudakov factor~\cite{Dokshitzer:1980hw},
of the form
\beq
{1 \over \sigma_0} {d \sigma \over d p_T^2}= {\as A \over 2\pi p_T^2}\ln
\left({Q^2 \over p_T^2} \right) \exp \left(-{\as A \over 4\pi }\ln^2
\left({Q^2 \over p_T^2} \right)\right) \, ,
\label{eq_sud;5qcd}
\eeq
where $A = 2 C_F$, and $\sigma_0$ is the total LO $q \bar{q} \lra V $ 
cross section. This approximation is commonly known as the
{\em Double Leading Logarithm Approximation} (DLLA).

The resummation in Eq.~(\ref{eq_sud;5qcd}) gives a finite but unphysically
{\em suppressed} result in the small $p_T$ limit. This suppression is
caused by the vanishing of strongly--ordered phase space, in which overall
transverse momentum conservation is ignored. The result in (\ref{eq_sud;5qcd})
corresponds to a configuration in which a {\em single} soft gluon balances
the vector boson transverse momentum, giving the overall
$\ln(Q^2 /p_T^2)/p_T^2$ term, while all other gluons have transverse momenta
$\ll p_T$. This is {\em not} the dominant configuration in the small $p_T$
limit. Equally important are non--strongly--ordered contributions
corresponding to the emission of soft ($\sim p_T$) gluons whose transverse
momenta add vectorially to give the overall $p_T$ of the vector boson.
Although such contributions are formally sub-leading order--by--order, they
do dominate the cross section in the region where the Sudakov form factor
suppresses the (formally) leading DLLA contributions.
The non--leading `kinematical' logarithms are correctly taken into account
by imposing transverse momentum conservation (rather than strong ordering),
and this is most easily achieved by means of a Fourier transform to impact
parameter ($b$--)space.

We next discuss analytic methods for resumming large logarithms in
$b$--space and $p_T$--space. As already mentioned, comparisons of resummed 
calculations with the predictions coming from parton shower Monte Carlo 
approaches are presented in 
Sects.~\ref{sec:corcella;3qcd} and \ref{sec:huston;3qcd}. 

\subsubsection{Analytic methods: $b$--space}

In the $b$--space method \cite{Parisi:1979se} one imposes transverse momentum 
conservation by Fourier transforming the $p_T$ distribution to impact 
parameter space and using the identity
\beq
\delta^{(2)} \left( \sum^N_{i=1} {\bf{k_{T_{i}}}} - {\bf{p_T}} \right) =
{ 1 \over 4 \pi^2} \int d^2 b {\rm e}^{- i {\bf {b \cdot p_T}}}
\prod^{N}_{i=1} {\rm e}^{i  {\bf {b \cdot k_{T_{i}}}}} \, .
\eeq
This allows for the derivation of  a general expression resumming
all terms of the perturbation series which are at least as singular
as ${1 /p_T^2}$ when $p_T \rightarrow 0$~\cite{Collins:1985kg,
Collins:1981uk,Collins:1982va}. The resummed expression is of the form
\beeq
\frac{d \sigma(AB \lra V(\lra {l {\bar{l'}}}) X )}{d p^2_T \,
dQ^2 \, dy \,  d\cos{\theta} \, d\phi} & = &
\frac{1}{256 \pi N_c s} \, \frac{Q^2}{(Q^2 - M_V^2)^2 + M_V^2 \Gamma_V^2}
\nonumber \\
& \times & \left[ Y_r(p_T^2, Q^2, y, \theta)
+ Y_f(p_T^2, Q^2,y, \theta, \phi) \right]  \, ,
\label{bspace;5qcd}
\eeeq
where $M_V$ and $\Gamma_V$ are the mass and the width of the vector boson,
and $\theta$ and $\phi$ stand for the lepton polar and azimuthal angles in
the Collins--Soper frame~\cite{Collins:1985kg,Collins:1981uk,Collins:1982va}.
$Y_r$ denotes the resummed part of the cross section, while $Y_f$ is the
remainder (that is, the fixed--order expression minus terms which are
already taken into account in $Y_r$, as in Eq.~(\ref{resmatch;5qcd})).
The exact expression for $Y_f$ can be found in~\cite{Ellis:1997sc}, whereas
\beeq
Y_r(p_T^2, Q^2, y, \theta) & = &  \Theta{(Q^2 - p_T^2)} ~{1 \over 2\pi}
\int_{0}^{\infty} db\;b  \, J_{0}(p_T  b)
\sum_{a,b}{} F^{NP}_{a b } (Q,b,x_A,x_B) \nn \\
& \times & H_{ab}(\theta) ~f'_{a/A}(x_A,\frac{b_0}{b_*}) ~f'_{b/B}
(x_B,\frac{b_0}{b_*}) ~\exp{[S(b,Q)]} \,.
\label{b_resum;5qcd}
\eeeq
Here $f'$ denotes a modified parton distribution, $H_{ab}(\theta)$ includes
coupling factors and the angular dependence of the lowest order cross
section~\cite{Ellis:1997sc}, and $b_*$ and $F_{ab}^{NP}$ are discussed below.
The Sudakov factor has the form
\beeq
\label{eq:abseries;5qcd}
S(b,Q^2) = - \int_{b_0^2 \over b^2}^{Q^2} \frac{d \mu^2}{\mu^2}
\bigg[ \ln \bigg ({Q^2 \over \mu^2} \bigg ) A(\as(\mu^2)) +
B(\as(\mu^2)) \bigg ] \, , \label{Sbs} \\
A(\as) = \sum^\infty_{i=1} \left(\frac{\as}{2 \pi}
\right)^i A^{(i)} \, , \quad
B(\as) = \sum^\infty_{i=1} \left(\frac{\as}{2 \pi}
\right)^i B^{(i)} \, ,
\label{AB;5qcd}
\eeeq
with $b_0=2\exp(-\gamma_E)$. The form in Eq.~(\ref{eq:abseries;5qcd}) is 
equally valid for processes initiated by $q{\bar q}$-annihilation 
(e.g. production of DY lepton pairs, $W$ and $Z$) and by $gg$-fusion 
(e.g. Higgs production). The coefficients $A^{(1)}, A^{(2)}$ and $B^{(1)}$
in each series~(\ref{AB;5qcd}) were computed in Ref.~\cite{Kodaira:1982nh} for
$q{\bar q}$-annihilation and in Ref.~\cite{Catani:1988vd} for $gg$-fusion.
These coefficients\footnote{In Ref.~\cite{Davies:1984hs} the coefficient
$B^{(2)}$ for $q{\bar q}$-annihilation was also computed. The coefficient
$B^{(2)}$ for $gg$-fusion is not yet known.} 
can also be obtained~\cite{Davies:1984hs} from the exact
fixed--order perturbative calculation in the high $p_T$ region by comparing
the logarithmic terms therein with the corresponding logarithms generated by
the first three terms of the expansion of $\exp(S(b,Q^2))$ 
in Eq.~(\ref{b_resum;5qcd}).

Although the $b$--space method succeeds in recovering a finite, positive
result in the $p_T \rightarrow 0$ limit, there are drawbacks associated with
the need to work in impact parameter space. The first is the difficulty of
matching the resummed and fixed--order predictions. Since the resummation is
performed in $b$--space one loses control over which logarithmic terms (in
$p_T$--space) are taken into account. Therefore there is no unambiguous
prescription for matching; existing prescriptions require switching from
resummed to fixed--order calculation at some value of $p_T$.
Secondly, since the integration in~(\ref{b_resum;5qcd}) extends from 0 to 
$\infty$, it is impossible to make predictions for {\it any} $p_T$ without 
having a prescription for how to deal with the non--perturbative regime of 
large $b$. One prescription is to artificially prevent $b$ from reaching 
large values by replacing it with a new variable $b_*$ and by parametrising the
non--perturbative large--$b$ region in terms of the form factor $F_{ab}^{NP}$.
The `freezing' of $b$ at $b_*$ is achieved by
\bann
b_*=\frac{b}{\sqrt{1+(b/b_{\rm \, lim})^2}}\; , \qquad
\qquad b_* < b_{\rm \, lim}\,,
\eann
with the parameter $b_{\rm \, lim} \sim 1/\Lambda_{\rm QCD}$ separating
perturbative and non--perturbative physics.
The detailed form of the non--perturbative function $F_{ab}^{NP}$ remains a
matter of theoretical dispute (for a review see~\cite{Ellis:1997sc}),
although it is assumed to have the general
form~\cite{Collins:1985kg,Collins:1981uk,Collins:1982va}
\bann
F_{ab}^{NP}(Q,b,x_A,x_B)=\exp\left\{-\left[h_Q(b)\ln\left({Q\over 2Q_0}\right)
    + h_a(b,x_A) +h_b(b,x_B)\right]\right\} \,.
\eann
In a very simple model in which the non--perturbative contribution arises from 
a Gaussian `intrinsic' $k_T$ distribution, one would have
$F \sim \exp(-\kappa b^2)$. The data are not inconsistent with such a form,
but suggest that the parameter $\kappa$ may have some dependence on $Q$
and $x$.

Phenomenological studies and numerical calculations based on the $b$--space
formalism are presented in 
Refs.~\cite{Altarelli:1984pt,Davies:1985sp,Arnold:1991yk,
Balazs:1997xd,Ellis:1997sc} (for DY lepton pair, $W$ and $Z$ production) 
and in Refs.~\cite{Hinchliffe:1988ap,Kauffman:1991jt,Kauffman:1992cx,
Balazs:2000wv} (for Higgs production).

\subsubsection{Analytic methods: $p_T$--space}

The difficulties mentioned above could in principle be overcome if one had a
method of performing the calculations directly in transverse momentum space.
Given an insight into which logarithmic terms are resummed, it should be
fairly straightforward to perform matching with the fixed--order result.
Moreover, the non--perturbative input would be required in (and would affect)
only the small $p_T$ region.

Three techniques have been proposed for carrying out resummation in
$p_T$--space~\cite{Ellis:1998ii,Frixione:1999dw,Kulesza:1999gm}. The main
difference lies in the selection of subsets of logarithmic terms which
each method resums; for a detailed discussion the reader is referred
to~\cite{Kulesza:2000sg}. The starting point for all techniques is the
general expression in impact parameter space for the vector boson transverse
momentum distribution in the DY
process~\cite{Collins:1985kg,Collins:1981uk,Collins:1982va},
at the quark level. To illustrate the results, we consider the approach
of \cite{Kulesza:1999gm}, and we give the expression for
the resummed part of the cross section $q \bar{q} \lra \gamma^* X$, in
the simplest case, with fixed coupling $\as$, at the parton level,
and retaining  only the leading coefficient $A^{(1)}$ in the series of
Eq.~(\ref{AB;5qcd}). It is of the form
\beeq
{1 \over \sigma_0} {d \sigma \over d p_T^2} =
{\lambda \over p_T^2} ~{\rm e}^{- {\lambda \over 2} L^2}
\sum_{N=1}^{\infty} {(-2 \lambda)^{(N-1)} \over (N-1)!}
\sum_{m=0}^{N-1} { \left( \begin{array}{c} N-1 \\ m \end{array} \right)}
L^{N-1-m}
\bigg[2\tau_{N+m}+ L \tau_{N+m-1}\bigg]\,.\nn \\
\label{qt_sum2;5qcd}
\eeeq
Here $L = \ln(Q^2/p_T^2)$, $\lambda = \as C_F /\pi$,
and the numbers $\tau_m$ are defined by
\beq
\tau_m \equiv \int_{0}^{\infty} d y J_{1}(y) \ln^m({y \over b_0}) \, .
\label{b_def;5qcd}
\eeq
The $\tau_m$ can be calculated explicitly using a generating
function~\cite{Kulesza:1999gm} so that e.g. $\tau_0=1$,
\mbox{$\tau_1=\tau_2=0$}, $\tau_3=-{1\over 2}\zeta(3)$, etc.
Notice that by setting all $\tau_m$ coefficients (except
$\tau_0$) to zero one would immediately recover the
DLLA form (\ref{qt_sum2;5qcd}). Since there are no explicit sub-leading
logarithms in~(\ref{qt_sum2;5qcd}), other than those related to kinematics,
the presence of the $\tau_m$ coefficients must correspond to relaxing the
strong--ordering condition. This can be checked explicitly by performing
the `exact' ${\cal O}(\as^2)$ calculation in transverse momentum space.
One finds
\beeq
\int d^2k_{T1} d^2k_{T2} \, \left[ {\ln(Q^2/k_{T1}^2) \over k_{T1}^2 }
\right]_+ \left[ { \ln(Q^2/k_{T2}^2) \over k_{T2}^2}  \right]_+
\delta^{(2)}({\bf{k_{T1}}}  + {\bf{k_{T2}}} - {\bf {p_T}} )
= {\pi \over p_T^2} \;  \left( - L^3 + 4 \zeta(3) \right).
\label{exact2;5qcd}
\eeeq
Strong ordering is equivalent to replacing the $\delta$ function by
$\delta^{(2)}({\bf{k_{T1}}} - {\bf{p_T}}) \times$ \mbox{$\theta(k_{T1}^2 -
k_{T2}^2)$} $ + (1 \leftrightarrow 2)$. This gives only the leading
$L^3$ term on the right--hand side. The $\zeta(3)$ term represents the
first appearance of the (kinematic) $\tau_3$ coefficient of 
Eq.~(\ref{qt_sum2;5qcd}).

\begin{figure}[t!]
\begin{center}
\includegraphics[width=0.5\textwidth,clip]{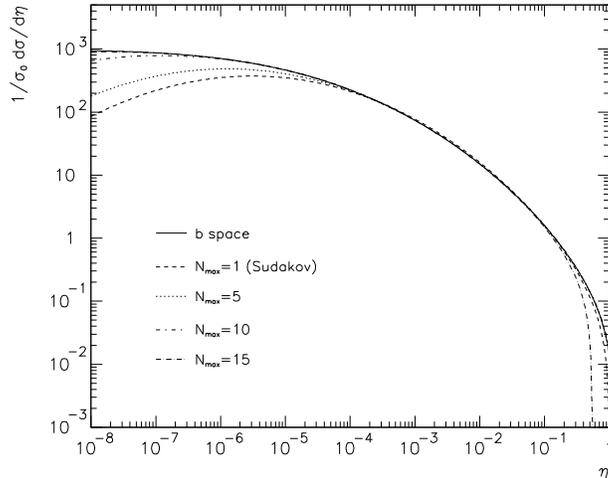}
\vskip-0.5cm
\caption{The $b$--space result (parton level, fixed coupling, only $A^{(1)}$)
  compared to the expression~(\ref{qt_sum2;5qcd}), calculated for various 
  values of $N_{\rm max}$. Here $\eta=p_T^2/Q^2$ and $N_{\rm max}$ is the 
  upper limit of the first summation in Eq.~\ref{qt_sum2;5qcd}.} 
\label{KSvb}
\end{center}
\end{figure}

In principle the formalism presented above allows for an inclusion of 
{\em any} number of such sub-leading kinematic logarithms. In practice, we use
Eq.~(\ref{qt_sum2;5qcd}) with a finite number of terms by introducing $N_{\rm 
max}$ as the upper limit of the first summation. $N_{\rm max}$ corresponds to 
the number of towers of logarithms which are fully resummed.
Figure~\ref{KSvb} shows that for small values of $p_T$ the approximation
of the $b$--space result improves with increasing $N_{\rm max}$.
Therefore by retaining sufficiently many terms one can obtain a good
approximation (i.e. adequate for phenomenological purposes) to the $b$--space
result by summing logarithms directly in $p_T$ space.\,\footnote{Notice
however that, due to the lack of knowledge of $A^{(3)},
\;B^{(3)}$, etc., it is only possible to obtain the complete result
for the first four `towers' of logarithms; subsequent towers can be included
only in the approximations leading to Eq.~(\ref{qt_sum2;5qcd}), see
\cite{Kulesza:1999gm}.}

The technique developed so far can be extended to include sub-leading $A$ and
$B$ coefficients, the running coupling and parton distributions, thus
yielding a `realistic' expression for the hadron--level cross section. The
result is too lengthy to reproduce here, but can be found in
\cite{Kulesza:1999gm, Kulesza:prep}.

Although the $p_T$--space method provides a simple matching prescription,
the form of the non--perturbative function in this approach (as well as
in $b$--space approach) remains an open theoretical issue. In particular,
the current lack of understanding of the $x$ and $Q^2$ dependence of the
non--perturbative contribution is a limiting factor in predicting the
$p_T \to 0 $ behaviour of the distribution at the LHC. However, it seems 
that the dependence on the amount of non--perturbative smearing weakens with 
increasing $Q$ (see Ref.~\cite{Balazs:2000sz} and 
the discussion in Sect.~\ref{sec:huston;3qcd}). 
It has also been shown~\cite{Kulesza:2000sg} that the quality
of the approximation to the $b$--space result achieved by various 
resummation approaches in $p_T$--space changes significantly only for small 
values of $p_T^2/Q^2$.
This in turn would suggest that the differences between these approaches
may become relevant for obtaining an accurate theoretical description of very
heavy boson (e.g. Higgs) production in the small $p_T$ regime.

\subsection{Small--$x$ resummations\protect\footnote{  
Contributing author: R.D.~Ball.}}
\label{sec:ressmallx;qcd}

If we are to make accurate predictions for LHC `background' processes with 
partonic centre--of--mass energy below 1~{\rm TeV}, we need to extrapolate 
cross sections measured at HERA and the Tevatron  
forward by between one and three orders of magnitude in $Q^2$, 
and back by between one and three orders of magnitude in $x$. 
Since away from thresholds these cross sections are generally rather smooth 
functions of $x$ and $Q^2$ one might try to do this by simply 
extrapolating parametric fits \cite{Desgrolard:1999ax,Cudell:2000tx}. 
However the uncertainties in such extrapolations are very difficult to 
quantify. Adding an assumption that the dominant singularities 
are Regge poles is not very helpful, since even with current 
data more than one `Pomeron' singularity is needed for a 
satisfactory fit \cite{Donnachie:1998gm,Donnachie:1999qv}. 
Moreover in this kind 
of approach it is not possible to relate all the various cross sections 
of interest, or for example calculate heavy quark production, or jet cross 
sections: each must be fitted individually. Clearly 
we need more dynamics. Strong interaction dynamics at high energies 
inevitably means perturbative QCD, and it is the current understanding of 
perturbative QCD at small $x$ that we summarise here. 

Provided there is a hard scale in the process, strong interaction processes
may generally be factorized into a hard partonic cross section, computable in 
perturbative QCD, and parton densities which must be determined empirically. 
At large scales $Q^2$ and not too small but fixed $x$ the QCD evolution 
equations~\cite{Gribov:1972ri,Georgi:1974sr,Gross:1974cs,Lipatov:1975qm,
Altarelli:1977zs} provide a reliable framework for the extrapolation of 
these parton densities from some initial scale $Q_0^2$ to higher 
values of $Q^2$. The complete AP splitting functions have been computed 
in perturbation theory at order $\as$ (LO) and 
$\as^2$ (NLO).
For the first
few moments the AP splitting functions at order $\as^3$ (NNLO) are also
known~\cite{Larin:1994vu,Larin:1997wd}. Once we have the parton 
distributions, it is straightforward to compute hadronic cross sections 
at LO or NLO: potentially large contributions of the form 
$(\as \ln Q^2/Q_0^2)^n$ (LLQ), $\as(\as \ln Q^2/Q_0^2)^n$ 
(NLLQ), \dots, have been resummed by solving the evolution equations,
so all that is necessary is the convolution of the evolved 
parton densities with the hard partonic cross section.

If we start with initial parton distributions that rise less steeply than  
a power in $1/x$ as $x$ decreases, then fixed order evolution to
higher $Q^2$ inevitably leads to distributions that become progressively 
steeper in $1/x$ as $Q^2$ increases\cite{DeRujula:1974rf}, in agreement with 
the rise in the $F_2$ data from HERA. More significantly 
the specific form and steepness of the rise is precisely 
\cite{Ball:1994du,Ball:1994kc,Forte:1995vs} as predicted. 
This is a major triumph for perturbative QCD, since it can be interpreted as 
direct evidence for asymptotic freedom \cite{Wilczek:1996bw}: the 
coefficient $\beta_0$ which determines the slope of the rise 
is the first coefficient of the QCD $\beta$--function. This 
has now been confirmed many times by successful NLO fits (see 
\cite{Botje:1999rp,Barone:2000yv} and Sect.~\ref{sec:pdf;qcd})
to increasingly precise HERA $F_2$ datasets. From 
these fits a gluon distribution may be extracted, 
and predictions made for $F_2^c$, di-jet production, and $F_L$, all of which
have now been confirmed by direct measurements 
\cite{Klein:1999yc,Marage:1999qa}.
Clearly fixed order perturbative QCD works well at HERA: none of these
predictions is trivial, and all are successful. Extrapolation to the
LHC region, and the calculation of relevant NLO cross sections, can then be
performed in the same way as at large $x$, with the added bonus that 
besides extrapolating up in $Q^2$ one can simultaneously extrapolate 
backwards in $x$. The errors in such predictions are the usual mix 
of experimental and parametrization uncertainties (see the discussion in
Sects.~\ref{sec:keller;2qcd}, \ref{sec:alekhin;2qcd} and in 
\cite{Ball:LesHouches}), 
and theoretical errors predominantly due to missing sub-leading corrections,
which may be estimated by partial calculations of NNLO terms 
\cite{Santiago:1999xb,vanNeerven:1999ca} (see also Sect.~\ref{sec:vogt;2qcd}).

However to obtain truly reliable predictions for processes at the LHC it is
not sufficient to confirm NLO QCD within errors at 
HERA: we must also be convinced that new sources of theoretical uncertainty do 
not arise as the kinematic region is extended.
In particular, as one goes to smaller values of $x$ it is not clear that  
retaining only the first few terms in the expansion (\ref{apexp;1qcd})
of the splitting 
functions in powers of $\as$ will be and remain a good approximation: 
as soon as $\xi=\ln{1/x}$ is sufficiently large that $\as \xi\sim 1$, 
terms of order $\as (\as \xi)^n$ (LLx), 
$\as^2 (\as \xi)^n$ (NLLx), \dots must also be considered in order 
to achieve a result which is reliable up to terms of order $\as^3$.
In fact $\as\xi\gsim 1$ throughout much of the kinematic 
region available at both HERA and the LHC, so one might naively expect these 
effects to be significant when extrapolating from one to the other. 
The fact that at HERA they seem to be small empirically is a 
mystery which must be solved if reliable predictions are to be made for the 
LHC.

Using the BFKL kernel $K(Q^2,k^2)$ 
\cite{Lipatov:1976zz,Fadin:1975cb,Kuraev:1976ge,
Kuraev:1977fs,Balitsky:1978ic} 
(see also Sect.~\ref{sec:smallx;qcd}) calculated to 
$O(\as)$ (LO) it is possible \cite{Jaroszewicz:1982gr,Catani:1990sg,
Catani:1991gu} to deduce the coefficients of the LLx singularities of the 
AP splitting function to all orders in perturbation theory. 
Summing these up, the splitting function (and thus the structure function) 
is predicted to grow as $x^{-\lambda}$ as $x\to 0$, where (at LLx) 
$\lambda=\lambda_0\equiv (12\as\ln 2)/\pi$. This 
procedure may be extended to NLLx singularities, using 
calculations of the coefficient function and gluon 
normalization \cite{Catani:1993rn,Catani:1994sq} and of the NLLx kernel 
\cite{Fadin:1998py,Fadin:1995xg,Fadin:1996tb,Fadin:1993wh,Fadin:1994fj,
Fadin:1996yv,Fadin:1996nw,Fadin:1998hr,Camici:1997ij,Ciafaloni:1998gs,
DelDuca:1996ki,DelDuca:1996me,DelDuca:1998kx,Bern:1998sc}, to give all
the NLLx terms in the splitting function 
\cite{Ball:1995vc,Ciafaloni:1995bn,Ball:1995tn,Camici:1997ta,
Ellis:1995gv,Catani:1996ze,Catani:1997sc}. 
It was known some time ago that reconciling these summed 
logarithms with the HERA data was going to be difficult 
\cite{Ball:1994du,Ball:1994kc,Forte:1995vs,Forte:1996xv,
Bojak:1997nr,Bojak:1997me}, simply because there is no evidence 
in the data for a rise with a fixed power $\lambda_0$.
Once all the NLLx corrections were known it became clearer why: 
the expansion in summed anomalous dimensions at LLx,NLLx,... 
is unstable \cite{Ball:1998be,Blumlein:1998pp,Ball:1999sh}, 
the ratio of NLLx to LLx contributions growing without bound as ${x}\to 0$. 
It follows that the previous theoretical estimates 
\cite{Ball:1995vc,Ciafaloni:1995bn,Ball:1995tn,Camici:1997ta,
Ellis:1995gv,Catani:1996ze,Catani:1997sc,Forte:1996xv,
Bojak:1997nr,Bojak:1997me} of the size of the effects of the 
small $x$ logarithms based on the fixed order BFKL equation, either 
at LO or NLO, were all hopelessly unreliable. Indeed any calculation 
which resums LO and NLO logs of $Q^2$, but sums up only 
LO and NLO logarithms of $x$ is seen to be 
insufficient: some sort of all order resummation of the small 
$x$ logarithms is necessary. Clearly there are many ways in which such
a resummation might be attempted: what are needed are guiding
principles to keep it under control.

There are two distinct strands to this problem. The first is the 
stability of the BFKL equation itself (see the discussion in 
Sect.~\ref{sec:smallxsalam;qcd}). Various proposals  
have been put forward: for example a particular choice of the renormalization 
scale~\cite{Brodsky:1999kn}, or a different 
identification of the large logs which are resummed 
\cite{Schmidt:1999mz,Forshaw:1999xm}.
However the root of the problem \cite{Salam:1998tj} 
is that the perturbative contributions to the kernel $K(Q^2,k^2)$
contain unresummed logarithms of the form $\as(\as t)^n$ (LLQ),
$\as^2(\as^n t)^n$ (NLLQ),\dots, where $t\equiv \ln Q^2/k^2$, 
which destabilise the fixed order expansion both in the ultraviolet 
region $Q^2\gg k^2$ and in the infrared $Q^2\ll k^2$.
These logarithmic contributions turn out to be so large that the 
fixed order expansion is useless, even in the small $x$ region, 
unless $\as$ is unrealistically small. In order to obtain 
a realistic approximation to the kernel, the large logarithms
of $Q^2$ must be resummed to all orders in perturbation theory.
Fortunately the ultraviolet logarithms not associated with the running of 
the coupling may be determined at LLQ and NLLQ from the LO and NLO 
Altarelli--Parisi splitting functions \cite{Altarelli:1999vw}. 
Summing them up, longitudinal momentum is automatically conserved: 
the relevant part of the kernel then satisfies the all order sum 
rule\cite{Altarelli:1999vw} $\int_{-\infty}^\infty \! dt K(t)=1$.
Furthermore, it turns out that when the LLQ and NLLQ contributions 
to the LO and NLO BFKL kernels are resummed, the expansion stabilises
in the perturbative ($Q^2 >> k^2$) region, and the residual
part of the kernel which resums the remaining small $x$ logarithms 
is relatively small.

However before we can use this resummed BFKL kernel to compute small $x$ 
resummation corrections we need to resolve a second issue: the 
inherent perturbative instability of the LLx and NLLx contributions 
to the splitting functions first noted in \cite{Ball:1998be,Blumlein:1998pp}. 
This is quite distinct from the previous problem: it can be shown (see
\cite{Ball:1999sh} and Sect.~\ref{sec:smallxsalam;qcd}) 
to follow inevitably from the shift in the 
value of $\lambda$ from its LLx value $\lambda_0$ to $\lambda_0 +
\Delta\lambda$ at NLLx. This shift must be accounted for exactly if a 
sensible resummed perturbative expansion is to be obtained. Since in practice 
the correction $\Delta\lambda$ is of the same order as $\lambda_0$, 
it seems probable that $\lambda=\lambda_0+\Delta\lambda$ is 
not calculable in perturbation theory: rather the 
value of $\lambda$ may be used to parameterise the
uncertainty in the value of the kernel $K(Q^2,k^2)$ when $Q^2\sim k^2$.

Putting together the two principles of momentum conservation and
perturbative stability, we can compute fully resummed NLO splitting 
functions \cite{Altarelli:1999vw}. The result depends on the unknown parameter
$\lambda$. Provided $\lambda \lsim 0$, the corrections to
conventional NLO evolution in the HERA region are tiny: 
this in itself is sufficient to explain the success of 
NLO evolution in describing the HERA data, 
and furthermore means that effect of resummed small $x$ 
logarithms on the extrapolation upwards in $Q^2$ from HERA 
to the LHC should also be rather small. More significant effects might be 
expected in the extrapolation down to smaller $x$, particularly if 
$Q^2$ is also small and $\lambda$ is positive. It should now be possible to 
quantify such uncertainties by a phenomenological analysis, using available 
HERA data to constrain $\lambda$.

One might have hoped that eventually it would be possible to compute $\lambda$
perturbatively. The main uncertainty in current calculations is due to
the unresummed infrared logarithms in the kernel $K(Q^2,k^2)$, which 
destabilise the fixed order perturbative expansion in the region $Q^2\ll k^2$.
In Refs.~\cite{Ciafaloni:1999iv,Ciafaloni:1999yw,Ciafaloni:1999au} an attempt 
is made to resum these logarithms 
through a symmetrization of $K(Q^2,k^2)$ in $Q^2$ and $k^2$: the 
idea is to deduce the infrared logarithms from the ultraviolet ones.  
The main shortcoming of this approach is that it makes implicit 
assumptions about the validity of perturbation 
theory when $Q^2$ is very small: symmetrization only works when 
running coupling effects are included, but making the coupling run with $Q^2$ 
or $k^2$ is not only very model dependent but seems inevitably to 
destabilise the small $x$ evolution \cite{Collins:1992nk,McDermott:1995jq,
Bartels:1996yk,Haakman:1998nu,Kovchegov:1998ae,Armesto:1998gt}, 
suggesting that effects beyond the reach of the usual perturbative expansion
become important in this region.

It seems that to make further progress we require either genuine
nonperturbative input, or a substantial extension of the perturbative
domain. A possible way in which this might be done through a 
new factorization procedure was explored in Ref.~\cite{Ball:1997vf}, from 
which the main conclusion
was that at small $x$ the coupling should run not with $Q^2$, but with
$W^2\sim Q^2/x$. Preliminary calculations \cite{Roberts:1999gb}  
suggest that this is not phenomenologically unacceptable. 
An alternative approach to factorization 
in high energy QCD based on Wilson lines may be found in 
Refs.~\cite{Balitsky:1998kc,Balitsky:1999ya}. 
Clearly much work remains to be done.

\section{PROMPT PHOTON PRODUCTION\protect\footnote{Session coordinators:
M.~Fontannaz, S.~Frixione and S.~Tapprogge.}}
\label{sec:photons;qcd}

\subsection{General features of photon production
\protect\footnote{Contributing authors: P.~Aurenche, 
M.~Fontannaz and S.~Frixione.}}

When mentioning the photon in the framework of high-energy collider
physics, one is immediately led to think -- with good reasons --
to Higgs searches through the gold-plated channel $H\to\gamma\gamma$. 
However, the production of photons also deserves attention on its own.
Firstly, a detailed understanding of the continuum two-photon production
is crucial in order to clearly disentangle any Higgs signals from the
background. Secondly, in hadronic collisions, where a very large number
of strong-interacting particles is produced, photon signals
are relatively clean, since the photon directly couples only to quarks.
Therefore, prompt-photon data can be used to study the underlying parton 
dynamics, in a complementary way with respect to analogous studies performed 
with hadrons or jets. For the same reason, these data represent a
very important tool in the determination of the gluon density in the 
proton, $f_g(x)$. Indeed, in recent years almost all the {\em direct} 
information (that is, not obtained through scaling violations as predicted 
by the DGLAP equations) on the intermediate- and high-$x$ behaviour
of $f_g(x)$ came from prompt-photon production, $pp\rightarrow \gamma X$ 
and $pN \rightarrow \gamma X$, in fixed-target experiments.
The main reason for this is that, at LO, a photon in the
final state is produced in the reactions $qg\to\gamma q$ and
$q\bar{q}\to\gamma g$, with the contribution of the former subprocess
being obviously sensitive to the gluon and usually dominant over that
of the latter. It is the `point-like' coupling of the photon to the
quark in these subprocesses that is responsible for a much cleaner
signal than, say, for the inclusive production of a $\pi^0$, which
proceeds necessarily through a fragmentation process.

There is, however, a big flaw in the arguments given above. In fact,
photons can also be produced through a fragmentation process, in which
a parton, scattered or produced in a QCD reaction, fragments into a
photon plus a number of hadrons. The problem with the fragmentation
component in the prompt-photon reaction is twofold: first, it
introduces in the cross section a dependence upon non-perturbative
fragmentation functions, similar to those relevant in the case of
single-hadron production, which are not calculable in perturbative QCD:
they depend on non-perturbative initial conditions~\cite{Gluck:1993zx,
Bourhis:1998yu}, and only their asymptotic behavior at very large scales 
is perturbatively calculable~\cite{Witten:1977ju}. These functions
are, at present, very poorly determined by the sparse LEP data available.
Secondly, {\em all} QCD partonic reactions contribute to the 
fragmentation component; thus, when addressing the problem of the
determination of the gluon density, the advantage of having a priori
only one partonic reaction ($q\bar{q}\to\gamma g$) competing with the
signal ($qg\to\gamma q$) is lost, even though some of the subprocesses
relevant to the fragmentation part at the same time result from a
gluon in the initial state.

The relative contribution of the fragmentation component with respect
to the direct component (where the photon participates in the
short-distance, hard-scattering process) is larger the larger the
centre-of-mass energy and the smaller the final-state transverse
momentum\footnote{Actually, in the fixed-target $pp\to\gamma X$
reaction, one can see the fragmentation component increasing
relatively to the direct one also at very {\em large} $\ptg$,
because of the direct cross section dying out very quickly at such
momenta. This effect is of no phenomenological relevance at the
LHC.}: at the LHC, for transverse momenta of the order of few tens of
GeV, it can become dominant.  However, here the situation is saved by
the so-called `isolation' cut, which is imposed on the photon signal
in experiments. Isolation is an experimental necessity: in a hadronic
environment the study of photons in the final state is complicated by
the abundance of $\pi^0$'s, eventually decaying into pairs of
$\gamma$'s. The isolation cut simply serves to improve the
signal-to-noise ratio: if a given neighbourhood of the photon is free
of energetic hadron tracks, the event is kept; it is rejected
otherwise. Fortunately, by requiring the photon to be isolated, one
also severely reduces the contribution of the fragmentation part to
the cross section. This is because fragmentation is an essentially
collinear process: therefore, photons resulting from parton
fragmentation are usually accompanied by hadrons, and are therefore
bound to be rejected after the imposition of an isolation cut.

It has to be stressed that, at fixed-target energies, the size of the
average transverse momentum allows to resolve the two photons coming
from $\pi^0$ decay and therefore to identify the $\pi^0$. It seems
therefore appropriate to recall some fixed target results before
turning to prompt photon production at the LHC. A recent review on the
comparisons between data and theory may be found in
\cite{Aurenche:1999gv}. Theory means NLO predictions including the
direct and the bremsstrahlung contributions \cite{Aurenche:1988fs,
  Aversa:1989vb, Aurenche:1993yc, Baer:1990ra, Gordon:1993qc}. A
Fortran code which puts together both contributions and allows simple
changes of parameters is now available \cite{werlen;6qcd}. The
conclusion reached in ref.~\cite{Aurenche:1999gv} is that some data
sets are incompatible with each other, or that theory must be
modified. A modification proposed in ref.~\cite{Apanasevich:1998hm}
consists in introducing transverse momentum of initial partons with a
large average value $<\kappa_{\bot}> \sim 1.4$~GeV. If this average
value varies with $\sqrt{s}$, then it is possible to adjust theory to
data. The resummation of threshold effects \cite{Catani:1999hs} (see
also Sect.~\ref{sec:rescal;qcd}) increases the cross section at large
$x_{\bot} = 2 p_{\bot}/\sqrt{s}$, but it cannot remove the discrepancy
between theory and data. Clearly an unsettled problem remains in this
fixed target energy range, which questions the possibility to
determine the gluon contents of the proton from prompt photon data
(see Sect.~\ref{sec:pdf;qcd}).

We now turn to the case of photon production at high-energy colliders;
after some general introductory remarks, we will present
phenomenological predictions relevant to the LHC; we remind the reader
that the production of prompt photons at LHC was first studied at the Aachen
workshop~\cite{pastlhcproc;6qcd}. No NLO corrections to the bremsstrahlung 
terms were available at that time, and the isolation prescriptions were 
implemented only at LO accuracy. Since then, theoretical computations
progressed toward a fully consistent NLO framework, which we will
discuss in the following.

\subsection{Isolation prescriptions\protect\footnote{Contributing 
author: S.~Frixione}}

As mentioned before, the fragmentation contribution, that threatened
to spoil the cleanliness of the photon signals at colliders, is
relatively well under control in the case of isolated-photon cross
sections. There is of course a price to pay for this gain: the
isolation condition poses additional problems in the theoretical
computations, which are not present in the case of fully-inclusive
photon cross sections. To be specific, we write the cross section for
the production of a single isolated photon in hadronic collisions as
follows\footnote{The production of pairs of isolated photons can be 
described in the very same manner; we will consider this case later. 
Here we stick to a simpler case in order to have as simple as notation 
as possible.}:
\beeq
&&d\sigma_{h_1h_2}(p_1,p_2;p_\gamma)=
\nonumber \\*&&\phantom{aa}
\int dx_1 dx_2 f_{a/h_1}(x_1,\mu_F) f_{b/h_2}(x_2,\mu_F) 
d\hat{\sigma}_{ab,\gamma}^{isol}(x_1 p_1,x_2 p_2;p_\gamma;\mu_R,\mu_F,\mug)
\nonumber \\*&&
+\int dx_1 dx_2 dz f_{a/h_1}(x_1,\mu_F) f_{b/h_2}(x_2,\mu_F) 
d\hat{\sigma}_{ab,c}^{isol}(x_1 p_1,x_2 p_2;p_\gamma/z;\mu_R,\mu_F,\mug) 
d_{\gamma/c}(z,\mug),\phantom{aaa}
\label{factth}
\eeeq
where $h_1$ and $h_2$ are the incoming hadrons, with momenta $p_1$ and $p_2$
respectively, and a sum over the parton indices $a$, $b$ and $c$ is 
understood. In the first term on the right hand side of eq.~(\ref{factth}) 
(the direct component) the subtracted partonic cross sections 
\mbox{$d\hat{\sigma}_{ab,\gamma}^{isol}$} get contributions from all 
the diagrams with a photon leg. On the other hand, the subtracted
partonic cross sections \mbox{$d\hat{\sigma}_{ab,c}^{isol}$}
appearing in the second term on the right hand side of eq.~(\ref{factth}) 
(the fragmentation component), get contribution from the pure 
QCD diagrams, with one of the partons eventually fragmenting 
in a photon, in a way described by the parton-to-photon fragmentation 
function $d_{\gamma/c}$. As the notation in eq.~(\ref{factth}) indicates, 
the isolation condition is embedded into the partonic cross sections. 

It is a well-known fact that, in perturbative QCD beyond LO,
and for all the isolation prescriptions known at present, with the
exception of that of ref.~\cite{Frixione:1998jh}, neither the direct nor
the fragmentation components are {\em separately} well defined
at any fixed order in perturbation theory: only their sum is
physically meaningful. In fact, the direct component is affected
by quark-to-photon collinear divergences, which are
subtracted by the bare fragmentation function that appears in
the unsubtracted fragmentation component. Of course, this subtraction 
is arbitrary as far as finite terms are concerned. This is formally 
expressed in eq.~(\ref{factth}) by the presence of the same scale 
$\mug$ in both the direct and fragmentation components: a finite piece
may be either included in the former or in the latter, without affecting
the physical predictions. The need for introducing a fragmentation 
contribution is physically better motivated from the fact that a QCD hard 
scattering process may produce, again through a fragmentation process, 
a $\rho$ meson that has the same quantum numbers as the photon and can 
thus convert into a photon, leading to the same signal. 

As far as the isolation prescriptions are concerned, here we will
restrict to those belonging to the class that can be denoted as `cone
isolations'~\cite{Baer:1990ra,Aurenche:1990gv,Berger:1991et,Glover:1992sf,%
Kunszt:1993ab,Gordon:1994ut}. In the framework of hadronic collisions,
where the need for invariance under longitudinal boosts (which is necessary for
collinear factorizability) suggests not to define physical quantities 
in terms of angles, the cone is drawn in the pseudorapidity--azimuthal 
angle plane, and corresponds to the set of points
\beq
{\cal C}_{R}=\left\{(\eta,\phi)\mid
\sqrt{(\eta-\etag)^2+(\phi-\phig)^2}\le R\right\},
\label{coneRz}
\eeq
where $\etag$ and $\phig$ are the pseudorapidity and azimuthal angle
of the photon, respectively, and $R$ is the aperture (or half-angle)
of the cone. After having drawn the cone, one has to actually impose
the isolation condition. We consider here two sub-classes of cone
isolation, whose difference lies mainly in the behaviour of the
fragmentation component. Prior to that, we need to define the total
amount of hadronic transverse energy deposited in a cone of half-angle
$R$ as
\beq
E_{T,had}(R)=\sum_{i=1}^n E_{Ti}\theta(R-R_{\gamma i}),
\eeq
where
\beq
R_{\gamma i}=\sqrt{(\eta_i-\etag)^2+(\phi_i-\phig)^2},
\eeq
and the sum runs over all the hadrons in the event (or, alternatively,
$i$ can be interpreted as an index running over the towers of a
hadronic calorimeter). For both the isolation prescriptions we are 
going to define below, the first step is to draw a cone of fixed 
half-angle $R_0$ around the photon axis, as given in eq.~(\ref{coneRz}). 
We will denote this cone as the isolation cone.

\begin{description}

\item[Definition A.] The photon is isolated if the total amount of 
hadronic transverse energy in the isolation cone fulfils the 
following condition:
\beq
E_{T,had}(R_0)\le \epc \ptg,
\label{iscondA}
\eeq
where $\epc$ is a fixed (generally small) parameter, and $\ptg$ is the 
transverse momentum of the photon.

\item[Definition B.] The photon is isolated if the following inequality 
is satisfied:
\beq
E_{T,had}(R)\le \epg\ptg {\cal Y}(R),
\label{iscondB}
\eeq
for {\it all} the cones lying inside the isolation cone, that is for
any $R\le R_0$. The function ${\cal Y}$ is arbitrary to a large extent, 
but must at least have the following property:
\beq
\lim_{R\to 0} {\cal Y}(R)=0,
\label{limY}
\eeq
and being different from zero everywhere except for $R=0$.

\end{description}

\noindent
Definition A was proven to lead to an infrared-safe cross section
at all orders of perturbation theory in ref.~\cite{Catani:1998yh}. 
The smaller $\epc$, the tighter the isolation. Loosely
speaking, for vanishing $\epc$ the direct component behaves like
\mbox{$\log\epc$}, while the fragmentation component behaves like
\mbox{$\epc\log\epc$}. Thus, for $\epc\to 0$ eq.~(\ref{factth})
diverges. This is obvious since the limit $\epc\to 0$ corresponds
to a fully-isolated-photon cross section, which cannot be a meaningful
quantity, whether experimentally (because of limited energy resolution)
or theoretically (because soft-particle emission inside the cone cannot be forbidden
without spoiling the infrared safety of the cross section).

Definition B was proposed and proven to lead to an infrared-safe cross 
section at all orders of perturbation theory in ref.~\cite{Frixione:1998jh}. 
Eq.~(\ref{limY}) implies that the energy of a parton falling into the isolation
cone ${\cal C}_{R_0}$ is correlated to its distance (in the $\eta$--$\phi$
plane) from the photon. In particular, a parton becoming collinear to
the photon is also becoming soft. When a quark is collinear to the photon,
there is a collinear divergence; however, if the quark is also soft,
this divergence is damped by the quark vanishing energy. When a gluon is
collinear to the photon, then either it is emitted from a quark, which is
itself collinear to the photon -- in which case, what was said previously
applies -- or the matrix element is finite. Finally, it is clear that
the isolation condition given above does not destroy the cancellation
of soft singularities, since a gluon with small enough energy can be 
emitted anywhere inside the isolation cone. The fact that this prescription
is free of final-state QED collinear singularities implies that the 
direct part of the cross section is finite. As far as the fragmentation
contribution is concerned, in QCD the fragmentation mechanism is purely
collinear. Therefore, by imposing eq.~(\ref{iscondB}), one forces the
hadronic remnants collinear to the photon to have zero energy. This
is equivalent to saying that the fragmentation variable $z$ is restricted
to the range $z=1$. Since the parton-to-photon fragmentation functions
do not contain any $\delta(1-z)$, this means that the fragmentation
contribution to the cross section is zero, because an integration over
a zero-measure set is carried out. Therefore, only the first term on the 
right hand side of eq.~(\ref{factth}) is different from zero, and it does 
not contain any $\mug$ dependence.

We stress again that the function ${\cal Y}$  can be rather freely 
defined. Any sufficiently well-behaved function, fulfilling
eq.~(\ref{limY}), could do the job, the key point being the correlation
between the distance of a parton from the photon and the parton energy,
which must be strong enough to cancel the quark-to-photon
collinear singularity. Throughout this paper, we will use
\beq
{\cal Y}(R)=\left(\frac{1-\cos R}{1-\cos R_0}\right)^n,\;\;\;\;\;\;
n=1.
\label{Yfun}
\eeq
We also remark that the traditional cone-isolation prescription, 
eq.~(\ref{iscondA}), can be formally recovered from eq.~(\ref{iscondB}) 
by setting ${\cal Y}=1$ and $\epg=\epc$.

\subsection{Single isolated photons at the LHC\protect\footnote{Contributing 
author: S.~Frixione}}

\begin{figure}[t!]
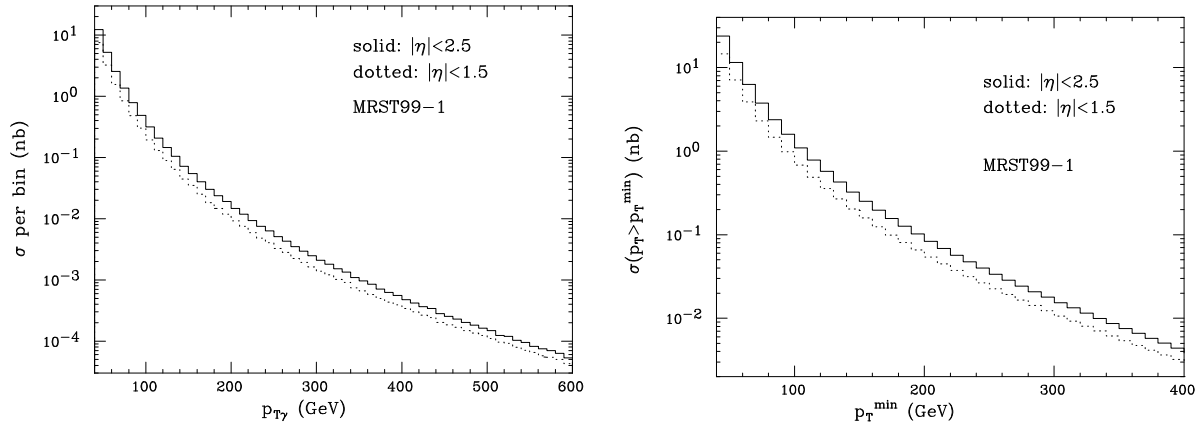

\centerline{
   \epsfig{figure=phoptdiff.ps,width=0.48\textwidth,clip=}
   \hfill
   \epsfig{figure=phoptint.ps,width=0.48\textwidth,clip=} }
\vskip-0.5cm
\caption{
Benchmark cross sections for isolated-photon production: differential
spectrum (left) and integrated spectrum (right).}
 \label{fig:phorates}
\end{figure}

In this section, we will present results for isolated-photon cross
sections in $pp$ collisions at 14 TeV. These results have been obtained
with the fully-exclusive NLO code of ref.~\cite{Frixione:1998hn}, and are
relevant to the isolation obtained with definition B; the actual
parameters used in the computation are given in eq.~(\ref{Yfun}),
together with $\epg=1$. We set $R_0=0.4$. We will comment in the 
following on the outcome of definition A.
Benchmark rates for isolated photons over different ranges of rapidity
are given in Fig.~\ref{fig:phorates}.

Any sensible perturbative computation should address the issue of
the perturbative stability of its results. A rigorous estimate of
the error affecting a cross section at a given order can be given
if the next order result is also available. If this is not the case,
it is customary to study the dependence of the physical observables
upon the renormalization ($\mu_R$) and factorization ($\mu_F$) scales.
It is important to stress that the resulting spread should not be
taken as the `theoretical error' affecting the cross section;
to understand this, it is enough to say that the range in which
$\mu_R$ and $\mu_F$ are varied is arbitrary. Rather, one should
compare the spread obtained at the various perturbative orders;
only if the scale dependence decreases when including higher
orders the cross section can be regarded as perturbatively stable
and sensibly compared to data.

Usually, $\mu_R$ and $\mu_F$ are imposed to have the same value, $\mu$,
which is eventually varied. However, this procedure might hide some
problems, because of a possible cancellation between the effects 
induced by the two scales. It is therefore desirable to vary $\mu_R$
and $\mu_F$ independently. Here, an additional problem arises at
the NLO. The expression of any cross section in terms of $\mu$
(that is, when $\mu_R=\mu_F$) is not ambiguous, 
while {\em it is} ambiguous if $\mu_R\ne\mu_F$. 
In fact, when $\mu_R\ne\mu_F$, the cross section can be written as the sum 
of a term corresponding to the contribution relevant to the case $\mu_R=\mu_F$, 
plus a term of the kind:
\beq
\as(\mu_A)\,{\cal B}(\as(\mu_R))\,\log\frac{\mu_R}{\mu_F},
\label{addterm}
\eeq
where ${\cal B}$ has the same power of $\as$ as the LO contribution,
say $\as^k$. The argument of the $\as$ in front of eq.~(\ref{addterm}),
$\mu_A$, can be chosen either equal to $\mu_R$ or equal to $\mu_F$, since
the difference between these two choices is of NNLO. Thus, it follows
that the dependence upon $\mu_R$ or $\mu_F$ of a NLO cross section
reflects the arbitrariness of the choice made in eq.~(\ref{addterm}),
which is negligible only if the NNLO ($\as^{k+2}$) corrections are
much smaller than the NLO ones ($\as^{k+1}$). This leads to the
conclusion that a study of the dependence upon $\mu_R$ or $\mu_F$ 
{\em only} can be misleading. In other words: ${\cal B}$ in
eq.~(\ref{addterm}) is determined through DGLAP equations in order to
cancel the scale dependence of the parton densities up to terms of order 
$\as^{k+2}$. This happens regardless of the choice made for $\mu_A$ in 
eq.~(\ref{addterm}). However, here we are not discussing the cancellation 
to a given perturbative order of the effects due to scale variations; we 
are concerned about the coefficient in front of the ${\cal O}(\as^{k+2})$
term induced by such variations, whose size is dependent upon the
choice made for $\mu_A$ and therefore, to some extent, arbitrary. We
have to live with this arbitrariness, if we decide to vary $\mu_R$ or
$\mu_F$ only. However, we can still vary $\mu_R$ and $\mu_F$
independently, but eventually putting together the results in some
sensible way, that reduces the impact of the choice made for $\mu_A$.
In this section, we will consider the quantities defined as follows:
\beeq
\left(\frac{\delta\sigma}{\sigma}\right)_\pm&=&
\pm\left\{\,\,
\left[\frac{\sigma(\mu_R=\muo,\mu_F=\muo)-\sigma(\mu_R=a_\pm\muo,\mu_F=\muo)}
           {\sigma(\mu_R=\muo,\mu_F=\muo)+\sigma(\mu_R=a_\pm\muo,\mu_F=\muo)}
\right]^2\right.
\nonumber \\&&\phantom{\pm}+\left.
\left[\frac{\sigma(\mu_R=\muo,\mu_F=\muo)-\sigma(\mu_R=\muo,\mu_F=a_\pm\muo)}
           {\sigma(\mu_R=\muo,\mu_F=\muo)+\sigma(\mu_R=\muo,\mu_F=a_\pm\muo)}
\right]^2\right\}^{\frac{1}{2}},
\label{delsigdef}
\eeeq
where $a_+$ and $a_-=1/a_+$ are two numbers of order one, which we
will take equal to 1/2 and 2 respectively; the $\pm$ sign in front of
the right hand side of eq.~(\ref{delsigdef}) is purely conventional. We can
evaluate \mbox{$(\delta\sigma/\sigma)_\pm$} by using $\mu_A=\mu_R$ or
$\mu_A=\mu_F$ in eq.~(\ref{addterm}). The reader can convince himself,
with the help of the renormalization group equation~(\ref{rgeq;1qcd}), 
that the difference between these two choices is of order $\as^4$ in the
expansion of {\em the contribution to}
\mbox{$(\delta\sigma/\sigma)_\pm^2$} {\em due to eq.~(\ref{addterm})};
on the other hand, this difference is only of order $\as^3$ in each of
the two terms under the square root in the right hand side of
eq.~(\ref{delsigdef}). This is exactly what we wanted to achieve: a
suitable combination of the cross sections resulting from independent
$\mu_R$ and $\mu_F$ variations is less sensitive to the choice for
$\mu_A$ made in eq.~(\ref{addterm}) than the results obtained by 
varying $\mu_R$ or $\mu_F$ {\em only}.

\begin{table}[t!]
\begin{center}
\begin{tabular}{|l||c|c|c|c|c||c|c||c|} \hline
& \multicolumn{5}{c||}{MRST99} 
& \multicolumn{2}{c||}{CTEQ5} 
& 
\\ \hline
& 1 & 2 & 3 & 4 & 5
& M & HJ
& $(\delta\sigma/\sigma)_\pm$
\\ \hline\hline
NLO, $\abs{\etag}<2.5$ 
  & 23.78 & 23.20 & 24.19 & 22.07 & 25.49 
  & 25.10 & 24.61 & $^{+0.068}_{-0.057}$
\\ \hline
 LO, $\abs{\etag}<2.5$ 
  & 10.34 & 10.07 & 10.52 & 9.875 & 10.78 
  & 10.91 & 10.66 & $^{+0.090}_{-0.072}$
\\ \hline
NLO, $\abs{\etag}<1.5$ 
  & 14.59 & 14.23 & 14.88 & 13.66 & 15.53 
  & 15.35 & 15.01 & $^{+0.068}_{-0.056}$
\\ \hline
 LO, $\abs{\etag}<1.5$ 
  & 6.457 & 6.270 & 6.583 & 6.212 & 6.657 
  & 6.771 & 6.596 & $^{+0.091}_{-0.073}$
\\ \hline
\end{tabular} 
\end{center} 
\vskip-0.25cm                                                           
\caption{
Isolated-photon cross sections (nb), with $40<\ptg<400$~GeV, in two
different rapidity ranges, for various MRST (MRST99-1/5) and CTEQ 
(CTEQ5M/HJ) parton densities. The scale dependence, evaluated according 
to eq.~(\ref{delsigdef}) and with the MRST99-1 set, is also shown.
}
\label{tab:xsec}
\end{table}                                                               
In table~\ref{tab:xsec} we present the results for the total isolated-photon
rates, both at NLO and at LO. The latter cross sections have been obtained
by retaining only the LO terms (${\cal O}(\aem\as)$)
in the short-distance cross section, and convoluting them with
NLO-evolved parton densities. Also, a two-loop expression for 
$\as$ has been used. There is of course a lot of freedom in
the definition of a Born-level result. However, we believe that
with this definition one has a better understanding of some 
issues related to the stability of the perturbative series.
To obtain the rates entering table~\ref{tab:xsec}, we
required the photon transverse momentum to be in the range
$40<\ptg<400$~GeV, and we considered the rapidity cuts $\abs{\etag}<1.5$ and
$\abs{\etag}<2.5$, in order to simulate a realistic geometrical acceptance 
of the LHC detectors. We first consider the scale dependence of our
results (last column), evaluated according to eq.~(\ref{delsigdef}).
We see that the NLO results are clearly more stable than the LO
ones; this is reassuring, and implies the possibility of a sensible
comparison between NLO predictions and the data. Notice that the 
size of the radiative corrections ($K$ factor, defined as the
ratio of the NLO result over the LO result) is quite large. 
From the table, we see that the cross sections obtained with
different parton densities differ by 6\% at the most (relative
to the result obtained with MRST99-1~\cite{Martin:1999ww}, which we take as 
the default set). MRST99 sets 2 and 3 are meant to give an estimate 
of the effects due to the current uncertainties affecting the gluon density
(see sect.~\ref{sec:pdf;qcd}), whereas sets 4 and 5 allow to study the
sensitivity of our predictions to the value of $\as(M_{\sss Z})$ (sets
1, 4 and 5 have $\lambdamsb=$220, 164 and 288~MeV respectively). On
the other hand, the difference between MRST99-1 and
CTEQ5M~\cite{Lai:2000wy} results is due to the inherent difference
between these two density sets (CTEQ5M has $\lambdamsb=$226~MeV, and
therefore the difference in the values of $\as(M_{\sss Z})$ plays only
a very minor role).

From inspection of table~\ref{tab:xsec}, we can conclude that
isolated-photon cross section at the LHC is under control, both 
in the sense of perturbation theory and of the dependence upon
non-calculable inputs, like $\as(M_{\sss Z})$ and parton densities.
The relatively weak dependence upon the parton densities, however,
is not a good piece of news if one aims at using photon data to
directly access the gluon density. On the other hand, the expected
statistics is large enough to justify attempts of a direct measurement 
of such a quantity. In the remainder of this section, we will
concentrate on this issue. We will consider 
\beq
{\cal R}_x=\frac{d\sigma_0/dx-d\sigma/dx}{d\sigma_0/dx+d\sigma/dx},
\eeq
where $x$ is any observable constructed with the kinematical variables
of the photon and, possibly, of the accompanying jets. $\sigma$ and
$\sigma_0$ are the cross sections obtained with two different sets
of parton densities, the latter of which is always the default
one (MRST99-1). We can imagine a gedanken experiment, where it is
possible to change at will the parton densities; in this way, we can assume
the relative statistical errors affecting $\sigma$ and $\sigma_0$ to 
decrease as $1/\sqrt{N}$ and $1/\sqrt{N_0}$, $N$ and $N_0$ being
the corresponding number of events. It is then straightforward to
calculate the statistical error affecting ${\cal R}_x$; by imposing
${\cal R}_x$ to be larger than its statistical error, one gets
\beq
{\cal R}_x\,>\,\left({\cal R}_x\right)_{min}\,\equiv\,
\frac{1}{\sqrt{2{\cal L}\ep\sigma(x,\Delta x)}},
\label{Rmin}
\eeq
where ${\cal L}$ is the integrated luminosity, $\ep\le 1$ collects
all the experimental efficiencies, and 
\beq
\sigma(x,\Delta x)=\int_{x-\Delta x/2}^{x+\Delta x/2} dx\,\frac{d\sigma}{dx}
\eeq
is the total cross section in a range of width $\Delta x$ around $x$.

\begin{figure}[t!]
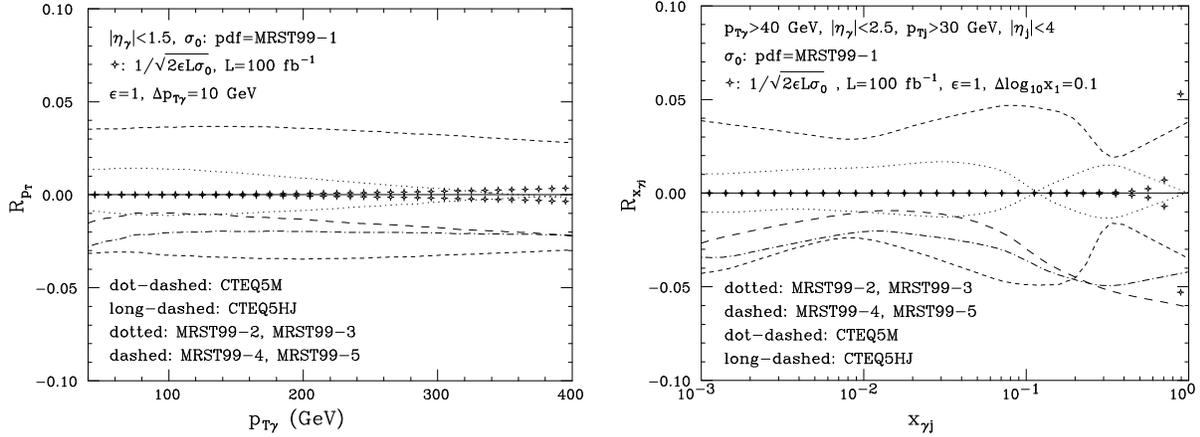

\centerline{
   \epsfig{figure=pt_pdf_e15.ps,width=0.48\textwidth,clip=}
   \hfill
   \epsfig{figure=xgj_pdf_cuts1.ps,width=0.48\textwidth,clip=} }
\vskip-0.5cm
\caption{
Dependence of isolated-photon and isolated-photon-plus-jet cross
section upon parton densities, as a function of $\ptg$ and $x_{\gamma j}$.
}
 \label{fig:pdfdep}
\end{figure}

In fig.~\ref{fig:pdfdep} we present our predictions for ${\cal R}_x$.
In the left panel of the figure we have chosen $x=\ptg$, while in the
right panel we have $x=x_{\gamma j}$, where
\beq
x_{\gamma j}=\frac{\ptg\exp(\etag)+\ptj\exp(\etaj)}{\sqrt{s}}.
\eeq
In this equation $\sqrt{s}$ is the centre-of-mass energy of the colliding
hadrons, and $\ptj$ and $\etaj$ are the transverse momentum and rapidity
of the hardest jet recoiling against the photon. In order to reconstruct
the jets, we adopted here a $k_{\perp}$-algorithm~\cite{Catani:1993hr}, in the
version of ref.~\cite{Ellis:1993tq} with $D=1$. Notice that $x_{\gamma j}$ 
exactly coincides at the LO with the longitudinal momentum fraction $x$ of 
the partons in one of the incoming hadrons; NLO corrections introduce only 
minor deviations.
For all the density sets considered, the dependence of ${\cal R}$
upon $\ptg$ is rather mild. The values in the low-$\ptg$ region
could also be inferred from table~\ref{tab:xsec}, since the cross
section is dominated by small $\ptg$'s. Analogously to what happens 
in the case of total rates, the sets MRST99-4 and MRST99-5 give rise
to extreme results for ${\cal R}_{\ptg}$, since the value of 
$\as(M_{\sss Z})$ is quite different from that of the default set. From the 
figure, it is apparent that, by studying the transverse momentum spectrum, 
it will not be easy to distinguish among the possible {\em shapes} of the
gluon density. On the other hand, it seems that, as far as the 
statistics is concerned, a distinction between any two sets
can be performed. Indeed, the symbols in the figure display
the quantity defined in eq.~(\ref{Rmin}), for ${\cal L}=100$~fb$^{-1}$,
$\Delta\ptg=10$~GeV and $\ep=1$. Of course, the latter value is not
realistic. However, a smaller value (leading to a larger 
$({\cal R})_{min}$), can easily be compensated by enlarging
$\Delta\ptg$ and by the fact that the total integrated luminosity 
is expected to be much larger than that adopted in fig.~\ref{fig:pdfdep}.

Turning to the right panel of fig.~\ref{fig:pdfdep}, we can see
a much more interesting situation. Actually, it can be shown that
the pattern displayed in the figure is rather faithfully reproduced
by plotting the analogous quantity, where one uses the gluon densities
instead of the cross sections. This does not come as a surprise.
First, $x_{\gamma j}$ is in an almost one-to-one correspondence
with the $x$ entering the densities. Secondly, photon production
is dominated by the gluon-quark channel, and therefore the cross
section has a linear dependence upon $f_g(x)$, which can be easily 
spotted. It does seem, therefore, to be rather advantageous to
look at more exclusive variables, like photon-jet correlations
(this is especially true if one considers the procedure of unfolding
the gluon density from the data: in the case of single-inclusive
variables, the unfolding requires a de-convolution, which is not
needed in the case of correlations). Of course, there is a price
to pay: the efficiency $\ep$ will be smaller in the case of
photon-jet correlations, with respect to the case of single-inclusive
photon observables, mainly because of the jet-tagging. However,
from the figure it appears that there should be no problem with 
statistics, except in the very large $x_{\gamma j}$ region.

Finally, we would like to comment on the fact that, for the case of
single-inclusive photon observables, we also computed the cross
section by isolating the photon according to definition A, using
\mbox{$\epc=2$~GeV$/\ptg$}. The two definitions return a $\ptg$
spectrum almost identical in shape, with definition B higher by a
factor of about 9\%. It is only at the smallest $\ptg$ values that we
considered, that definition B returns a slightly steeper spectrum.
The fact that such different definitions produce very similar cross
sections may be surprising. This happens because, prior to applying
the isolation condition, partons tend to be radiated close to the
photon; therefore, most of them are rejected when applying the
isolation, no matter of which type. This situation has already been
encountered in the production of photons at much smaller energies. The
reader can find a detailed discussion on this point in
ref.~\cite{Frixione:1999gr}.

In the previous paragraphs, we concentrated on the possibility that
isolated-photon data can be used to constrain or measure the gluon
density in the proton. However, it is well known that $f_g(x)$ is
rather strongly correlated to $\as$. This is not a problem if one
is interested in observables that only depend upon the quantity
$\as f_g(x)$. On the other hand, the determination of the gluon
density alone is important in many respects. Thus, one has to assume 
an accurate knowledge of $\as$ to extract $f_g(x)$ from the data.
It is of course possible to turn this argument the other way round:
that is, to assume a good knowledge of $f_g(x)$ to measure $\as$.
The sensitivity of the isolated-photon cross section at the LHC
upon the value of $\as$ can be inferred from table~\ref{tab:xsec}
and fig.~\ref{fig:pdfdep}, looking at the results obtained with
the sets MRST99-4 and MRST99-5. Unfortunately, since the 
gluon-initiated processes dominate the cross section, and 
the gluon is the least known among the parton densities, this
procedure will probably result in sizeable systematic errors;
on the other hand, thanks to the size of the production rate,
we should expect a precise result on a statistical basis.
These considerations should encourage us to find alternative
ways of measuring $\as$ by using photon data. Since the main
problem is in the dependence of the cross section upon $f_g(x)$,
the guide line is that of considering observables that are less 
sensitive to the parton densities than the isolated-photon cross 
section.

In what follows, we will argue that an observable of this
kind is given by the ratio
\beq
{\cal X}(p_T)=
\frac{d\sigma_j}{d\ptj}(p_T)\, \Big/
\frac{d\sigma_\gamma}{d\ptg}(p_T)\,.
\label{Xdef}
\eeq
Here, \mbox{$d\sigma_j/d\ptj$} is the single-inclusive jet 
transverse momentum spectrum, while \mbox{$d\sigma_\gamma/d\ptg$}
is the transverse momentum spectrum of the isolated photon.

\begin{table}[t!]
\begin{center}
\begin{tabular}{|l||c|c|c|} \hline
$\ptmin$ (GeV) & 40 & 100 & 200 
\\ \hline
& \multicolumn{3}{c|}{$\abs{\etag}<1.5$} 
\\ \hline
MRST99-2
  & $1.006\pm 0.009$ & $1.003\pm 0.025$ & $0.991\pm 0.051$
\\ \hline
MRST99-3
  & $1.002\pm 0.009$ & $1.009\pm 0.023$ & $1.007\pm 0.048$
\\ \hline
& \multicolumn{3}{c|}{$\abs{\etag}<2.5$}  
\\ \hline
MRST99-2
  & $1.003\pm 0.008$ & $1.002\pm 0.023$ & $0.998\pm 0.042$
\\ \hline
MRST99-3
  & $1.009\pm 0.008$ & $1.009\pm 0.023$ & $0.999\pm 0.046$
\\ \hline
\end{tabular} 
\end{center}                                                            
\vskip-0.25cm                                                           
\caption{
NLO predictions for the double ratio $D$ defined in eq.~(\ref{Dratio}),
for various $\ptmin$ and two ranges in rapidity.
}
\label{tab:Dres}
\end{table}                                                               
It is immediate to see that, at the LO, ${\cal X}$ is proportional to 
$\as$. In the ratio that defines ${\cal X}$, one expects that the dependence
upon the parton densities cancel to a good extent, thus giving an
observable suited to measure $\as$, regardless of the precision to which
$f_g(x)$ is known. In hadronic physics, the trick of considering ratios of 
cross sections (instead of the cross sections themselves) in order to
reduce the dependence on the parton densities is frequently used.
In particular, for the measurement of $\as$ at hadron colliders, one 
can think to the $W+1$-jet over $W+0$-jet ratio (${\cal A}$), and to 
the 3-jet over 2-jet ratio (${\cal B}$). We have to stress an important 
difference between these two quantities and ${\cal X}$: in the ratio
that defines ${\cal A}$ and ${\cal B}$, the numerator requires the
definition (through final-state cuts) of an hard object in addition
to those already present in the denominator. This implies that the
kinematical configurations in the numerator and denominator can be
sizably different. Therefore, one faces the following problem: even if 
${\cal A}$ and ${\cal B}$ are formally proportional (at the LO) to $\as$, 
it is not straightforward to determine the scale at which $\as$ is calculated.
Furthermore, since the numerator and the denominator have different hard 
scales, the parton densities appearing in these two quantities will be 
probed at different momenta: this of course will partially destroy the
cancellation that one is willing to achieve when considering such ratios.
One the other hand, this problem does not affect ${\cal X}$: both the 
isolated-photon and the single-inclusive cross sections are dominated
by two-body, back-to-back configurations: it is therefore pretty
intuitive that $\as$ will be evaluated at a scale equal to the
transverse momentum of the observed photon and jet. On the other
hand, the partonic subprocesses contributing to the numerator and
the denominator of ${\cal A}$ and ${\cal B}$ are basically the same.
This is not true for ${\cal X}$, because of the different hard
production processes involved. Therefore, one might argue that in the 
latter case the cancellation of the dependence on parton densities will
not take place. We can however observe the following: at the LHC, and
if one does not consider too large values in $p_T$, the average momentum 
fraction $x$ probed is small: thus, the quark densities are dominated 
by the sea, which is in turn related to $f_g(x)$. In this way, we can 
expect to recover the cancellation.

Of course, there is no way to tell beforehand which observable displays
the smallest dependence upon the parton density choice. In order to
study this issue in the case of ${\cal X}$, we will consider in
the following the double ratio
\beq
D(\ptmin)=\overline{\cal X}(\ptmin)\,/\,\overline{\cal X}_0(\ptmin),
\label{Dratio}
\eeq
where
\beq
\overline{\cal X}(\ptmin)=
\int_{\ptmin}^{\ptmax} d\ptj\frac{d\sigma_j}{d\ptj}\, \Bigg/
\int_{\ptmin}^{\ptmax} d\ptg\frac{d\sigma_\gamma}{d\ptg}\,.
\label{Xbardef}
\eeq
In eq.~(\ref{Dratio}), $\overline{\cal X}_0$ is computed with our
default parton density set (MRST99-1), while $\overline{\cal X}$ is 
computed with the other sets. Notice that we considered $\overline{\cal X}$
instead of ${\cal X}$ just because we collected a limited amount
of statistics in the MC runs performed so far, and $\overline{\cal X}$
stands a better chance than ${\cal X}$ to be insensitive to fluctuations.
Notice, however, that the relevant transverse momentum spectra are
quite steep, and therefore $\overline{\cal X}(\ptmin)$ is dominated
by ${\cal X}(\ptmin)$. In eq.~(\ref{Xbardef}), the upper limit $\ptmax$ 
can be chosen at will. A possible choice is to set it equal to the 
kinematical limit; in the results presented in this section, we have set 
$\ptmax=400$~ GeV.

\begin{table}[t!]
\begin{center}
\begin{tabular}{|l||c|c|c|} \hline
$\ptmin$ (GeV) & 40 & 100 & 200 
\\ \hline
& \multicolumn{3}{c|}{$\abs{\etag}<1.5$} 
\\ \hline
MRST99-2
  & $0.974\pm 0.003$ & $0.966\pm 0.010$ & $0.984\pm 0.027$
\\ \hline
MRST99-3
  & $1.019\pm 0.003$ & $1.016\pm 0.010$ & $1.012\pm 0.025$
\\ \hline
& \multicolumn{3}{c|}{$\abs{\etag}<2.5$}  
\\ \hline
MRST99-2
  & $0.976\pm 0.002$ & $0.973\pm 0.008$ & $0.987\pm 0.019$
\\ \hline
MRST99-3
  & $1.017\pm 0.002$ & $1.010\pm 0.008$ & $1.010\pm 0.018$
\\ \hline
\end{tabular} 
\end{center}                                                            
\vskip-0.25cm                                                           
\caption{
NLO predictions for the ratio defined in eq.~(\ref{gammarat}).
This table has to be compared to table~\ref{tab:Dres}.
}
\label{tab:xsecrat}
\end{table}                                                               
Our NLO predictions for the double ratio $D$ are presented in 
table~\ref{tab:Dres}. By inspection of the table, we can see that
$D$ is remarkably stable with respect to the choice of the
density set; it has to be stressed, however, that an increase of 
the statistics is mandatory at the highest $\ptmin$ considered. 
In the table, we limited ourselves to considering only the sets
MRST99-2 and MRST99-3. The reason is the following: by construction,
these sets gauge the current uncertainty affecting the determination
of $f_g(x)$, with MRST99-1 being assumed to return the ``true'' densities.
Thus, since $D$ is compatible with one, we are indeed checking that
the dependence upon the parton densities in ${\cal X}$ (actually,
$\overline{\cal X}$) almost perfectly cancels. If we were considering
other sets, like MRST99-4, we would expect 
\mbox{$D\simeq \as(\Lambda_{\rm MRST99-4})/\as(\Lambda_{\rm MRST99-1})$}.
However, the strong correlation between $\as$ and $f_g(x)$ might
spoil this naive expectation. The same can be said when considering
the sets of the CTEQ group: in this case, a further bias can be
introduced by the fact that MRST and CTEQ use different parametrizations
and evolution codes. We postpone a more careful analysis of this
problem to a forthcoming work.

It can be argued that the results displayed in table~\ref{tab:Dres}
are due to the fact that the densities used are actually not that 
different in the $x$ range of interest. This, however, is not true. 
In fact, at the level of cross sections, the differences between
the predictions obtained with the default set or with the other
sets are much larger. This can be seen from table~\ref{tab:xsec}. 
More precisely, we can consider the ratio
\beq
\int_{\ptmin}^{\ptmax} d\ptg\frac{d\sigma_\gamma}{d\ptg}\, \Bigg/
\int_{\ptmin}^{\ptmax} d\ptg\frac{d\sigma_{0\gamma}}{d\ptg}\,,
\label{gammarat}
\eeq
where $d\sigma_{0\gamma}$ is calculated using MRST99-1, and
$d\sigma_\gamma$ with all the other density sets. The results
for this quantity are presented in table~\ref{tab:xsecrat}. Each
entry of this table has to be compared with the corresponding
entry in table~\ref{tab:Dres}. From this exercise, it is indeed
evident that ${\cal X}$ is much less sensitive than the isolated-%
photon cross section to the choice of the density set, at least
at small $\ptmin$. When $\ptmin$ approaches larger values, no
firm conclusion can be reached, given the statistics collected;
as mentioned before, one can suspect that, the higher $\ptmin$,
the larger the dependence of $\overline{\cal X}$ upon the densities.
One the other hand, it can be observed that smaller momenta allow
an easier observation of the running of $\as$.

\subsection{Pairs of isolated photons: infrared sensitivity with 
standard cone isolation\protect\footnote{Contributing authors:
T.~Binoth, J.P.~Guillet and E.~Pilon.}}

In the discussion given before, we restricted to the case of the
production of a single isolated photons. Of course, the considerations
we made can be extended with obvious modification in eq.~(\ref{factth})
to the case of the production of photon pairs. In such a case, the
cross section splits naturally in three {\em unphysical} components:
direct, single-fragmentation and double-fragmentation, corresponding
to the processes where both photons, one photon and none of the
photons are directly entering the hard subprocess. As far as the 
isolation prescription is concerned, things are unchanged: this
cut has to be imposed on both photons, and possibly supplemented
by the requirement that the photons be isolated from each other.

\begin{figure}[t!]
\begin{center}
\includegraphics[width=0.5\textwidth,clip]{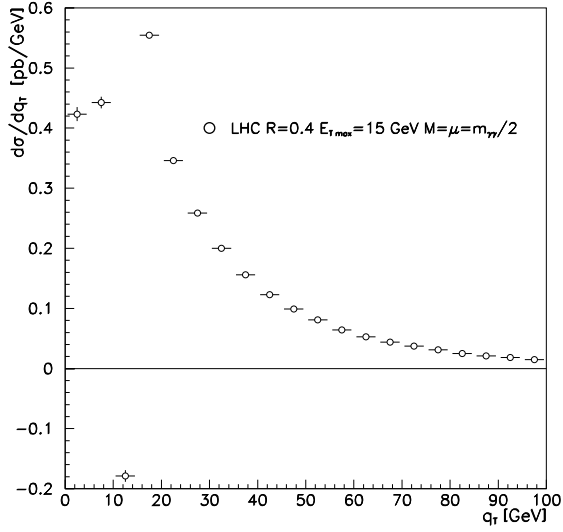}
\end{center}
\vskip-0.8cm
\caption{
 Di-Photon differential cross section $d\sigma/dq_T$ at LHC, $\sqrt{s}=14$ TeV,
 with the kinematic cuts $p_{T}(\gamma_{1}) \geq 40$ GeV, $p_{T}(\gamma_{2}) 
 \geq 25$ GeV, $|y(\gamma_{1,2})| \leq 2.5$, and with isolation criterion 
 $E_{T max}=15$ GeV in $R=0.4$. The scale choice for initial state
 factorization scale ($M$), fragmentation scale ($M_{f}$) and renormalization
 scale ($\mu$)  is  $M = M_{f} = \mu = m_{\gamma \gamma}/2$. 
 }
\label{flc13}
\end{figure}

In Sect.~\ref{sec:backg;qcd},
the production of photon pairs is described with a 
special emphasis on its role as a background to Higgs searches.
Here we would like to concentrate on a different,
more technical aspect, which is more relevant
to pure-QCD studies. We investigate appearance of infrared 
divergences {\em inside} the physical spectrum. An example of such divergences 
appears in the transverse momentum ($q_{T}$) spectrum of a pair of isolated
photons - or of a jet+isolated photon system. This can be seen in 
Fig.~\ref{flc13}, which shows $d \sigma/dq_{T}$ vs. $q_{T}$  for isolated photon
pairs, computed at NLO accuracy \cite{Binoth:1999qq}. The rather large  
value of isolation cut used here , $E_{T max}=15$ GeV, is not motivated by any 
phenomenological consideration: it instead allows to split the well known 
infrared issue in the vicinity of $q_{T} \rightarrow 0$ from the new one at 
$q_{T} \rightarrow E_{T max}$.

The trouble comes from the ``single fragmentation" contribution (the
contribution where only one photon comes from the fragmentation of a hard
parton, the other being emitted by the partonic sub-process). In the QCD 
improved parton model framework, the fragmentation is a strictly collinear
process, hence all the hadronic debris of the parton-to-photon fragmentation 
fall inside the cone of the photon from fragmentation. At LO, both photons are 
back-to-back in the transverse plane, so, due to transverse momentum
conservation, $q_{T} = E_{T}^{had}$. Since the transverse hadronic energy
deposited in the isolation cone has to be less that $E_{T \, max}$, the 
LO ``single
fragmentation" contribution of the $q_{T}$ distribution has a stepwise behavior.
Then, as shown in \cite{Catani:1997xc}, at NLO such an observable gets an 
infrared double logarithmic divergence at the critical point
$q_{T} = E_{T \, max}$. The details of this infrared structure are very 
sensitive to the kinematic constraints and the observable considered. In the
case at hand, the NLO contribution to $d \sigma/d q_{T}$ gets a double
logarithm below the critical point, which is produced by the convolution  of
the  lowest order stepwise term with the probability distribution for emitting
a soft and collinear gluon, yielding:
\begin{equation}
 \left( \frac{d \sigma}{d q_{T}} \right)_{NLO} 
 \sim - 
 \left( \frac{d \sigma}{d q_{T}} \right)_{LO} 
 \; \Theta \left( E_{T \, max} - q_{T} \right)
 \mbox{} \times \as  \ln^{2}
 \left( 1 - \frac{q_{T}^{2}}{E_{T \, max}^{2}} \right) \; + \cdots
\end{equation}
More generally, at each order in
$\alpha_{s}$, up to two powers of such logarithms will appear, making any fixed
order calculation diverge at $q_{T} = E_{T \, max}$, so  that the spectrum
computed by any fixed order calculation is unreliable in the vicinity of this
critical value. In principle, an all order resummation has to be carried out 
if possible, in order to restore any predictability. In practice, the
phenomenologically relevant values of $E_{T \, max}$ are fairly lower than 15
GeV, so that this problem may affect only the very first bins of the $q_{T}$
distribution.

\subsubsection{Mismatch theory/experiment with very severe isolation cuts}

Another issue deserves some care, when isolated photons are selected by mean of
the above standard cone criterion. In an actual prompt photon event the
transverse energy deposited inside the isolation cone has several physical
origins. One is when hadrons coming from the hadronization of hard partons
involved in the subprocess fall into the cone. A second one is given by the
debris of the fragmentation producing the photon, when the latter comes from
such a mechanism. A third source of accompanying transverse energy is provided
by ``minimum bias". Moreover at high luminosity, piled-up events may also 
contaminate the hadronic environment of a previous photon event. From an
experimental point of view, the value of $E_{T \, max}$ has to be as low as
possible in order to suppress background events and events with photons from
fragmentation, while retaining most of the ``true" direct photons. The goal is
thus to use an experimental value of $E_{T \, max}$ basically saturated by 
``minimum bias" - and pile-up. For example this is nearly achieved by CDF at
the Tevatron requiring $E_{T \, max} = 1$ GeV in $R = 0.4$.  In partonic
calculations, the first two sources of accompanying transverse energy are taken
into account, whereas the last two are ignored. However if  the accompanying
$E_{T}^{had}$ is to be saturated by ``minimum bias" and  pile-up, then in a
partonic calculation, this leaves almost no room for  accompanying partonic
$E_{T}$ coming from the hard subprocess itself. Therefore, a partonic
calculation meant to incorporate the effect of such an experimental  cut should
use an effective value for $E_{T \, max}$ in the calculation, which is much
smaller than the one experimentally used, e.g. at most a few hundred MeV for
CDF. The correspondence between the values used in experiments, or full Monte
Carlo simulations (which model the ``minimum bias"), and their counterparts
in  higher order partonic calculations has to be further studied. Such a
comparison is worthwhile especially because the actual isolation cuts used by
colliders experiments are more exclusive and sophisticated than the schematic
criterion defined above.

However when the experimental value of $E_{T \, max}$ is nearly saturated by 
``minimum bias", such a study is complicated by an infrared problem. Indeed, an
infrared divergence appears in partonic calculations, when photons are required
to be absolutely isolated, i.e. accompanied by a {\it vanishing amount} of
partonic transverse energy inside a cone of finite size, because this
amputation of gluon phase space prevents the cancellation of the infrared
singularities associated with soft gluon emission. With a finite value $E_{T \,
max}$, this would translate into the appearance of $\ln(E_{T \, max}/Q)$ (where
$Q$ is some large scale, of the order of the photon's $p_{T}$)  which would
become large with a tiny $E_{T \, max}$. Whereas the ``fragmentation"
contribution to, e.g. the $p_{T}$ distribution of direct photons
\cite{Gordon:1994ut,CFGP;6qcd}, or the invariant mass distribution of photon
pairs, is roughly 
\begin{equation}
\sigma^{fragm} \sim \varepsilon \left( \ln^{2} \varepsilon + 
                    \ln \varepsilon \ln R + \cdots\right)
\end{equation}
(with $\varepsilon = E_{T \, max}/Q$), the ``direct" contribution behaves as
\begin{equation}
\sigma^{dir} \sim R^{2} \ln \varepsilon + {\cal O}(1)
\end{equation}      
The theoretical partonic calculation would then become unstable and unreliable,
when $\varepsilon \ll 1$ with finite $R$. Moreover, this problem is not
localized in the sole vicinity of some isolated point, at the border of or
inside the spectrum, but in principle it plagues the calculation over the whole
spectrum - at least some extended range of it - for observables
such  as, e.g., the $p_{T}$ distribution of direct photons, or the invariant 
mass distribution of photon pairs. The dependence of theoretical partonic
calculations on the isolation parameters, especially on $E_{T \, max}$, has
still to be studied in detail \cite{bgp;7qcd} in order to fix this puzzle.


\section{SMALL X PHYSICS\protect\footnote{Section
    coordinators: R. Ball, V. Del Duca and A. de Roeck.}}
\label{sec:smallx;qcd}

\subsection{Jet physics at large rapidity intervals and the BFKL 
equation\protect\footnote{
Contributing authors: V. Del Duca and W.J. Stirling.}}
\label{sec:vddwjs}

The LHC offers a unique opportunity to explore
semi-hard strong-interaction processes, which are characterized by two large
and disparate kinematic scales. In inclusive jet production, jets of
transverse energy $E_\perp = 50$ GeV can span a
kinematic range of up to 11 units of rapidity.
Processes with two large and disparate kinematic scales
typically lead to cross sections 
containing large logarithms.  Examples of this type of process 
are di-jet production in hadron collisions at large rapidity 
intervals~\cite{Abachi:1996et}, forward jet production in 
DIS~\cite{Aid:1995we,Breitweg:1999ed,Adloff:1999fa}, and
$\gamma^*\gamma^*$ collisions in double-tag events, 
$e^+e^- \to e^+e^- +$ hadrons~\cite{Acciarri:1999ix}.
In large-rapidity di-jet production the large 
logarithm is the rapidity interval between the jets,
$\Delta y\simeq\ln(\hat s/|\hat t|)$, with
$\hat s$ the squared parton center-of-mass energy and $|\hat t|$ of the
order of the squared jet transverse energy. In forward jet production
in DIS the large logarithm is $\ln(x/x_{bj})$, where $x_{bj}$ is the Bjorken
scaling variable and $x$ the momentum fraction of the parton entering the
hard scattering. These logarithms will 
arise in a perturbative calculation at each order in the coupling 
constant $\as$.  Alternatively, if the logarithms are large 
enough, it is possible to include them through an all-order resummation 
in the leading logarithmic (LL) approximation performed by means of the
Balitsky-Fadin-Kuraev-Lipatov (BFKL) 
equation~\cite{Kuraev:1976ge,Kuraev:1977fs,Balitsky:1978ic}.

In the high-energy limit, $\hat s\gg |\hat t|$, the BFKL theory assumes 
that any scattering
process is dominated by gluon exchange in the crossed channel~\footnote{The
crossed-channel gluon dominance
is also used as a diagnostic tool for 
discriminating between different dynamical models for parton scattering.
In the measurement of di-jet angular distributions,
models which feature gluon exchange in the crossed channel, like QCD, 
predict a characteristic $\sin^{-4}(\theta^\star/2)$
di-jet angular distribution~\cite{Ellis:1992qq,Abe:1992sj,Weerts:1994ui}, 
while models featuring contact-term interactions, which do not have gluon 
exchange in the crossed channel,
predict a flattening of the di-jet angular distribution
at large $\hat s/|\hat t|$~\cite{Abe:1996mj,Abbott:1998nf}.}
which for a given scattering occurs at ${\cal O}(\as^2)$. 
This constitutes the leading-order (LO) term of the BFKL resummation.
The corresponding QCD amplitude
factorizes into a gauge-invariant effective amplitude formed by 
two scattering centers, the LO impact factors, connected by the gluon 
exchanged in the crossed channel. The impact factors are characteristic
of the scattering process at hand. The BFKL equation then resums the universal
LL corrections, of ${\cal O}(\as^n\ln^n(\hat s/|\hat t|))$,
to the gluon exchange in the crossed channel. These are obtained in the
limit of a strong rapidity ordering of the emitted gluon radiation, i.e.
for $n$ gluons produced in the scattering,
\begin{equation}
y_1\gg y_2\gg \ldots \gg y_{n-1}\gg y_n\, .\label{mrk}
\end{equation}

Di-Jet production in hadron collisions at large rapidity intervals
is the simplest process to which to apply the BFKL resummation, and one
of the topical BFKL processes at the LHC, thus we shall use it as the
paradigm process.
Since di-jet production at large rapidity intervals
is dominated by gluon exchange in the crossed channel,
the functional form of the QCD amplitudes for gluon-gluon, gluon-quark 
or quark-quark scattering at LO is the
same; they differ only by the colour strength in the parton-production
vertices.  We can then write the cross section
in the following factorized form 
\cite{Mueller:1987ey,DelDuca:1994mn,Stirling:1994zs}
\begin{equation}
{d\sigma\over d^2 p_{a'\perp} d^2 p_{b'\perp} dy_{a'} dy_{b'}}\, =\,
x_a^0 f_{\rm eff}(x_a^0,\mu_F^2)\, x_b^0 f_{\rm eff}(x_b^0,\mu_F^2)\,
{d\hat\sigma_{gg}\over d^2 p_{a'\perp} d^2 p_{b'\perp}}\, ,\label{mrfac}
\end{equation}
where $\mu_F$ is the factorisation scale, 
$a'$ and $b'$ label the forward and backward outgoing jet, respectively,
and $p_\perp$ are two-dimensional vectors in the plane transverse to the
collision axis, the {\em azimuthal} plane.
$x_a^0,\, x_b^0$ are the parton momentum fractions in the high-energy limit,  
\begin{equation}
x_a^0 = {|p_{a'\perp}|\over\sqrt{s}} e^{y_{a'}}\qquad \qquad
x_b^0 = {|p_{b'\perp}|\over\sqrt{s}} e^{-y_{b'}}\, ,\label{nkin0}
\end{equation}
and the effective parton distribution functions  
are~\cite{Combridge:1984jn} 
\begin{equation}
f_{\rm eff}(x,\mu_F^2) = f_g(x,\mu_F^2) + {4\over 9}\sum_f
\left[ \, f_{q_f}(x,\mu_F^2) + f_{{\bar q}_f}(x,\mu_F^2)\right], \label{effec}
\end{equation}
where the sum is over the quark flavours. In the high-energy limit, 
the gluon-gluon scattering cross section becomes~\cite{Mueller:1987ey}
\begin{equation}
{d\hat\sigma_{gg}\over d^2 p_{a'\perp} d^2 p_{b'\perp}}\ =\
\biggl[{C_A\as\over p_{a'\perp}^2}\biggr] \,
f(q_{a\perp},q_{b\perp},\Delta y) \,
\biggl[{C_A\as\over p_{b'\perp}^2}\biggr] \ ,
\label{cross}
\end{equation}
with $C_{A}=N_{c}=3$, $\Delta y = y_{a'}-y_{b'}$ and $q_{i\perp}$ the 
momenta transferred in the $t$-channel, with $q_{a\perp}=-p_{a'\perp}$ and 
$q_{b\perp}=p_{b'\perp}$, and where we use the shorthand for the magnitude 
squared, $|p_{\perp}|^2\equiv p_{\perp}^2$.
The quantities in square brackets
are the LO impact factors for jet production. 
The function $f(q_{a\perp},q_{b\perp}, \Delta y)$ is the
Green's function associated with the gluon exchanged in the 
crossed channel. It is process independent and given in the LL 
approximation by the solution of the BFKL equation. 
This equation is a two-dimensional
integral equation which describes the evolution in transverse momentum
 of the gluon propagator exchanged in the crossed channel.  
If we transform to moment space via
\begin{equation}
f(q_{a\perp},q_{b\perp},\Delta y) \ =\ \int {d\omega\over 2\pi i}\, 
e^{\omega\Delta y}\, 
f_{\omega}(q_{a\perp},q_{b\perp})\label{moment}
\end{equation}
we can write the BFKL equation as
\begin{equation}
\omega\, f_{\omega}(q_{a\perp},q_{b\perp})\, =
{1\over 2}\,\delta^2(q_{a\perp}-q_{b\perp})\, +\, {\bar \as \over \pi} 
\cK\,[f_{\omega}(q_{a\perp},q_{b\perp})]\, ,\label{bfklb}
\end{equation}
with $\bar \as = \as N_c/\pi$, and where the kernel $\cK$ is given by
\begin{equation}
\cK\,[f_{\omega}(q_{a\perp},q_{b\perp})] = 
\int {d^2k_{\perp}\over k_{\perp}^2}\, \left[
f_{\omega}(q_{a\perp}+k_{\perp},
q_{b\perp}) - {q_{a\perp}^2\over k_{\perp}^2 + 
(q_{a\perp}+k_{\perp})^2}\, f_{\omega}(q_{a\perp},q_{b\perp}) \right]
\, .\label{kern}
\end{equation}
The first term in the kernel accounts for the emission of a real gluon of 
transverse momentum $k_{\perp}$ and the second term accounts 
for the virtual radiative corrections, which {\em reggeise} the gluon 
exchanged in the crossed channel. The solution to the BFKL equation is,
\begin{equation}
f(q_{a\perp},q_{b\perp},\Delta y)\, = {1\over (2\pi)^2 
\sqrt{q_{a\perp}^2 q_{b\perp}^2}} 
\sum_{n=-\infty}^{\infty} e^{in\phi_{ab}}\, \int_{-\infty}^{\infty} d\nu\, 
e^{\omega(\nu,n)\Delta y}\, \left(q_{a\perp}^2\over q_{b\perp}^2
\right)^{i\nu}\, ,\label{solc}
\end{equation}
with $\phi_{ab}$ the azimuthal angle between $q_{a\perp}$ and $q_{b\perp}$,
and $\omega(\nu,n)$ the eigenvalue of the BFKL equation 
\begin{equation}
\omega(\nu,n)\, =\, -{\bar \as}\, \left[\psi\left({|n|+1
\over 2} +i\nu\right) + \psi\left({|n|+1
\over 2} -i\nu\right) + 2\gE\right]\, ,\label{om}
\end{equation}
with $\psi$ the digamma function, $\gE= -\psi(1)$ the Euler constant,
and with maximum at $\omega(0,0)\equiv \lambda = 4\bar \as\ln{2}$. 
Thus the solution of the BFKL equation resums powers of $\Delta y$.
The resulting gluon-gluon cross section
grows with $\Delta y$ as $f(q_{a\perp},q_{b\perp},\Delta y) \sim 
\exp(\lambda \Delta y)$~\cite{Kuraev:1977fs,Balitsky:1978ic}, in contrast
to the leading-order (${\cal O}(\as^2)$) 
cross section which is constant at large $\Delta y$.

In order to detect evidence of a BFKL-type behaviour in a scattering
process, we need to have $\Delta y$ as large as possible.
In di-jet production it can be done by minimizing 
the jet transverse energy, and maximizing $\hat s$. Since $\hat s = x_a^0
x_b^0 s$, in a fixed-energy collider this is achieved by increasing the
parton momentum fractions $x_{a,b}$, and then measuring e.g. 
the di-jet production rate $d\sigma/ d\Delta y$. However,
as the $x$'s grow the parton luminosity falls off, making it
difficult to disentangle the eventual BFKL-driven rise of the
parton cross section from the \pdfs\ fall 
off~\cite{DelDuca:1994mn,Stirling:1994zs}. One way to circumvent this problem
is to use a variable-energy collider: 
the increase in $\hat s$ can then be achieved by fixing the $x$'s (and hence the 
\pdfs) and by 
letting the hadron 
center-of-mass energy $s$ grow.  The advantage 
of this set-up is that variations in the \pdfs\
are minimised, while variations in the parton dynamics, and thus in
the eventual underlying BFKL behaviour, are 
stressed~\cite{Mueller:1987ey,Orr:1998hc}. 
The D0 collaboration have recently attempted to uncover BFKL behavior
in this way  by comparing di-jet cross sections measured at $\sqrt{s} = 630$~GeV
and $1.8$~TeV~\cite{Abbott:1999ai}. In a contribution to this 
Workshop~\cite{Peschanski:2000am},
the possibility of testing for BFKL-type behaviour by comparing di-jet cross sections
at the Tevatron (2~TeV) and the LHC (14~TeV) has been investigated.        
The difficulty here is that one is comparing jets measured in two very different
detectors, with resulting systematic uncertainties in the relative cross sections.
One could also, of course,  contemplate running the LHC at a lower collision energy.
Note that a variable-energy configuration can 
be more easily realised: in forward-jet production in DIS,
since a fixed-energy $ep$ collider is nonetheless a variable-energy collider 
in the photon-proton frame~\cite{Mueller:1991er,Tang:1992am,Bartels:1992tf,
Kwiecinski:1992vf,Bartels:1996gr,Orr:1998tf};
in $\gamma^*\gamma^*$ collisions in double-tag events, 
$e^+e^- \to e^+e^- +$ hadrons, by varying the energy in the photon-photon 
frame~\cite{Brodsky:1997sd,Bartels:1996ke}.

As a more practical alternative to varying the collider energy, one can study
less inclusive observables. 
In particular,  the correlation between the tagging jets, 
which at LO are supposed to be back to back, is smeared by gluon
radiation induced by parton showers and by hadronization. However, if we
look at the correlation also as a function of $\Delta y$, we expect
the (BFKL) gluon radiation in the rapidity interval 
between the jets to further blur the information on 
the mutual position in transverse momentum space, and thus the decorrelation
to grow with $\Delta y$. 
Accordingly, the transverse 
momentum imbalance \cite{DelDuca:1994mn,DelDuca:1994xy}, and the azimuthal 
angle decorrelation \cite{DelDuca:1994mn,Stirling:1994zs,DelDuca:1995fx,
DelDuca:1995ng,Orr:1997im} have been proposed as BFKL observables.
In particular, it is straightforward to derive from (\ref{solc})
the prediction for the dependence of $\langle \cos\phi_{ab}\rangle$
on $\Delta y$:\footnote{In practice one integrates the di-jet
 transverse momenta above some threshold, $|p_{a'\perp}|, |p_{b'\perp}| 
> p_{\perp}^{\rm min}$.} $\langle \cos\phi_{ab}\rangle \approx 0 $.
One finds~\cite{DelDuca:1994mn,Stirling:1994zs,DelDuca:1995fx,
DelDuca:1995ng,Orr:1997im}  that $\langle \cos\phi_{ab}\rangle $
decreases rapidly from 1 at small $\Delta y$ (back-to-back jets), and
approaches zero as $\Delta y \to \infty$.
Such an  azimuthal angle decorrelation has indeed been
observed at the Tevatron
Collider~\cite{Abachi:1996et}.
However, the LL BFKL formalism predicts 
a much stronger decorrelation than that observed in 
the data. On the other hand a NLO partonic Monte Carlo generator
(JETRAD~\cite{Giele:1993dj,Giele:1994gf}), in which the exact $2\to 2$ 
and $2\to 3$ matrix 
elements are taken into account, predicts too little decorrelation. 
In fact the data are well described by the HERWIG Monte Carlo 
generator~\cite{Marchesini:1988cf,Knowles:1988vs,Marchesini:1992ch}, 
which `dresses' the basic $2\to 2$ parton scattering  with parton showers and
also includes hadronization.
Thus the present conclusion is that at least for di-jets with transverse momenta
$ > 20 $~GeV and with  rapidity intervals $< 6$~units, 
as analysed by the D0 Collaboration at the 
Tevatron,  there is no evidence for LL
BFKL-induced gluon radiation in the azimuthal angle decorrelation.

A possible explanation of the failure of the LL BFKL prediction
to describe the Tevatron data is that the sub-leading corrections are large.
There are various sources of such corrections: next-to-leading order
corrections to the BFKL kernel in (\ref{kern}), 
which have recently been calculated (see Sect.~\ref{sec:smallxsalam;qcd}), 
related running coupling effects\footnote{Note that the solution
given in (\ref{solc}) assumes a fixed value for $\as$.}, and finally
kinematic corrections that take into account the limited phase space available
for BFKL-type gluon emission. In the derivation leading to 
the result (\ref{solc}), the transverse momentum of each emitted gluon is 
unbounded, and it is  this unrestricted emission of gluons with
transverse momenta $\sim |p_{a'\perp}|, |p_{b'\perp}|$ that leads
to the strong decorrelation in azimuthal angle.

In an attempt to go beyond the analytic LL BFKL results,
a Monte Carlo approach has been 
adopted~\cite{Schmidt:1997fg,Orr:1997im,Orr:1998hc}. By solving the BFKL
equation (\ref{bfklb}) by iteration, which amounts to
`unfolding' the summation over the intermediate radiated
gluons and making their contributions explicit,
it is possible to include the effects of both the 
running coupling and  the overall kinematic constraints.
It is also straightforward
to implement the resulting iterated solution in an event generator.

The first step in this procedure is to separate the $k_{\perp}$
integral in (\ref{bfklb}) 
into `resolved' and `unresolved' contributions,
according to whether they lie above or below a small transverse energy
scale $\mu$.  The scale $\mu$ is assumed to be
small compared to the other relevant scales in the problem (the minimum
transverse momentum $p_{\perp}^{\rm min}$ for example).  
The virtual and unresolved 
contributions are then combined into a single, finite integral.  The BFKL
equation  becomes 
\begin{eqnarray}
\omega\, f_{\omega}(q_{a\perp},q_{b\perp}) & =&
{1\over 2}\,\delta^2(q_{a\perp}-q_{b\perp})\, +\, {\bar\as\over \pi} 
\int_{k_{\perp}^2 > \mu^2} {d^2k_{\perp}\over k_{\perp}^2}\,
f_{\omega}(q_{a\perp}+k_{\perp},q_{b\perp}) \nonumber \\
&+&  {\bar\as\over \pi} \int {d^2k_{\perp}\over k_{\perp}^2}
\left[  f_{\omega}(q_{a\perp}+k_{\perp},q_{b\perp})\, \theta(\mu^2 -k_{\perp}^2)\, - \, {q_{a\perp}^2 \, f_{\omega}(q_{a\perp},q_{b\perp}) 
 \over k_{\perp}^2 + (q_{a\perp}+k_{\perp})^2}
\right] .\label{bfklbx}
\end{eqnarray}
The combined unresolved/virtual integral can be simplified by noting 
that since $ k_{\perp}^2 \ll q_{a\perp}^2,q_{b\perp}^2$ by construction,
the $k_{\perp}$ term in the argument of $f_{\omega}$ can be neglected,
giving
\begin{equation}
(\omega - \omega_0)\, f_{\omega}(q_{a\perp},q_{b\perp}) \, =\,
{1\over 2}\,\delta^2(q_{a\perp}-q_{b\perp})\, +\, {\bar\as\over \pi} 
\int_{k_{\perp}^2 > \mu^2} {d^2k_{\perp}\over k_{\perp}^2}\,
f_{\omega}(q_{a\perp}+k_{\perp},q_{b\perp}) \nonumber \\
\, ,\label{bfklcx}
\end{equation}
where
\begin{equation}
\omega_0  = {\bar\as\over \pi} 
\int {d^2k_{\perp}\over k_{\perp}^2}
\left[   \theta(\mu^2 -k_{\perp}^2)\, - \, {q_{a\perp}^2 
 \over k_{\perp}^2 + (q_{a\perp}+k_{\perp})^2}
\right] 
= {\bar\as}\, \ln\left( { \mu^2 \over q_{a\perp}^2 }   \right) 
\, .    \label{eq:omega0}
\end{equation}
The virtual and unresolved contributions are now contained
in $\omega_0$ and we are left with an integral over resolved real
gluons. We can now solve (\ref{bfklcx}) iteratively, 
and performing the inverse transform
we have 
\begin{equation}
f(q_{a\perp},q_{b\perp},\Delta y)\, =  \, 
  \sum_{n=0}^{\infty} f^{(n)}(q_{a\perp},q_{b\perp},\Delta y) \; .
\label{eq:b7}
\end{equation}
where
\begin{eqnarray}
f^{(0)}(q_{a\perp},q_{b\perp},\Delta y) &  =  &  
\left[ \frac{\mu^2}{q_{a\perp}^2} \right]^{\bar\as\Delta y}\,
\,\frac{1}{2}\, \delta^2(q_{a\perp}-q_{b\perp} ) \nonumber \\
f^{(n\geq 1)}(q_{a\perp},q_{b\perp},\Delta y) &  =  &  
\left[ \frac{\mu^2}{q_{a\perp}^2} \right]^{\bar\as\Delta y}\,
\left\{ \prod_{i=1}^{n}  \int d^2 k_{i\perp}\, dy_i \, {\cal F}_i \right\}
\,\frac{1}{2}\, \delta^2(q_{a\perp}-q_{b\perp} - \sum_{i=1}^n k_{i\perp})
\nonumber \\
{\cal F}_i &=& \frac{\bar\as}{\pi k_{i\perp}^2}\, 
\theta(k_{i\perp}^2 -\mu^2)\, \theta(y_{i-1}-y_i)\,
\left[ { (q_{a\perp} +\sum_{j=1}^{i-1}k_{j\perp}  )^2
 \over (q_{a\perp} +\sum_{j=1}^{i}k_{j\perp}  )^2 }\right]^{\bar\as y_i}
\label{eq:b8}
\end{eqnarray}
Thus the solution to the BFKL equation is recast in terms
of phase space integrals for resolved gluon emissions, with form factors
representing the net effect of unresolved and virtual emissions.
Unlike in the case of DGLAP evolution, there is no strong ordering
of the transverse momenta $k_{i\perp}$.
 Strictly speaking, the derivation given above
only applies for fixed coupling because we have left $\as$
outside the integrals.  The modifications necessary to account for 
a running coupling $\as (k_{\perp}^2)$ are 
straightforward~\cite{Orr:1997im}.

The expression for $f$  in (\ref{eq:b7},\ref{eq:b8}) above is amenable to numerical integration, 
and one can for example reproduce the analytic result given in (\ref{solc}). 
More importantly, having made explicit the BFKL gluon emission 
phase space, we can impose overall
energy and momentum conservation. In particular the parton momentum fractions 
in the presence of BFKL gluon emission become
\begin{eqnarray}
x_a &=&   \frac{e^{y_{a'}}}{\sqrt{s}}
   \left( |p_{a'\perp}| \,     +\,  |p_{b'\perp}| e^{-\Delta y} \,   
+ \, \sum_i |k_{i\perp}| e^{y_i-y_{a'} } \right) \, ,  \nonumber \\
x_b &=&   \frac{e^{-y_{b'}}}{\sqrt{s}}
   \left(  |p_{b'\perp}|  \,  + \,          |p_{a'\perp}| e^{-\Delta y} \,  
+ \, \sum_i |k_{i\perp}| e^{-y_i+y_{b'}} \right) \, .  
\end{eqnarray}
The momentum fractions in the high-energy limit
given in (\ref{nkin0}) are recovered by imposing strong rapidity ordering, 
eq.~(\ref{mrk}). 
Note that the requirement $x_a,x_b \leq 1$ effectively imposes
an upper limit on the transverse momentum ($k_{i\perp}$) integrals.
This in turn means that the analytic result (\ref{solc})
is {\it not} reproduced in the presence of such a constraint,
since they require the internal transverse momenta integrals
to extend to infinity. Formally, the kinematic constraints $x_a,x_b \leq 1$ induce an infinite sequence of
sub-leading logarithms $\as^n\Delta y^{n-1},\ \as^n\Delta y^{n-2},
\ \ldots$ that suppress the growth of the parton scattering cross section 
with $\Delta y$.
 
Applying kinematic constraints and including the running coupling 
suppresses the emission of energetic BFKL gluons, and therefore weakens the azimuthal
decorrelation predicted at LL level~\cite{Schmidt:1997fg,Orr:1997im}. 
As a result, reasonable agreement with the D0 decorrelation data is recovered.
It is clear, therefore, that one needs a higher-energy collider such as the LHC
in order to discriminate between the BFKL and parton shower (DGLAP) dynamics.

\begin{figure}[t!]
\begin{center}
\includegraphics[width=0.55\textwidth,clip]{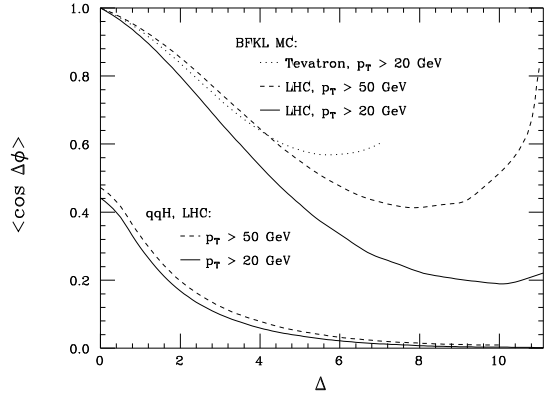}
\vskip-0.5cm
\caption{The azimuthal angle decorrelation in di-jet production at the Tevatron 
($\sqrt{s}=1.8$~GeV) and LHC ($\sqrt{s}=14$~TeV)
as a function of di-jet rapidity difference $\Delta y$~\cite{Orr:1998ps}.  
The upper curves are computed using the improved BFKL MC with running $\as$;
they are: (i) Tevatron, $p_T>20$~GeV (dotted curve),
(ii) LHC, $p_T>20$~GeV (solid curve), and (iii) LHC, $p_T>50$~GeV
(dashed curve).  The lower curves are for di-jet production in the process
$qq\to qqH$ for $p_T>20$~GeV (solid curve) and $p_T>50$~GeV
(dashed curve).} \label{fig:lhcdecorr}
\end{center}
\end{figure}

Figure~\ref{fig:lhcdecorr}  shows \
the mean value of $\cos{\Delta\phi}$ in di-jet production in an improved BFKL
MC approach \cite{Orr:1998ps} that includes kinematic constraints and 
running couplings (upper curves).
The jets are completely correlated (i.e. 
back-to-back in the azimuthal plane) at $\Delta y  =0$, and as $\Delta y$
increases
we see the characteristic BFKL decorrelation, 
followed by a flattening out and then an increase in 
$\langle\cos{\Delta\phi}\rangle$
as  the kinematic limit is approached\footnote{For any given
transverse momentum threshold, there is some $\Delta y$ at which the jet
pair ($a',b'$)alone saturates the kinematic limit, and emission of additional 
(real) gluons is completely suppressed and the azimuthal correlation
returns.  As we approach that limiting value of $\Delta y$
we therefore expect to see a transition back towards correlated jets.}.
 Not surprisingly, the kinematic
constraints have a much stronger effect when the $p_{\perp}^{\rm min}$ 
threshold is  set at $50$~GeV (dashed curve) than at $20$~GeV (solid curve); 
in the latter case
more phase space is available to radiate gluons.  We also show for 
comparison the decorrelation for di-jet production at the Tevatron
for $p_T>20$~GeV.  There we see that the lower collision energy (1.8~TeV)
limits the allowed rapidity difference and 
substantially suppresses the decorrelation at large $\Delta y$.  
 Note that the larger center-of-mass energy compared to  transverse momentum threshold 
at the LHC would seem to give it a significant advantage as far 
as observing BFKL effects is concerned.

The lower set of curves in Fig.~\ref{fig:lhcdecorr} refer to Higgs production
via the $WW,\; ZZ$ fusion process $qq \to qq H$, and are included for comparison
\cite{Orr:1998ps}.
This process automatically provides a `BFKL-like' di-jet sample with large
rapidity separation, although evidently the jets are significantly less correlated
in azimuthal angle. 

In summary, the LHC offers an important test of BFKL dynamics in the production
of relatively low transverse momentum jet pairs with a large rapidity separation.
In this section we have given an overview of the relevant theory. An important next step
is to include the effects of the next-to-leading  order contributions to the BFKL kernel,
and to consider other related processes with gluon exchange in the crossed 
channel\footnote{Examples include $qg \to Wqg$, $gg \to b\bar b g$ etc.}.
On the experimental side, it remains a challenge to trigger on such low $p_{\perp}$
jets in the far forward regions of the detector.

\subsection{Small-$x$ Effects in Final States\protect\footnote{
Contributing authors: C.~Ewerz and B.R.~Webber.} }

To understand the special features of QCD dynamics at small $x$, it
will be essential not only to study the fully inclusive cross sections
for small-$x$ processes at the LHC, such as the Drell-Yan process at
dilepton mass-squared $Q^2$ much smaller than the 
c.m.\ energy-squared, but also to investigate the structure
of the associated final states. One important aspect of the final state
is the number of mini-jets produced. By mini-jets we mean jets with
transverse momenta above some resolution scale $\mR$, where
$\mR^2\ll Q^2$. Thus the mini-jet multiplicity at small $x$ involves
not only $\ln x\gg 1$ but also another large logarithm,
$T=\ln(Q^2/\mR^2)$, which needs to be resummed.  The results
presented below include all terms of the form $(\as\ln x)^n T^m$
where $1\le m\le n$.  Terms with $m=n$ are called double-logarithmic (DL)
while those with $1\le m<n$ give single-logarithmic (SL) corrections.
The DL contributions to the mini-jet multiplicity have been obtained in 
\cite{Webber:1998we}, and the SL terms have been included in 
\cite{Ewerz:1999fn,Ewerz:1999tt}. 
In these calculations the BFKL formalism \cite{Fadin:1975cb,Balitsky:1978ic} 
has been used, but the results are expected to hold \cite{Salam:1999ft}
also in the CCFM 
formalism \cite{Ciafaloni:1988ur,Catani:1990yc,Catani:1990sg,Catani:1991gu} 
based on angular ordering of gluon emissions. 

We start by considering the gluon structure function at the 
momentum scale $Q^2$, $F(x,Q^2)$. 
It is the sum of contributions $F^{(r\res)}(x,Q^2,\mR^2)$ 
in which different numbers $r$ of final-state mini-jets 
are resolved with transverse momentum greater than $\mR$, 
\begin{equation}\label{Frx}
F^{(r\res)}(x,Q^2,\mR^2) = F(x,\mR^2)\otimes G^{(r)}(x,T)
\equiv\int_x^1\frac{dz}{z} F(z,\mR^2) G^{(r)}(x/z,T)\,.
\end{equation}
To determine the coefficient function $G^{(r)}$ to leading 
logarithmic order in $x$, it is convenient to apply a Mellin 
transformation, 
\begin{equation}\label{mellin}
f_\om(\ldots) = \int_0^1 dx\,x^\om f(x,\ldots)\;.
\end{equation}
In $\omega$-space the evolution of the structure function 
is $F_\om(Q^2)=\exp[\glip(\alb/\om)T]F_\om(\mR^2)$, 
where $\glip$ is the Lipatov anomalous dimension, i.e. the solution
obtained from eq.~(\ref{om})  by setting $n=0$ and $\gamma=1/2 + i\nu$,
\begin{equation}\label{omlip}
\omega = -\alb\left[\psi(\gamma)+\psi(1-\gamma) + 2\gE \right]
\equiv \alb\,\chi(\gamma)\;. 
\end{equation}
The Lipatov anomalous dimension 
 can be written as an expansion in powers of $\as/\om$, 
\begin{equation}\label{glip}
\glip(\alb/\om) = \alom+
2\zeta(3)\left(\alom\right)^4+
2\zeta(5)\left(\alom\right)^6+
\ldots\;.
\end{equation}
In \cite{Ewerz:1999tt} it has been shown that 
the generating function
$G_\om(u,T)=\sum_{r=0}^\infty u^r G^{(r)}_\om(T)$
can be written as 
\begin{equation}\label{Gomsol}
G_\om(u,T)=\frac{I_\om(u,0)}{I_\om(u,T)}\;,
\end{equation}
where 
\begin{equation}\label{Idef}
I_\om(u,T) =\int_\Gamma \frac{d\gamma}{\gamma}
\,e^{-\gamma T+\phi_\om(u,\gamma)}\,,
\end{equation}
$\Gamma$ being a contour parallel to the imaginary axis on
the left of all singularities of the integrand, and
\begin{equation}\label{phidef}
\phi_\om(u,\gamma) = \frac{u}{u-1}\int_{\frac{1}{2}}^{\gamma}
d\gamma'\,\left[\frac{\om}{\alb u}-\chi(\gamma')\right]
\;.
\end{equation}
One can obtain the moments of the  jet multiplicity distribution
from the generating function as follows:
\begin{equation}\label{rmoms}
\VEV{r(r-1)\ldots(r-s+1)}_\om 
=\exp[-\glip(\alb/\om)T]
\left.\frac{\partial^s G_\om}{\partial u^s}\right|_{u=1}\;.
\end{equation}
Using the expressions (\ref{Gomsol})-(\ref{phidef}) 
we thus find for the mean number of jets 
\begin{equation}\label{meanres}
\VEV{r}_\om = 
- \frac{1}{\chi'} 
\left(\frac{1}{\glip} + \frac{\chi''}{2 \chi'}+ \chi \right) T
- \frac{1}{2 \chi'} T^2
\end{equation}
where $\chi'$ means the derivative of $\chi(\gamma)$ evaluated at
$\gamma=\glip$. The corresponding expression for the variance 
in the number of jets,
$\sig^2_\om\equiv\VEV{r^2}_\om -\VEV{r}^2_\om$,
is more complicated~\cite{Ewerz:1999tt}. 
Interestingly, the variance is a polynomial of third degree 
in $T$. This implies that the distribution in the number of 
jets remains narrow for large $T$ in the sense that its width 
grows slower than its mean. 

Considered as functions of $\om$ 
the coefficients of the powers of $T$ in eq.\ (\ref{meanres}) 
and in the corresponding expression for $\sig^2_\om$~\cite{Ewerz:1999tt}
exhibit bad behaviour at large values of $\alb/\om$. 
This is associated with the singularity of the leading-order Lipatov
anomalous dimension $\glip$ at $\alb/\om = (4\ln 2)^{-1}$. We would expect this
behaviour to be modified strongly by higher order corrections. 
Although the next-to-leading corrections to $\glip$ are known 
\cite{Fadin:1998py,Camici:1997ij,Ciafaloni:1998gs} a full 
calculation of the corresponding corrections to the associated jet
multiplicity has not been performed and would appear very difficult.

For practical purposes it is necessary to determine the 
multiplicity moments as functions of $x$. This can be done 
using (\ref{omlip}) and the perturbative expansion (\ref{glip}) 
of the anomalous dimension. The inverse Mellin transformation 
can then be applied to this series term by term using 
\begin{equation}\label{melom}
\frac{1}{2\pi i}\int_C d\om\,x^{-\om-1}\left(\alom\right)^n
=\frac{\alb}{x}\frac{[\alb\ln(1/x)]^{n-1}}{(n-1)!}\;.
\end{equation}
In this way one easily finds a series for the inverse Mellin 
transform $\VEV{r}(x)$ of $\VEV{r}_\om$, for example. 
We note that the factorial in the denominator makes the resulting series 
in $x$-space converge very rapidly. 
It is then straightforward to compute the 
mini-jet multiplicity associated with point-like scattering on
the gluonic component of the proton at small $x$ 
using 
\begin{equation} \label{nx}
n(x)=\frac{F(x,Q^2) \otimes \VEV{r}(x)}{F(x,Q^2)} \,. 
\end{equation}

To illustrate the effects of BFKL resummation we compute 
the number of associated jets in central Higgs production at 
the LHC. 
The dominant production process for a SM Higgs boson 
at the LHC is expected to be gluon-gluon fusion. 
The production cross section for a Higgs boson of mass $M_H$ and rapidity $y$
by gluon-gluon fusion in proton-proton collisions at 
centre-of mass energy $\sqrt{s}$ takes the form 
\begin{equation}\label{higgsxsec}
\frac{d\sigma}{dy} = F(x_1,M_H^2)\,F(x_2,M_H^2)\,C(M_H^2)\,,
\end{equation}
where for central production of the Higgs ($y=0$) we have 
$x_1 = x_2=M_H/\sqrt{s}$, 
and for LHC $\sqrt{s}=14\,\mbox{TeV}$. 
$C$ represents the $gg\to H$ vertex, which is perturbatively 
calculable as an intermediate top-quark loop. 
A more careful treatment would involve
replacing the Higgs production vertex $C(M_H^2)$ by an 
impact factor $C(M_H^2, k_1^2,k_2^2)$ and convoluting
it with unintegrated gluon densities taken at the off-shell 
gluon virtualities $k_1^2$ and $k_2^2$, respectively. 
The dependence of the impact factor $C(M_H^2, k_1^2,k_2^2)$ 
on these virtualities is expected to be weak, and we have neglected
it to arrive at eq.~(\ref{higgsxsec}). Then $C$ cancels in 
the mean number of mini-jets and its dispersion, and we do
not need to know its detailed form.  

Since the gluon emissions in the regions of positive and negative 
rapidity are independent, we can simply 
add the numbers $n_1=n(x_1)$ and $n_2=n(x_2)$ of mini-jets 
produced in these regions. 
The mean multiplicity $N$ of associated mini-jets 
becomes\footnote{We do not count any jets emerging from 
the proton remnants.} 
\begin{equation}
N(x) = n_1+ n_2 = 2 n(x) \,,
\end{equation}
where $n(x)$ can be calculated as in (\ref{nx}) after 
replacing $Q^2$ by $M_H^2$. 
Similarly, the variance is
\begin{equation}
\sigma^2_N(x) = \sigma^2_n(x_1)+\sigma^2_n(x_2)
= 2 \sigma^2_n(x)
\;.
\end{equation}
The variance $\sig^2_n$ can be obtained in a similar 
way as the mean (for details, see ref.~\cite{Ewerz:1999tt}). 

  \begin{figure}[t!]
\input{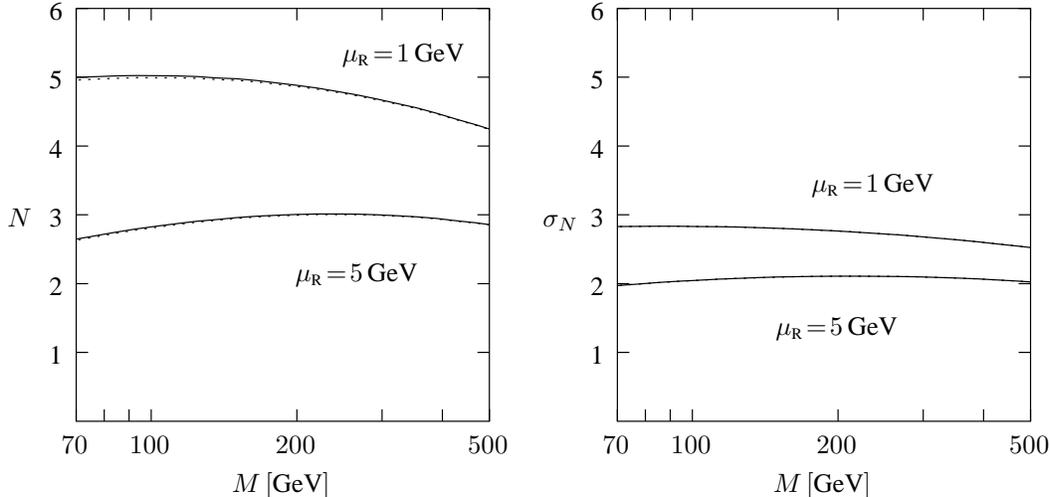}
\vskip-0.5cm
  \caption{The mean value and dispersion of the number of (mini-)jets
    in central Higgs production 
    at LHC for two different resolution scales $\mR$. Solid lines 
    show the SL results up to the 15th order in perturbation theory, 
    dashed lines correspond to the DL approximation.}
\label{fig_higgsjets}
  \end{figure}

We have calculated the dependence of 
$N$ and $\sig_N$ on the Higgs mass $M_H$ 
using the leading-order MRST gluon distribution 
\cite{Martin:1998sq}. 
Our numerical results are shown in fig.~\ref{fig_higgsjets}. 
The DL results, obtained by keeping only the first term 
in eq.\ (\ref{glip}), give an excellent approximation and 
the SL terms are less significant.
We see that the mini-jet multiplicity and its dispersion are rather
insensitive to the Higgs mass at the energy of the LHC.
The mean number of associated mini-jets is rather low, such 
that the identification of the Higgs boson should not be seriously 
affected by them. 
In view of the rapid convergence of the perturbative series in 
$x$-space we do not expect the result for the mini-jet multiplicity 
to be strongly modified by higher order corrections. 

\subsection{The next-to-leading corrections\protect\footnote{
Contributing author: G.P. Salam.} }
\label{sec:smallxsalam;qcd}

As has already been discussed, in practically all experimental
contexts, the LL BFKL equations fails to reproduce
the data. It is likely that the problem is due to the presence of
significant sub-leading corrections. 

The next-to-leading logarithmic (NLL) correction terms $\as (\as \ln
s)^n$ are therefore of particular interest. Such terms can arise for
example from configurations containing a pair of particles which are
close in rapidity, or due to the running of the coupling.  We write the
kernel of the BFKL equation~(\ref{bfklb}) as
\begin{equation}
\cK\,[f_{\omega}(q_{a\perp},q_{b\perp})] =
\cK_0[f_{\omega}(q_{a\perp},q_{b\perp})] +
  \alb\,\cK_1[f_{\omega}(q_{a\perp},q_{b\perp})] + \cO{\alb^2}\, ,
\end{equation}
where $\cK_0$ is the LL kernel~(\ref{kern}), and $\cK_1$ contains the
NLL corrections. A number of different pieces contribute to $\cK_1$:
the emission of two close-in-rapidity partons (two gluons~\cite{Fadin:1989kf,
DelDuca:1996ki}
or a $q \bar q$ pair~\cite{Catani:1990xk,Catani:1991eg,Fadin:1996nw,
DelDuca:1996me,Fadin:1998hr}) from the gluon ladder; the one-loop 
corrections~\cite{Fadin:1993wh,Fadin:1994fj,Fadin:1996yv,Fadin:1998sh,
DelDuca:1999cx} to the emission of a gluon from the ladder; the NLL
corrections to a reggeised gluon~\cite{Fadin:1995xg,Fadin:1996tb,Fadin:1996km,
Blumlein:1998ib}. The various pieces were put together in
\cite{Fadin:1998py,Camici:1997ij,Ciafaloni:1998gs}.

The resulting corrections have a number of interesting features, such
as the fact that they imply the emitted transverse momentum
as being the appropriate scale for $\as$, and
certain parts of the resulting kernel can be associated with
physical contributions such as the finite-$z$ part of the DGLAP
splitting functions.  However from the point of view of their direct
use in phenomenology, the NLL corrections present problems: applying
the NLL kernel to the LL eigenfunctions, $(k_\perp^2)^\gamma$, with
$\gamma$ as in eq.~(\ref{omlip}), the BFKL
exponent becomes~\cite{Fadin:1998py,Ciafaloni:1998gs}
\begin{equation}
  \label{eq:NLLpower}
  \lambda   \simeq 4\ln2 \alb (1 - 6.2\as )\,,
\end{equation}
and inserting a value of $\as=0.2$ relevant for many BFKL studies
leads to a negative power. A detailed study of the resummation of the
kernel reveals the even worse property that for $\as > 0.05$ the NLL
corrections lead to negative cross sections~\cite{Ross:1998xw}.

\subsubsection{Beyond NLL}

At first sight one might therefore conclude that the NLL corrections
remove all predictive power from BFKL physics. Various groups have
however proposed rather different approaches for the inclusion and
resummation of higher-order terms with a view to stabilising the
perturbative series. Three basic strategies have been suggested: BLM
resummation together with an appropriate scheme change, a rapidity
veto, and resummation of collinearly enhanced terms.

A standard approach in situations where the perturbative series
converges slowly is to apply a scale change. One such procedure is BLM
scale setting \cite{Brodsky:1983gc}, where it is argued that for any
given observable, some of the NLL corrections come from the natural
scale being different from $Q^2$, and that the appropriate scale can
be deduced from the coefficient of the $\nf$-dependent part of the NLL
correction. In \cite{Brodsky:1999kn} the procedure is applied to the
BFKL NLL corrections. The authors find that in the $\msbar$ scheme,
BLM scale setting makes little difference to the poor convergence of
the series.  They then show that in certain other schemes, notably the
MOM (based on the symmetric triple-gluon vertex) and $\Upsilon$ (based
on $\Upsilon \to ggg$ decay) schemes, the coefficient of the $\nf$
dependence is significantly modified --- the BLM resummation then has
a much larger effect leading to an estimate for the exponent,
$\lambda\simeq0.15$ fairly independently of $Q^2$. The problem of negative
cross sections still persists however, albeit to a lesser extent.
There are also questions regarding the naturalness of the particular
scheme choices that are required in order to obtain a stable answer,
there being arguments both for and against.

The rapidity veto approach has been studied in detail in
\cite{Schmidt:1999mz}. The background of this approach is that the
BFKL kernel is formally valid only for branchings separated by a large
rapidity --- but to obtain the high-energy power-growth one then
normally integrates over all possible rapidity intervals between
successive branchings, including small rapidities. One can equally
well place a rapidity veto, i.e. integrate only over rapidities beyond
some cut, $\Delta y$, of order 1 or 2. This corresponds to introducing
a set of corrections at NLL and beyond, and one argues that part of
the actual NLL corrections may come from something akin to such a
rapidity veto. One then studies the effect of the rapidity veto at all
orders (while fixing the NLL corrections). This was done in
\cite{Schmidt:1999mz} where it was found that for large rapidity
vetoes ($\Delta y > 2.2$) the exponent $\lambda$ is quite stable against
variations in $\Delta y$ and that the problems of negative cross
sections disappear. But for smaller rapidity vetoes, the usual
problems persist.

\begin{figure}[t!]
  \begin{center}
    \includegraphics[width=0.55\textwidth,clip]{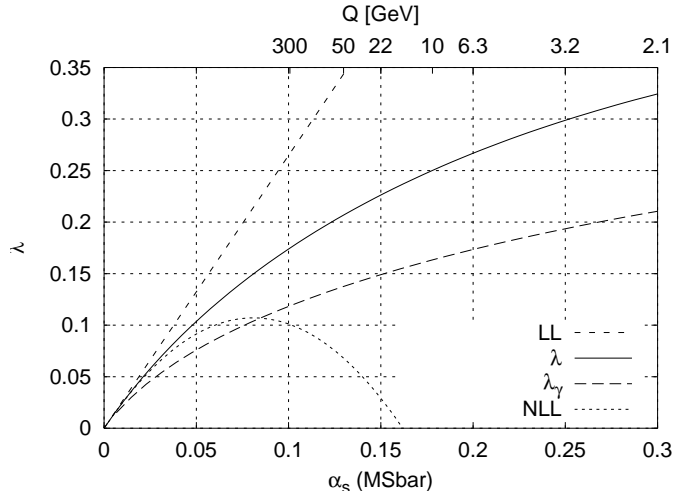}
\vskip-0.5cm
    \caption{The high-energy exponent in various approaches; $\lambda$ is
      the exponent relevant to processes such as Mueller-Navelet jets,
      including the NLL corrections and collinear
      improvements; the equivalent exponent relevant to anomalous
      dimensions is $\lambda_\gamma$.}
    \label{fig:NLLIntercept}
  \end{center}
\end{figure}

The two above approaches conjecture some new physical effect (natural
non-Abelian scheme, rapidity veto). The third approach is a little
different in that it takes the small-$x$ kernel and supplements it in
such a way as to render it consistent with DGLAP evolution in the
collinear and anti-collinear limits, i.e. where one of the interacting
objects has a much larger transverse scale than the other. The
motivation for doing this comes from the observation that while the
convergence of the small-$x$ expansion is poor for normal high-energy
scattering (both objects of the same transverse scale), for
(anti)collinear high-energy scattering the expansion becomes far worse
and so \emph{must} be resummed: technically speaking, the LL
characteristic function\footnote{In the notation of Sect.
  \ref{sec:vddwjs} and generalising eq.~(\ref{omlip}), 
$\om(\nu,0) = \alb \chi_0(1/2+i\nu) +
  \alb^2\chi_1(1/2+i\nu)+\ldots$. Higher azimuthal components
  $\om(\nu,n\ge1)$ are not included. However, they contribute only
to azimuthal angle correlations such as those discussed in 
Sect.~\ref{sec:vddwjs}.}
$\chi_0(\gamma)$ diverges as $1/\gamma$ in the collinear limit $\gamma\to0$,
while the NLL function, $\chi_1(\gamma)$, diverges as $1/\gamma^3$. Since
the structure of these divergences is governed by collinear physics,
it can be calculated at all orders. It turns out that there are double
and single collinear logs and alone they are responsible for most of
the NLL correction even outside the collinear region. They have been
resummed respectively in \cite{Salam:1998tj,Andersson:1996ju} and
\cite{Ciafaloni:1999iv,Ciafaloni:1999yw}, leading to a stable result
for the exponent $\lambda$, free of the problem of negative cross
sections. The dependence of $\lambda$ on $\as$ is shown in
figure~\ref{fig:NLLIntercept}, together with the leading and
next-to-leading results, for comparison.  There is relatively little
dependence on changes of scheme and scale \cite{Ciafaloni:1999yw} and
on the additional introduction of a rapidity veto
\cite{Forshaw:1999xm}. This approach therefore seems to be the most
likely candidate for practical phenomenology.

\subsubsection{Spin-offs from the NLL results: understanding running coupling} 

One of the spin-offs of the NLL corrections was that they identified
the correct scale to be used in the kernel: $\as(q^2)$, where $q$ is
the emitted transverse momentum. However to understand the effects of
running coupling in high-energy cross sections it is necessary to
understand the \emph{iteration} of the kernel with running
coupling. The two contexts of interest are for quantities such as
Mueller-Navelet jets, and for anomalous dimensions. 

In the former, one has a situation where diffusion takes place both
above and below the scale set by the jets. The running of the coupling
causes diffusion below the typical scale $E_t^2$ of the jets to be
enhanced compared to that above --- as a result, as the rapidity
separation increases and diffusion increases, evolution below $E_t^2$
is increasingly favoured, and the cross section grows faster than
$e^{\lambda(E_t^2) Y}$: an extra term appears in the exponent,
proportional to $\as^5(E_t^2)Y^3$
\cite{Kovchegov:1998ae,Levin:1998pk}. This causes the effective power
growth to increase gradually. A second, recently hypothesised effect
called \emph{tunneling} \cite{Ciafaloni:1999au}, should at a certain
point cause a sudden increase in the observed power growth, as the
contribution from very-low-scale evolution becomes larger than that
from evolution at scales of order $E_t^2$. This happens at a rapidity
of $Y\simeq \ln{Q^2}/\lambda_P$, where $\lambda_P$ is the exponent
characteristic of low scales. It remains to be seen whether such an
effect will be phenomenologically observable.

Another quantity for which running coupling effects turn out to be
very important is anomalous dimensions, or equivalently small-$x$
splitting functions.  Very schematically, anomalous dimensions at a
scale $Q^2$ seem to involve small-$x$ branching only above $Q^2$:
branching below that scale has already been factorized out.
Consequently they sample a region where the running coupling is
smaller than $\as(Q^2)$.  Thus the observed small-$x$ exponent of
the anomalous dimension, $\lambda_\gamma(Q^2)$, is smaller than the
exponent $\lambda(Q^2)$ relevant in say Mueller-Navelet jets with
scale $E_t^2=Q^2$
\cite{Ciafaloni:1999yw,Ciafaloni:1999au,Thorne:1999sg}. An alternative
point of view \cite{Ball:1999sh,Altarelli:1999vw} is discussed
in Sect.~\ref{sec:ressmallx;qcd}.



\section{DOUBLE PARTON SCATTERING\protect\footnote{Section
    coordinator: D. Treleani.}$^{,}$~\protect\footnote{Contributing 
authors:A.~Del
    Fabbro and D.~Treleani.}}
\label{sec:doubpar;qcd}

\subsection{Introduction}

The large flux of partons, which becomes available for hard collisions
at high energies,
justifies the expectation, at the LHC, of sizeable effects 
due to the unitarization of the 
hard component of the interaction. In fact it is not difficult to foresee hard 
collision processes
with a cross section larger than the total cross
section itself~\cite{Pancheri:1986qg, Lomatch:1989uc}. Such a result is not 
inconsistent, if one keeps into
account that the inclusive cross section, described by the
single scattering expression of the QCD-parton model, includes a multiplicity
factor which keeps into account the possibility of having several partonic 
interactions in the same
hadronic inelastic event~\cite{Ametller:1988ru, Brown:1989gp}. The possibility 
of hard processes
with multiple parton
interactions, namely different pair of partons interacting independently 
with a large momentum transfer in
the same hadronic collision, was on the other hand foreseen long ago by several
authors
\cite{Landshoff:1978fq, Takagi:1979wn, Goebel:1980mi, Paver:1982yp,
Humpert:1983pw, Mekhfi:1985dv, Mekhfi:1985az, Humpert:1985ay,
Sjostrand:1987su, Halzen:1987ue, Mangano:1989sq,
Godbole:1990ti, Drees:1996rw}. In a multi-parton interaction the different pairs of
interacting partons are separated in transverse space by a distance of the
order of the hadron radius. As a consequence the
transverse momenta have to be balanced 
independently in the different partonic collisions, giving in this way a well
defined characterization to the process.
The simplest event of that kind, the double parton
scattering, has been a topic of 
experimental search of all high energy hadron collider
experiments since several years~\cite{Akesson:1987iv, Alitti:1991rd,
Abe:1993rv}. While initially the results
have been sparse and 
not very consistent, recently CDF has reported the observation
of a large number of events with double parton scatterings~\cite{Abe:1997bp,
Abe:1997xk}. 
\begin{figure}[t!]
\begin{center}
\begin{tabular}{lr}
\begin{minipage}[t]{0.47\textwidth}
    \includegraphics[width=\textwidth,clip]{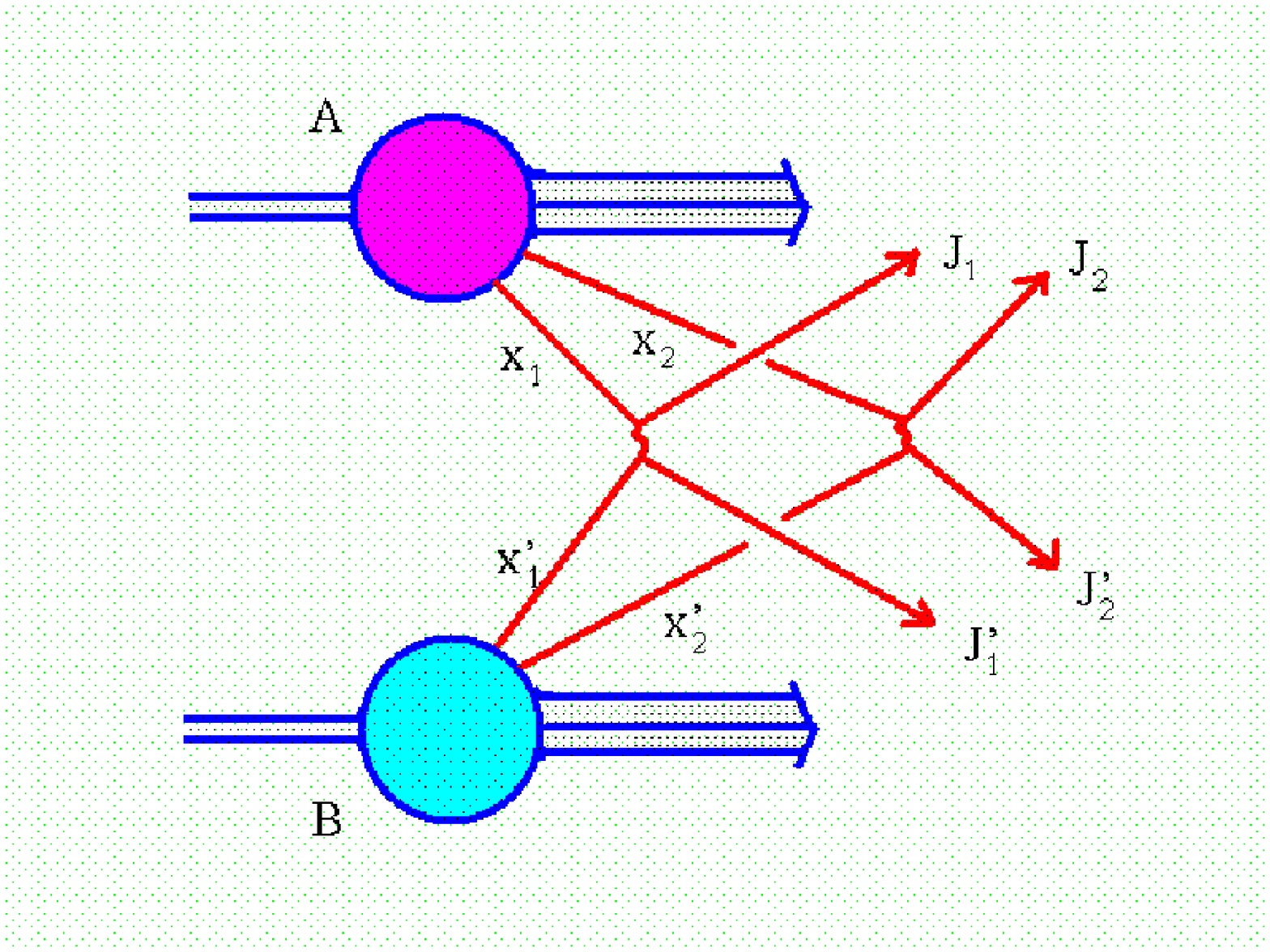}
\vskip-0.5cm
    \caption{Double parton scattering.}
    \label{fig:double;9QCD}
\end{minipage}
&
\begin{minipage}[t]{0.47\textwidth}
    \includegraphics[width=\textwidth,clip]{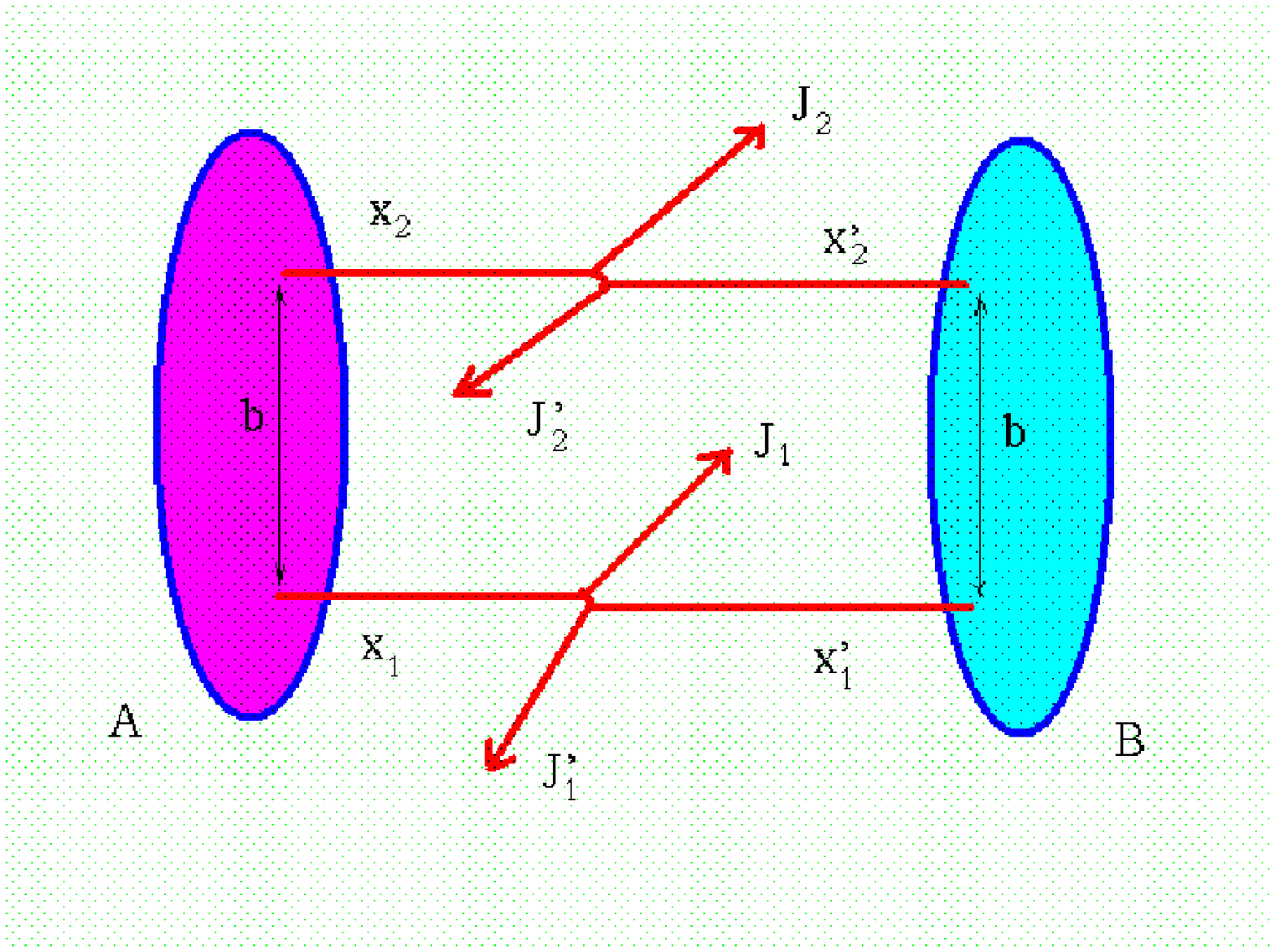}
\vskip-0.5cm
    \caption{Graphical representation of Eq.~\ref{sigmad;9QCD}.}
    \label{fig:doubleb;9QCD}
\end{minipage}
\\
\end{tabular}
  \end{center}
\vspace{-0.5cm}
\end{figure}


\subsection{Cross section for double parton scattering}

The inclusive cross section of a double parton scattering has a simple
probabilistic expression. 
Interference effects between the two partonic collisions are in fact
negligible, since the
partonic interactions are localized in a 
much smaller region, with a size of the order of the inverse of the momentum
transfer, as compared to 
the distance in transverse space between the different partonic
interactions. The
non-perturbative component of the process gets factorized, as a consequence,
into a function 
which depends on the fractional momenta 
of the partons taking part the interaction and on their distance in transverse
space,
which has to be the same for both the target and
the projectile partons, in order to have the alignment which is needed
for the interaction to occur. One obtains therefore for the double
parton scattering cross section the expression (see 
fig.~\ref{fig:doubleb;9QCD})
\begin{equation}
\sigma_D={1\over2}\int_{p_T^{cut}}\Gamma_A(x_1,x_2;b)\hat{\sigma}(x_1,x_1')
\hat{\sigma}(x_2,x_2')\Gamma_B(x_1',x_2';b)dx_1dx_1'dx_2dx_2'd^2b \;,
\label{sigmad;9QCD}
\end{equation}
where the non perturbative input is the
two-body parton distribution $\Gamma(x_1, x_2;b)$, whose arguments are the two
fractional momenta, $x_1$ and $x_2$, and the distance of the two partons
in transverse space $b$. The partonic cross sections,
$\hat{\sigma}(x,x')$, are integrated on the momentum transfer, at a fixed value
of the partonic center of mass
energy, and the cutoff $p_T^{cut}$ is introduced to regularize the
singularity at small $p_T$ and at small $x$ values. The two-body parton 
distributions $\Gamma(x_1, x_2;b)$
represent the new property of the hadron structure which becomes
accessible through the observation of the double parton collision
processes. It is a non perturbative quantity which is independent on
the one-body parton distributions, namely on the non-perturbative input
to the large $p_T$ processes usually considered. The two-body
parton distributions are in fact
related directly to the two-body parton correlations in the hadron structure.  


If the two pairs of partons undergoing the hard interactions 
are not correlated in $x$ and if the dependence on $b$
can be factorized, the two-body parton distributions are nevertheless expressed
as $\Gamma(x_1, x_2;b)=f(x_1) f(x_2) F(b)$, where $f(x)$ is the usual one-body
parton distribution, appearing in large $p_T$ inclusive processes, and
$F(b)$ is a function which describes the distribution of the partons in
transverse space. With these assumptions the cross
section for a double parton collision leads, in
the case of two indistinguishable parton interactions, to the simplest
factorized expression
\begin{equation}
\sigma_D(p_T^{cut})={\bigl[\sigma_S(p_T^{cut})\bigr]^2\over2\sigma_{eff}} \;,
\label{fact;9QCD}
\end{equation}
where $\sigma_S$ is the usual inclusive cross section of the
perturbative QCD, i.e. the convolution of parton distributions with
the partonic cross section, $p_T^{cut}$ is the lower integration
threshold and
$\sigma_{eff}$ is a scale factor, with dimensions of a cross
section. It is the result of the integration on the transverse distance $b$, 
actually $1/\sigma_{eff}=\int
d^2bF^2(b)$. All the information on the parton correlation in
transverse space is summarized in $\sigma_{eff}$~\cite{Calucci:1998ii}. The
geometrical origin of $\sigma_{eff}$ justifies the expectation that its
value is both a energy and cutoff independent quantity. 

The double parton scattering process has been measured at Fermilab by
CDF by looking at final states with 
three mini-jets and one photon~\cite{Abe:1997bp, Abe:1997xk}. The measured 
value of the scale
factor is: 
\begin{equation}
\sigma_{eff}=14.5\pm1.7^{+1.7}_{-2.3} \;{\rm mb} .
\end{equation}
In the limited range of $x$
experimentally accessible, $\sigma_{eff}$ does not show evidence of dependence
on the fractional momenta, which indicates that the
simplest hypotheses above are not in contradiction with the experiment.  

The qualitative features of the double parton scattering process are
easily read in the factorized expression in Eq.~(\ref{fact;9QCD}). As a
consequence of the proportionality of $\sigma_D$ with $\sigma_S^2$, the
double parton scattering cross section is characterized by a rapid
decrease for $p_T\to\infty$ and by a rapid growth for 
$p_T\to0$. As for the energy behavior, $\sigma_D$ increases faster with $s$ as
compared to the single scattering
cross section (it goes as $\sigma_S^2$). Multiple parton collisions are
therefore enhanced at the LHC.

\subsection{Four jet production}

The most obvious case where multiple parton collisions play a role at
high energy is in the 
production of jets, since the integrated cross section can easily
exceed the unitarity limit at large energies and with a fixed value of
$p_T^{cut}$. One has in fact that, for any value of
$p_T^{cut}$, when $s$ is sufficiently large $\sigma_S>\sigma_{inel}$.
The simplest case to consider is the  production of four large $p_T$
jets, where 
one can compare the leading $(2\to4)$ process with the power suppressed
$(2\to2)^2$
double parton collision. 

\begin{figure}[t!]
\begin{center}
    \includegraphics[width=0.5\textwidth,clip]{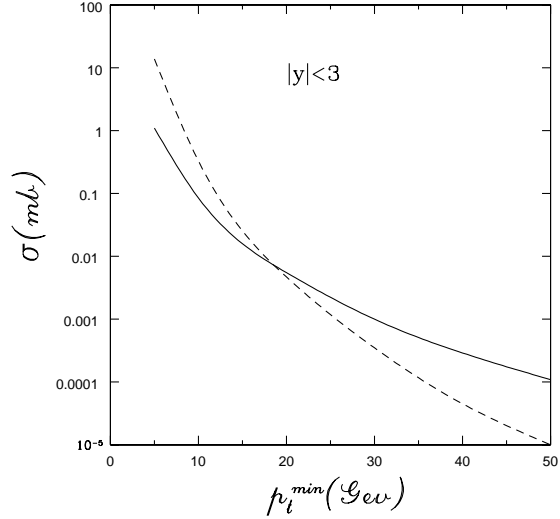}
\vskip-0.5cm
    \caption{Integrated cross section for production of four jets with $|y|<3$
    as a function of the lowest transverse momentum of the jets $p_T^{min}$. 
The continuous curve
    is the expected cross section as from the leading QCD production
    mechanism $2\to4$, the dashed curve is the expected cross section due to 
the
    contribution of double parton collisions $(2\to2)^2$.}
    \label{fig:jets;9QCD}
\end{center}
\end{figure}

In fig.~(\ref{fig:jets;9QCD}) we show the expected rates of production of four
large $p_T$ jets in the central rapidity region ($|y|<3$) with the two
different production mechanisms, as a function of the lowest value
of the transverse momenta of the produced jets $p_T^{min}$. The continuous 
curve
is the expected cross section as from the leading QCD production
mechanism $(2\to4)$~\cite{Berends:1989ie, Berends:1991vx}. The dashed curve is 
the double parton
collisions $(2\to2)^2$ cross section. The curve representing the double 
parton collision in fig.~(\ref{fig:jets;9QCD}) has to be regarded as a lower 
limit,
rather than as the expected rate of the double parton collision process, since
no factor $K$,
accounting for higher order correction terms in $\alpha_S$, has been included
in the evaluation.
Notice that higher order corrections in $\alpha_S$
will contribute with a factor $K^2$ in the double parton collision cross
section. The overall qualitative feature is that, at the LHC, the double parton
collision dominates, with 
respect to the leading QCD single scattering interaction, when one of the jets 
has a transverse
momentum which becomes as low as $20$~GeV.

\subsection{$
l + b{\bar b}$ production}

\begin{figure}[t!]
\begin{center}
\begin{tabular}{lr}
\begin{minipage}[t]{0.47\textwidth}
    \includegraphics[width=\textwidth,clip]{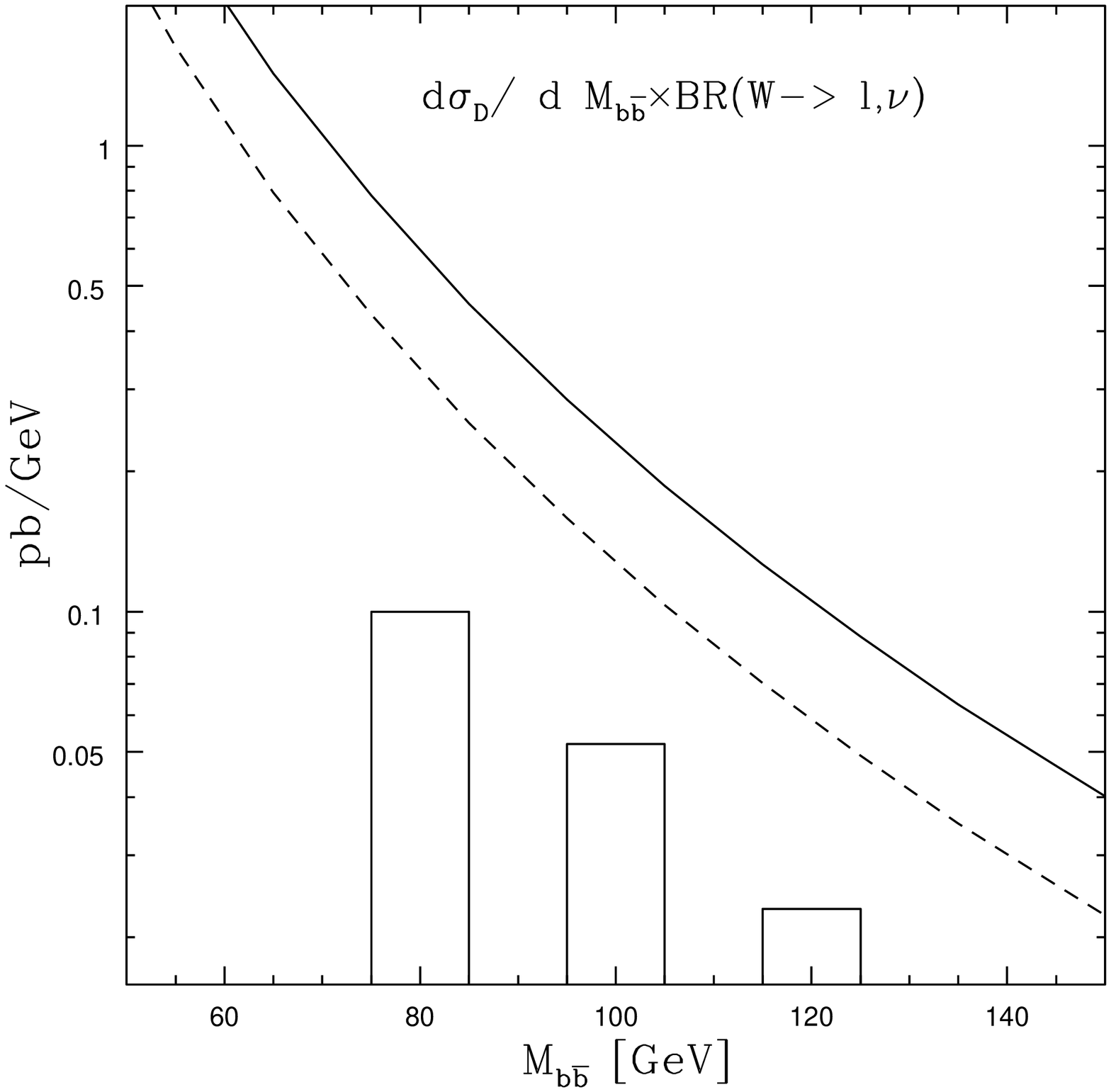}
 \vskip-0.5cm
   \caption{ 
Double parton scattering background to Higgs boson production in
    association with a $W$
as a function of the $b{\bar b}$ invariant mass. The expected Higgs
signal is for three possible values of the Higgs mass, 80, 100 and 120~GeV. 
The dashed line is the background at the LO in perturbation theory.
The continuous line is the result for the double parton scattering background
when computing the $b{\bar b}$ cross section at order $\alpha_S^3$~\cite{Mangano:1992jk}.}
    \label{fig:bh;9QCD}
\end{minipage}
&
\begin{minipage}[t]{0.47\textwidth}
    \includegraphics[width=\textwidth,clip]{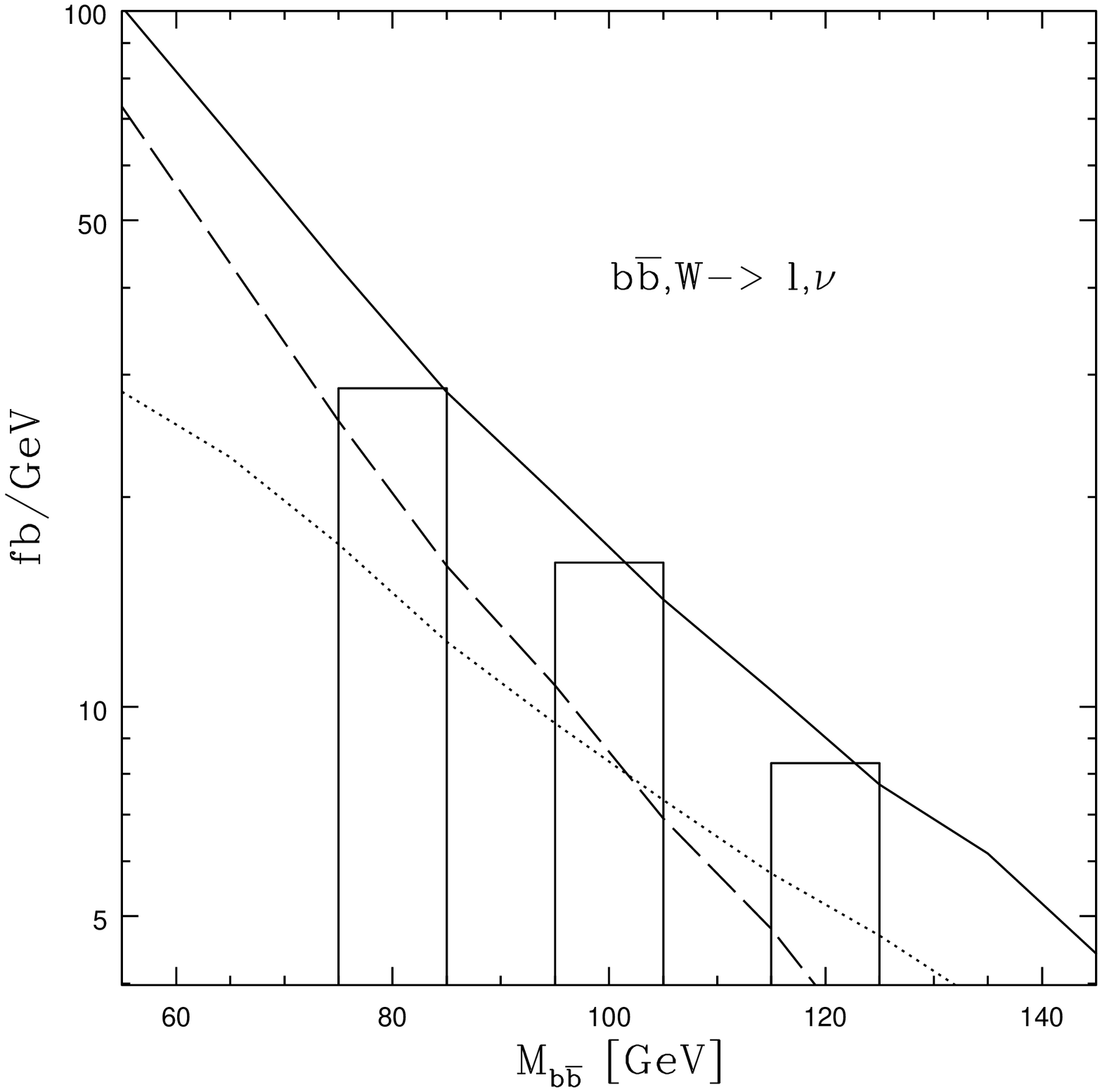}
\vskip-0.5cm
    \caption{
The backgrounds to Higgs boson production is compared with the signal after 
the cuts (see main text).
Dotted line: 
single scattering contribution to the $Wb{\bar b}$ channel. Dashed line:
double parton scattering background. 
Continuous line: total estimated background.}
    \label{fig:bw;9QCD}
\end{minipage}
\\
\end{tabular}
  \end{center}
\vspace{-0.5cm}
\end{figure}

Although multi-parton collisions have been mostly considered to describe
the multiplicity distributions in high energy hadronic interactions 
(for a discussion of multi-parton interactions at LHCb,  
we refer the reader to the Bottom Production Chapter of this Report), 
the role of multi-parton collisions is not limited to the case of
production of large or relatively large $p_T$ jets. One may find in fact 
various other processes of
interest at the LHC where multiple parton collisions are
relevant~\cite{DelFabbro:1999tf, Kulesza:1999zh}.  
While $\sigma_{eff}$ may depend in principle on the different species of 
partons involved
in the interaction, $\sigma_{eff}$ should not vary much in the different
processes and one would 
expect that it is, to a large extent, a process independent
quantity~\cite{Calucci:1999yz}. We will therefore consider it, in the 
following, as a
universal quantity and we will use for $\sigma_{eff}$ the value which has been
measured in the CDF experiment. 
The cross section of a double
parton interaction, resulting from the two distinguishable parton collisions 
$A$
and $B$, is therefore expressed as
\begin{equation}
\sigma_D={\sigma_A\sigma_B\over\sigma_{eff}} \;.
\end{equation}
As a meaningful example we have 
considered the production of an isolated lepton and of a $b{\bar b}$
pair \cite{DelFabbro:2000tf}, which represents 
an interesting channel to detect the Higgs boson
production at the LHC in the intermediate Higgs mass range, $80{\rm
GeV}<M_H<150{\rm GeV}$.
A background to the process 
$p+p\to WH+X$, with $W\to l\nu_l$ and $H\to b\bar{b}$, is represented
by the double parton scattering interaction where the
intermediate vector boson $W$ and the $b\bar{b}$ pair are 
created in two independent parton interactions. 
If one uses $\sigma(W)\times BR(W\to l\nu_l)\simeq40{\rm nb}\nonumber$ 
~\cite{Martin:1999ww}
and  
$\sigma(b\bar{b})\simeq 5\times10^2 \mu{\rm b}$, one obtains for
the double collision cross section the value of $1.4$~nb. The Higgs production 
cross sections,
$p+p\to WH+X$, with $W\to
l\nu_l$ and $H\to b\bar{b}$, has been estimated to be rather of order of 
$1$~pb~\cite{Moretti:1996wa, Kunszt:1997yp}. 
Obviously the three
orders of magnitude of difference in the integrated cross section are mainly
due to the 
configurations where the $b{\bar b}$ pair is produced with an invariant mass 
close to the
threshold of $b{\bar b}$ production. The expected background to the
Higgs production signal, caused by the double parton collision process, is
shown in fig.~(\ref{fig:bh;9QCD}) as a function of the invariant mass of the
$b{\bar b}$ pair. 

In fig.~(\ref{fig:bh;9QCD}) we have plotted the expected signal in the ${b 
\bar{b}}$ invariant mass
due to the Higgs boson production for three possible values of the Higgs mass,
80, 100 and 120~GeV. The dashed line is the double parton scattering
background at the LO
in perturbation theory.
The continuous line is the result for the double parton scattering background
when computing the $b{\bar b}$ cross section at order 
$\alpha_S^3$\cite{Mangano:1992jk}.

%

In fig.~(\ref{fig:bw;9QCD}) we compare the signal and the background after 
applying all the
typical cuts considered to select the Higgs signal in this
channel~\cite{Moretti:1996wa}:  

- - for the lepton we require: 
$p_T^l>20$~GeV, $|\eta^l|<2.5$ and isolation from the $b$'s,
$\Delta R_{l,b}>.7$

- - for the two $b$ partons: 
$p_T^b>15$~GeV, $|\eta^b|<2$ and $\Delta
R_{b,\bar b}>.7$


As in the previous figure the Higgs signal in the $b{\bar b}$ invariant mass
corresponds to three possible values for the mass of the Higgs boson, $80$,
$100$ and $120$~GeV. The dotted line is the single parton scattering
background, where the $Wb{\bar b}$ state is created directly in a single
partonic interaction. The dashed line is the expected background originated by 
the double parton
scattering process, evaluated by estimating the $b{\bar b}$ production cross
section at ${\cal O}(\alpha_S^3)$. 
The continuous line is the total expected background.
In the calculations of the background and signal  we used, for the LO
matrix elements, the 
packages MadGraph ~\cite{Stelzer:1994ta} and HELAS~\cite{Murayama:1992gi}. The 
integration was
performed by VEGAS~\cite{Lepage:1978sw} with the parton distributions 
MRS99~\cite{Martin:1999ww}.

Also after using the more realistic cuts just
described, the double parton scatterings process remains a rather
substantial component of the background, as one may see by comparing in
fig.~(\ref{fig:bw;9QCD}) the total background
estimate (continuous curve) with the more conventional single scattering 
background
estimate (dotted curve).

\subsection{Summarizing remarks}
At the LHC one has to expect large effects from 
multiple parton collisions in various
processes of interest.  
To the purpose of illustration, we have
presently studied 
the production of a $b{\bar b}$ pair in association with a $W$ boson, 
followed by the decay  
$W\to l\nu$, in the mass range $M_{b{\bar b}}\simeq100$~GeV. The
channel is of interest for the observation of the Higgs boson production when
the 
Higgs mass is below the threshold of $W^+W^-$ production. We find that, if one
applies the 
standard cuts to the final state usually considered to isolate the Higgs
signal in this channel, the
background due to double parton
scatterings ($b{\bar b}$ pair and $W$ boson produced in two different
partonic interactions) is comparable to the more traditional 
background, where the 
$b{\bar b}$ pair and the $W$ boson
are produced in a single parton collision. A similar situation can be
expected with several other final states:
\begin{itemize}
\item{ } $Zb{\bar b}$,
\item{ } $W+{\rm jets}$, $Wb+{\rm jets}$ and $Wb{\bar b}+{\rm jets}$, 
\item{ } $t{\bar t}\to llb{\bar b}$, 
\item{ } $t{\bar b}\to b{\bar b}l\nu$,
\item{ } $b{\bar b}+{\rm jets}$,
\item{ } final states with many jets when $p_T^{min}\simeq20$, $30$~GeV.
\end{itemize}
The well definite characterization of the states produced by the
multiple parton scattering processes
allows nevertheless one to figure out more efficient selection criteria to get
rid of this further background source,
or to measure it in a
precise way. The present analysis however points out that, as
a consequence of the enhanced role of multiple parton collisions at
high energy, a
detailed and systematic study of the expected rates and  
backgrounds, due to 
multiple parton collision processes, is of great importance at the LHC and it
represents one of the topics which have to
be addressed seriously in the next future.


%


\section{BACKGROUNDS TO NEUTRAL HIGGS BOSONS SEARCHES\protect\footnote{Section
    coordinators: J.-P. Guillet, E. Pilon and E. Richter-Was.}$^{,}$~\protect\footnote{
    Contributing authors: T.~Binoth, D.~de Florian, M.~Grazzini, J.-P.~Guillet,
    J.~Huston, V.~Ilyin, Z.~Kunszt, Ph.~Min\'e, E.~Pilon, E.~Richter-Was
    and M.~Werlen.}}
\label{sec:backg;qcd}

\subsection{Introduction} \label{intro;11qcd} 

The most important goal of the physics programme of the LHC experiments ATLAS
\cite{ATLAS-TDR;11qcd} and CMS \cite{CMS-TP;11qcd} is to perform measurements which lead to
the  understanding of the mechanism of electroweak symmetry breaking. In the
framework of the SM, as well as its extensions, e.g. 
super-symmetric (SUSY), it translates into the major topic of Higgs boson
searches. The  SM assumes one doublet of scalar fields, implying the existence
of one neutral scalar particle. In SUSY models, the Higgs sector is extended to
contain at least two doublets of scalar fields leading to the prediction of
five Higgs particles, three electrically neutral and two charged. The following
discussion focuses on neutral bosons.

The Higgs boson mass remains largely unconstrained in the SM. From perturbative
unitarity arguments an upper limit of $\sim$~1 TeV can be derived.  The
requirements of stability of electroweak vacuum, and of perturbative validity of
the SM seen as an effective theory, allow to set upper and lower bounds
depending on the cut-off value chosen for the energy scale up to which the SM
is assumed to be valid
\cite{Maiani:1978cg,Cabibbo:1979ay,Dashen:1983ts,Callaway:1984zd,Beg:1984tu,Lindner:1986uk,Altarelli:1994rb,Casas:1995qy,Casas:1996aq,Grzadkowski:1986zw,Hambye:1997wb}.
If the cut-off is assumed to be about the Planck mass, which means that no new
physics appears up to that scale, the Higgs boson is predicted to be in the
range 130~-190~GeV. This bound becomes weaker if new physics appears at a lower
mass scale. A global fit to all electroweak data in the SM framework seems to
favour a rather light Higgs boson:  $m_{H} = 76^{+85}_{-47}$ GeV \cite{lep1;11qcd}.
Moreover, SUSY extensions of the SM generically predict the existence of one
rather light neutral Higgs boson (e.g. roughly $m_{H} \leq 130$ GeV in the
minimal SUSY extension). The LEP2 experiments are searching Higgs bosons with
masses up to about 110 GeV \cite{lep2;11qcd}. Assuming that no Higgs boson will be
found at LEP, the above indications raise even more interest in the Higgs boson
searches at LHC in the intermediate mass range from 95~GeV to $2m_Z$.

The Higgs boson searches scenarios prepared by the ATLAS \cite{ATLAS-TDR;11qcd} and
CMS \cite{CMS-TP;11qcd} Collaborations cover a large spectrum of final state
signatures in this mass range. The rare $H \rightarrow \gamma \gamma$ decay
mode is expected to be accessible in inclusive Higgs production in the mass
range 90~-140 GeV already for an integrated luminosity of
$100~fb^{-1}$. This observability can be also complemented by looking at an
additional jet (production in association with jets) or lepton in the final
state ($t \bar{t} H$, $WH$, $ZH$ associated production). The additional
isolated lepton in the final state will also allow to access the dominant $H
\rightarrow b \bar{b}$ decay mode, and such observability has been established
in the ATLAS searches scenarios for the  $t \bar{t} H$ production channel.
Higgs decay into $WW$in inclusive or associated production lead to the clean
signature of 2 or 3 leptons in the final state. A signature with even higher
lepton multiplicity is provided by the $H \rightarrow Z Z^*$ channel in the
inclusive and associated production. The possible observability of the latest
one is still under investigation, as presented below. A rich spectrum of final
state signatures was proposed recently, which explored $WW$ and $ZZ$ fusion
mechanisms producing a Higgs boson in association with two forward/backward
jets. The observability of the $H \rightarrow \gamma \gamma$, $H \rightarrow
\tau^{+} \tau^{-}$ and $H \rightarrow W W^*$ as established so far in
\cite{Rainwater:1997dg,Rainwater:1999kj,Plehn:1999xi,Rainwater:1999sd} at the particle level seems very promising. 

Given the very large spectrum of final state signatures which have become of
interest in the intermediate mass range, this section will be focused on 
recent progress in the evaluation of backgrounds to two-photon and
multi-lepton signatures, and in the observability of the latter in associated
production. Recent results concerning the two-photon background in the mass
range 90~-~140 GeV, together with the NLO contribution to the signal of
associated production $H$ + jet, are given in Sect.~\ref{2gam;11qcd}. A recent
investigation on $WH$ associated production for $m_{H} \geq 140$ GeV is
presented in Sect.~\ref{wh;11qcd}.

\subsection{The two-photon channel in the mass range 90~-~140 GeV} 
\label{2gam;11qcd}

In this range, the most promising channel is $H \rightarrow \gamma  \gamma$.
The branching ratio is however small\footnote{The cross section for the
production of a SM Higgs  boson at the Tevatron in this range is $\sim 1 pb$,
not enough to allow a search in this mode given its small branching ratio. A
search for a non SM Higgs Boson in this mode has been carried out by both CDF
and D$\not{\! 0}$ with negative conclusions \cite{cdfhiggs;11qcd,Abbott:1999vv}.} , typically
$B(H \rightarrow  \gamma \gamma) \sim {\cal O}(10^{-3})$, and initially the
background is eight orders  of magnitude larger than the signal. This
background is splitted into two  components, called {\it irreducible} and {\it
reducible}. 

\subsubsection{Irreducible background: prompt photon pairs.}\label{irred}

This class of background comes from prompt photon pair production, where
``prompt" means that the photons do not come from the decay of high-$p_{T}$
$\pi^{0}$ or $\eta$, but from hard partonic interactions. A large amount of
this background, which we therefore call {\it irreducible}, passes the photon
isolation cuts. Further kinematic cuts have to be used to suppress it.
Regarding the efficiency of background rejection, one may distinguish between
the signal processes of  {\it inclusive} production, and of {\it associated}
production (and corresponding backgrounds). The first class yields higher
rates  than the second one. On the other hand, kinematical cuts are more
efficient in the case of associated production, and the background may be
theoretically better controlled than in the inclusive case. These issues are
discussed in the following.

\noindent
{\it Mechanisms of prompt photon pair production.}\label{mech} \\
Schematically, three mechanisms produce prompt photon pairs with a large
invariant mass: the ``direct" mechanism produces both photons directly from the
hard subprocess; the ``single-fragmentation" mechanism, instead, involves 
precisely one photon resulting from the fragmentation of a hard parton; the 
``double-fragmentation" mechanism
yields both photons by fragmentation. 
Topologically, a photon from fragmentation is most probably accompanied by a
jet of hadrons, therefore will be more strongly rejected by the isolation
criterion. From a calculational point of view, this schematic classification
emerges from the QCD factorization procedure described in 
Sect.~\ref{sec:intro;qcd} (see \cite{Binoth:1999qq} for more details). Although
this classification is convenient,  one has to keep in mind that the splitting
between these different contributions  is arbitrary: none of these
contributions is separately measurable, only their sum is. Due to the high
gluon density at LHC, ``single-fragmentation"  dominates the inclusive production
of prompt photon pairs. Beyond NLO, a new process of the ``direct" type
appears, the so-called box $g g \rightarrow \gamma \gamma$ contribution. 
Strictly speaking, it is a NNLO contribution. However, the large gluon luminosity
at LHC magnifies it to a size comparable with the Born term $q \bar{q}
\rightarrow \gamma \gamma$ in the invariant mass range  90~-~140 GeV. Therefore
it is usually included in LHC phenomenological studies 
\cite{Binoth:1999qq,Aurenche:1985yk,Aurenche:1990yf,Bailey:1992br,Bailey:1993jz,Bailey:1994ae,Balazs:1998xd,Balazs:1999bm}.

\noindent
{\it Recent improvements}\label{improv} \\
Early calculations \cite{Aurenche:1985yk,Aurenche:1990yf} 
of photon-pair production were not suitable to estimate the background to 
Higgs boson production. A first improvement \cite{Bailey:1992br,Bailey:1993jz,Bailey:1994ae} 
implemented these results
in a more flexible way by combining analytical and Monte-Carlo
techniques. Following a similar approach, recent work goes further along two
directions. 

In \cite{Balazs:1998xd,Balazs:1999bm}, multiple soft gluons effects in the ``direct"
contribution are summed to next-to-leading logarithmic accuracy in the
Collins-Soper framework. This provides a prediction for semi-inclusive
observables such as the  transverse momentum ($q_{T}$) distribution of photon
pairs that extends over the whole spectrum, thanks to a matching between the
resummed part (suited for the low $q_{T}$ peak) and a fixed order calculation
for the high $q_{T}$ tail. These features are encoded in the computer program 
{\it RESBOS} \cite{Balazs:1998xd,Balazs:1999bm}. In this calculation, 
the ``single-fragmentation" contribution is evaluated at LO and 
``double-fragmentation" is neglected.

Another recent improvement is the
computation of the NLO corrections to both fragmentation contributions (using
the set of NLO fragmentation functions of \cite{Bourhis:1998yu}), which provides a
consistent NLO approximation suitable for inclusive observables. This 
calculation, also  implemented in a computer code {\it DIPHOX} of Monte Carlo
type, is described in \cite{Binoth:1999qq}. No soft gluon summation has so far
been implemented in \cite{Binoth:1999qq}.

\noindent
{\it Effects of isolation}\label{isolation}\\ 
Actually, the isolation requirements, imposed experimentally to suppress the
reducible background, severely reduce the fragmentation components, too (which,
properly speaking, are thus not really irreducible\footnote{This misleading
terminology sometimes \cite{Hjet1;11qcd,Abdullin:1998er} leads to call
irreducible only the ``direct" component, and reducible the $\pi^{0}$, $\eta$,
etc {\it plus} the ``fragmentation" components. Although it seems 
intuitive at LO, this alternative  classification is ill defined beyond LO, as
the splitting between  ``fragmentation" components and higher order corrections
to the ``direct" one  is theoretically ambiguous.}\label{def-reducible}). The
isolation criterion commonly used is schematically the 
following\footnote{This isolation criterion for single prompt photon production
is discussed in the theoretical
literature in Refs.~\cite{Kunszt:1993ab,Berger:1996cc,Berger:1996vy,Catani:1998yh} 
($e^{+}e^{-}$ collisions)
and in Refs.~\cite{Baer:1990ra,Aurenche:1990gv,Berger:1991et,Gordon:1994ut,CFGP;6qcd}
(hadronic collisions). 
An alternative  criterion has been recently proposed in \cite{Frixione:1998jh}.
More discussion on the issue of isolation can be found in 
Sects.~\ref{sec:photons;qcd}}.
A photon is called isolated if, inside a cone about the photon, defined in
rapidity and azimuthal angle by  $(\eta - \eta_{\gamma})^{2} + (\phi
-~\phi_{\gamma})^{2}~\leq~R^{2}$, the  deposited transverse hadronic energy 
$E_{T}^{had}$ is less than some specified value $E_{T \, max}$. Severe
isolation requirements, as $E_{T \, max} = 5$ GeV inside a cone of radius 
$R =0.4$,
suppress the "single-fragmentation" component by a factor 20 to 50, and kill the
``double-fragmentation" contribution, so that the production of {\it isolated}
photon pairs is dominated by the ``direct" mechanism\footnote{The situation is
essentially  the same for a less severe cut as $E_{T \, max} = 10$ GeV.
Note however that such a partonic calculation ignores the hadronic transverse
energy splashed in by underlying events. The value of $E_{T \, max}$ used in
this type of calculation may then be considered as an effective parameter,
smaller than  the actual value used experimentally. This issue has still to be
clarified, especially when the experimental value is nearly saturated by
underlying events and pile-up effects.}. Isolation implies however  that one is
not really dealing with inclusive quantities anymore. Although the
factorization property  of collinear singularities still holds in this case 
\cite{Catani:1998yh,CFGP;6qcd}, infrared divergences can appear {\it inside}
the physical  spectrum for some distributions calculated at fixed order, e.g.
NLO, accuracy, due to isolation. The appearance and the pattern of these
singularities depend strongly on the kinematics and on the type of isolation
criterion used. Moreover, potential infrared instabilities may affect the
reliability of the predictions, when a very low value of $E_{T \, max}$
compared to the $p_{T}$ of the isolated photon, is used. A better understanding
of these problems is required ( see \cite{Binoth:1999qq} and 
Sect.~\ref{sec:photons;qcd} for a more detailed discussion).

\begin{figure}[t!]
\begin{center}
\includegraphics[width=0.6\textwidth,clip]{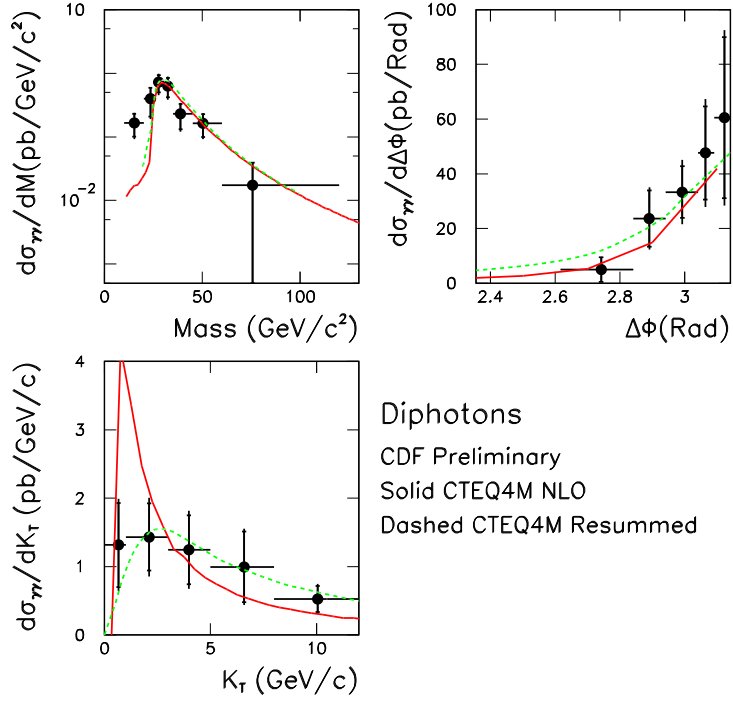}
\end{center}
\vskip-0.5cm
\caption{
A comparison of the NLO and ResBos predictions for di-photon production at
the Tevatron for the di-photon mass, the di-photon azimuthal angle (denoted here
by
$\Delta\phi$) and the di-photon transverse momentum (denoted here by $K_T$).} 
\label{fig:makemyplot;11qcd}
\end{figure}

\noindent
{\it Phenomenology}\label{pheno}\\
Our understanding of photon pair production is already tested at the Tevatron
\cite{Abe:1993cy,cdf2;11qcd,d0;11qcd}. A comparison of the CDF di-photon cross section to NLO and
resummed predictions is shown in Fig.~\ref{fig:makemyplot;11qcd} (for a recent
comparison with \mbox{D$\not{\! 0}$} data see, e.g., \cite{Binoth:1999qq}). Measured
inclusive observables, such as the invariant mass distribution, each photon's
$p_{T}$ distribution, the azimuthal angle ($\phi_{\gamma \gamma}$)
distribution of pairs, agree reasonably well with NLO calculations
\cite{Aurenche:1985yk,Aurenche:1990yf,Bailey:1992br,Bailey:1993jz,Bailey:1994ae,Binoth:1999qq}. 
However, the measured di-photon $q_T$
distribution is noticeably broader than the NLO
prediction, but it is in agreement with the resummed prediction of
\cite{Balazs:1998xd,Balazs:1999bm}. This is 
expected since the $q_T$ distribution is particularly sensitive to soft gluon 
effects\footnote{Infrared sensitive distributions, such as the
$q_{T}$  distribution near $q_{T} \rightarrow 0$, and  the $\phi_{\gamma \gamma}$
distribution near  $\phi_{\gamma \gamma} \rightarrow
\pi$, can be reliably estimated only with resummed calculations. Note that, for the
$\phi_{\gamma \gamma}$ distribution near  $\phi_{\gamma \gamma} \rightarrow
\pi$, not only the ``direct" component diverges order by order and requires a
soft gluon summation, but also both fragmentation contributions. This much more
complicated case has not been treated yet.} \cite{les-houches;11qcd}.

\begin{figure}[t!]
\begin{center}
\includegraphics[width=0.8\linewidth,height=10cm]{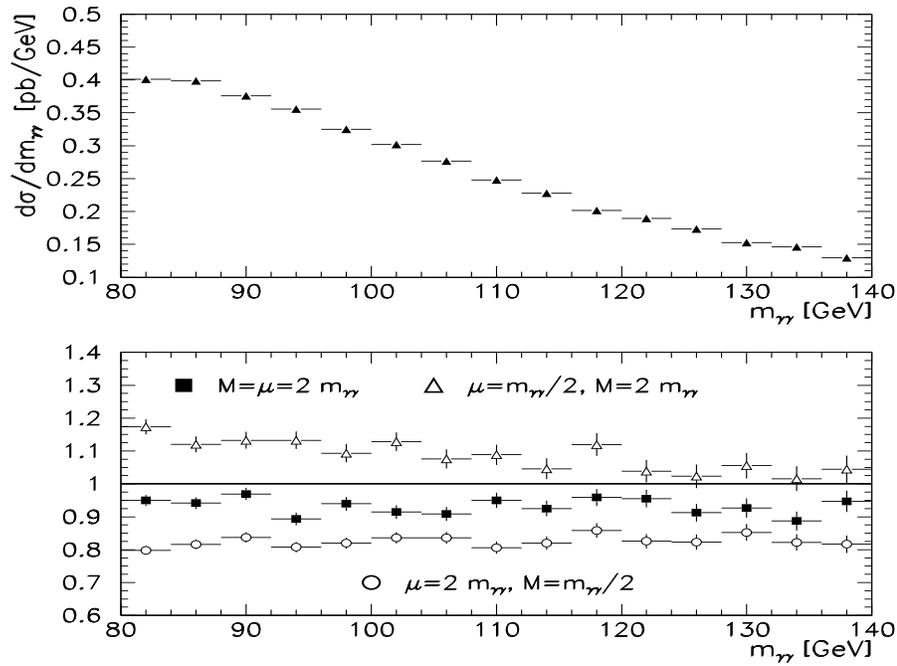}
\end{center}
\vskip-0.5cm
\caption{
 Top: di-photon differential cross section $d\sigma/d m_{\gamma \gamma}$ vs.
 $m_{\gamma \gamma}$ at LHC,  with isolation criterion
 $E_{T \, max}=5$~GeV in $R=0.4$, for the scale choice $M = \mu = m_{\gamma 
 \gamma}/2$. 
 Bottom: factorization ($M$) and renormalization ($\mu$) scale dependences of 
 the NLO cross section $d\sigma/d m_{\gamma \gamma}$ vs. $m_{\gamma \gamma}$,
 normalized by  $d\sigma/d m_{\gamma \gamma}|_{M = \mu = m_{\gamma \gamma}/2}$. 
 }
\label{fig.scale.dep.isol;11qcd}
\end{figure}

The results from Run 1 at the Tevatron were obtained with less than
100 $pb^{-1}$ of data. During Run 2, a data sample  approximately 20 times as
large will be available, allowing both the di-photon signal and its background
to be studied in detail. In particular, the di-photon $q_T$ distribution will be
measured to much greater precision, allowing a study of the $q_T$ resummation
techniques for a $g g$ initial state, necessary for both Higgs and di-photon
production at the LHC~\cite{les-houches;11qcd}.
 
On the theoretical side, scale ambiguities as well as the uncertainties from
unknown beyond NLO corrections plague the predictions. A study of scale
uncertainties has been performed \cite{Binoth:1999qq} for inclusive observables such as
the invariant mass distribution of photon pairs at LHC in the range 90~-~140 
GeV, (Fig. \ref{fig.scale.dep.isol;11qcd}). In the isolated case with
$E_{T \, max} = 5$
GeV inside a cone with $R =0.4$, the scale uncertainties are dominated by the
dependences on the factorization and renormalization scales $M$ and $\mu$; 
while the fragmentation scale ($M_{f}$) dependence is negligible due to the
strong suppression of the fragmentation contribution. The scale uncertainties
are rather small (less than 5\%) when the factorization and renormalization
scales 
are set to be equal and are  varied between 
$m_{\gamma \gamma}/2$ and $2 m_{\gamma
\gamma}$. On the other hand,  anti-correlated
variations of $M$ and $\mu$ in the same range lead to still  rather large (up
to 20\%) uncertainties. In summary, the higher order corrections in prompt
photon pair production are {\it not} fully under control yet.  The consistent
calculation at full NNLO accuracy would involve, in particular, two-loop  $q
\bar{q} \rightarrow \gamma \gamma$ amplitudes and the NNLO evolution of the
parton distributions. Despite recent progress \cite{Smirnov:1999gc,Tausk:1999vh,Smirnov:1999wz,vanNeerven:1999ca}
in this direction\footnote{For more details, see also 
Sect.~\ref{sec:focal;qcd}}, such a NNLO description is not yet available.
Furthermore, the box contribution $gg \rightarrow \gamma \gamma$ is the 
lowest order term of a new subprocess.
Reducing its scale dependence would involve the calculation of N$^3$LO
corrections\footnote{Although incomplete, the N$^{3}$LO corrections to sole
$g g$
initiated subprocesses, especially the first correction to the box, might
already reduce the scale uncertainties. A complete N$^3$LO calculation goes
beyond the scope of available techniques.}. Meanwhile, preliminary numerical
comparisons have been initiated between these new NLO (and resummed) partonic
calculations, and Monte Carlo event generators \cite{les-houches;11qcd}. They 
have to be pushed further.

\subsubsection{Reducible background}
Before any cut is applied, most of the $H \rightarrow \gamma \gamma$ background
comes from large-$p_{T}$ $\pi^{0}$, $\eta$ or $\omega$, decaying into photons. 
It can be severely reduced by imposing combined geometric and calorimetric
isolation criteria. A small fraction of this huge background, consisting in
large-$p_{T}$ isolated  $\pi^{0}$ or $\eta$ may still pass such cuts. Earlier
estimations of this background rely on Monte Carlo event generators, in which
the tails of fragmentation distributions near the end point are rather poorly
known. An improvement can be provided by using isolated $\pi^{0}$ pairs  and
$\gamma \pi^{0}$ Tevatron data, compared with Monte-Carlo type NLO 
calculations, such as \cite{Chiappetta:1996wp}, to improve NLO fragmentation functions at
large $z$. 

Like continuum di-photon production, its background from $\gamma\pi^0$ and
$\pi^0\pi^0$ production has been extensively studied at the
Tevatron~\cite{Abe:1993cy,cdf2;11qcd,d0;11qcd}. This study can serve as a useful benchmark for the
reducible background prediction, as well as for very  useful tests of QCD. The
inclusive $\pi^0\pi^0$ and $\gamma\pi^0$ cross sections are orders of
magnitude larger than the $\gamma\gamma$ cross sections, making an extraction
of the latter difficult, unless additional selection criteria are applied. As
in essentially all collider photon measurements, an isolation cut needs to be
applied to each of the di-photon candidates\footnote{Other cuts  are applied as
well but the main impact on the background is from the isolation cut.}. In the
case of CDF (in Run 1B), the isolation cut requires that any additional energy 
in a cone of radius R = 0.4 ($R = \sqrt{\Delta\eta^2 + \Delta\phi^2}$) 
around the photon direction be less than 1 GeV. This requirement is basically
saturated by the energy deposited by the  di-photon underlying event and any
additional minimum bias interactions that may have occurred during the same
crossing.  Such a strict isolation requirement rejects the majority of the
$\gamma\pi^0$ and $\pi^0\pi^0$ backgrounds while retaining the true di-photon
events with 80\% efficiency\footnote{For the sake of  compactness, only
$\pi^0$ backgrounds are listed, but other backgrounds, for example, from $\eta$
and $\omega$ production, are also considered.}.

The isolation cut suppresses the di-photon backgrounds to the point  where they
are comparable to the di-photon signal. One still needs a  technique that allows
for the separation of the di-photon signal from the background, in a Monte Carlo
independent manner. CDF uses two such  techniques: a measurement of the
electro-magnetic shower width using a wire chamber placed at the EM shower
maximum position, and a measurement of the fraction of the photon candidates
that have converted in the magnet coil. The two photons from the $\pi^0$ can
not be separately  reconstructed given the tower granularity, but  they do have
a different shower width distribution and a different conversion probability
than single photons. These differences allow the extraction of the di-photon
signal, not on an event-by-event basis, but on a statistical basis, at each
kinematic point being considered. The latter consideration is important since
the background fraction does vary with the  kinematics of the events being
considered.  

With the 1 GeV isolation cut for each photon, the di-photon signal fraction
varies from about 30\% at low $E_T$ to essentially 100\% at high $E_T$ (50
GeV). The dominant source of background was determined to be from $\pi^0\pi^0$
production.\footnote{A study of the di-photon backgrounds at ATLAS found the
$\gamma\pi^0$ and $\pi^0\pi^0$ backgrounds to be of roughly equal size in the
low mass Higgs signal region, with each of the backgrounds being of the order
of 20\% of the di-photon continuum \cite{ATLAS-TDR;11qcd}.} Note that if the leakage 
of the electro-magnetic shower energy into the isolation cone is correctly
accounted for, there is no reason to have a fractional isolation scale (some
fixed fraction of the photon energy) rather than a fixed amount of energy
allowed in the isolation cone. A fixed energy isolation cut provides a
discrimination against jet backgrounds that increases in  effectiveness as the
energy of the photon candidate increases. At higher transverse energies, the
isolation cut requires the jet to fragment into a $\pi^0$ at larger values of
the fragmentation  variable $z$, a process greatly suppressed by the steeply
falling fragmentation function. The large $z$ ( $z > 0.95$) region is poorly 
known since inclusive measurements of jet fragmentation \cite{Webber:1999ui} have few
statistics in this region. This statement is even more true for the  case of
gluon jets, which form the bulk of the background source at the LHC. The
di-photon trigger at the Tevatron selects those rare jets that fragment into
isolated $\pi^0$'s. Thus, it would be useful to try to normalize the
predictions of the event generators  such as PYTHIA \cite{Sjostrand:1994yb}, which are
used for background studies at the LHC to the background data at the Tevatron.
Such a comparison is now in progress \cite{background;11qcd}.

\subsubsection{Production in association with jets}
In order to improve the signal/background ratio, it has been suggested 
\cite{Hjet1;11qcd,Abdullin:1998er} to study the associated production of $H(\rightarrow \gamma
\gamma)+$ jet. For this process, both signal $S$ and background $B$ are reduced
but still remain at the level of $\sim 100$ signal events at low LHC
luminosity.   The LO estimate has shown that the $S/B$ ratio is
improved critically with the same level of significance $S/\sqrt{B}$.
Furthermore, higher order corrections to the background have been shown
recently \cite{deFlorian:1999tp} to be under better control than in the inclusive case.

\noindent
{\it Background: associated vs. inclusive} \label{avsi}\\
Indeed, the situations in the inclusive and associated channels are quite 
different. In the inclusive case, the main reason why the magnitude of the NNLO
box contribution is comparable to the LO cross section is that the latter is
initiated by $q \bar{q}$, whereas the former involves $g g$. The $g g$
luminosity, much larger than the $q \bar{q}$ one, compensates numerically the
extra $\alpha_s^2$ factor of the box. This is not the case in the channel
$\gamma \gamma$ + jet, since the LO cross section is dominated instead by a $q
g$ initiated subprocess. The $q g$ luminosity is sizably larger than the $q
\bar{q}$ one, which guarantees that the corresponding NNLO contribution remains small (less
than  20\% for $p_T>30$ GeV) compared to the LO result \cite{deFlorian:1999tp}. Thus, 
expecting that the subprocess $g g \rightarrow \gamma \gamma  g$ gives the main
NNLO correction, a quantitative description of the background with an accuracy
better than 20\% could be achieved already at NLO in the $\gamma \gamma$+ jet
channel for a high-$p_{T}$ jet. All the helicity amplitudes needed for the
implementation of the (``direct" contribution to the) background to NLO 
accuracy
are now available 
\cite{Bern:1995fz,Signer:1995nk,DelDuca:1999pa}.

\noindent
{\it Signal vs. background} \label{sb}\\ 
The 3-body kinematics of the process allows more refined cuts to improve the
$S/B$ ratio up to $1/2-1/3$ \cite{Hjet1;11qcd,Abdullin:1998er} (to be compared with   $S/B \ge
1/7$  for the inclusive channel). Due to helicity and total angular momentum
conservation the $s$-wave state does not contribute to the dominant signal
subprocess $g g \rightarrow H g$. On the contrary, all angular momentum states
contribute to the subprocesses $g q  \rightarrow \gamma \gamma q$ and $q
\bar{q} \rightarrow \gamma \gamma g$. Therefore, the signal has a more
suppressed threshold behaviour compared to the background. The $S/B$ ratio can
thus be improved by increasing the partonic c.m.s. energy $\sqrt{\hat s}$ far
beyond threshold. Indeed, a cut $\sqrt{\hat s}>300$ GeV has been found to give
the best S/B ratio for the LHC. The effect can not be fully explained by the
threshold behavior only, since that would result in a uniform suppression
factor. It was shown in \cite{Hjet1;11qcd,Abdullin:1998er} (see Figs. 5 and 6
there) that the dependences of the background and the signal on the c.m.s.
angular variables are quite different, therefore, the strong $\hat s$ cut
affects them with different suppression factors (see
\cite{Hjet1;11qcd,Abdullin:1998er} for more details). This effect can be
exploited to enhance the significance $S/\sqrt{B}$  at the same level as $S/B$.
If the cut $\cos({\vartheta^*)({j \gamma})}<-0.87$  on the jet-photon angle in 
the partonic c.m.s. is applied for $\sqrt{\hat s}<300$ GeV and  combined with
the cut $\sqrt{\hat s}>300$  GeV, the change on $S/B$ is rather small, while
the significance is improved by a factor $\sim$ 1.3. The same effect can be
observed with the cut on the jet angle $\vartheta^*({j})$
in the partonic c.m.s. (cf. the Fig.~5 mentioned above), but one
should notice that the two variables, $\vartheta^*({j \gamma})$ and
$\vartheta^*({j})$, are correlated. Therefore, it is desirable to perform a
multi-variable optimization of the event selection. Notice that the present
discussion is based on a LO analysis, and concerns only what was defined above
as the ``direct" component of the  irreducible background. One now has to 
understand how this works at NLO.

Other, reducible, sources of background are potentially dangerous. The 
above-defined  ``single-fragmentation" component to the so-called irreducible
background, and the reducible background coming from misidentification of jet
events were treated on a similar footing in the LO analysis of 
\cite{Hjet1;11qcd,Abdullin:1998er} as a  {\it de facto} reducible background (see footnote
\ref{def-reducible}). In \cite{Hjet1;11qcd,Abdullin:1998er}, a rough analysis found that this
reducible background is less than 20\% of  the irreducible one after cuts are
imposed.  The misidentification rate is given mainly by the subprocesses $g q
\rightarrow \gamma g q$, $g g \rightarrow \gamma q \bar{q}$ and $q
q'\rightarrow \gamma q(g) q'(g)$, when the final state parton produces an
energetic isolated photon but other products of the hadronization escape the
detection as a jet. There, a $\gamma(\pi^0)/\mbox{jet}$ rejection factor equal to 2500
for a jet misidentified as a photon and 5000 for a well separated
$\gamma(\pi^0)$ production by a jet were used. No additional $\pi^0$ rejection
algorithms were applied. Furthermore, this reducible background is expected to
be suppressed even more strongly than the irreducible background of ``direct"
type when a cut on $\sqrt{\hat s}$ is applied.

In summary, the associated channel $H(\rightarrow \gamma \gamma) +$ jet with
jet transverse energy $E_T>30$ GeV and rapidity $|\eta|<4.5$ (thus
involving forward hadronic calorimeters) opens a promising possibility for
discovering the Higgs boson with a mass of 100-140 GeV at LHC even at low 
luminosity. However, to perform a quantitative analysis, the NLO calculations 
of the background
have to be completed and included in a more realistic
final state analysis.

\noindent
{\it Signal at NLO} $\mbox{}$ \\
The exact calculation of the NLO corrections to the signal is very complex,
since the gluons interact with the Higgs boson via virtual quark loops.
Fortunately, the effective field theory approach \cite{Dawson:1991zj,Djouadi:1991tk} applicable in the
large top mass limit with effective gluon-gluon-Higgs boson coupling  gives an
accurate approximation with an error less than 5\%, provided $m_{H} \le 2m_{t}$. 
Recently, in this approximation and using the helicity method,  the
transition amplitudes relevant to the NLO corrections have been analytically
calculated for all contributing subprocesses (loop corrections \cite{Schmidt:1997wr} and
bremsstrahlung \cite{Dawson:1992au,Kauffman:1997ix}). The subtraction method of \cite{Frixione:1996ms,Frixione:1997np} has been used
to cancel analytically the soft and collinear singularities and to implement
the amplitudes into a numerical program of Monte-Carlo type which allows to
calculate any infrared-safe observable for the production of a Higgs boson with
one jet at NLO accuracy \cite{deFlorian:1999zd}. 

\begin{figure}[htb]
\begin{center}
\includegraphics[width=0.8\linewidth,height=6cm]{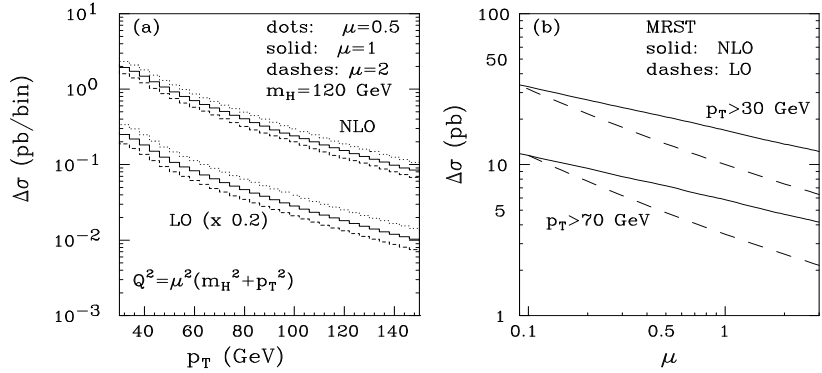}
\end{center}
\vskip-0.5cm
\caption{
Scale dependence of LO and NLO distributions for Higgs boson production. 
(a) $p_{T}$ distributions at different scales and (b) the scale dependence
of the integrated cross sections for $p_{T}>30$ and  $70$ GeV. The MRST
parton distributions are used.
} 
\label{kdf;11qcd}
\end{figure}

One of the main results of the calculation is that the NLO corrections are
large and increase considerably the cross section, with a $K$ factor $\sim$ 
1.5-1.6 ($K=\sigma^{NLO}/\sigma^{LO}$) and almost constant for a large
kinematical range of $p_T$ and rapidity of the Higgs boson. Furthermore, the
NLO result is less dependent on variations of the factorization and
renormalization scales. Fig. \ref{kdf;11qcd}(a) displays the $p_T$ distribution at
both LO and NLO for a Higgs boson with $m_H=120$ GeV. The curves correspond to
three different  renormalization/factorization scale choices $Q=\mu \,
(m_H^2+p_T^2)^{1/2}$,  with $\mu=0.5,1,2$, and show that  the scale dependence
is reduced at NLO.   The same features can be observed in more detail in Fig.
\ref{kdf;11qcd}(b), where the LO and NLO cross sections integrated  for $p_{T}$
larger than 30 and 70 GeV are shown  as a function of the
renormalization/factorization scale. Both  the LO and NLO  cross sections
increase monotonically with decreasing  $\mu$,  down to the limiting value 
where perturbative QCD can still  be applied,  indicating that the stability of
the NLO result is not completely satisfactory. However, in the usual range of
variation of $\mu$ from 0.5 to 2, the LO scale uncertainty amounts to $\pm$
35\%, whereas at NLO it is reduced to $\pm$ 20\%.

\subsection{Multi-lepton channels in the mass range $m_{H} \geq 140$ GeV.}
\label{wh;11qcd}

Above 140 GeV, the most promising channel is $H \rightarrow Z Z^{(*)} 
\rightarrow 4$ leptons. The corresponding irreducible background comes mainly
from the non resonant $Z Z^{(*)}$ production. Severe isolation cuts are needed
to suppress reducible $t \bar{t}$ and $Z b \bar{b}$ backgrounds for Higgs boson
masses below the $ZZ$ threshold. The topic of  weak boson pair production is
presented in a dedicated Section of the Electroweak Physics Chapter of this
Report. In
particular, the latter gathers the effects of NLO contributions to
distributions of invariant mass, or transverse momentum of weak boson pairs,
and comparisons between Monte Carlo event generators and recent NLO partonic
calculations. 

The $H \rightarrow WW \rightarrow 2 l + \not{\!\! E}_{T}$ channels was
recently found \cite{Dittmar:1997ss,Dittmar:1996sp} to be very promising in this mass range
around 170 GeV, where the significance of the $H \rightarrow ZZ ^{*}
\rightarrow 4 l$ channel is relatively small due to the suppression of
$ZZ^{*}$ branching ratio as the $WW$ mode opens up. In this mass range, the
leptonic branching ration of the $WW$ mode is approximately 100 times
larger than the $Z Z^{*}\rightarrow 4 l$ mode. Although the Higgs boson mass
peak cannot be directly reconstructed in this case, the transverse mass
distribution can be used to sign the Higgs boson and extract information on its
mass. 

The multileptonic channels $H \rightarrow WW^{(*)}$ and 
$H \rightarrow Z Z^{(*)}$  are also of great interest for the associated $WH$
production. Although the cross section for the associated production is a 
factor 50 to 100 lower than for the inclusive production, the $S/B$ ratio is 
substantially improved. They are also interesting to determine the Higgs boson
couplings, since only the couplings to gauge bosons appear in the production
and decay chain. The observability of $WH$ with $H \rightarrow WW^{*} 
\rightarrow 2 l \; 2 \nu$ has been recently proposed in \cite{Baer:1998cm} and 
experimentally studied in \cite{ATLAS-TDR;11qcd}.  The observability of the
associated production $WH$, $H \rightarrow Z Z^{*}  \rightarrow 4 l$ has been
recently considered in \cite{NOTE006;11qcd} and is sketched below. Due to the small
number of events expected for $ZH$ and $t \bar{t}H$ production, only the $WH$
process has been  investigated.

\subsubsection{Associated $WH$ production, five lepton channel}

{\it Selection criteria} \\ 
All simulations of Higgs boson and background events have been made with the 
PYTHIA 5.702 and JETSET 7.408 Monte Carlo programs implemented in the 
CMSIM/CMANA package \cite{NOTE002;11qcd}. The processes implemented in PYTHIA were
simulated with parton showers, with the exception of internal bremsstrahlung,
generated by PHOTOS \cite{Barberio:1991ms}. No $K$ factors were used, so the final
numbers of signal events may be underestimated by about a factor 1.3
\cite{Spira:1998dg}. The experimental resolution of CMS for lepton reconstruction was 
simulated by a Gaussian smearing:
\begin{center}
 $\Delta p_{T}/p_{T} = 4.5 \% \sqrt{p_{T}/1000}$  \quad for muons, \\   
 $(\Delta E / E)^2 = (4\%/\sqrt{E})^2 + (0.230/E)^2 + (0.55\%)^2$ \quad for electrons, 
\end{center}
where $p_{T}$ and $E$ are expressed in GeV. Dedicated programs  calculate the
dependence on $\eta$ and $p_{T}$ of the geometrical and kinematical
acceptances, the invariant mass cuts to select the $Z$ or $Z^*$,  and the
rejection of non isolated leptons in jets with cuts selecting leptons without
charged tracks above $p_{T} >$ 2 GeV in a cone $R < 0.1$  ($R^2 = \Delta \eta^2
+ \Delta \phi^2$).  A few events were also fully generated and visualized in
CMS by CMSIM. The reactions $W^{\pm} H \rightarrow \mu^{\pm} \nu_{\mu} Z
Z^{(*)} \rightarrow  5 \mu^{\pm} \nu_{\mu}$ and $W^{\pm} H \rightarrow e^{\pm}
\nu_{e} Z Z^{(*)}  \rightarrow 5 e^{\pm} \nu_{e}$ have been studied in details.
Although the branching ratios are identical, some differences between these
channels are expected due to differences in acceptances and trigger
efficiencies. The generated leptons are sorted in decreasing $p_{T}$ order,
from 1 to 5, then the following cuts are applied. \\ 
For muon events :
\begin{itemize}
 \item $\mid \eta \mid  < 2.1$ for $\mu_1$ and $\mu_2$
 $\;\;\;\;\;    \mid \eta \mid  < 2.5$ for $\mu_3$ to $\mu_5$
 \item $    p_{T} > 20$ GeV for $\mu_1$
 $\;\;\;\;\;\;    p_{T} > 10$ GeV for $\mu_2$
 $\;\;\;\;\;\;    p_{T} > 5$ or $10$ GeV for $\mu_3 ,  \mu_4$  and  $\mu_5$
\end{itemize}
For electron events :
\begin{itemize}
 \item $\mid \eta \mid  < 2.5$ for $e_1$ to $e_5$
 \item $ p_{T} > 20$ GeV for $e_1$
 $\;\;\;\;\;\; p_{T} > 15$ GeV for $e_2$
 $\;\;\;\;\;\; p_{T} > 7, 10$ or $15$ GeV for $e_3, e_4$  and  $e_5$
\end{itemize} 
Leptons 1 and 2 are the ones used to trigger events, leptons 3 to 5 $p_{T}$
thresholds can be set at lower values. Almost no difference is observed when
the trigger threshold is set at a higher value (30 and 20~GeV), as expected
since leptons 1 and 2 produced by $W$ and  $Z$ decays are very energetic. The
other possible final states: $2 e + 3\mu$,  $2 \mu + 3 e$, $4 e + 1 \mu$ and $4
\mu + 1 e$ are also good candidates. Since only small numerical differences
were found in the results between the pure electronic and muonic final states,
the 4 mixed ones were not simulated and the total number of expected events was
multiplied by a factor 8. As the expected cross section is very low, the
present search would be meaningful at high luminosity only. The pile-up at high
luminosity has a minor impact for the detection of leptons. Nevertheless it has
to be taken into account when using  the isolation cuts.

\begin{table}[t!]
\begin{center}
\begin{tabular}{||l|r|r|r|r||}  \hline \hline
                        &  no cut& isolation cut & $Z$ mass cut  & all cuts \\ \hline \hline
$WH , M_{H} = 150 GeV$        &  3.56  &  3.42     &  2.89  &  2.69  \\ \hline
$t\bar{t}$ background  &  141.  &  3.10     &  26.1  &  0.098  \\ \hline
$Zb\bar{b}$ background &  17.3  &  3.46     &  13.8  &  3.46  \\ \hline \hline
$WH , M_{H} = 200 GeV$        &  5.92  &  5.55     &  3.95  &  3.76  \\ \hline
$WH , M_{H} = 300 GeV$        &  1.45  &  1.30     &  0.91  &  0.86  \\ \hline
$t\bar{t}$ background  &  141.  &  3.10     &  0.098  &  0.098  \\ \hline
$Zb\bar{b}$ background &  17.3  &  3.46     &  1.73  &  0     \\ \hline \hline
\end{tabular}
\caption {Number of events in the 5 leptons channel for $L = 10^{5}
pb^{-1}$, $p_{T}$ cut = 10 GeV. No mass window on 4 leptons is applied.} 
\end{center}
\end{table}

\noindent
$H \rightarrow Z Z^*$ \\ 
This channel concerns the mass range $m_{H} <2 m_{Z}$. The irreducible
background, due to the non resonant $W Z Z^*$ production, is not included in
PYTHIA. In order to get a rough order of magnitude, the $S/B$
ratio was  then assumed to be of the same order as the one of direct production
of $H \rightarrow Z Z^*$, compared to non resonant $Z Z^*$. This ratio has been
estimated in \cite{NOTE004;11qcd} to be lower than 10 \% for $m_{H}$ = 150 GeV.  The
reducible background comes from the $t \bar{t}$ and $Z b \bar{b}$ channels with
three leptons coming from semi-leptonic decays of $B$ and $D$ mesons. The
initial cross sections of these processes are very high and, without cut, this
background is much higher than the irreducible one. 

\begin{figure}[t!] 
 \begin{center}
    \includegraphics[width=10cm,height=10cm]{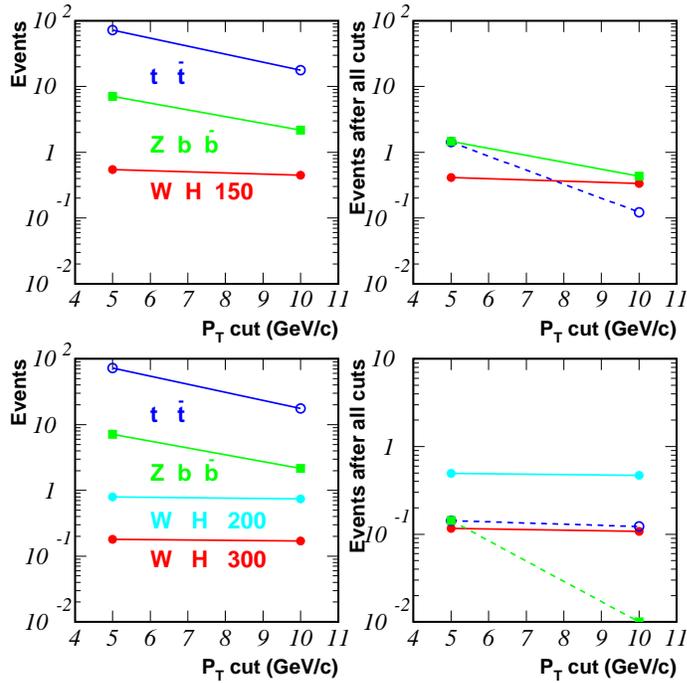}
\vskip-0.5cm
    \caption{Number of events at $L = 10^5 pb^{-1}$
    for $W H \rightarrow 5 \mu $ channel and background
    for M$_{H}$ = 150  GeV (top), 200 and 300~GeV (bottom)
    after kinematical cuts (left), isolation
    and $Z Z$ mass cuts combined (right). 
    $P_{T}$ cut refers to the softest of the five muons.
    Dotted line is an upper limit (no Monte Carlo event survive the cuts).}
    \label{evt;11qcd}
  \end{center}
\end{figure}

The selection requests one pair of opposite sign muons or electrons with a mass equal to
$m_{Z} \pm 5$ GeV, and one pair of opposite sign muons or electrons with a mass
below $m_{Z}$. This removes only 19 \% of the signal events which fall in the
tails of the mass distributions. An additional effect of widening the Z mass
would come from the $e^{\pm}$ bremsstrahlung in the tracker material
\cite{NOTEIvica;11qcd} and contribute to decrease the acceptance. The lepton pair
mass spectra of the $t \bar{t}$ and $Z b \bar{b}$ backgrounds exhibit a peak at
low mass. A cut at  $m_{Z^{*}} > 10$ GeV would further reduce these backgrounds
by 20 \% without affecting the signal. No detector reconstruction inefficiency
was considered at this level. The isolation cut is used to reject leptons from
$b$ or $c$ quark decay, in the reducible background channels. The events
exhibiting tracks with $p_{T} > 2$ GeV contained in a cone $R <$ 0.1 around
any of the five leptons are rejected (Fig. \ref{evt;11qcd}). Actually a better
rejection is expected in the CMS detector when using the information from the
$b$ vertex position \cite{NOTE95-059;11qcd}.

Another reducible background was considered: the non resonant production of $Z
Z^{*}$ where one of the $Z^{(*)}$ decays into two leptons and the other decays
into $b \bar{b}$, the $b$ quarks decaying semi-leptonically. The number of
events before acceptance, mass and isolation cut is about 70 \% of the signal,
but as we expect the leptons from the $b$'s to be very soft and non isolated, 
that this background can be considered as negligible.  

\noindent
$H \rightarrow Z Z$ \\ 
This channel is similar to the previous one except that we request two pairs of
opposite sign muons or electrons with masses equal to $m_{Z} \pm 5$ GeV. This
cut removes 32 to 34 \% of the signal events. It is now much more efficient
against the $t \bar{t}$ background than against the  $Z b \bar{b}$, because the
$Z b \bar{b}$ channel involves a real $Z$. The calculations were made for Higgs
bosons with  $m_{H} = 200$ and 300 GeV (Fig. \ref{evt;11qcd}). The acceptances of the
signal vary only slightly as a function of $p_{T}$ cut and other selection
cuts. The four leptons mass spectrum for the background is a wide distribution 
centered around 150 GeV. A cut on this spectrum can be used to obtain an
additional rejection factor of the order of 10 to 50, after the Higgs boson
mass has been previously measured in a more sensitive channel, like the
inclusive $H \rightarrow 4l$ \cite{NOTE95-059;11qcd}.

\noindent
{\it Results} \\  
The number of expected 5 muons or 5 electrons events for one year of running at
high luminosity  $100 fb^{-1}$ is low: 0.34 for a Higgs boson mass of 150 GeV,
0.47 for 200 GeV and 0.11 for 300 GeV/c. Considering all the possible 5 leptons
channels, these numbers must be multiplied by a factor 8. They are summarized
in table 1, together with the corresponding backgrounds (not including the cut
on the four leptons mass spectrum described above). The $S/B$ ratio is better
for $m_{H} = 200$ GeV and is unacceptable for $m_{H} = 150$ GeV. Thus this
channel can be considered almost hopeless for the discovery of the Higgs boson
below the $ZZ$ threshold. On the other hand, if the Higgs boson is in the mass
range 200 to 300 GeV, the detection of these rare 5 lepton events above a low
background would be a valuable information for the study of the Higgs boson
couplings. 

However, before drawing any definitive conclusion, several issues should be
improved concerning the backgrounds. Firstly the irreducible $W Z Z^*$
background has to be calculated, e.g. using an automatized calculation like
\cite{Pukhov:1999gg} and included in the analysis.  Moreover the reducible $Z b
\bar{b}$ process should be revisited with another Monte Carlo generator, as the
implementation in PYTHIA 5.7 for the $Zb \bar{b}$ process is known to suffer
from an instability  in the phase space generation (this implementation has
been removed from the version PYTHIA 6.1 for this reason). Finally, another
source of 5 leptons events, not evaluated with enough statistics so far is
the semi-leptonic decay of $b \bar{b}$ or $c \bar{c}$ generated by initial
or final gluon radiation.   

An extension of this study would also be the investigation of the associated
production of a higher mass  Higgs boson using other decay modes with larger
branching ratios like $Z \rightarrow$ jet jet.


\end{document}